\documentclass[nolinenum]{jfp}
\usepackage[]{stmaryrd}
\usepackage[]{amsthm}
\usepackage[]{mathpartir}
\usepackage[]{polytable}
\usepackage[]{subcaption}
\usepackage[]{tikz}
\usepackage[capitalize]{cleveref}
\newtheorem{definition}{Definition}
\newtheorem{theorem}{Theorem}

\newwrite\boxesfile\immediate\openout\boxesfile=tensors.boxes
\newsavebox{\marxupbox}\usepackage{newunicodechar}\usepackage[scaled=0.94]{helvet}\newcommand{\tikzstar}[5]{% inner radius, outer radius, tips, rot angle, options
\begin{tikzpicture}[baseline={([yshift=-\axisht]current bounding box.center)},x=1ex,y=1ex,line width=.1ex]
  \pgfmathsetmacro{\starangle}{360/#3}
  \draw[#5] (#4:#1)
  \foreach \x in {1,...,#3}
  { -- (#4+\x*\starangle-\starangle/2:#2) -- (#4+\x*\starangle:#1)
  }
  -- cycle;
\end{tikzpicture}
}

\newcommand{\varsqdiamond}{\mathbin{\text{\tikz [x=1ex,y=1ex,line width=.1ex,line join=round] \draw (0,0) rectangle (1,1) (.5\pgflinewidth,.5) -- (.5,1ex-.5\pgflinewidth) -- (1ex-.5\pgflinewidth,.5) -- (.5,.5\pgflinewidth) -- (.5\pgflinewidth,.5) -- cycle;}}}
\newcommand{\varsquare}{\mathbin{\text{\tikz [x=1ex,y=1ex,line width=.1ex,line join=round] \draw (0,0) rectangle (1,1);}}}
\newcommand{\smalltriangleleft}{\mathbin{\text{\tikz [baseline={([yshift=-\axisht]current bounding box.center)},x=1ex,y=1ex,line width=.1ex] \draw (0,0.3) -- (-0.52,0) -- (0,-0.3) -- cycle;}}}

\def\axisht{\dimexpr.5\fontcharht\font`)-.5\fontchardp\font`)\relax} % https://tex.stackexchange.com/questions/386288
\def\doubleequals{\mathrel{\unitlength 0.01em
  \begin{picture}(78,40)
    \put(7,34){\line(1,0){25}} \put(45,34){\line(1,0){25}}
    \put(7,14){\line(1,0){25}} \put(45,14){\line(1,0){25}}
  \end{picture}}}

\newunicodechar{ }{\,}
\newunicodechar{¬}{\ensuremath{\neg}} % 
\newunicodechar{²}{^2}
\newunicodechar{³}{^3}
\newunicodechar{·}{\ensuremath{\cdot}}
\newunicodechar{¹}{^1}
\newunicodechar{×}{\ensuremath{\times}} % 
\newunicodechar{÷}{\ensuremath{\div}} % 
\newunicodechar{ʲ}{^j}
\newunicodechar{ʳ}{^r}
\newunicodechar{ˡ}{^l}
\newunicodechar{ˢ}{^s}
\newunicodechar{̷}{\not} % 
\newunicodechar{Γ}{\ensuremath{\Gamma}}   %
\newunicodechar{Δ}{\ensuremath{\Delta}} % 
\newunicodechar{Η}{\ensuremath{\textrm{H}}} % 
\newunicodechar{Θ}{\ensuremath{\Theta}} % 
\newunicodechar{Λ}{\ensuremath{\Lambda}} % 
\newunicodechar{Ξ}{\ensuremath{\Xi}} % 
\newunicodechar{Π}{\ensuremath{\Pi}}   %
\newunicodechar{Σ}{\ensuremath{\Sigma}} % 
\newunicodechar{Φ}{\ensuremath{\Phi}} % 
\newunicodechar{Ψ}{\ensuremath{\Psi}} % 
\newunicodechar{Ω}{\ensuremath{\Omega}} % 
\newunicodechar{α}{\ensuremath{\mathnormal{\alpha}}}
\newunicodechar{β}{\ensuremath{\beta}} % 
\newunicodechar{γ}{\ensuremath{\mathnormal{\gamma}}} % 
\newunicodechar{δ}{\ensuremath{\mathnormal{\delta}}} % 
\newunicodechar{ε}{\ensuremath{\mathnormal{\varepsilon}}} % 
\newunicodechar{ζ}{\ensuremath{\mathnormal{\zeta}}} % 
\newunicodechar{η}{\ensuremath{\mathnormal{\eta}}} % 
\newunicodechar{θ}{\ensuremath{\mathnormal{\theta}}} % 
\newunicodechar{ι}{\ensuremath{\mathnormal{\iota}}} % 
\newunicodechar{κ}{\ensuremath{\mathnormal{\kappa}}} % 
\newunicodechar{λ}{\ensuremath{\mathnormal{\lambda}}} % 
\newunicodechar{μ}{\ensuremath{\mathnormal{\mu}}} % 
\newunicodechar{ν}{\ensuremath{\mathnormal{\nu}}} % 
\newunicodechar{ξ}{\ensuremath{\mathnormal{\xi}}} % 
\newunicodechar{π}{\ensuremath{\mathnormal{\pi}}}
\newunicodechar{ρ}{\ensuremath{\mathnormal{\rho}}} % 
\newunicodechar{σ}{\ensuremath{\mathnormal{\sigma}}} % 
\newunicodechar{τ}{\ensuremath{\mathnormal{\tau}}} % 
\newunicodechar{φ}{\ensuremath{\mathnormal{\phi}}} % 
\newunicodechar{χ}{\ensuremath{\mathnormal{\chi}}} % 
\newunicodechar{ψ}{\ensuremath{\mathnormal{\psi}}} % 
\newunicodechar{ω}{\ensuremath{\mathnormal{\omega}}} %  
\newunicodechar{ϕ}{\ensuremath{\mathnormal{\phi}}} % 
\newunicodechar{ϵ}{\ensuremath{\mathnormal{\epsilon}}} % 
\newunicodechar{ᵀ}{\ensuremath{^T}}
\newunicodechar{ᵏ}{^k}
\newunicodechar{ᵐ}{^m}
\newunicodechar{ᵢ}{\ensuremath{_i}} % 
\newunicodechar{ᵣ}{_r}
\newunicodechar{ᶿ}{^\theta}
\newunicodechar{ }{\quad}
\newunicodechar{†}{\dagger}
\newunicodechar{ }{\,}
\newunicodechar{′}{\ensuremath{^\prime}}  % 
\newunicodechar{″}{\ensuremath{^\second}} % 
\newunicodechar{‴}{\ensuremath{^\third}}  % 
\newunicodechar{ⁱ}{^i}
\newunicodechar{⁵}{\ensuremath{^5}}
\newunicodechar{⁺}{\ensuremath{^+}} %% \newunicodechar{⁺}{^+}
\newunicodechar{⁻}{\ensuremath{^-}} %% 
\newunicodechar{ⁿ}{^n}
\newunicodechar{₀}{\ensuremath{_0}} % 
\newunicodechar{₁}{\ensuremath{_1}} % \newunicodechar{₁}{_1}
\newunicodechar{₂}{\ensuremath{_2}} % \newunicodechar{₂}{_2}
\newunicodechar{₃}{\ensuremath{_3}}
\newunicodechar{₊}{\ensuremath{_+}} %% 
\newunicodechar{₋}{\ensuremath{_-}} %% 
\newunicodechar{ₖ}{\ensuremath{_k}} % 
\newunicodechar{ₗ}{\ensuremath{_l}}
\newunicodechar{ₘ}{_m}
\newunicodechar{ₙ}{_n} % 
\newunicodechar{ₛ}{_s}
\newunicodechar{⃰}{\ensuremath{^{\ast}}}
\newunicodechar{ℂ}{\ensuremath{\mathbb{C}}} % 
\newunicodechar{ℒ}{\ensuremath{\mathscr{L}}}
\newunicodechar{ℕ}{\mathbb{N}} % 
\newunicodechar{ℚ}{\ensuremath{\mathbb{Q}}}
\newunicodechar{ℝ}{\ensuremath{\mathbb{R}}} % 
\newunicodechar{ℤ}{\ensuremath{\mathbb{Z}}} % 
\newunicodechar{ℳ}{\mathscr{M}}
\newunicodechar{⅋}{\ensuremath{\parr}} % 
\newunicodechar{←}{\ensuremath{\leftarrow}} % 
\newunicodechar{↑}{\ensuremath{\uparrow}} % 
\newunicodechar{→}{\ensuremath{\rightarrow}} %
\newunicodechar{↔}{\ensuremath{\leftrightarrow}} % 
\newunicodechar{↖}{\nwarrow} %
\newunicodechar{↗}{\nearrow} %
\newunicodechar{↝}{\ensuremath{\leadsto}}
\newunicodechar{↦}{\ensuremath{\mapsto}}
\newunicodechar{⇆}{\ensuremath{\leftrightarrows}} % 
\newunicodechar{⇐}{\ensuremath{\Leftarrow}} % 
\newunicodechar{⇒}{\ensuremath{\Rightarrow}} % 
\newunicodechar{⇔}{\ensuremath{\Leftrightarrow}} % 
\newunicodechar{∀}{\ensuremath{\forall}}   %
\newunicodechar{∂}{\ensuremath{\partial}}
\newunicodechar{∃}{\ensuremath{\exists}} % 
\newunicodechar{∅}{\ensuremath{\varnothing}} % 
\newunicodechar{∇}{\ensuremath{\nabla}} % 
\newunicodechar{∈}{\ensuremath{\in}}
\newunicodechar{∉}{\ensuremath{\not\in}} % 
\newunicodechar{∋}{\ensuremath{\ni}}  % 
\newunicodechar{∎}{\ensuremath{\qed}}%\newunicodechar{∎}{\ensuremath{\blacksquare}} % end of proof
\newunicodechar{∏}{\prod}
\newunicodechar{∑}{\sum}
\newunicodechar{∗}{\ensuremath{\ast}} % 
\newunicodechar{∘}{\ensuremath{\circ}} % 
\newunicodechar{∙}{\ensuremath{\bullet}} % \newunicodechar{∙}{\ensuremath{\cdot}}
\newunicodechar{∝}{\propto}
\newunicodechar{∞}{\ensuremath{\infty}} % 
\newunicodechar{∣}{\ensuremath{\mid}} % 
\newunicodechar{∧}{\ensuremath{\wedge}}% 
\newunicodechar{∨}{\vee}% 
\newunicodechar{∩}{\ensuremath{\cap}} % 
\newunicodechar{∪}{\ensuremath{\cup}} % 
\newunicodechar{∫}{\int}
\newunicodechar{∷}{::} % 
\newunicodechar{∼}{\ensuremath{\sim}} % 
\newunicodechar{≃}{\ensuremath{\simeq}} % 
\newunicodechar{≅}{\ensuremath{\cong}} % 
\newunicodechar{≈}{\ensuremath{\approx}} % 
\newunicodechar{≜}{\ensuremath{\stackrel{\scriptscriptstyle {\triangle}}{=}}} % \newunicodechar{≜}{\triangleq}
\newunicodechar{≝}{\ensuremath{\stackrel{\scriptscriptstyle {\text{def}}}{=}}}
\newunicodechar{≟}{\ensuremath{\stackrel {_\text{\textbf{?}}}{\text{\textbf{=}}\negthickspace\negthickspace\text{\textbf{=}}}}} % or {\ensuremath{\stackrel{\scriptscriptstyle ?}{=}}}
\newunicodechar{≠}{\ensuremath{\neq}}% 
\newunicodechar{≡}{\ensuremath{\equiv}}% 
\newunicodechar{≤}{\ensuremath{\le}} % 
\newunicodechar{≥}{\ensuremath{\ge}} % 
\newunicodechar{⊂}{\ensuremath{\subset}} % 
\newunicodechar{⊃}{\ensuremath{\supset}} %  
\newunicodechar{⊆}{\ensuremath{\subseteq}} %  
\newunicodechar{⊇}{\ensuremath{\supseteq}} % 
\newunicodechar{⊎}{\ensuremath{\uplus}} %
\newunicodechar{⊑}{\ensuremath{\sqsubseteq}} % 
\newunicodechar{⊒}{\ensuremath{\sqsupseteq}} % 
\newunicodechar{⊓}{\ensuremath{\sqcap}} % 
\newunicodechar{⊔}{\ensuremath{\sqcup}} % 
\newunicodechar{⊕}{\ensuremath{\oplus}} % 
\newunicodechar{⊗}{\ensuremath{\otimes}} % 
\newunicodechar{⊙}{\ensuremath{\odot}} % 
\newunicodechar{⊛}{\ensuremath{\circledast}}
\newunicodechar{⊢}{\ensuremath{\vdash}} % 
\newunicodechar{⊤}{\ensuremath{\top}}
\newunicodechar{⊥}{\ensuremath{\bot}} % bottom
\newunicodechar{⊧}{\models} % 
\newunicodechar{⊨}{\models} % 
\newunicodechar{⊩}{\Vdash}
\newunicodechar{⊬}{\not \vdash}
\newunicodechar{⊸}{\ensuremath{\multimap}} % 
\newunicodechar{⋁}{\ensuremath{\bigvee}}
\newunicodechar{⋃}{\ensuremath{\bigcup}} % 
\newunicodechar{⋄}{\ensuremath{\diamond}} % 
\newunicodechar{⋅}{\ensuremath{\cdot}}
\newunicodechar{⋆}{\ensuremath{\star}} % 
\newunicodechar{⋮}{\ensuremath{\vdots}} % 
\newunicodechar{⋯}{\ensuremath{\cdots}} % 
\newunicodechar{─}{---}
\newunicodechar{■}{\ensuremath{\blacksquare}} % black square
\newunicodechar{□}{\ensuremath{\varsquare}} % 
\newunicodechar{▴}{\ensuremath{\blacktriangledown}}
\newunicodechar{▵}{\ensuremath{\triangle}}
\newunicodechar{▹}{\ensuremath{\triangleright}} % 
\newunicodechar{▻}{\ensuremath{\rhd}} % 
\newunicodechar{▾}{\ensuremath{\blacktriangle}}
\newunicodechar{▿}{\ensuremath{\triangledown}}
\newunicodechar{◃}{\ensuremath{\triangleleft}}
\newunicodechar{◅}{\ensuremath{\lhd}}
\newunicodechar{◇}{\ensuremath{\diamond}} % 
\newunicodechar{◽}{\ensuremath{\varsquare}}
\newunicodechar{★}{\ensuremath{\star}}   %
\newunicodechar{♭}{\ensuremath{\flat}} % 
\newunicodechar{♯}{\ensuremath{\sharp}} % 
\newunicodechar{⛋}{\ensuremath{\varsqdiamond}} % 
\newunicodechar{✓}{\ensuremath{\checkmark}} % 
\newunicodechar{⟂}{\ensuremath{^\bot}} % PERPENDICULAR 
\newunicodechar{⟦}{\ensuremath{\llbracket}}
\newunicodechar{⟧}{\ensuremath{\rrbracket}}
\newunicodechar{⟨}{\ensuremath{\langle}} % 
\newunicodechar{⟩}{\ensuremath{\rangle}} % 
\newunicodechar{⟶}{{\longrightarrow}}
\newunicodechar{⟷} {\ensuremath{\leftrightarrow}}
\newunicodechar{⟹}{\ensuremath{\Longrightarrow}} % 
\newunicodechar{⦇}{\rlparenthesis}
\newunicodechar{⦈}{\llparenthesis}
\newunicodechar{⪅}{\ensuremath{\lessapprox}}
\newunicodechar{⪆}{\ensuremath{\gtrapprox}}
\newunicodechar{ⱼ}{\ensuremath{_j}} % 
\newunicodechar{𝒟}{\ensuremath{\mathcal{D}}} % 
\newunicodechar{𝒢}{\ensuremath{\mathcal{G}}}
\newunicodechar{𝒦}{\ensuremath{\mathcal{K}}} % 
\newunicodechar{𝒫}{\ensuremath{\mathcal{P}}}
\newunicodechar{𝔸}{\ensuremath{\mathds{A}}} % 
\newunicodechar{𝔹}{\ensuremath{\mathds{B}}} % 
\newunicodechar{𝔼}{\mathds{E}}
\newunicodechar{𝟙}{\ensuremath{\mathds{1}}} % 
\begin{document}\begin{authgrp}\author{Jean-Philippe Bernardy}\affiliation{Chalmers University of Technology and University of Gothenburg, Gothenburg, Sweden(\email{jean-philippe.bernardy@gu.se})}\author{Patrik Jansson}\affiliation{Chalmers University of Technology and University of Gothenburg, Gothenburg, Sweden(\email{patrikj@chalmers.se})}\end{authgrp}\journaltitle{JFP}\cpr{The Author(s),}\doival{10.1017/xxxxx}\lefttitle{Bernardy, Jansson}\righttitle{Domain-Specific Tensor Languages}\totalpg{\pageref{lastpage01}}\jnlDoiYr{2025}\def\dual#1{{#1}^*}
\title{Domain-Specific Tensor Languages}\begin{abstract} The tensor notation used in several areas of mathematics is a useful
one, but it is not widely available to the functional programming
community. In a practical sense, the (embedded) domain-specific
languages ({\sc{}dsl}s) that are currently in use for tensor algebra are either 1.
array-oriented languages that do not enforce or take advantage of
tensor properties and algebraic structure or 2. follow the categorical
structure of tensors but require the programmer to manipulate tensors
in an unwieldy point-free notation.

A deeper issue is that for tensor calculus, the dominant pedagogical paradigm
assumes an audience which is either
comfortable with notational liberties which programmers cannot afford,
or focus on the applied mathematics of tensors, largely leaving their
linguistic aspects (behaviour of variable binding, syntax and
semantics, etc.) for the reader to figure out by themselves.
This state of affairs is hardly surprising, because, as we highlight,
several properties of standard tensor notation are somewhat
exotic from the perspective of lambda calculi.

 We bridge the gap by defining a {\sc{}dsl}, embedded in Haskell, whose
syntax closely captures the index notation for tensors in wide use in
the literature.  The semantics of this {\sc{}edsl} is defined in terms of the
algebraic structures which define tensors in their full generality.
This way, we believe that our {\sc{}edsl} can be used both as a tool for
scientific computing, but also as a vehicle to express and present
the theory and applications of tensors.
 \end{abstract}\maketitle{} \section{Introduction and Motivation}\label{0} 
Tensor calculus is an essential tool in physics and applied
mathematics.  It was instrumental already a century ago in the
formulation of Einstein's general relativity, and its usage has spread
to many areas of science. At its heart lies linear algebra, which is
defined as the study of linear maps between vector spaces. In
applications, one commonly manipulates \emph{representations} of
linear algebraic objects: vectors as 1-dimensional arrays and linear
maps as matrices (2-dimensional arrays). Indeed, assuming a given
basis, the representations are equivalent to the algebraic objects.
Likewise, tensors are often thought of as a higher-dimensional
version of matrices: their algebraic formulation is as a
category of linear maps between vector spaces.

Viewing the above situation through the lens of programming language
theory, the algebraic formulation forms a set of combinators and the
array-based representations is a possible semantics for them.  Even
though the praxis is to blur the distinction between algebraic objects
and their coefficient representations, it is a source of confusion in
the case of tensor calculus, which studies tensor fields, in the sense
of tensor-valued functions defined over a manifold. (We provide some
evidence in \cref{400}.) Notably, difficulties arise because the basis
varies over the manifold.  The first contribution of this paper is to
provide a clear conceptual picture by highlighting the
syntax-semantics distinction.

On the practical side, the situation is similar. One can find a
plethora of languages and libraries purportedly geared towards tensor
manipulation, but they inevitably focus on their multi-dimensional
array representations. There is nearly no support for algebraic tensor
field expressions.  In this paper we work towards bridging this gap,
by applying programming-language methodology to the notations of
tensor algebra and tensor calculus--- thus viewing them as
domain-specific languages.
 For the readership with a programming language background, we aim to
provide a down-to-earth presentation of tensor notations.  We capture
all their important properties, in particular by making use of linear
types.  We also aim to attract a readership that already has a working
knowledge of tensors. For them we aim to fully formalise the
relationship between the representation-oriented notation for tensor
fields and its linear-algebraic semantics. We do so by viewing this
syntax as terms in a (linear-typed) lambda calculus.  As usual with
{\sc{}dsl}s, this presentation comes with an executable semantics. This
means we end up with a usable tool to manipulate tensor fields, which is the second contribution of this paper.

\subsection{Overview}\label{1} 
To make the presentation more pedagogical, we delay the introduction of tensor fields over manifolds until \cref{135}.
Until then, the reader can think of each tensor as ``just'' an element of a
certain vector space. This allows us to present the core concepts in a
simpler setting, even though they will apply unchanged in the more general context.
As hinted above, we will use an algebraic semantics for tensors,
following a categorical structure (\cref{37}). Together, the combinators
forming this categorical structure form a point-free {\sc{}edsl}, which we
refer to as {\sc{}Roger} in reference to Roger Penrose (see \cref{405} as for why).

Every {\sc{}Roger} program can be evaluated to morphisms in any suitable
tensor category. This includes matrices, but also string diagrams with
the appropriate structure as well.  {\sc{}Roger} is useful in its own right,
but has all the downsides of a point-free language, and thus is not in
wide use in the mathematics community, where the so-called \emph{Einstein notation} is
preferred.  The Einstein notation mimics the usual notation to access
components of matrices, but speaks about these components in a
wholesale manner, that is, with index variables that range over all
the dimensions. We formalise this notation in an index-based {\sc{}edsl} (\cref{114}).  We refer to this {\sc{}edsl} as {\sc{}Albert} in the rest of the
paper.  Expressions in {\sc{}Albert} evaluate to morphisms in {\sc{}Roger}, and
thus in any tensor category.

\begin{figure}[]\begin{center}{\begin{tikzpicture}\path[-,line width=0.4pt,line cap=butt,line join=miter,dash pattern=](-131.91pt,44.6953pt)--(-54pt,44.6953pt)--(-54pt,5.1097pt)--(-131.91pt,5.1097pt)--cycle;
\node[anchor=north west,inner sep=0] at (-131.91pt,44.6953pt){\savebox{\marxupbox}{{\(\begin{array}{c}\text{Einstein notation}\\\text{\cref{33}}\\\text{\(tᵢʲ\)}\end{array}\)}}\immediate\write\boxesfile{2}\immediate\write\boxesfile{\number\wd\marxupbox}\immediate\write\boxesfile{\number\ht\marxupbox}\immediate\write\boxesfile{\number\dp\marxupbox}\box\marxupbox};
\path[-,line width=0.4pt,line cap=butt,line join=miter,dash pattern=](-135.91pt,48.6953pt)--(-50pt,48.6953pt)--(-50pt,1.1097pt)--(-135.91pt,1.1097pt)--cycle;
\path[-,draw=black,line width=0.4pt,line cap=butt,line join=miter,dash pattern=](-135.91pt,48.6953pt)--(-50pt,48.6953pt)--(-50pt,1.1097pt)--(-135.91pt,1.1097pt)--cycle;
\path[-,line width=0.4pt,line cap=butt,line join=miter,dash pattern=](5.7774pt,126.4025pt)--(53.2774pt,126.4025pt)--(53.2774pt,87.4025pt)--(5.7774pt,87.4025pt)--cycle;
\node[anchor=north west,inner sep=0] at (5.7774pt,126.4025pt){\savebox{\marxupbox}{{\(\begin{array}{c}\text{{\sc{}Albert}}\\\text{\cref{114}}\\\text{\(λ\mskip 3.0mu_{\mathsf{i}}\mskip 3.0mu^{\mathsf{j}}\mskip 3.0mu\mathnormal{\rightarrow }\mskip 3.0mu\mathsf{t}\mskip 3.0mu_{\mathsf{i}}\mskip 3.0mu^{\mathsf{j}}\)}\end{array}\)}}\immediate\write\boxesfile{3}\immediate\write\boxesfile{\number\wd\marxupbox}\immediate\write\boxesfile{\number\ht\marxupbox}\immediate\write\boxesfile{\number\dp\marxupbox}\box\marxupbox};
\path[-,line width=0.4pt,line cap=butt,line join=miter,dash pattern=](1.7774pt,130.4025pt)--(57.2774pt,130.4025pt)--(57.2774pt,83.4025pt)--(1.7774pt,83.4025pt)--cycle;
\path[-,draw=black,line width=0.4pt,line cap=butt,line join=miter,dash pattern=](1.7774pt,130.4025pt)--(57.2774pt,130.4025pt)--(57.2774pt,83.4025pt)--(1.7774pt,83.4025pt)--cycle;
\path[-,line width=0.4pt,line cap=butt,line join=miter,dash pattern=](4pt,44.4025pt)--(55.0548pt,44.4025pt)--(55.0548pt,5.4025pt)--(4pt,5.4025pt)--cycle;
\node[anchor=north west,inner sep=0] at (4pt,44.4025pt){\savebox{\marxupbox}{{\(\begin{array}{c}\text{{\sc{}Roger}}\\\text{\cref{37}}\\\text{\(\mathsf{t}\mskip 3.0mu\mathnormal{:}\mskip 3.0muT_{z}\mskip 3.0mu\allowbreak{}\mathnormal{\leadsto }\allowbreak{}\mskip 3.0muT_{z}\)}\end{array}\)}}\immediate\write\boxesfile{4}\immediate\write\boxesfile{\number\wd\marxupbox}\immediate\write\boxesfile{\number\ht\marxupbox}\immediate\write\boxesfile{\number\dp\marxupbox}\box\marxupbox};
\path[-,line width=0.4pt,line cap=butt,line join=miter,dash pattern=](0pt,48.4025pt)--(59.0548pt,48.4025pt)--(59.0548pt,1.4025pt)--(0pt,1.4025pt)--cycle;
\path[-,draw=black,line width=0.4pt,line cap=butt,line join=miter,dash pattern=](0pt,48.4025pt)--(59.0548pt,48.4025pt)--(59.0548pt,1.4025pt)--(0pt,1.4025pt)--cycle;
\path[-,line width=0.4pt,line cap=butt,line join=miter,dash pattern=](-120.455pt,-37.5975pt)--(-65.455pt,-37.5975pt)--(-65.455pt,-89.5975pt)--(-120.455pt,-89.5975pt)--cycle;
\node[anchor=north west,inner sep=0] at (-120.455pt,-37.5975pt){\savebox{\marxupbox}{{\(\begin{array}{c}\text{Matrices}\\\text{\cref{110}}\\\text{\(\left[\begin{array}{cc}t^1_1&t^1_2\\t^2_1&t^2_2\end{array}\right]\)}\end{array}\)}}\immediate\write\boxesfile{5}\immediate\write\boxesfile{\number\wd\marxupbox}\immediate\write\boxesfile{\number\ht\marxupbox}\immediate\write\boxesfile{\number\dp\marxupbox}\box\marxupbox};
\path[-,line width=0.4pt,line cap=butt,line join=miter,dash pattern=](-124.455pt,-33.5975pt)--(-61.455pt,-33.5975pt)--(-61.455pt,-93.5975pt)--(-124.455pt,-93.5975pt)--cycle;
\path[-,draw=black,line width=0.4pt,line cap=butt,line join=miter,dash pattern=](-124.455pt,-33.5975pt)--(-61.455pt,-33.5975pt)--(-61.455pt,-93.5975pt)--(-124.455pt,-93.5975pt)--cycle;
\path[-,line width=0.4pt,line cap=butt,line join=miter,dash pattern=](113.0548pt,45.805pt)--(193.4249pt,45.805pt)--(193.4249pt,4pt)--(113.0548pt,4pt)--cycle;
\node[anchor=north west,inner sep=0] at (113.0548pt,45.805pt){\savebox{\marxupbox}{{\(\begin{array}{c}\text{Diagram notation}\\\text{\cref{37}}\\\text{{\begin{tikzpicture}\path[-,draw=black,line width=0.4pt,line cap=butt,line join=miter,dash pattern=](-16.32pt,5.7525pt)--(-9.32pt,5.7525pt);
\path[-,draw=black,line width=0.4pt,line cap=butt,line join=miter,dash pattern=](0pt,5.7525pt)--(7pt,5.7525pt);
\path[-,line width=0.4pt,line cap=butt,line join=miter,dash pattern=](-6.32pt,8.505pt)--(-3pt,8.505pt)--(-3pt,3pt)--(-6.32pt,3pt)--cycle;
\node[anchor=north west,inner sep=0] at (-6.32pt,8.505pt){\savebox{\marxupbox}{{\(t\)}}\immediate\write\boxesfile{12}\immediate\write\boxesfile{\number\wd\marxupbox}\immediate\write\boxesfile{\number\ht\marxupbox}\immediate\write\boxesfile{\number\dp\marxupbox}\box\marxupbox};
\path[-,draw=black,line width=0.4pt,line cap=butt,line join=miter,dash pattern=](-9.32pt,9.005pt)..controls(-9.32pt,10.3857pt)and(-8.2007pt,11.505pt)..(-6.82pt,11.505pt)--(-2.5pt,11.505pt)..controls(-1.1193pt,11.505pt)and(0pt,10.3857pt)..(0pt,9.005pt)--(0pt,2.5pt)..controls(0pt,1.1193pt)and(-1.1193pt,0pt)..(-2.5pt,0pt)--(-6.82pt,0pt)..controls(-8.2007pt,0pt)and(-9.32pt,1.1193pt)..(-9.32pt,2.5pt)--cycle;
\path[-,draw=black,line width=0.4pt,line cap=butt,line join=miter,dash pattern=](0pt,5.7525pt)--(0pt,5.7525pt);
\path[-,draw=black,line width=0.4pt,line cap=butt,line join=miter,dash pattern=](-9.32pt,5.7525pt)--(-9.32pt,5.7525pt);
\path[-,line width=0.4pt,line cap=butt,line join=miter,dash pattern=](-20.9516pt,8.0921pt)--(-18.32pt,8.0921pt)--(-18.32pt,3.4128pt)--(-20.9516pt,3.4128pt)--cycle;
\node[anchor=north west,inner sep=0] at (-20.9516pt,8.0921pt){\savebox{\marxupbox}{{\({\scriptstyle i}\)}}\immediate\write\boxesfile{13}\immediate\write\boxesfile{\number\wd\marxupbox}\immediate\write\boxesfile{\number\ht\marxupbox}\immediate\write\boxesfile{\number\dp\marxupbox}\box\marxupbox};
\path[-,line width=0.4pt,line cap=butt,line join=miter,dash pattern=](-22.9516pt,10.0921pt)--(-16.32pt,10.0921pt)--(-16.32pt,1.4128pt)--(-22.9516pt,1.4128pt)--cycle;
\path[-,line width=0.4pt,line cap=butt,line join=miter,dash pattern=](9pt,8.8313pt)--(12.8617pt,8.8313pt)--(12.8617pt,2.6737pt)--(9pt,2.6737pt)--cycle;
\node[anchor=north west,inner sep=0] at (9pt,8.8313pt){\savebox{\marxupbox}{{\({\scriptstyle j}\)}}\immediate\write\boxesfile{14}\immediate\write\boxesfile{\number\wd\marxupbox}\immediate\write\boxesfile{\number\ht\marxupbox}\immediate\write\boxesfile{\number\dp\marxupbox}\box\marxupbox};
\path[-,line width=0.4pt,line cap=butt,line join=miter,dash pattern=](7pt,10.8313pt)--(14.8617pt,10.8313pt)--(14.8617pt,0.6737pt)--(7pt,0.6737pt)--cycle;
\end{tikzpicture}}}\end{array}\)}}\immediate\write\boxesfile{6}\immediate\write\boxesfile{\number\wd\marxupbox}\immediate\write\boxesfile{\number\ht\marxupbox}\immediate\write\boxesfile{\number\dp\marxupbox}\box\marxupbox};
\path[-,line width=0.4pt,line cap=butt,line join=miter,dash pattern=](109.0548pt,49.805pt)--(197.4249pt,49.805pt)--(197.4249pt,0pt)--(109.0548pt,0pt)--cycle;
\path[-,draw=black,line width=0.4pt,line cap=butt,line join=miter,dash pattern=](109.0548pt,49.805pt)--(197.4249pt,49.805pt)--(197.4249pt,0pt)--(109.0548pt,0pt)--cycle;
\path[-to,draw=black,line width=0.4pt,line cap=butt,line join=miter,dash pattern=](29.5274pt,83.4023pt)..controls(29.5274pt,69.7948pt)and(29.5274pt,62.0106pt)..(29.5274pt,48.4032pt);
\path[-,line width=0.4pt,line cap=butt,line join=miter,dash pattern=](33.5274pt,69.415pt)--(80.5174pt,69.415pt)--(80.5174pt,62.39pt)--(33.5274pt,62.39pt)--cycle;
\node[anchor=north west,inner sep=0] at (33.5274pt,69.415pt){\savebox{\marxupbox}{{evaluates to}}\immediate\write\boxesfile{7}\immediate\write\boxesfile{\number\wd\marxupbox}\immediate\write\boxesfile{\number\ht\marxupbox}\immediate\write\boxesfile{\number\dp\marxupbox}\box\marxupbox};
\path[-,line width=0.4pt,line cap=butt,line join=miter,dash pattern=](29.5274pt,73.415pt)--(84.5174pt,73.415pt)--(84.5174pt,58.39pt)--(29.5274pt,58.39pt)--cycle;
\path[-to,draw=black,line width=0.4pt,line cap=butt,line join=miter,dash pattern=](-92.955pt,1.1096pt)..controls(-92.955pt,-12.8979pt)and(-92.955pt,-21.1501pt)..(-92.955pt,-33.5961pt);
\path[-,line width=0.4pt,line cap=butt,line join=miter,dash pattern=](-88.955pt,-14.815pt)--(-44.775pt,-14.815pt)--(-44.775pt,-23.88pt)--(-88.955pt,-23.88pt)--cycle;
\node[anchor=north west,inner sep=0] at (-88.955pt,-14.815pt){\savebox{\marxupbox}{{generalises}}\immediate\write\boxesfile{8}\immediate\write\boxesfile{\number\wd\marxupbox}\immediate\write\boxesfile{\number\ht\marxupbox}\immediate\write\boxesfile{\number\dp\marxupbox}\box\marxupbox};
\path[-,line width=0.4pt,line cap=butt,line join=miter,dash pattern=](-92.955pt,-10.815pt)--(-40.775pt,-10.815pt)--(-40.775pt,-27.88pt)--(-92.955pt,-27.88pt)--cycle;
\path[to-to,draw=black,line width=0.4pt,line cap=butt,line join=miter,dash pattern=](109.0544pt,24.9025pt)..controls(91.8337pt,24.9025pt)and(80.0613pt,24.9025pt)..(59.0552pt,24.9025pt);
\path[-,line width=0.4pt,line cap=butt,line join=miter,dash pattern=](79.4486pt,20.9025pt)--(103.3186pt,20.9025pt)--(103.3186pt,11.8475pt)--(79.4486pt,11.8475pt)--cycle;
\node[anchor=north west,inner sep=0] at (79.4486pt,20.9025pt){\savebox{\marxupbox}{{equiv.}}\immediate\write\boxesfile{9}\immediate\write\boxesfile{\number\wd\marxupbox}\immediate\write\boxesfile{\number\ht\marxupbox}\immediate\write\boxesfile{\number\dp\marxupbox}\box\marxupbox};
\path[-,line width=0.4pt,line cap=butt,line join=miter,dash pattern=](75.4486pt,24.9025pt)--(107.3186pt,24.9025pt)--(107.3186pt,7.8475pt)--(75.4486pt,7.8475pt)--cycle;
\path[to-to,draw=black,line width=0.4pt,line cap=butt,line join=miter,dash pattern=](-49.9996pt,24.9025pt)..controls(-32.597pt,24.9025pt)and(-20.9327pt,24.9025pt)..(-0.0004pt,24.9025pt);
\path[-,line width=0.4pt,line cap=butt,line join=miter,dash pattern=](-43.6488pt,20.9025pt)--(-19.7788pt,20.9025pt)--(-19.7788pt,11.8475pt)--(-43.6488pt,11.8475pt)--cycle;
\node[anchor=north west,inner sep=0] at (-43.6488pt,20.9025pt){\savebox{\marxupbox}{{equiv.}}\immediate\write\boxesfile{10}\immediate\write\boxesfile{\number\wd\marxupbox}\immediate\write\boxesfile{\number\ht\marxupbox}\immediate\write\boxesfile{\number\dp\marxupbox}\box\marxupbox};
\path[-,line width=0.4pt,line cap=butt,line join=miter,dash pattern=](-47.6488pt,24.9025pt)--(-15.7788pt,24.9025pt)--(-15.7788pt,7.8475pt)--(-47.6488pt,7.8475pt)--cycle;
\path[-to,draw=black,line width=0.4pt,line cap=butt,line join=miter,dash pattern=](1.7773pt,88.3242pt)..controls(-23.189pt,71.6098pt)and(-35.0462pt,63.6715pt)..(-57.416pt,48.6953pt);
\path[-,line width=0.4pt,line cap=butt,line join=miter,dash pattern=](-77.1738pt,76.8975pt)--(-35.7138pt,76.8975pt)--(-35.7138pt,69.9025pt)--(-77.1738pt,69.9025pt)--cycle;
\node[anchor=north west,inner sep=0] at (-77.1738pt,76.8975pt){\savebox{\marxupbox}{{formalises}}\immediate\write\boxesfile{11}\immediate\write\boxesfile{\number\wd\marxupbox}\immediate\write\boxesfile{\number\ht\marxupbox}\immediate\write\boxesfile{\number\dp\marxupbox}\box\marxupbox};
\path[-,line width=0.4pt,line cap=butt,line join=miter,dash pattern=](-81.1738pt,80.8975pt)--(-31.7138pt,80.8975pt)--(-31.7138pt,65.9025pt)--(-81.1738pt,65.9025pt)--cycle;
\end{tikzpicture}}\end{center}\caption{Tensor notations, {\sc{}edsl}s and relationships between them. Even though
the index notation, the morphism notation and the string diagram
notation are all equivalent mathematically, in our implementation
{\sc{}Roger} is coded as a (set of) type-classes, and the index and diagram
notations are instances of it.}\label{15}\end{figure} 
In sum, because the index-notation, diagram notation and matrices are
instances of tensor categories, programs written in any of our {\sc{}edsl}s can be executed as tensor programs using the matrix instance or can
generate index or diagram notation for the code in question.  The
relationships between these notations and {\sc{}edsl}s are depicted in
\cref{15}.

\newpage{} This means that a function in {\sc{}Albert}, say
\begin{list}{}{\setlength\leftmargin{1.0em}}\item\relax
\ensuremath{\begin{parray}\column{B}{@{}>{}l<{}@{}}\column[0em]{1}{@{}>{}l<{}@{}}\column{E}{@{}>{}l<{}@{}}%
\>[1]{\mathsf{example}\mskip 3.0mu\mathsf{t}\mskip 3.0mu\mathsf{u}\mskip 3.0mu\mathnormal{=}\mskip 3.0mu\mathsf{contract}\mskip 3.0mu\allowbreak{}\mathnormal{(}\mskip 0.0muλ\mskip 3.0mu^{\mathsf{i}}\mskip 3.0mu_{\mathsf{i}}\mskip 3.0mu\mathnormal{\rightarrow }\mskip 3.0mu\mathsf{t}\mskip 3.0mu^{\mathsf{i}}\mskip 3.0mu{\tikzstar{0.11}{0.25}{5}{-18}{fill=black}}\mskip 3.0mu\mathsf{deriv}\mskip 3.0mu_{\mathsf{i}}\mskip 3.0mu\mathsf{u}\mskip 0.0mu\mathnormal{)}\allowbreak{}}\<[E]\end{parray}}\end{list} will, depending on the type, either:
\begin{enumerate}\item{}render itself in Einstein notation as \(t{^i}∇{_i}u\);\item{}render itself as the diagram \({\begin{tikzpicture}[baseline=(current bounding box.center)]\path[-,draw=lightgray,line width=0.4pt,line cap=butt,line join=miter,dash pattern=](-70.31pt,-2.4837pt)--(-63.31pt,-2.4837pt);
\path[-,draw=lightgray,line width=0.4pt,line cap=butt,line join=miter,dash pattern=](7pt,0pt)--(14pt,0pt);
\path[-,draw=lightgray,line width=0.4pt,line cap=butt,line join=miter,dash pattern=](-63.31pt,-2.4837pt)--(-56.31pt,-2.4837pt);
\path[-,draw=lightgray,line width=0.4pt,line cap=butt,line join=miter,dash pattern=](-63.31pt,-16.2256pt)--(-56.31pt,-16.2256pt);
\path[-,draw=lightgray,line width=0.4pt,line cap=butt,line join=miter,dash pattern=](-50.31pt,-2.4837pt)--(-50.31pt,-2.4837pt);
\path[-,draw=black,line width=0.4pt,line cap=butt,line join=miter,dash pattern=](-50.31pt,-12.4837pt)--(-50.31pt,-12.4837pt);
\path[-,draw=black,line width=0.4pt,line cap=butt,line join=miter,dash pattern=](-50.31pt,-19.9675pt)--(-50.31pt,-19.9675pt);
\path[-,draw=lightgray,line width=0.4pt,line cap=butt,line join=miter,dash pattern=](-44.31pt,-2.4837pt)--(-44.31pt,-2.4837pt);
\path[-,draw=black,line width=0.4pt,line cap=butt,line join=miter,dash pattern=](-44.31pt,-7.4837pt)--(-44.31pt,-7.4837pt);
\path[-,draw=black,line width=0.4pt,line cap=butt,line join=miter,dash pattern=](-44.31pt,-19.9675pt)--(-44.31pt,-19.9675pt);
\path[-,draw=black,line width=0.4pt,line cap=butt,line join=miter,dash pattern=](-38.31pt,0pt)--(-38.31pt,0pt);
\path[-,draw=lightgray,line width=0.4pt,line cap=butt,line join=miter,dash pattern=](-38.31pt,-9.9675pt)--(-38.31pt,-9.9675pt);
\path[-,draw=black,line width=0.4pt,line cap=butt,line join=miter,dash pattern=](-38.31pt,-19.9675pt)--(-38.31pt,-19.9675pt);
\path[-,draw=black,line width=0.4pt,line cap=butt,line join=miter,dash pattern=](-32.31pt,0pt)--(-31.31pt,0pt);
\path[-,draw=lightgray,line width=0.4pt,line cap=butt,line join=miter,dash pattern=](-32.31pt,-14.9675pt)--(-31.31pt,-14.9675pt);
\path[-,draw=black,line width=0.4pt,line cap=butt,line join=miter,dash pattern=](-32.31pt,-19.9675pt)--(-31.31pt,-19.9675pt);
\path[-,draw=lightgray,line width=0.4pt,line cap=butt,line join=miter,dash pattern=](0pt,0pt)--(7pt,0pt);
\path[-,draw=lightgray,line width=0.4pt,line cap=butt,line join=miter,dash pattern=](0pt,-19.9675pt)--(7pt,-19.9675pt);
\path[-,draw=lightgray,line width=0.4pt,line cap=butt,line join=miter,dash pattern=](7pt,0pt)--(7pt,0pt);
\path[-,line width=0.4pt,line cap=butt,line join=miter,dash pattern=on 0.4pt off 1pt](7pt,0pt)--(7pt,0pt)--(7pt,-19.9675pt)--(7pt,-19.9675pt)--cycle;
\path[-,draw=black,line width=0.4pt,line cap=butt,line join=miter,dash pattern=on 0.4pt off 1pt](4pt,3pt)--(10pt,3pt)--(10pt,-22.9675pt)--(4pt,-22.9675pt)--cycle;
\path[-,fill=lightgray,line width=0.4pt,line cap=butt,line join=miter,dash pattern=](8pt,-19.9675pt)..controls(8pt,-19.4152pt)and(7.5523pt,-18.9675pt)..(7pt,-18.9675pt)..controls(6.4477pt,-18.9675pt)and(6pt,-19.4152pt)..(6pt,-19.9675pt)..controls(6pt,-20.5198pt)and(6.4477pt,-20.9675pt)..(7pt,-20.9675pt)..controls(7.5523pt,-20.9675pt)and(8pt,-20.5198pt)..(8pt,-19.9675pt)--cycle;
\path[-,line width=0.4pt,line cap=butt,line join=miter,dash pattern=](-17.315pt,2.7525pt)--(-13.995pt,2.7525pt)--(-13.995pt,-2.7525pt)--(-17.315pt,-2.7525pt)--cycle;
\node[anchor=north west,inner sep=0] at (-17.315pt,2.7525pt){\savebox{\marxupbox}{{\(t\)}}\immediate\write\boxesfile{16}\immediate\write\boxesfile{\number\wd\marxupbox}\immediate\write\boxesfile{\number\ht\marxupbox}\immediate\write\boxesfile{\number\dp\marxupbox}\box\marxupbox};
\path[-,draw=black,line width=0.4pt,line cap=butt,line join=miter,dash pattern=](-20.315pt,3.2525pt)..controls(-20.315pt,4.6332pt)and(-19.1957pt,5.7525pt)..(-17.815pt,5.7525pt)--(-13.495pt,5.7525pt)..controls(-12.1143pt,5.7525pt)and(-10.995pt,4.6332pt)..(-10.995pt,3.2525pt)--(-10.995pt,-3.2525pt)..controls(-10.995pt,-4.6332pt)and(-12.1143pt,-5.7525pt)..(-13.495pt,-5.7525pt)--(-17.815pt,-5.7525pt)..controls(-19.1957pt,-5.7525pt)and(-20.315pt,-4.6332pt)..(-20.315pt,-3.2525pt)--cycle;
\path[-,draw=lightgray,line width=0.4pt,line cap=butt,line join=miter,dash pattern=](0pt,0pt)--(-10.995pt,0pt);
\path[-,draw=black,line width=0.4pt,line cap=butt,line join=miter,dash pattern=](-31.31pt,0pt)--(-20.315pt,0pt);
\path[-,draw=black,line width=0.4pt,line cap=butt,line join=miter,dash pattern=](-25.31pt,-14.9675pt)--(-25.31pt,-14.9675pt);
\path[-,draw=lightgray,line width=0.4pt,line cap=butt,line join=miter,dash pattern=](-25.31pt,-19.9675pt)--(-25.31pt,-19.9675pt);
\path[-,draw=black,line width=0.8pt,line cap=butt,line join=miter,dash pattern=](-21.31pt,-13.9675pt)..controls(-21.31pt,-12.3106pt)and(-19.9669pt,-10.9675pt)..(-18.31pt,-10.9675pt)--(-7pt,-10.9675pt)..controls(-5.3431pt,-10.9675pt)and(-4pt,-12.3106pt)..(-4pt,-13.9675pt)--(-4pt,-25.1825pt)..controls(-4pt,-26.8393pt)and(-5.3431pt,-28.1825pt)..(-7pt,-28.1825pt)--(-18.31pt,-28.1825pt)..controls(-19.9669pt,-28.1825pt)and(-21.31pt,-26.8393pt)..(-21.31pt,-25.1825pt)--cycle;
\path[-,draw=lightgray,line width=0.4pt,line cap=butt,line join=miter,dash pattern=](-21.31pt,-19.9675pt)--(-18.31pt,-19.9675pt);
\path[-,draw=black,line width=0.4pt,line cap=butt,line join=miter,dash pattern=](-21.31pt,-14.9675pt)--(-21.31pt,-14.9675pt);
\path[-,draw=lightgray,line width=0.4pt,line cap=butt,line join=miter,dash pattern=](-7pt,-19.9675pt)--(-4pt,-19.9675pt);
\path[-,draw=black,line width=0.4pt,line cap=butt,line join=miter,dash pattern=](-25.31pt,-14.9675pt)--(-21.31pt,-14.9675pt);
\path[-,draw=lightgray,line width=0.4pt,line cap=butt,line join=miter,dash pattern=](-25.31pt,-19.9675pt)--(-21.31pt,-19.9675pt);
\path[-,draw=lightgray,line width=0.4pt,line cap=butt,line join=miter,dash pattern=](-4pt,-19.9675pt)--(0pt,-19.9675pt);
\path[-,line width=0.4pt,line cap=butt,line join=miter,dash pattern=](-15.31pt,-17.7525pt)--(-10pt,-17.7525pt)--(-10pt,-22.1825pt)--(-15.31pt,-22.1825pt)--cycle;
\node[anchor=north west,inner sep=0] at (-15.31pt,-17.7525pt){\savebox{\marxupbox}{{\(u\)}}\immediate\write\boxesfile{17}\immediate\write\boxesfile{\number\wd\marxupbox}\immediate\write\boxesfile{\number\ht\marxupbox}\immediate\write\boxesfile{\number\dp\marxupbox}\box\marxupbox};
\path[-,draw=black,line width=0.4pt,line cap=butt,line join=miter,dash pattern=](-18.31pt,-17.2525pt)..controls(-18.31pt,-15.8718pt)and(-17.1907pt,-14.7525pt)..(-15.81pt,-14.7525pt)--(-9.5pt,-14.7525pt)..controls(-8.1193pt,-14.7525pt)and(-7pt,-15.8718pt)..(-7pt,-17.2525pt)--(-7pt,-22.6825pt)..controls(-7pt,-24.0632pt)and(-8.1193pt,-25.1825pt)..(-9.5pt,-25.1825pt)--(-15.81pt,-25.1825pt)..controls(-17.1907pt,-25.1825pt)and(-18.31pt,-24.0632pt)..(-18.31pt,-22.6825pt)--cycle;
\path[-,draw=lightgray,line width=0.4pt,line cap=butt,line join=miter,dash pattern=](-7pt,-19.9675pt)--(-7pt,-19.9675pt);
\path[-,draw=lightgray,line width=0.4pt,line cap=butt,line join=miter,dash pattern=](-18.31pt,-19.9675pt)--(-18.31pt,-19.9675pt);
\path[-,draw=lightgray,line width=0.4pt,line cap=butt,line join=miter,dash pattern=](-31.31pt,-14.9675pt)..controls(-27.31pt,-14.9675pt)and(-29.31pt,-19.9675pt)..(-25.31pt,-19.9675pt);
\path[-,draw=black,line width=0.4pt,line cap=butt,line join=miter,dash pattern=](-31.31pt,-19.9675pt)..controls(-27.31pt,-19.9675pt)and(-29.31pt,-14.9675pt)..(-25.31pt,-14.9675pt);
\path[-,draw=black,line width=0.4pt,line cap=butt,line join=miter,dash pattern=](-38.31pt,0pt)--(-32.31pt,0pt);
\path[-,draw=lightgray,line width=0.4pt,line cap=butt,line join=miter,dash pattern=](-38.31pt,-9.9675pt)..controls(-34.31pt,-9.9675pt)and(-36.31pt,-14.9675pt)..(-32.31pt,-14.9675pt);
\path[-,draw=black,line width=0.4pt,line cap=butt,line join=miter,dash pattern=](-38.31pt,-19.9675pt)--(-32.31pt,-19.9675pt);
\path[-,draw=lightgray,line width=0.4pt,line cap=butt,line join=miter,dash pattern=](-44.31pt,-2.4837pt)..controls(-40.31pt,-2.4837pt)and(-42.31pt,-9.9675pt)..(-38.31pt,-9.9675pt);
\path[-,draw=black,line width=0.4pt,line cap=butt,line join=miter,dash pattern=](-44.31pt,-7.4837pt)..controls(-40.31pt,-7.4837pt)and(-42.31pt,0pt)..(-38.31pt,0pt);
\path[-,draw=black,line width=0.4pt,line cap=butt,line join=miter,dash pattern=](-44.31pt,-19.9675pt)--(-38.31pt,-19.9675pt);
\path[-,draw=lightgray,line width=0.4pt,line cap=butt,line join=miter,dash pattern=](-50.31pt,-2.4837pt)--(-44.31pt,-2.4837pt);
\path[-,draw=black,line width=0.4pt,line cap=butt,line join=miter,dash pattern=](-50.31pt,-12.4837pt)..controls(-46.31pt,-12.4837pt)and(-48.31pt,-7.4837pt)..(-44.31pt,-7.4837pt);
\path[-,draw=black,line width=0.4pt,line cap=butt,line join=miter,dash pattern=](-50.31pt,-19.9675pt)--(-44.31pt,-19.9675pt);
\path[-,draw=lightgray,line width=0.4pt,line cap=butt,line join=miter,dash pattern=](-56.31pt,-2.4837pt)--(-50.31pt,-2.4837pt);
\path[to-,draw=black,line width=0.4pt,line cap=butt,line join=miter,dash pattern=](-53.37pt,-12.8195pt)..controls(-52.53pt,-12.5955pt)and(-51.51pt,-12.4837pt)..(-50.31pt,-12.4837pt);
\path[-,draw=black,line width=0.4pt,line cap=butt,line join=miter,dash pattern=](-56.31pt,-16.2256pt)..controls(-56.31pt,-14.4756pt)and(-55.33pt,-13.3421pt)..(-53.37pt,-12.8195pt);
\path[-,draw=black,line width=0.4pt,line cap=butt,line join=miter,dash pattern=](-53.37pt,-19.6317pt)..controls(-52.53pt,-19.8557pt)and(-51.51pt,-19.9675pt)..(-50.31pt,-19.9675pt);
\path[-,draw=black,line width=0.4pt,line cap=butt,line join=miter,dash pattern=](-56.31pt,-16.2256pt)..controls(-56.31pt,-17.9756pt)and(-55.33pt,-19.1091pt)..(-53.37pt,-19.6317pt);
\path[-,draw=lightgray,line width=0.4pt,line cap=butt,line join=miter,dash pattern=](-63.31pt,-2.4837pt)--(-63.31pt,-2.4837pt);
\path[-,line width=0.4pt,line cap=butt,line join=miter,dash pattern=on 0.4pt off 1pt](-63.31pt,-2.4837pt)--(-63.31pt,-2.4837pt)--(-63.31pt,-16.2256pt)--(-63.31pt,-16.2256pt)--cycle;
\path[-,draw=black,line width=0.4pt,line cap=butt,line join=miter,dash pattern=on 0.4pt off 1pt](-66.31pt,0.5163pt)--(-60.31pt,0.5163pt)--(-60.31pt,-19.2256pt)--(-66.31pt,-19.2256pt)--cycle;
\path[-,fill=lightgray,line width=0.4pt,line cap=butt,line join=miter,dash pattern=](-62.31pt,-16.2256pt)..controls(-62.31pt,-15.6733pt)and(-62.7577pt,-15.2256pt)..(-63.31pt,-15.2256pt)..controls(-63.8623pt,-15.2256pt)and(-64.31pt,-15.6733pt)..(-64.31pt,-16.2256pt)..controls(-64.31pt,-16.7779pt)and(-63.8623pt,-17.2256pt)..(-63.31pt,-17.2256pt)..controls(-62.7577pt,-17.2256pt)and(-62.31pt,-16.7779pt)..(-62.31pt,-16.2256pt)--cycle;
\end{tikzpicture}}\) or\item{}run on matrix representations of the tensors \(t\) and \(u\) and compute the result (a
  scalar field in this case, representing the directional derivative of \(u\) in the direction of \(t\)).\end{enumerate} 
Together, {\sc{}Albert} and {\sc{}Roger} form a Haskell library for expressing
tensors.\footnote{The code is available at \url{https://github.com/jyp/linear-smc/}.\label{18}}  This library
leverages linear types as implemented in {\sc{}ghc} 9.  This
implementation defines an executable semantics of {\sc{}Albert}, and is presented in
\cref{394}. All the examples presented in this paper were prepared using our library. In particular, the diagrams are
generated with it.

In \cref{135}, we move to deal with tensor fields
proper. Essentially this means that every expression in either {\sc{}edsl} corresponds to a tensor field, and that we can manipulate
derivatives of such fields. With this addition, the {\sc{}edsl}s can be used
for symbolic calculations of tensor fields. We can, for example, apply
covariant derivatives to tensor expressions, re-express them in terms
of partial derivatives and Christoffel symbols, and instantiate those to
concrete coordinates systems. We demonstrate this workflow in
\cref{265}, where we express Einstein's General Relativity equation for
the curvature of space-time and verify that the Schwarzschild metric
tensor is a solution.

We start in \cref{19} with a summary of the notions of linear algebra and tensors.

\section{Background: linear algebra and tensors}\label{19} 
The goal of this section is both to provide the canonical
presentation as reference, and to expose its abstruse character. The
summary does not replace a proper introduction to the topic, and we
urge the reader to turn to an appropriate reference if necessary (see references in \cref{400}).

A typical definition of tensor that one might find is the following:
\begin{quote}An \(n\)th-rank\footnote{What is called here \emph{rank} is referred to as \emph{order} in our text. We choose this terminology because rank has
another meaning in linear algebra. Namely, it is the minimum number of
simple tensors that sum to it. A simple tensor is a tensor product of
a number of non-zero vectors and co-vectors: \(\mathsf{t}\mskip 3.0mu\mathnormal{=}\mskip 3.0mu\mathsf{v}_{1}\mskip 3.0mu{⊗}\mskip 3.0mu\mathnormal{...}\mskip 3.0mu{⊗}\mskip 3.0mu\mathsf{v}_{n}\). Most
proper tensors are not simple. That is, their rank is more than
one.
  \label{20}} tensor in \(m\)-dimensional
space is a mathematical object that has \(n\) indices and
\(m^n\) components and obeys certain transformation
rules.{\unskip\nobreak\hfil\penalty50 \hskip2em\hbox{}\nobreak\hfil\citep{rowland_tensor_2023}\parfillskip=0pt \finalhyphendemerits=0 \par}\end{quote} (The transformations in question relate to change of basis, as we will see.) This kind of definition is heavily geared
towards coordinate representations, rather than their algebraic definition.
Why do pedagogical accounts widely refer to coordinate representations
rather than semantics?  One answer is that calculations are eventually
always performed using coordinates. Another answer is that the kind of
algebraic thinking required to grasp tensors may be too abstract to form
an intuition. Our point of view is that it is indeed at the wrong abstraction level,
and that the categorical structures are better suited to reasoning about tensors than
the pure linear-algebraic ones. Nonetheless, we will have to refer to the algebraic
definitions of tensors down the road, so we provide  a minimal recap below.

\subsection{Pure algebraic point of view}\label{21} 
The main object of study are homomorphisms between vector spaces:
linear transformations, also called linear maps. We will later see that tensors are such maps.

\begin{definition}[vector space] A vector space (over a field \(S\)) is a commutative group
\(\mathsf{v}\) equipped with a compatible notion of scaling by elements of \(S\).
\begin{list}{}{\setlength\leftmargin{1.0em}}\item\relax
\begin{tabular}{c@{\quad\quad}c}\ensuremath{\begin{parray}\column{B}{@{}>{}l<{}@{}}\column[0em]{1}{@{}>{}l<{}@{}}\column[1em]{2}{@{}>{}l<{}@{}}\column{3}{@{}>{}l<{}@{}}\column{E}{@{}>{}l<{}@{}}%
\>[1]{\mathbf{class}\mskip 3.0mu\mathsf{Group}\mskip 3.0mu\mathsf{v}\mskip 3.0mu\mathbf{where}}\<[E]\\
\>[2]{\allowbreak{}\mathnormal{(}\mskip 0.0mu\mathnormal{+}\mskip 0.0mu\mathnormal{)}\allowbreak{}\mskip 3.0mu}\>[3]{\mathnormal{::}\mskip 3.0mu\mathsf{v}\mskip 3.0mu\mathnormal{\rightarrow }\mskip 3.0mu\mathsf{v}\mskip 3.0mu\mathnormal{\rightarrow }\mskip 3.0mu\mathsf{v}}\<[E]\\
\>[2]{0\mskip 3.0mu}\>[3]{\mathnormal{::}\mskip 3.0mu\mathsf{v}}\<[E]\\
\>[2]{\mathsf{negate}\mskip 3.0mu}\>[3]{\mathnormal{::}\mskip 3.0mu\mathsf{v}\mskip 3.0mu\mathnormal{\rightarrow }\mskip 3.0mu\mathsf{v}}\<[E]\end{parray}}& \ensuremath{\begin{parray}\column{B}{@{}>{}l<{}@{}}\column[0em]{1}{@{}>{}l<{}@{}}\column[1em]{2}{@{}>{}l<{}@{}}\column{3}{@{}>{}l<{}@{}}\column{4}{@{}>{}l<{}@{}}\column{5}{@{}>{}l<{}@{}}\column{6}{@{}>{}l<{}@{}}\column{7}{@{}>{}l<{}@{}}\column{E}{@{}>{}l<{}@{}}%
\>[1]{\mathbf{class}\mskip 3.0mu\allowbreak{}\mathnormal{(}\mskip 0.0mu\mathsf{Group}\mskip 3.0mu\mathsf{v}\mskip 0.0mu}\>[6]{\mathnormal{)}\allowbreak{}\mskip 3.0mu\mathnormal{\Rightarrow }\mskip 3.0mu\mathsf{VectorSpace}\mskip 3.0mu\mathsf{v}\mskip 3.0mu}\>[7]{\mathbf{where}}\<[E]\\
\>[2]{\allowbreak{}\mathnormal{(}\mskip 0.0mu\smalltriangleleft \mskip 0.0mu\mathnormal{)}\allowbreak{}\mskip 3.0mu}\>[3]{\mathnormal{::}\mskip 3.0mu}\>[4]{\mathsf{S}\mskip 3.0mu\mathnormal{\rightarrow }\mskip 3.0mu}\>[5]{\mathsf{v}\mskip 3.0mu\mathnormal{\rightarrow }\mskip 3.0mu\mathsf{v}}\<[E]\end{parray}}\end{tabular}\end{list} A vector space must additionally satisfy a number of laws, including
that scaling is a linear operation:
\(s \smalltriangleleft (x+y) = s \smalltriangleleft x+ s \smalltriangleleft y\).\label{22}\end{definition}  
The exact nature of this field of scalars (\(S\)) has little bearing on the
algebraic development\footnote{Symmetrisation and antisymmetrisation (\cref{265}) require the field to have characteristic zero, which is true for real and complex fields.\label{23}}, but we assume throughout that they are real numbers. Note that \(S\) is itself
a vector space, with scaling \(\allowbreak{}\mathnormal{(}\mskip 0.0mu\smalltriangleleft \mskip 0.0mu\mathnormal{)}\allowbreak{}\) then being scalar multiplication.

\begin{definition}[linear map] A function \(f: V ⟶ W\) is a linear map iff. for all collections of scalars \(cᵢ\) and vectors \(\vec vᵢ\) we have \begin{displaymath}f \left(∑ᵢ cᵢ \smalltriangleleft \vec vᵢ\right) = ∑ᵢ cᵢ \smalltriangleleft (f(\vec vᵢ))\end{displaymath} \label{24}\end{definition}For a fixed domain and codomain, linear maps themselves form a vector space.
\begin{list}{}{\setlength\leftmargin{1.0em}}\item\relax
  \ensuremath{\begin{parray}\column{B}{@{}>{}l<{}@{}}\column[0em]{1}{@{}>{}l<{}@{}}\column[1em]{2}{@{}>{}l<{}@{}}\column{3}{@{}>{}l<{}@{}}\column{4}{@{}>{}l<{}@{}}\column{5}{@{}>{}l<{}@{}}\column{6}{@{}>{}l<{}@{}}\column{7}{@{}>{}l<{}@{}}\column{E}{@{}>{}l<{}@{}}%
\>[1]{\mathbf{instance}\mskip 3.0mu\allowbreak{}\mathnormal{(}\mskip 0.0mu\mathsf{Group}\mskip 3.0mu\mathsf{w}\mskip 0.0mu\mathnormal{)}\allowbreak{}\mskip 3.0mu\mathnormal{\Rightarrow }\mskip 3.0mu\mathsf{Group}\mskip 3.0mu\allowbreak{}\mathnormal{(}\mskip 0.0mu\mathsf{v}\mskip 3.0mu\mathnormal{⟶}\mskip 3.0mu\mathsf{w}\mskip 0.0mu\mathnormal{)}\allowbreak{}\mskip 3.0mu\mathbf{where}}\<[E]\\
\>[2]{\mathsf{negate}\mskip 3.0mu\mathsf{f}\mskip 3.0mu\mathnormal{=}\mskip 3.0mu\allowbreak{}\mathnormal{(}\mskip 0.0mu}\>[6]{λ\mskip 3.0mu\mathsf{v}\mskip 3.0mu\mathnormal{\rightarrow }\mskip 3.0mu\mathsf{negate}\mskip 3.0mu\allowbreak{}\mathnormal{(}\mskip 0.0mu\mathsf{f}\mskip 3.0mu\mathsf{v}\mskip 0.0mu\mathnormal{)}\allowbreak{}\mskip 0.0mu\mathnormal{)}\allowbreak{}}\<[E]\\
\>[2]{\mathsf{f}\mskip 3.0mu}\>[3]{\mathnormal{+}\mskip 3.0mu\mathsf{f'}\mskip 3.0mu}\>[5]{\mathnormal{=}\mskip 3.0mu\allowbreak{}\mathnormal{(}\mskip 0.0mu}\>[7]{λ\mskip 3.0mu\mathsf{v}\mskip 3.0mu\mathnormal{\rightarrow }\mskip 3.0mu\mathsf{f}\mskip 3.0mu\mathsf{v}\mskip 3.0mu\mathnormal{+}\mskip 3.0mu\mathsf{f'}\mskip 3.0mu\mathsf{v}\mskip 0.0mu\mathnormal{)}\allowbreak{}}\<[E]\\
\>[2]{0\mskip 3.0mu\mathnormal{=}\mskip 3.0mu\allowbreak{}\mathnormal{(}\mskip 0.0mu}\>[4]{λ\mskip 3.0mu\mathsf{v}\mskip 3.0mu\mathnormal{\rightarrow }\mskip 3.0mu0\mskip 0.0mu\mathnormal{)}\allowbreak{}}\<[E]\end{parray}} 

 \ensuremath{\begin{parray}\column{B}{@{}>{}l<{}@{}}\column[0em]{1}{@{}>{}l<{}@{}}\column[1em]{2}{@{}>{}l<{}@{}}\column{3}{@{}>{}l<{}@{}}\column{4}{@{}>{}l<{}@{}}\column{5}{@{}>{}l<{}@{}}\column{6}{@{}>{}l<{}@{}}\column{E}{@{}>{}l<{}@{}}%
\>[1]{\mathbf{instance}\mskip 3.0mu\allowbreak{}\mathnormal{(}\mskip 0.0mu\mathsf{VectorSpace}\mskip 3.0mu\mathsf{v}\mskip 0.0mu}\>[4]{\mathnormal{,}\mskip 3.0mu\mathsf{VectorSpace}\mskip 3.0mu\mathsf{w}\mskip 0.0mu}\>[5]{\mathnormal{)}\allowbreak{}\mskip 3.0mu\mathnormal{\Rightarrow }\mskip 3.0mu\mathsf{VectorSpace}\mskip 3.0mu\allowbreak{}\mathnormal{(}\mskip 0.0mu}\>[6]{\mathsf{v}\mskip 3.0mu\mathnormal{⟶}\mskip 3.0mu\mathsf{w}\mskip 0.0mu\mathnormal{)}\allowbreak{}\mskip 3.0mu\mathbf{where}}\<[E]\\
\>[2]{\mathsf{c}\mskip 3.0mu\smalltriangleleft \mskip 3.0mu\mathsf{f}\mskip 3.0mu}\>[3]{\mathnormal{=}\mskip 3.0mu\allowbreak{}\mathnormal{(}\mskip 0.0muλ\mskip 3.0mu\mathsf{v}\mskip 3.0mu\mathnormal{\rightarrow }\mskip 3.0mu\mathsf{c}\mskip 3.0mu\smalltriangleleft \mskip 3.0mu\mathsf{f}\mskip 3.0mu\mathsf{v}\mskip 0.0mu\mathnormal{)}\allowbreak{}}\<[E]\end{parray}} \end{list} The eager reader should be warned that, for now, indices are used to range of over arbitrary sets of vectors and scalars (and bound by \(\sum\)), in a usual way.
Indices take a special meaning only when we get to coordinates and the Einstein notation (from \cref{30} and \cref{33}).

\begin{definition}[covector space] Given a vector space \(V\), the
covector space \(\dual V\) is defined as the set of linear maps
\(V ⟶ S\).\label{25}\end{definition} Since covector spaces are special cases of linear maps, they form vector spaces too.
In a similar vein, the set of linear maps \(f: S ⟶ W\) is isomorphic to \(W\).
(Indeed \(f(s) = f (s \smalltriangleleft 1) = s \smalltriangleleft f(1)\), and thus the vector \(f(1)\) in \(W\) fully determines the linear function \(f\).)

 \begin{definition}[bilinear map] A function \(f: V×W ⟶ U\) is a bilinear map iff. for all \(cᵢ, dⱼ : S\), \(\vec vᵢ : V\), and \(\vec wⱼ : W\) we have
\begin{displaymath}f \left(∑ᵢ cᵢ \smalltriangleleft \vec vᵢ, ∑ⱼ dⱼ \smalltriangleleft \vec wⱼ\right) = ∑_{i,j} cᵢdⱼ \smalltriangleleft (f(\vec vᵢ,\vec wⱼ))\end{displaymath} \label{26}\end{definition} 
\begin{definition}[Tensor product of vector spaces] Given two vector spaces \(V\) and \(W\),
their tensor product is a vector space,
denoted by \(V⊗W\),
together with a bilinear map \(φ:(V×W) ⟶ (V⊗W)\) with the following universal property.
For every vector space \(Z\) and every bilinear map \(h:(V×W) ⟶ Z\), there exists
a unique linear map \(h' : (V⊗W) ⟶ Z\) such that \(h = h' ∘ φ\).
The output \(φ(v,w)\) is often denoted by \(v⊗w\), overloading the same symbol.
(We let the reader check that the tensor product always exists.)\label{27}\end{definition} 
 
\paragraph*{Examples:}\hspace{1.0ex}\label{28} Here is an attempt at providing an intuition
for what is, and is not, a bilinear function. Consider the simplest
case of the definition of bilinear map where there is just one vector
\(\vec v\) as the first argument and one vector \(\vec w\) as
the second argument to \(f\).
We then have \(f(\vec v,0) = f(1\smalltriangleleft \vec v,0\smalltriangleleft \vec w) = (1×0)\smalltriangleleft f(\vec v,\vec w) = 0\).  This
means that vector addition is \emph{not} bilinear
because \(\vec v+0=\vec v ≠ 0\). Similarly, \(f\) cannot be first or
second projection, because they are also linear, not bilinear.

We also have that we can ``move constant factors'' between \(\vec v\) and
\(\vec w\): \(f(c \smalltriangleleft \vec v, 1 \smalltriangleleft \vec w) = (c×1) \smalltriangleleft f(\vec v,\vec w) = (1×c) \smalltriangleleft f(\vec v,\vec w) =
f(1 \smalltriangleleft \vec v, c \smalltriangleleft \vec w)\). In connection with the tensor product this means
that even though, for any two vectors \(\vec v : V\) and \(\vec w : W\),
we can construct a tensor \(u = \phi(\vec v,\vec w) : V⊗W\) which looks like
we have embedded a pair, we cannot extract \(\vec v\) and \(\vec w\) again
– they are mixed up together (entangled).

What a bilinear function can (and must) do, as we can see from the
definition, when given two linear combinations, is to compute a linear
combination based on all pairwise products of the coefficients,
without depending on the coefficients themselves.

\paragraph*{Order of a tensor}\hspace{1.0ex}\label{29} Often, tensors are used in a context where there is a single (atomic) underlying
vector space \(\mathsf{T}\) which isn't just the scalars.
Then the complexity of a vector space built from \(\mathsf{T}\) can be measured by its order.
The order of \(\mathsf{T}\) is defined to be 1 and the order of the scalar space is 0.
The order of a tensor space \(\mathsf{V}\mskip 3.0mu\mathnormal{⊗}\mskip 3.0mu\mathsf{W}\) is the sum of the order of
spaces \(\mathsf{V}\) and \(\mathsf{W}\), and this way we can build spaces of
arbitrarily large order. The order of a linear map can be defined
either as the pair of the orders of its input and output spaces, or as
their sum (depending on convention). For example a linear operator on an atomic vector space has
order (1,1) or 2 in the respective conventions. Morphisms of order three or more are properly called
tensors. Conversely, tensors of any order (including 0, 1 and 2) are
linear maps, of the appropriate domain and codomain.
When there is more that one underlying vector space, the order is not enough to
characterise a tensor space: the full type needs to be specified, as in \cref{37}.
(Yet this level of complexity won't be exercised in this paper.)

\subsection{Coordinate representations}\label{30} 
In practice, the algebraic definitions are not easy to manipulate for
concrete problems, thus one most commonly works with coordinate
representations instead. (Our goal will be to break free of those eventually.)  As a reminder, given a basis \(\vec eᵢ\),
any vector \(\vec x∈V\) can be uniquely expressed as \(\vec x = ∑ᵢ xⁱ
\smalltriangleleft \vec eᵢ\) where each \(xⁱ\) coordinate is a
scalar. In this way, given a basis \(\vec eᵢ\), a vector space is
isomorphic to its set of coordinate representations.  Note that a superscript is used for the index of such coordinates. The general convention
that governs whether one should write indices in low or high positions
is explained in \cref{33}; for now it is enough to know that
they are indexing notations.

Like vectors, linear maps are also commonly manipulated as matrices of coefficients.
For a linear map \(f\) from a vector space with basis \(\vec dᵢ\) to
a space with basis \(\vec eⱼ\), each column is given by the
coefficients of \(f(\vec dᵢ)\). Indeed, using \(F_i^j\) to denote the coefficients we have:

\begin{displaymath}
  f(\vec x) =
  f\left(∑ᵢ xⁱ \smalltriangleleft \vec dᵢ\right) =
  ∑ᵢ xⁱ \smalltriangleleft f(\vec dᵢ) =
  ∑ᵢ xⁱ \smalltriangleleft \left(∑_j Fᵢʲ \smalltriangleleft \vec eⱼ\right) =
  ∑ⱼ \left(∑_i Fᵢʲ xⁱ\right) \smalltriangleleft \vec eⱼ
\end{displaymath} In general, the values of the matrix coefficients \(Fᵢʲ\) depend on
the choice of bases \(\vec dᵢ\) and \(\vec eⱼ\), but to reduce
the number of moving parts one usually works with a coherent set of
bases. 
\paragraph*{Coherent bases}\hspace{1.0ex}\label{31} Starting from an atomic vector space
\(T\), one can build a collection of more complicated tensor spaces
using tensor product, dual, and the unit (the scalar field \(S\)). For coordinate representations each such space could, in
general, have its own basis, but it is standard to work with a
collection of coherent bases.  Given a basis \(\vec eᵢ\) for a finite-dimensional
atomic vector space \(T\), the coherent basis for \(\dual T\) is
the set of covectors \(\tilde eʲ\) such that \(\tilde eʲ(\vec
eᵢ) = δᵢʲ\). (It is usually called the dual basis.) Likewise, given two coherent bases \(\vec dᵢ\) and
\(\vec eᵢ\) respectively for \(V\) and \(W\), the coherent
basis for \(V⊗W\) is \(b_{i,j} = φ(\vec dᵢ,\vec eⱼ)\), where
\(φ\) is given by \cref{27}. Note that this basis is indexed
by a pair. Accordingly, if the dimension of \(V\) is \(m\) and
the dimension of \(W\) is \(n\), the dimension of \(V⊗W\) is
\(m×n\). Additionally, re-associating tensor spaces do not change
coherent bases (\(\vec e_{(i,j),k}\) is the same as \(\vec d_{i,(j,k)}\), up to applying the corresponding associator).
Finally the scalar vector space has dimension one, and thus has a
single base vector, which is coherently chosen to be the unit of the
scalar field (the number \(1 : S\)).   
\paragraph*{Coordinate transformations}\hspace{1.0ex}\label{32} 
Assuming one basis \(\vec eⱼ\) and
another basis      \(\vec dᵢ\) for
the same vector space \(V\) such that \(\vec dᵢ = ∑ⱼ Fᵢʲ \vec eⱼ\),
then the coordinates in basis \(\vec dᵢ\) for \(\vec x\) are
\(\hat xʲ = ∑ᵢ Fᵢʲ xⁱ\).
 We say that the matrix \(F\) is the
transformation matrix for \(V\) given the choice of bases made above, and denote it \(J(V)\).
Then the transformation matrices
for vector spaces built from an atomic space \(T\) using the coherent set of bases
defined above are given by the following structural rules:

\begin{align*}J(V⊗W)&= J(V)⊗J(W)&\text{Kronecker product of matrices}\\J(\dual V)&= J(V)^{-1}&\text{matrix inverse}\\J(S)&= 1&\text{scalar unit}\end{align*} Furthermore, the matrix representation \(G\) of a linear map \(g : V ⟶ W\) is
transformed to: \(J(W) · G · J(\dual V)\), where \((·)\) is matrix multiplication.
These are the ``transformation rules'' that
\citet{rowland_tensor_2023} allude to in the above quote.
 
\subsection{Einstein Notation}\label{33} 
The previous section showed how to deal with concrete matrices, using
a concrete choice of bases. The next step is to manipulate symbolic
expressions involving matrices. The language of such expressions
(together with a couple of simple conventions) is colloquially
referred to as Einstein notation.

In this notation, every index ranges over the dimensions of an atomic
vector space.\footnote{That is, the components of tensor spaces are
always indexed separately; and the tensor unit (scalar) is never
indexed. Previously, indices were used to range over arbitrary sets of vectors and scalars--- without regard for the dimensions of the spaces which they inhabit.\label{34}} Consequently, the total number of free (non repeated) indices indicates the order of
a tensor expression in index notation.  An index can be written as a
subscript (and called a low index) or as a superscript (and called a high index).

The location (high or low) of the index is dictated by which
coordinate transformation applies to it. That is, if a high index
ranges over the dimensions of \(V\), then \(J(V)\) applies, whereas
\(J(\dual V)\) applies for a low index.  Additionally, every
reference to a tensor is fully saturated, in the sense that a symbolic
tensor is always applied to as many indices as its order. Thus, for
instance \(x^i\) denote (the components of) a vector, and
\(y_j\) denote (the components of) a covector. The expression
\(t{_i}{^j}\) refers to (components of) a linear transformation of
order (1,1).  In the absence of contraction (see below),
multiplication increases the order of tensors. For instance
\(x^i y_j\) also has order (1,1).  In general, if \(t\) and
\(u\) are expressions denoting tensors of order \(m\) and
\(n\) respectively, then their product \(t u\) denotes a tensor of order
\(m+n\).

\paragraph*{Contraction}\hspace{1.0ex}\label{35} In Einstein notation, the convention is that, within a
term, a repeated index is implicitly summed over. (In terms familiar
to this journal: such indices are implicitly bound by a summation
operator.)  Because summation is a linear operator, within a term all the
well-scoped locations of the summation operator are equivalent--- so it
makes a lot of sense to omit them.
 Additionally, when an index is
repeated, it must be repeated exactly twice; once as a high index and
once as a low index. Mentioning an index twice is called
\emph{contraction}.  Viewing tensors as higher dimensional matrices of
coefficients, contraction consists in summing coefficients along a
diagonal.
Therefore a contraction reduces the order of the tensor by
two.\footnote{As such, it is the generalisation of a trace operation. We come back to the connection between contraction and trace in \cref{86}.\label{36}} The indices which are contracted are sometimes called ``dummy''
and those that are not contracted are called ``live'' (In terms
familiar to the functional programming community, dummies
are bound variables and live indices are free variables.)  To be
well-scoped, every term in a sum must use the same live indices.  For
instance the expression \(t{_l}{^j}u{_m}{^k}v{_i}{^l}{^m} + v{_i}{^j}{^k}\) denotes a tensor of order
(1,2). Its live indices are \({}_i,{}^j,{}^k\), and the indices \(m\) and
\(l\) are dummies.

At this stage, one can see the Einstein notation as a convenient way
to notate expressions which manipulate coordinates of tensors. The
high/low index convention makes it clear which transformations
apply.  Yet it may be mysterious why indices must be repeated
exactly twice, and why (live) indices cannot be omitted from a term.
The answer lies in the following observation. Even though the Einstein notation may originate as a
convenient way to express coordinates, it really is intended to describe algebraic objects. The physicists
\citet{thorne15:ModernClassicalPhysics} put it this way: \begin{quote}[we suggest to]
momentarily think of [Einstein notation] as a relationship between components of
tensors in a specific basis; then do a quick mind-flip and regard it
quite differently, as a relationship between geometric,
basis-independent tensors with the indices playing the roles of slot
names.\end{quote} \noindent{} (A ``slot'' is a component of input or output tensor space.)
\noindent{} The key of this ``mind flip'' is that live indices correspond to
inputs (or outputs) of linear functions, and contraction corresponds
to connecting inputs to outputs.
The main contribution of this paper is to work out this connection in full as a pair of two {\sc{}edsl}s.

\section{Categorical structures}\label{37} 
The key concepts needed to understand the essence of Einstein notation are the
categorical structures that tensors inhabit. Besides, these structures
will be instrumental in our design: we will let the user of {\sc{}Albert} write expressions which are (close to) Einstein notation, but they
will be evaluated to morphisms in the appropriate category. The
underlying category can then be specialised according to the
application at hand.

The categorical approach consists in raising the abstraction level,
and focusing on the ways that linear maps are combined to construct more
complex ones. The first step is to view linear maps as morphisms of a
category whose objects are vector spaces. We
render the type of morphisms from \(\mathsf{a}\) to \(\mathsf{b}\) as \(\mathsf{a}\mskip 1.0mu\overset{z}{\leadsto }\mskip 1.0mu\mathsf{b}\), corresponding to \(z\mskip 3.0mu\mathsf{a}\mskip 3.0mu\mathsf{b}\) in Haskell code.
\begin{list}{}{\setlength\leftmargin{1.0em}}\item\relax
\ensuremath{\begin{parray}\column{B}{@{}>{}l<{}@{}}\column[0em]{1}{@{}>{}l<{}@{}}\column[1em]{2}{@{}>{}l<{}@{}}\column{3}{@{}>{}l<{}@{}}\column{4}{@{}>{}l<{}@{}}\column{5}{@{}>{}l<{}@{}}\column{E}{@{}>{}l<{}@{}}%
\>[1]{\mathbf{class}\mskip 3.0mu\mathsf{Category}\mskip 3.0muz\mskip 3.0mu\mathbf{where}}\<[E]\\
\>[2]{\mathsf{id}\mskip 3.0mu}\>[3]{\mathnormal{::}\mskip 3.0mu}\>[4]{\mathsf{a}\mskip 1.0mu\overset{z}{\leadsto }\mskip 1.0mu\mathsf{a}}\<[E]\\
\>[2]{\allowbreak{}\mathnormal{(}\mskip 0.0mu\allowbreak{}\mathnormal{∘}\allowbreak{}\mskip 0.0mu\mathnormal{)}\allowbreak{}\mskip 3.0mu}\>[3]{\mathnormal{::}\mskip 3.0mu}\>[5]{\allowbreak{}\mathnormal{(}\mskip 0.0mu\mathsf{b}\mskip 1.0mu\overset{z}{\leadsto }\mskip 1.0mu\mathsf{c}\mskip 0.0mu\mathnormal{)}\allowbreak{}\mskip 3.0mu\mathnormal{\rightarrow }\mskip 3.0mu\allowbreak{}\mathnormal{(}\mskip 0.0mu\mathsf{a}\mskip 1.0mu\overset{z}{\leadsto }\mskip 1.0mu\mathsf{b}\mskip 0.0mu\mathnormal{)}\allowbreak{}\mskip 3.0mu\mathnormal{\rightarrow }\mskip 3.0mu\allowbreak{}\mathnormal{(}\mskip 0.0mu\mathsf{a}\mskip 1.0mu\overset{z}{\leadsto }\mskip 1.0mu\mathsf{c}\mskip 0.0mu\mathnormal{)}\allowbreak{}}\<[E]\end{parray}} \end{list} 
\begin{figure}[]\begin{center}\begin{tabular}{cc}{\begin{tikzpicture}[baseline=(current bounding box.center)]\path[-,draw=black,line width=0.4pt,line cap=butt,line join=miter,dash pattern=](-1pt,0pt)--(0pt,0pt);
\path[-,draw=black,line width=0.4pt,line cap=butt,line join=miter,dash pattern=](6pt,0pt)--(7pt,0pt);
\path[-,draw=black,line width=0.4pt,line cap=butt,line join=miter,dash pattern=](0pt,0pt)--(6pt,0pt);
\path[-,line width=0.4pt,line cap=butt,line join=miter,dash pattern=](-5.6316pt,2.3396pt)--(-3pt,2.3396pt)--(-3pt,-2.3396pt)--(-5.6316pt,-2.3396pt)--cycle;
\node[anchor=north west,inner sep=0] at (-5.6316pt,2.3396pt){\savebox{\marxupbox}{{\({\scriptstyle i}\)}}\immediate\write\boxesfile{38}\immediate\write\boxesfile{\number\wd\marxupbox}\immediate\write\boxesfile{\number\ht\marxupbox}\immediate\write\boxesfile{\number\dp\marxupbox}\box\marxupbox};
\path[-,line width=0.4pt,line cap=butt,line join=miter,dash pattern=](-7.6316pt,4.3396pt)--(-1pt,4.3396pt)--(-1pt,-4.3396pt)--(-7.6316pt,-4.3396pt)--cycle;
\path[-,line width=0.4pt,line cap=butt,line join=miter,dash pattern=](9pt,3.0788pt)--(12.8617pt,3.0788pt)--(12.8617pt,-3.0788pt)--(9pt,-3.0788pt)--cycle;
\node[anchor=north west,inner sep=0] at (9pt,3.0788pt){\savebox{\marxupbox}{{\({\scriptstyle j}\)}}\immediate\write\boxesfile{39}\immediate\write\boxesfile{\number\wd\marxupbox}\immediate\write\boxesfile{\number\ht\marxupbox}\immediate\write\boxesfile{\number\dp\marxupbox}\box\marxupbox};
\path[-,line width=0.4pt,line cap=butt,line join=miter,dash pattern=](7pt,5.0788pt)--(14.8617pt,5.0788pt)--(14.8617pt,-5.0788pt)--(7pt,-5.0788pt)--cycle;
\end{tikzpicture}}&{\begin{tikzpicture}[baseline=(current bounding box.center)]\path[-,draw=black,line width=0.4pt,line cap=butt,line join=miter,dash pattern=](-16.32pt,0pt)--(-9.32pt,0pt);
\path[-,draw=black,line width=0.4pt,line cap=butt,line join=miter,dash pattern=](18.31pt,0pt)--(25.31pt,0pt);
\path[-,draw=black,line width=0.4pt,line cap=butt,line join=miter,dash pattern=](0pt,0pt)--(7pt,0pt);
\path[-,line width=0.4pt,line cap=butt,line join=miter,dash pattern=](10pt,2.215pt)--(15.31pt,2.215pt)--(15.31pt,-2.215pt)--(10pt,-2.215pt)--cycle;
\node[anchor=north west,inner sep=0] at (10pt,2.215pt){\savebox{\marxupbox}{{\(u\)}}\immediate\write\boxesfile{40}\immediate\write\boxesfile{\number\wd\marxupbox}\immediate\write\boxesfile{\number\ht\marxupbox}\immediate\write\boxesfile{\number\dp\marxupbox}\box\marxupbox};
\path[-,draw=black,line width=0.4pt,line cap=butt,line join=miter,dash pattern=](7pt,2.715pt)..controls(7pt,4.0957pt)and(8.1193pt,5.215pt)..(9.5pt,5.215pt)--(15.81pt,5.215pt)..controls(17.1907pt,5.215pt)and(18.31pt,4.0957pt)..(18.31pt,2.715pt)--(18.31pt,-2.715pt)..controls(18.31pt,-4.0957pt)and(17.1907pt,-5.215pt)..(15.81pt,-5.215pt)--(9.5pt,-5.215pt)..controls(8.1193pt,-5.215pt)and(7pt,-4.0957pt)..(7pt,-2.715pt)--cycle;
\path[-,draw=black,line width=0.4pt,line cap=butt,line join=miter,dash pattern=](18.31pt,0pt)--(18.31pt,0pt);
\path[-,draw=black,line width=0.4pt,line cap=butt,line join=miter,dash pattern=](7pt,0pt)--(7pt,0pt);
\path[-,line width=0.4pt,line cap=butt,line join=miter,dash pattern=](-6.32pt,2.7525pt)--(-3pt,2.7525pt)--(-3pt,-2.7525pt)--(-6.32pt,-2.7525pt)--cycle;
\node[anchor=north west,inner sep=0] at (-6.32pt,2.7525pt){\savebox{\marxupbox}{{\(t\)}}\immediate\write\boxesfile{41}\immediate\write\boxesfile{\number\wd\marxupbox}\immediate\write\boxesfile{\number\ht\marxupbox}\immediate\write\boxesfile{\number\dp\marxupbox}\box\marxupbox};
\path[-,draw=black,line width=0.4pt,line cap=butt,line join=miter,dash pattern=](-9.32pt,3.2525pt)..controls(-9.32pt,4.6332pt)and(-8.2007pt,5.7525pt)..(-6.82pt,5.7525pt)--(-2.5pt,5.7525pt)..controls(-1.1193pt,5.7525pt)and(0pt,4.6332pt)..(0pt,3.2525pt)--(0pt,-3.2525pt)..controls(0pt,-4.6332pt)and(-1.1193pt,-5.7525pt)..(-2.5pt,-5.7525pt)--(-6.82pt,-5.7525pt)..controls(-8.2007pt,-5.7525pt)and(-9.32pt,-4.6332pt)..(-9.32pt,-3.2525pt)--cycle;
\path[-,draw=black,line width=0.4pt,line cap=butt,line join=miter,dash pattern=](0pt,0pt)--(0pt,0pt);
\path[-,draw=black,line width=0.4pt,line cap=butt,line join=miter,dash pattern=](-9.32pt,0pt)--(-9.32pt,0pt);
\path[-,line width=0.4pt,line cap=butt,line join=miter,dash pattern=](-20.9516pt,2.3396pt)--(-18.32pt,2.3396pt)--(-18.32pt,-2.3396pt)--(-20.9516pt,-2.3396pt)--cycle;
\node[anchor=north west,inner sep=0] at (-20.9516pt,2.3396pt){\savebox{\marxupbox}{{\({\scriptstyle i}\)}}\immediate\write\boxesfile{42}\immediate\write\boxesfile{\number\wd\marxupbox}\immediate\write\boxesfile{\number\ht\marxupbox}\immediate\write\boxesfile{\number\dp\marxupbox}\box\marxupbox};
\path[-,line width=0.4pt,line cap=butt,line join=miter,dash pattern=](-22.9516pt,4.3396pt)--(-16.32pt,4.3396pt)--(-16.32pt,-4.3396pt)--(-22.9516pt,-4.3396pt)--cycle;
\path[-,line width=0.4pt,line cap=butt,line join=miter,dash pattern=](27.31pt,3.0788pt)--(31.1717pt,3.0788pt)--(31.1717pt,-3.0788pt)--(27.31pt,-3.0788pt)--cycle;
\node[anchor=north west,inner sep=0] at (27.31pt,3.0788pt){\savebox{\marxupbox}{{\({\scriptstyle j}\)}}\immediate\write\boxesfile{43}\immediate\write\boxesfile{\number\wd\marxupbox}\immediate\write\boxesfile{\number\ht\marxupbox}\immediate\write\boxesfile{\number\dp\marxupbox}\box\marxupbox};
\path[-,line width=0.4pt,line cap=butt,line join=miter,dash pattern=](25.31pt,5.0788pt)--(33.1717pt,5.0788pt)--(33.1717pt,-5.0788pt)--(25.31pt,-5.0788pt)--cycle;
\end{tikzpicture}}\\\(\mathsf{id}\)&\(\mathsf{u}\mskip 3.0mu\allowbreak{}\mathnormal{∘}\allowbreak{}\mskip 3.0mu\mathsf{t}\)\\\(δ{_i}{^j}\)&\(u{_k}{^j}t{_i}{^k}\)\end{tabular}\end{center}\caption{Diagram, categorical, and index notations for identity and composition.}\label{44}\end{figure} 
Vector spaces form a commutative monoid under tensor product. Hence, linear maps form
a symmetric monoidal category, or {\sc{}smc}, whose combinators are as follows.
\begin{list}{}{\setlength\leftmargin{1.0em}}\item\relax
 \ensuremath{\begin{parray}\column{B}{@{}>{}l<{}@{}}\column[0em]{1}{@{}>{}l<{}@{}}\column[1em]{2}{@{}>{}l<{}@{}}\column{3}{@{}>{}l<{}@{}}\column{4}{@{}>{}l<{}@{}}\column{5}{@{}>{}l<{}@{}}\column{6}{@{}>{}l<{}@{}}\column{7}{@{}>{}l<{}@{}}\column{8}{@{}>{}l<{}@{}}\column{9}{@{}>{}l<{}@{}}\column{E}{@{}>{}l<{}@{}}%
\>[1]{\mathbf{class}\mskip 3.0mu}\>[4]{\mathsf{Category}\mskip 3.0muz\mskip 3.0mu}\>[7]{\mathnormal{\Rightarrow }\mskip 3.0mu\mathsf{SymmetricMonoidal}\mskip 3.0muz\mskip 3.0mu\mathbf{where}}\<[E]\\
\>[2]{\allowbreak{}\mathnormal{(}\mskip 0.0mu{⊗}\mskip 0.0mu\mathnormal{)}\allowbreak{}\mskip 3.0mu}\>[3]{\mathnormal{::}\mskip 3.0mu}\>[9]{\allowbreak{}\mathnormal{(}\mskip 0.0mu\mathsf{a}\mskip 1.0mu\overset{z}{\leadsto }\mskip 1.0mu\mathsf{b}\mskip 0.0mu\mathnormal{)}\allowbreak{}\mskip 3.0mu\mathnormal{\rightarrow }\mskip 3.0mu\allowbreak{}\mathnormal{(}\mskip 0.0mu\mathsf{c}\mskip 1.0mu\overset{z}{\leadsto }\mskip 1.0mu\mathsf{d}\mskip 0.0mu\mathnormal{)}\allowbreak{}\mskip 3.0mu\mathnormal{\rightarrow }\mskip 3.0mu\allowbreak{}\mathnormal{(}\mskip 0.0mu\mathsf{a}\mskip 3.0mu\mathnormal{⊗}\mskip 3.0mu\mathsf{c}\mskip 0.0mu\mathnormal{)}\allowbreak{}\mskip 1.0mu\overset{z}{\leadsto }\mskip 1.0mu\allowbreak{}\mathnormal{(}\mskip 0.0mu\mathsf{b}\mskip 3.0mu\mathnormal{⊗}\mskip 3.0mu\mathsf{d}\mskip 0.0mu\mathnormal{)}\allowbreak{}}\<[E]\\
\>[2]{σ\mskip 3.0mu}\>[3]{\mathnormal{::}\mskip 3.0mu}\>[6]{\allowbreak{}\mathnormal{(}\mskip 0.0mu\mathsf{a}\mskip 3.0mu\mathnormal{⊗}\mskip 3.0mu\mathsf{b}\mskip 0.0mu\mathnormal{)}\allowbreak{}\mskip 1.0mu\overset{z}{\leadsto }\mskip 1.0mu\allowbreak{}\mathnormal{(}\mskip 0.0mu\mathsf{b}\mskip 3.0mu\mathnormal{⊗}\mskip 3.0mu\mathsf{a}\mskip 0.0mu\mathnormal{)}\allowbreak{}}\<[E]\\
\>[2]{α\mskip 3.0mu}\>[3]{\mathnormal{::}\mskip 3.0mu}\>[8]{\allowbreak{}\mathnormal{(}\mskip 0.0mu\allowbreak{}\mathnormal{(}\mskip 0.0mu\mathsf{a}\mskip 3.0mu\mathnormal{⊗}\mskip 3.0mu\mathsf{b}\mskip 0.0mu\mathnormal{)}\allowbreak{}\mskip 3.0mu\mathnormal{⊗}\mskip 3.0mu\mathsf{c}\mskip 0.0mu\mathnormal{)}\allowbreak{}\mskip 1.0mu\overset{z}{\leadsto }\mskip 1.0mu\allowbreak{}\mathnormal{(}\mskip 0.0mu\mathsf{a}\mskip 3.0mu\mathnormal{⊗}\mskip 3.0mu\allowbreak{}\mathnormal{(}\mskip 0.0mu\mathsf{b}\mskip 3.0mu\mathnormal{⊗}\mskip 3.0mu\mathsf{c}\mskip 0.0mu\mathnormal{)}\allowbreak{}\mskip 0.0mu\mathnormal{)}\allowbreak{}}\<[E]\\
\>[2]{\bar{α}\mskip 3.0mu}\>[3]{\mathnormal{::}\mskip 3.0mu}\>[8]{\allowbreak{}\mathnormal{(}\mskip 0.0mu\mathsf{a}\mskip 3.0mu\mathnormal{⊗}\mskip 3.0mu\allowbreak{}\mathnormal{(}\mskip 0.0mu\mathsf{b}\mskip 3.0mu\mathnormal{⊗}\mskip 3.0mu\mathsf{c}\mskip 0.0mu\mathnormal{)}\allowbreak{}\mskip 0.0mu\mathnormal{)}\allowbreak{}\mskip 1.0mu\overset{z}{\leadsto }\mskip 1.0mu\allowbreak{}\mathnormal{(}\mskip 0.0mu\allowbreak{}\mathnormal{(}\mskip 0.0mu\mathsf{a}\mskip 3.0mu\mathnormal{⊗}\mskip 3.0mu\mathsf{b}\mskip 0.0mu\mathnormal{)}\allowbreak{}\mskip 3.0mu\mathnormal{⊗}\mskip 3.0mu\mathsf{c}\mskip 0.0mu\mathnormal{)}\allowbreak{}}\<[E]\\
\>[2]{ρ\mskip 3.0mu}\>[3]{\mathnormal{::}\mskip 3.0mu}\>[5]{\mathsf{a}\mskip 1.0mu\overset{z}{\leadsto }\mskip 1.0mu\allowbreak{}\mathnormal{(}\mskip 0.0mu\mathsf{a}\mskip 3.0mu\mathnormal{⊗}\mskip 3.0mu\mathbf{1}\mskip 0.0mu\mathnormal{)}\allowbreak{}}\<[E]\\
\>[2]{\bar{ρ}\mskip 3.0mu}\>[3]{\mathnormal{::}\mskip 3.0mu}\>[5]{\allowbreak{}\mathnormal{(}\mskip 0.0mu\mathsf{a}\mskip 3.0mu\mathnormal{⊗}\mskip 3.0mu\mathbf{1}\mskip 0.0mu\mathnormal{)}\allowbreak{}\mskip 1.0mu\overset{z}{\leadsto }\mskip 1.0mu\mathsf{a}}\<[E]\end{parray}} \end{list} In the above {\sc{}smc} class definition we follow the usual convention of using the
same symbol \(\allowbreak{}\mathnormal{(}\mskip 0.0mu\mathnormal{⊗}\mskip 0.0mu\mathnormal{)}\allowbreak{}\) both for the product of objects and the
parallel composition of morphisms. In fact, this morphism operator is
also called a tensor product in the literature.
An {\sc{}smc} comes with a number of
laws, which are both unsurprising and extensively documented elsewhere
\citep{barr_category_1999}. We omit them here.
The operations \(σ\) (swap), \(α\), \(\bar{α}\) (associators) witness the
commutative monoidal structure which tensor products possess.  The
unit of the tensor product, written \(\mathbf{1}\), is the scalar vector space (\(\mathsf{S}\)),
which is witnessed by the isomorphisms  \(ρ\) and \(\bar{ρ}\), called unitors.

\begin{figure}[]\begin{center}\begin{tabular}{cccc}{\begin{tikzpicture}\path[-,draw=black,line width=0.4pt,line cap=butt,line join=miter,dash pattern=](-18.31pt,5.7525pt)--(-11.31pt,5.7525pt);
\path[-,draw=black,line width=0.4pt,line cap=butt,line join=miter,dash pattern=](-18.31pt,-11.215pt)--(-11.31pt,-11.215pt);
\path[-,draw=black,line width=0.4pt,line cap=butt,line join=miter,dash pattern=](0pt,5.7525pt)--(7pt,5.7525pt);
\path[-,draw=black,line width=0.4pt,line cap=butt,line join=miter,dash pattern=](0pt,-11.215pt)--(7pt,-11.215pt);
\path[-,line width=0.4pt,line cap=butt,line join=miter,dash pattern=](-7.315pt,8.505pt)--(-3.995pt,8.505pt)--(-3.995pt,3pt)--(-7.315pt,3pt)--cycle;
\node[anchor=north west,inner sep=0] at (-7.315pt,8.505pt){\savebox{\marxupbox}{{\(t\)}}\immediate\write\boxesfile{45}\immediate\write\boxesfile{\number\wd\marxupbox}\immediate\write\boxesfile{\number\ht\marxupbox}\immediate\write\boxesfile{\number\dp\marxupbox}\box\marxupbox};
\path[-,draw=black,line width=0.4pt,line cap=butt,line join=miter,dash pattern=](-10.315pt,9.005pt)..controls(-10.315pt,10.3857pt)and(-9.1957pt,11.505pt)..(-7.815pt,11.505pt)--(-3.495pt,11.505pt)..controls(-2.1143pt,11.505pt)and(-0.995pt,10.3857pt)..(-0.995pt,9.005pt)--(-0.995pt,2.5pt)..controls(-0.995pt,1.1193pt)and(-2.1143pt,0pt)..(-3.495pt,0pt)--(-7.815pt,0pt)..controls(-9.1957pt,0pt)and(-10.315pt,1.1193pt)..(-10.315pt,2.5pt)--cycle;
\path[-,draw=black,line width=0.4pt,line cap=butt,line join=miter,dash pattern=](0pt,5.7525pt)--(-0.995pt,5.7525pt);
\path[-,draw=black,line width=0.4pt,line cap=butt,line join=miter,dash pattern=](-11.31pt,5.7525pt)--(-10.315pt,5.7525pt);
\path[-,line width=0.4pt,line cap=butt,line join=miter,dash pattern=](-8.31pt,-9pt)--(-3pt,-9pt)--(-3pt,-13.43pt)--(-8.31pt,-13.43pt)--cycle;
\node[anchor=north west,inner sep=0] at (-8.31pt,-9pt){\savebox{\marxupbox}{{\(u\)}}\immediate\write\boxesfile{46}\immediate\write\boxesfile{\number\wd\marxupbox}\immediate\write\boxesfile{\number\ht\marxupbox}\immediate\write\boxesfile{\number\dp\marxupbox}\box\marxupbox};
\path[-,draw=black,line width=0.4pt,line cap=butt,line join=miter,dash pattern=](-11.31pt,-8.5pt)..controls(-11.31pt,-7.1193pt)and(-10.1907pt,-6pt)..(-8.81pt,-6pt)--(-2.5pt,-6pt)..controls(-1.1193pt,-6pt)and(0pt,-7.1193pt)..(0pt,-8.5pt)--(0pt,-13.93pt)..controls(0pt,-15.3107pt)and(-1.1193pt,-16.43pt)..(-2.5pt,-16.43pt)--(-8.81pt,-16.43pt)..controls(-10.1907pt,-16.43pt)and(-11.31pt,-15.3107pt)..(-11.31pt,-13.93pt)--cycle;
\path[-,draw=black,line width=0.4pt,line cap=butt,line join=miter,dash pattern=](0pt,-11.215pt)--(0pt,-11.215pt);
\path[-,draw=black,line width=0.4pt,line cap=butt,line join=miter,dash pattern=](-11.31pt,-11.215pt)--(-11.31pt,-11.215pt);
\path[-,line width=0.4pt,line cap=butt,line join=miter,dash pattern=](-22.9416pt,8.0921pt)--(-20.31pt,8.0921pt)--(-20.31pt,3.4128pt)--(-22.9416pt,3.4128pt)--cycle;
\node[anchor=north west,inner sep=0] at (-22.9416pt,8.0921pt){\savebox{\marxupbox}{{\({\scriptstyle i}\)}}\immediate\write\boxesfile{47}\immediate\write\boxesfile{\number\wd\marxupbox}\immediate\write\boxesfile{\number\ht\marxupbox}\immediate\write\boxesfile{\number\dp\marxupbox}\box\marxupbox};
\path[-,line width=0.4pt,line cap=butt,line join=miter,dash pattern=](-24.9416pt,10.0921pt)--(-18.31pt,10.0921pt)--(-18.31pt,1.4128pt)--(-24.9416pt,1.4128pt)--cycle;
\path[-,line width=0.4pt,line cap=butt,line join=miter,dash pattern=](-24.1717pt,-8.1362pt)--(-20.31pt,-8.1362pt)--(-20.31pt,-14.2938pt)--(-24.1717pt,-14.2938pt)--cycle;
\node[anchor=north west,inner sep=0] at (-24.1717pt,-8.1362pt){\savebox{\marxupbox}{{\({\scriptstyle j}\)}}\immediate\write\boxesfile{48}\immediate\write\boxesfile{\number\wd\marxupbox}\immediate\write\boxesfile{\number\ht\marxupbox}\immediate\write\boxesfile{\number\dp\marxupbox}\box\marxupbox};
\path[-,line width=0.4pt,line cap=butt,line join=miter,dash pattern=](-26.1717pt,-6.1362pt)--(-18.31pt,-6.1362pt)--(-18.31pt,-16.2938pt)--(-26.1717pt,-16.2938pt)--cycle;
\path[-,line width=0.4pt,line cap=butt,line join=miter,dash pattern=](9pt,8.2436pt)--(13.3398pt,8.2436pt)--(13.3398pt,3.2614pt)--(9pt,3.2614pt)--cycle;
\node[anchor=north west,inner sep=0] at (9pt,8.2436pt){\savebox{\marxupbox}{{\({\scriptstyle k}\)}}\immediate\write\boxesfile{49}\immediate\write\boxesfile{\number\wd\marxupbox}\immediate\write\boxesfile{\number\ht\marxupbox}\immediate\write\boxesfile{\number\dp\marxupbox}\box\marxupbox};
\path[-,line width=0.4pt,line cap=butt,line join=miter,dash pattern=](7pt,10.2436pt)--(15.3398pt,10.2436pt)--(15.3398pt,1.2614pt)--(7pt,1.2614pt)--cycle;
\path[-,line width=0.4pt,line cap=butt,line join=miter,dash pattern=](9pt,-8.6892pt)--(11.5769pt,-8.6892pt)--(11.5769pt,-13.7408pt)--(9pt,-13.7408pt)--cycle;
\node[anchor=north west,inner sep=0] at (9pt,-8.6892pt){\savebox{\marxupbox}{{\({\scriptstyle l}\)}}\immediate\write\boxesfile{50}\immediate\write\boxesfile{\number\wd\marxupbox}\immediate\write\boxesfile{\number\ht\marxupbox}\immediate\write\boxesfile{\number\dp\marxupbox}\box\marxupbox};
\path[-,line width=0.4pt,line cap=butt,line join=miter,dash pattern=](7pt,-6.6892pt)--(13.5769pt,-6.6892pt)--(13.5769pt,-15.7408pt)--(7pt,-15.7408pt)--cycle;
\end{tikzpicture}}&{\begin{tikzpicture}\path[-,draw=black,line width=0.4pt,line cap=butt,line join=miter,dash pattern=](-1pt,17.4184pt)--(0pt,17.4184pt);
\path[-,draw=black,line width=0.4pt,line cap=butt,line join=miter,dash pattern=](-1pt,10pt)--(0pt,10pt);
\path[-,draw=black,line width=0.4pt,line cap=butt,line join=miter,dash pattern=](-1pt,0pt)--(0pt,0pt);
\path[-,draw=black,line width=0.4pt,line cap=butt,line join=miter,dash pattern=](6pt,17.4184pt)--(7pt,17.4184pt);
\path[-,draw=black,line width=0.4pt,line cap=butt,line join=miter,dash pattern=](6pt,7.4184pt)--(7pt,7.4184pt);
\path[-,draw=black,line width=0.4pt,line cap=butt,line join=miter,dash pattern=](6pt,0pt)--(7pt,0pt);
\path[-,draw=black,line width=0.4pt,line cap=butt,line join=miter,dash pattern=](0pt,17.4184pt)--(6pt,17.4184pt);
\path[-,draw=black,line width=0.4pt,line cap=butt,line join=miter,dash pattern=](0pt,10pt)..controls(4pt,10pt)and(2pt,7.4184pt)..(6pt,7.4184pt);
\path[-,draw=black,line width=0.4pt,line cap=butt,line join=miter,dash pattern=](0pt,0pt)--(6pt,0pt);
\path[-,line width=0.4pt,line cap=butt,line join=miter,dash pattern=](-5.6316pt,19.758pt)--(-3pt,19.758pt)--(-3pt,15.0788pt)--(-5.6316pt,15.0788pt)--cycle;
\node[anchor=north west,inner sep=0] at (-5.6316pt,19.758pt){\savebox{\marxupbox}{{\({\scriptstyle i}\)}}\immediate\write\boxesfile{51}\immediate\write\boxesfile{\number\wd\marxupbox}\immediate\write\boxesfile{\number\ht\marxupbox}\immediate\write\boxesfile{\number\dp\marxupbox}\box\marxupbox};
\path[-,line width=0.4pt,line cap=butt,line join=miter,dash pattern=](-7.6316pt,21.758pt)--(-1pt,21.758pt)--(-1pt,13.0788pt)--(-7.6316pt,13.0788pt)--cycle;
\path[-,line width=0.4pt,line cap=butt,line join=miter,dash pattern=](-6.8617pt,13.0788pt)--(-3pt,13.0788pt)--(-3pt,6.9212pt)--(-6.8617pt,6.9212pt)--cycle;
\node[anchor=north west,inner sep=0] at (-6.8617pt,13.0788pt){\savebox{\marxupbox}{{\({\scriptstyle j}\)}}\immediate\write\boxesfile{52}\immediate\write\boxesfile{\number\wd\marxupbox}\immediate\write\boxesfile{\number\ht\marxupbox}\immediate\write\boxesfile{\number\dp\marxupbox}\box\marxupbox};
\path[-,line width=0.4pt,line cap=butt,line join=miter,dash pattern=](-8.8617pt,15.0788pt)--(-1pt,15.0788pt)--(-1pt,4.9212pt)--(-8.8617pt,4.9212pt)--cycle;
\path[-,line width=0.4pt,line cap=butt,line join=miter,dash pattern=](-7.3398pt,2.4911pt)--(-3pt,2.4911pt)--(-3pt,-2.4911pt)--(-7.3398pt,-2.4911pt)--cycle;
\node[anchor=north west,inner sep=0] at (-7.3398pt,2.4911pt){\savebox{\marxupbox}{{\({\scriptstyle k}\)}}\immediate\write\boxesfile{53}\immediate\write\boxesfile{\number\wd\marxupbox}\immediate\write\boxesfile{\number\ht\marxupbox}\immediate\write\boxesfile{\number\dp\marxupbox}\box\marxupbox};
\path[-,line width=0.4pt,line cap=butt,line join=miter,dash pattern=](-9.3398pt,4.4911pt)--(-1pt,4.4911pt)--(-1pt,-4.4911pt)--(-9.3398pt,-4.4911pt)--cycle;
\path[-,line width=0.4pt,line cap=butt,line join=miter,dash pattern=](9pt,19.9442pt)--(11.5769pt,19.9442pt)--(11.5769pt,14.8926pt)--(9pt,14.8926pt)--cycle;
\node[anchor=north west,inner sep=0] at (9pt,19.9442pt){\savebox{\marxupbox}{{\({\scriptstyle l}\)}}\immediate\write\boxesfile{54}\immediate\write\boxesfile{\number\wd\marxupbox}\immediate\write\boxesfile{\number\ht\marxupbox}\immediate\write\boxesfile{\number\dp\marxupbox}\box\marxupbox};
\path[-,line width=0.4pt,line cap=butt,line join=miter,dash pattern=](7pt,21.9442pt)--(13.5769pt,21.9442pt)--(13.5769pt,12.8926pt)--(7pt,12.8926pt)--cycle;
\path[-,line width=0.4pt,line cap=butt,line join=miter,dash pattern=](9pt,9.0262pt)--(15.1685pt,9.0262pt)--(15.1685pt,5.8106pt)--(9pt,5.8106pt)--cycle;
\node[anchor=north west,inner sep=0] at (9pt,9.0262pt){\savebox{\marxupbox}{{\({\scriptstyle m}\)}}\immediate\write\boxesfile{55}\immediate\write\boxesfile{\number\wd\marxupbox}\immediate\write\boxesfile{\number\ht\marxupbox}\immediate\write\boxesfile{\number\dp\marxupbox}\box\marxupbox};
\path[-,line width=0.4pt,line cap=butt,line join=miter,dash pattern=](7pt,11.0262pt)--(17.1685pt,11.0262pt)--(17.1685pt,3.8106pt)--(7pt,3.8106pt)--cycle;
\path[-,line width=0.4pt,line cap=butt,line join=miter,dash pattern=](9pt,1.6078pt)--(13.3581pt,1.6078pt)--(13.3581pt,-1.6078pt)--(9pt,-1.6078pt)--cycle;
\node[anchor=north west,inner sep=0] at (9pt,1.6078pt){\savebox{\marxupbox}{{\({\scriptstyle n}\)}}\immediate\write\boxesfile{56}\immediate\write\boxesfile{\number\wd\marxupbox}\immediate\write\boxesfile{\number\ht\marxupbox}\immediate\write\boxesfile{\number\dp\marxupbox}\box\marxupbox};
\path[-,line width=0.4pt,line cap=butt,line join=miter,dash pattern=](7pt,3.6078pt)--(15.3581pt,3.6078pt)--(15.3581pt,-3.6078pt)--(7pt,-3.6078pt)--cycle;
\end{tikzpicture}}&{\begin{tikzpicture}\path[-,draw=black,line width=0.4pt,line cap=butt,line join=miter,dash pattern=](-1pt,7.2177pt)--(0pt,7.2177pt);
\path[-,draw=black,line width=0.4pt,line cap=butt,line join=miter,dash pattern=](-1pt,-0.2008pt)--(0pt,-0.2008pt);
\path[-,draw=black,line width=0.4pt,line cap=butt,line join=miter,dash pattern=](6pt,7.0169pt)--(7pt,7.0169pt);
\path[-,draw=black,line width=0.4pt,line cap=butt,line join=miter,dash pattern=](6pt,0pt)--(7pt,0pt);
\path[-,draw=black,line width=0.4pt,line cap=butt,line join=miter,dash pattern=](0pt,7.2177pt)..controls(4pt,7.2177pt)and(2pt,0pt)..(6pt,0pt);
\path[-,draw=black,line width=0.4pt,line cap=butt,line join=miter,dash pattern=](0pt,-0.2008pt)..controls(4pt,-0.2008pt)and(2pt,7.0169pt)..(6pt,7.0169pt);
\path[-,line width=0.4pt,line cap=butt,line join=miter,dash pattern=](-5.6316pt,9.5573pt)--(-3pt,9.5573pt)--(-3pt,4.878pt)--(-5.6316pt,4.878pt)--cycle;
\node[anchor=north west,inner sep=0] at (-5.6316pt,9.5573pt){\savebox{\marxupbox}{{\({\scriptstyle i}\)}}\immediate\write\boxesfile{57}\immediate\write\boxesfile{\number\wd\marxupbox}\immediate\write\boxesfile{\number\ht\marxupbox}\immediate\write\boxesfile{\number\dp\marxupbox}\box\marxupbox};
\path[-,line width=0.4pt,line cap=butt,line join=miter,dash pattern=](-7.6316pt,11.5573pt)--(-1pt,11.5573pt)--(-1pt,2.878pt)--(-7.6316pt,2.878pt)--cycle;
\path[-,line width=0.4pt,line cap=butt,line join=miter,dash pattern=](-6.8617pt,2.878pt)--(-3pt,2.878pt)--(-3pt,-3.2795pt)--(-6.8617pt,-3.2795pt)--cycle;
\node[anchor=north west,inner sep=0] at (-6.8617pt,2.878pt){\savebox{\marxupbox}{{\({\scriptstyle j}\)}}\immediate\write\boxesfile{58}\immediate\write\boxesfile{\number\wd\marxupbox}\immediate\write\boxesfile{\number\ht\marxupbox}\immediate\write\boxesfile{\number\dp\marxupbox}\box\marxupbox};
\path[-,line width=0.4pt,line cap=butt,line join=miter,dash pattern=](-8.8617pt,4.878pt)--(-1pt,4.878pt)--(-1pt,-5.2795pt)--(-8.8617pt,-5.2795pt)--cycle;
\path[-,line width=0.4pt,line cap=butt,line join=miter,dash pattern=](9pt,9.508pt)--(13.3398pt,9.508pt)--(13.3398pt,4.5258pt)--(9pt,4.5258pt)--cycle;
\node[anchor=north west,inner sep=0] at (9pt,9.508pt){\savebox{\marxupbox}{{\({\scriptstyle k}\)}}\immediate\write\boxesfile{59}\immediate\write\boxesfile{\number\wd\marxupbox}\immediate\write\boxesfile{\number\ht\marxupbox}\immediate\write\boxesfile{\number\dp\marxupbox}\box\marxupbox};
\path[-,line width=0.4pt,line cap=butt,line join=miter,dash pattern=](7pt,11.508pt)--(15.3398pt,11.508pt)--(15.3398pt,2.5258pt)--(7pt,2.5258pt)--cycle;
\path[-,line width=0.4pt,line cap=butt,line join=miter,dash pattern=](9pt,2.5258pt)--(11.5769pt,2.5258pt)--(11.5769pt,-2.5258pt)--(9pt,-2.5258pt)--cycle;
\node[anchor=north west,inner sep=0] at (9pt,2.5258pt){\savebox{\marxupbox}{{\({\scriptstyle l}\)}}\immediate\write\boxesfile{60}\immediate\write\boxesfile{\number\wd\marxupbox}\immediate\write\boxesfile{\number\ht\marxupbox}\immediate\write\boxesfile{\number\dp\marxupbox}\box\marxupbox};
\path[-,line width=0.4pt,line cap=butt,line join=miter,dash pattern=](7pt,4.5258pt)--(13.5769pt,4.5258pt)--(13.5769pt,-4.5258pt)--(7pt,-4.5258pt)--cycle;
\end{tikzpicture}}&{\begin{tikzpicture}\path[-,draw=black,line width=0.4pt,line cap=butt,line join=miter,dash pattern=](-7pt,5pt)--(0pt,5pt);
\path[-,draw=black,line width=0.4pt,line cap=butt,line join=miter,dash pattern=](0pt,5pt)--(7pt,5pt);
\path[-,draw=lightgray,line width=0.4pt,line cap=butt,line join=miter,dash pattern=](0pt,0pt)--(7pt,0pt);
\path[-,draw=black,line width=0.4pt,line cap=butt,line join=miter,dash pattern=](0pt,5pt)--(0pt,5pt);
\path[-,line width=0.4pt,line cap=butt,line join=miter,dash pattern=on 0.4pt off 1pt](0pt,5pt)--(0pt,5pt)--(0pt,0pt)--(0pt,0pt)--cycle;
\path[-,draw=black,line width=0.4pt,line cap=butt,line join=miter,dash pattern=on 0.4pt off 1pt](-3pt,8pt)--(3pt,8pt)--(3pt,-3pt)--(-3pt,-3pt)--cycle;
\path[-,fill=lightgray,line width=0.4pt,line cap=butt,line join=miter,dash pattern=](1pt,0pt)..controls(1pt,0.5523pt)and(0.5523pt,1pt)..(0pt,1pt)..controls(-0.5523pt,1pt)and(-1pt,0.5523pt)..(-1pt,0pt)..controls(-1pt,-0.5523pt)and(-0.5523pt,-1pt)..(0pt,-1pt)..controls(0.5523pt,-1pt)and(1pt,-0.5523pt)..(1pt,0pt)--cycle;
\path[-,line width=0.4pt,line cap=butt,line join=miter,dash pattern=](-11.6316pt,7.3396pt)--(-9pt,7.3396pt)--(-9pt,2.6604pt)--(-11.6316pt,2.6604pt)--cycle;
\node[anchor=north west,inner sep=0] at (-11.6316pt,7.3396pt){\savebox{\marxupbox}{{\({\scriptstyle i}\)}}\immediate\write\boxesfile{61}\immediate\write\boxesfile{\number\wd\marxupbox}\immediate\write\boxesfile{\number\ht\marxupbox}\immediate\write\boxesfile{\number\dp\marxupbox}\box\marxupbox};
\path[-,line width=0.4pt,line cap=butt,line join=miter,dash pattern=](-13.6316pt,9.3396pt)--(-7pt,9.3396pt)--(-7pt,0.6604pt)--(-13.6316pt,0.6604pt)--cycle;
\path[-,line width=0.4pt,line cap=butt,line join=miter,dash pattern=](9pt,8.0788pt)--(12.8617pt,8.0788pt)--(12.8617pt,1.9212pt)--(9pt,1.9212pt)--cycle;
\node[anchor=north west,inner sep=0] at (9pt,8.0788pt){\savebox{\marxupbox}{{\({\scriptstyle j}\)}}\immediate\write\boxesfile{62}\immediate\write\boxesfile{\number\wd\marxupbox}\immediate\write\boxesfile{\number\ht\marxupbox}\immediate\write\boxesfile{\number\dp\marxupbox}\box\marxupbox};
\path[-,line width=0.4pt,line cap=butt,line join=miter,dash pattern=](7pt,10.0788pt)--(14.8617pt,10.0788pt)--(14.8617pt,-0.0788pt)--(7pt,-0.0788pt)--cycle;
\end{tikzpicture}}\\\(\mathsf{t}\mskip 3.0mu{⊗}\mskip 3.0mu\mathsf{u}\)&\(α\)&\(σ\)&\(ρ\)\\\(t{_i}{^k}u{_j}{^l}\)&\(δ{_i}{^l}δ{_j}{^m}δ{_k}{^n}\)&\(δ{_i}{^l}δ{_j}{^k}\)&\(δ{_i}{^j}\)\\\(\mathsf{T}\mskip 3.0mu\mathnormal{⊗}\mskip 3.0mu\mathsf{T}\mskip 1.0mu\overset{z}{\leadsto }\mskip 1.0mu\mathsf{T}\mskip 3.0mu\mathnormal{⊗}\mskip 3.0mu\mathsf{T}\)&\(\allowbreak{}\mathnormal{(}\mskip 0.0mu\mathsf{T}\mskip 3.0mu\mathnormal{⊗}\mskip 3.0mu\mathsf{T}\mskip 0.0mu\mathnormal{)}\allowbreak{}\mskip 3.0mu\mathnormal{⊗}\mskip 3.0mu\mathsf{T}\mskip 1.0mu\overset{z}{\leadsto }\mskip 1.0mu\mathsf{T}\mskip 3.0mu\mathnormal{⊗}\mskip 3.0mu\allowbreak{}\mathnormal{(}\mskip 0.0mu\mathsf{T}\mskip 3.0mu\mathnormal{⊗}\mskip 3.0mu\mathsf{T}\mskip 0.0mu\mathnormal{)}\allowbreak{}\)&\(\mathsf{T}\mskip 3.0mu\mathnormal{⊗}\mskip 3.0mu\mathsf{T}\mskip 1.0mu\overset{z}{\leadsto }\mskip 1.0mu\mathsf{T}\mskip 3.0mu\mathnormal{⊗}\mskip 3.0mu\mathsf{T}\)&\(\mathsf{T}\mskip 1.0mu\overset{z}{\leadsto }\mskip 1.0mu\mathsf{T}\mskip 3.0mu\mathnormal{⊗}\mskip 3.0mu\mathbf{1}\)\end{tabular}\end{center}\caption{Diagram, categorical, and Einstein notations for
morphisms of symmetric monoidal categories. They are in general polymorphic, but we display them here as
acting on an atomic vector space \(T\), or the simplest allowable
combination thereof (see the last row in the figure for the monomorphic type of the
respective morphisms). The morphisms \(\bar{α}\) and \(\bar{ρ}\) are not shown, but are drawn symmetrically to \(α\) and
\(ρ\) respectively.}\label{63}\end{figure} 
As an example, take the morphism \(\mathsf{ex}\) =
\(\allowbreak{}\mathnormal{(}\mskip 0.0mu\mathsf{id}\mskip 3.0mu{⊗}\mskip 3.0muσ\mskip 0.0mu\mathnormal{)}\allowbreak{}\mskip 3.0mu\allowbreak{}\mathnormal{∘}\allowbreak{}\mskip 3.0mu\bar{α}\mskip 3.0mu\allowbreak{}\mathnormal{∘}\allowbreak{}\mskip 3.0mu\allowbreak{}\mathnormal{(}\mskip 0.0mu\mathsf{id}\mskip 3.0mu{⊗}\mskip 3.0muα\mskip 3.0mu\allowbreak{}\mathnormal{∘}\allowbreak{}\mskip 3.0mu\allowbreak{}\mathnormal{(}\mskip 0.0muσ\mskip 3.0mu{⊗}\mskip 3.0mu\mathsf{id}\mskip 0.0mu\mathnormal{)}\allowbreak{}\mskip 3.0mu\allowbreak{}\mathnormal{∘}\allowbreak{}\mskip 3.0mu\bar{α}\mskip 0.0mu\mathnormal{)}\allowbreak{}\mskip 3.0mu\allowbreak{}\mathnormal{∘}\allowbreak{}\mskip 3.0muα\mskip 3.0mu\allowbreak{}\mathnormal{∘}\allowbreak{}\mskip 3.0muα\). It is polymorphic, but has in particular type
\(\allowbreak{}\mathnormal{(}\mskip 0.0mu\allowbreak{}\mathnormal{(}\mskip 0.0mu\mathsf{T}\mskip 3.0mu\mathnormal{⊗}\mskip 3.0mu\mathsf{T}\mskip 0.0mu\mathnormal{)}\allowbreak{}\mskip 3.0mu\mathnormal{⊗}\mskip 3.0mu\mathsf{T}\mskip 0.0mu\mathnormal{)}\allowbreak{}\mskip 3.0mu\mathnormal{⊗}\mskip 3.0mu\mathsf{T}\mskip 1.0mu\overset{z}{\leadsto }\mskip 1.0mu\allowbreak{}\mathnormal{(}\mskip 0.0mu\mathsf{T}\mskip 3.0mu\mathnormal{⊗}\mskip 3.0mu\mathsf{T}\mskip 0.0mu\mathnormal{)}\allowbreak{}\mskip 3.0mu\mathnormal{⊗}\mskip 3.0mu\allowbreak{}\mathnormal{(}\mskip 0.0mu\mathsf{T}\mskip 3.0mu\mathnormal{⊗}\mskip 3.0mu\mathsf{T}\mskip 0.0mu\mathnormal{)}\allowbreak{}\). Its input and output orders are both
4, for a total order of (4,4) or 8. It is written \(δ{_i}{^m}δ{_j}{^p}δ{_k}{^n}δ{_l}{^o}\) in
Einstein notation, which makes more explicit the connection between inputs and output.
An even more explicit notation is its rendering as a diagram: 
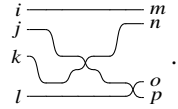
{\begin{tikzpicture}[baseline={([yshift=-0.8ex]current bounding box.center)}]\path[-,draw=black,line width=0.4pt,line cap=butt,line join=miter,dash pattern=](-31pt,32.4184pt)--(-30pt,32.4184pt);
\path[-,draw=black,line width=0.4pt,line cap=butt,line join=miter,dash pattern=](-31pt,25pt)--(-30pt,25pt);
\path[-,draw=black,line width=0.4pt,line cap=butt,line join=miter,dash pattern=](-31pt,15pt)--(-30pt,15pt);
\path[-,draw=black,line width=0.4pt,line cap=butt,line join=miter,dash pattern=](-31pt,0pt)--(-30pt,0pt);
\path[-,draw=black,line width=0.4pt,line cap=butt,line join=miter,dash pattern=](12pt,32.4184pt)--(13pt,32.4184pt);
\path[-,draw=black,line width=0.4pt,line cap=butt,line join=miter,dash pattern=](12pt,27.2028pt)--(13pt,27.2028pt);
\path[-,draw=black,line width=0.4pt,line cap=butt,line join=miter,dash pattern=](12pt,5.46pt)--(13pt,5.46pt);
\path[-,draw=black,line width=0.4pt,line cap=butt,line join=miter,dash pattern=](12pt,-0.46pt)--(13pt,-0.46pt);
\path[-,draw=black,line width=0.4pt,line cap=butt,line join=miter,dash pattern=](-24pt,32.4184pt)--(-24pt,32.4184pt);
\path[-,draw=black,line width=0.4pt,line cap=butt,line join=miter,dash pattern=](-24pt,25pt)--(-24pt,25pt);
\path[-,draw=black,line width=0.4pt,line cap=butt,line join=miter,dash pattern=](-24pt,5pt)--(-24pt,5pt);
\path[-,draw=black,line width=0.4pt,line cap=butt,line join=miter,dash pattern=](-24pt,0pt)--(-24pt,0pt);
\path[-,draw=black,line width=0.4pt,line cap=butt,line join=miter,dash pattern=](-18pt,32.4184pt)--(-18pt,32.4184pt);
\path[-,draw=black,line width=0.4pt,line cap=butt,line join=miter,dash pattern=](-18pt,15pt)--(-18pt,15pt);
\path[-,draw=black,line width=0.4pt,line cap=butt,line join=miter,dash pattern=](-18pt,5pt)--(-18pt,5pt);
\path[-,draw=black,line width=0.4pt,line cap=butt,line join=miter,dash pattern=](-18pt,0pt)--(-18pt,0pt);
\path[-,draw=black,line width=0.4pt,line cap=butt,line join=miter,dash pattern=](0pt,32.4184pt)--(0pt,32.4184pt);
\path[-,draw=black,line width=0.4pt,line cap=butt,line join=miter,dash pattern=](0pt,15pt)--(0pt,15pt);
\path[-,draw=black,line width=0.4pt,line cap=butt,line join=miter,dash pattern=](0pt,5pt)--(0pt,5pt);
\path[-,draw=black,line width=0.4pt,line cap=butt,line join=miter,dash pattern=](0pt,0pt)--(0pt,0pt);
\path[-,draw=black,line width=0.4pt,line cap=butt,line join=miter,dash pattern=](6pt,32.4184pt)--(6pt,32.4184pt);
\path[-,draw=black,line width=0.4pt,line cap=butt,line join=miter,dash pattern=](6pt,27.2028pt)--(6pt,27.2028pt);
\path[-,draw=black,line width=0.4pt,line cap=butt,line join=miter,dash pattern=](6pt,5pt)--(6pt,5pt);
\path[-,draw=black,line width=0.4pt,line cap=butt,line join=miter,dash pattern=](6pt,0pt)--(6pt,0pt);
\path[-,draw=black,line width=0.4pt,line cap=butt,line join=miter,dash pattern=](6pt,32.4184pt)--(12pt,32.4184pt);
\path[-,draw=black,line width=0.4pt,line cap=butt,line join=miter,dash pattern=](6pt,27.2028pt)--(12pt,27.2028pt);
\path[-,draw=black,line width=0.4pt,line cap=butt,line join=miter,dash pattern=](6pt,5pt)..controls(10pt,5pt)and(8pt,-0.46pt)..(12pt,-0.46pt);
\path[-,draw=black,line width=0.4pt,line cap=butt,line join=miter,dash pattern=](6pt,0pt)..controls(10pt,0pt)and(8pt,5.46pt)..(12pt,5.46pt);
\path[-,draw=black,line width=0.4pt,line cap=butt,line join=miter,dash pattern=](0pt,32.4184pt)--(6pt,32.4184pt);
\path[-,draw=black,line width=0.4pt,line cap=butt,line join=miter,dash pattern=](0pt,15pt)..controls(4pt,15pt)and(2pt,27.2028pt)..(6pt,27.2028pt);
\path[-,draw=black,line width=0.4pt,line cap=butt,line join=miter,dash pattern=](0pt,5pt)--(6pt,5pt);
\path[-,draw=black,line width=0.4pt,line cap=butt,line join=miter,dash pattern=](0pt,0pt)--(6pt,0pt);
\path[-,draw=black,line width=0.4pt,line cap=butt,line join=miter,dash pattern=](-18pt,32.4184pt)--(0pt,32.4184pt);
\path[-,draw=black,line width=0.4pt,line cap=butt,line join=miter,dash pattern=](-12pt,15pt)--(-12pt,15pt);
\path[-,draw=black,line width=0.4pt,line cap=butt,line join=miter,dash pattern=](-12pt,10pt)--(-12pt,10pt);
\path[-,draw=black,line width=0.4pt,line cap=butt,line join=miter,dash pattern=](-12pt,0pt)--(-12pt,0pt);
\path[-,draw=black,line width=0.4pt,line cap=butt,line join=miter,dash pattern=](-6pt,15pt)--(-6pt,15pt);
\path[-,draw=black,line width=0.4pt,line cap=butt,line join=miter,dash pattern=](-6pt,10pt)--(-6pt,10pt);
\path[-,draw=black,line width=0.4pt,line cap=butt,line join=miter,dash pattern=](-6pt,0pt)--(-6pt,0pt);
\path[-,draw=black,line width=0.4pt,line cap=butt,line join=miter,dash pattern=](-6pt,15pt)--(0pt,15pt);
\path[-,draw=black,line width=0.4pt,line cap=butt,line join=miter,dash pattern=](-6pt,10pt)..controls(-2pt,10pt)and(-4pt,5pt)..(0pt,5pt);
\path[-,draw=black,line width=0.4pt,line cap=butt,line join=miter,dash pattern=](-6pt,0pt)--(0pt,0pt);
\path[-,draw=black,line width=0.4pt,line cap=butt,line join=miter,dash pattern=](-12pt,15pt)..controls(-8pt,15pt)and(-10pt,10pt)..(-6pt,10pt);
\path[-,draw=black,line width=0.4pt,line cap=butt,line join=miter,dash pattern=](-12pt,10pt)..controls(-8pt,10pt)and(-10pt,15pt)..(-6pt,15pt);
\path[-,draw=black,line width=0.4pt,line cap=butt,line join=miter,dash pattern=](-12pt,0pt)--(-6pt,0pt);
\path[-,draw=black,line width=0.4pt,line cap=butt,line join=miter,dash pattern=](-18pt,15pt)--(-12pt,15pt);
\path[-,draw=black,line width=0.4pt,line cap=butt,line join=miter,dash pattern=](-18pt,5pt)..controls(-14pt,5pt)and(-16pt,10pt)..(-12pt,10pt);
\path[-,draw=black,line width=0.4pt,line cap=butt,line join=miter,dash pattern=](-18pt,0pt)--(-12pt,0pt);
\path[-,draw=black,line width=0.4pt,line cap=butt,line join=miter,dash pattern=](-24pt,32.4184pt)--(-18pt,32.4184pt);
\path[-,draw=black,line width=0.4pt,line cap=butt,line join=miter,dash pattern=](-24pt,25pt)..controls(-20pt,25pt)and(-22pt,15pt)..(-18pt,15pt);
\path[-,draw=black,line width=0.4pt,line cap=butt,line join=miter,dash pattern=](-24pt,5pt)--(-18pt,5pt);
\path[-,draw=black,line width=0.4pt,line cap=butt,line join=miter,dash pattern=](-24pt,0pt)--(-18pt,0pt);
\path[-,draw=black,line width=0.4pt,line cap=butt,line join=miter,dash pattern=](-30pt,32.4184pt)--(-24pt,32.4184pt);
\path[-,draw=black,line width=0.4pt,line cap=butt,line join=miter,dash pattern=](-30pt,25pt)--(-24pt,25pt);
\path[-,draw=black,line width=0.4pt,line cap=butt,line join=miter,dash pattern=](-30pt,15pt)..controls(-26pt,15pt)and(-28pt,5pt)..(-24pt,5pt);
\path[-,draw=black,line width=0.4pt,line cap=butt,line join=miter,dash pattern=](-30pt,0pt)--(-24pt,0pt);
\path[-,line width=0.4pt,line cap=butt,line join=miter,dash pattern=](-35.6316pt,34.758pt)--(-33pt,34.758pt)--(-33pt,30.0788pt)--(-35.6316pt,30.0788pt)--cycle;
\node[anchor=north west,inner sep=0] at (-35.6316pt,34.758pt){\savebox{\marxupbox}{{\({\scriptstyle i}\)}}\immediate\write\boxesfile{64}\immediate\write\boxesfile{\number\wd\marxupbox}\immediate\write\boxesfile{\number\ht\marxupbox}\immediate\write\boxesfile{\number\dp\marxupbox}\box\marxupbox};
\path[-,line width=0.4pt,line cap=butt,line join=miter,dash pattern=](-37.6316pt,36.758pt)--(-31pt,36.758pt)--(-31pt,28.0788pt)--(-37.6316pt,28.0788pt)--cycle;
\path[-,line width=0.4pt,line cap=butt,line join=miter,dash pattern=](-36.8617pt,28.0788pt)--(-33pt,28.0788pt)--(-33pt,21.9212pt)--(-36.8617pt,21.9212pt)--cycle;
\node[anchor=north west,inner sep=0] at (-36.8617pt,28.0788pt){\savebox{\marxupbox}{{\({\scriptstyle j}\)}}\immediate\write\boxesfile{65}\immediate\write\boxesfile{\number\wd\marxupbox}\immediate\write\boxesfile{\number\ht\marxupbox}\immediate\write\boxesfile{\number\dp\marxupbox}\box\marxupbox};
\path[-,line width=0.4pt,line cap=butt,line join=miter,dash pattern=](-38.8617pt,30.0788pt)--(-31pt,30.0788pt)--(-31pt,19.9212pt)--(-38.8617pt,19.9212pt)--cycle;
\path[-,line width=0.4pt,line cap=butt,line join=miter,dash pattern=](-37.3398pt,17.4911pt)--(-33pt,17.4911pt)--(-33pt,12.5089pt)--(-37.3398pt,12.5089pt)--cycle;
\node[anchor=north west,inner sep=0] at (-37.3398pt,17.4911pt){\savebox{\marxupbox}{{\({\scriptstyle k}\)}}\immediate\write\boxesfile{66}\immediate\write\boxesfile{\number\wd\marxupbox}\immediate\write\boxesfile{\number\ht\marxupbox}\immediate\write\boxesfile{\number\dp\marxupbox}\box\marxupbox};
\path[-,line width=0.4pt,line cap=butt,line join=miter,dash pattern=](-39.3398pt,19.4911pt)--(-31pt,19.4911pt)--(-31pt,10.5089pt)--(-39.3398pt,10.5089pt)--cycle;
\path[-,line width=0.4pt,line cap=butt,line join=miter,dash pattern=](-35.5769pt,2.5258pt)--(-33pt,2.5258pt)--(-33pt,-2.5258pt)--(-35.5769pt,-2.5258pt)--cycle;
\node[anchor=north west,inner sep=0] at (-35.5769pt,2.5258pt){\savebox{\marxupbox}{{\({\scriptstyle l}\)}}\immediate\write\boxesfile{67}\immediate\write\boxesfile{\number\wd\marxupbox}\immediate\write\boxesfile{\number\ht\marxupbox}\immediate\write\boxesfile{\number\dp\marxupbox}\box\marxupbox};
\path[-,line width=0.4pt,line cap=butt,line join=miter,dash pattern=](-37.5769pt,4.5258pt)--(-31pt,4.5258pt)--(-31pt,-4.5258pt)--(-37.5769pt,-4.5258pt)--cycle;
\path[-,line width=0.4pt,line cap=butt,line join=miter,dash pattern=](15pt,34.0262pt)--(21.1685pt,34.0262pt)--(21.1685pt,30.8106pt)--(15pt,30.8106pt)--cycle;
\node[anchor=north west,inner sep=0] at (15pt,34.0262pt){\savebox{\marxupbox}{{\({\scriptstyle m}\)}}\immediate\write\boxesfile{68}\immediate\write\boxesfile{\number\wd\marxupbox}\immediate\write\boxesfile{\number\ht\marxupbox}\immediate\write\boxesfile{\number\dp\marxupbox}\box\marxupbox};
\path[-,line width=0.4pt,line cap=butt,line join=miter,dash pattern=](13pt,36.0262pt)--(23.1685pt,36.0262pt)--(23.1685pt,28.8106pt)--(13pt,28.8106pt)--cycle;
\path[-,line width=0.4pt,line cap=butt,line join=miter,dash pattern=](15pt,28.8106pt)--(19.3581pt,28.8106pt)--(19.3581pt,25.595pt)--(15pt,25.595pt)--cycle;
\node[anchor=north west,inner sep=0] at (15pt,28.8106pt){\savebox{\marxupbox}{{\({\scriptstyle n}\)}}\immediate\write\boxesfile{69}\immediate\write\boxesfile{\number\wd\marxupbox}\immediate\write\boxesfile{\number\ht\marxupbox}\immediate\write\boxesfile{\number\dp\marxupbox}\box\marxupbox};
\path[-,line width=0.4pt,line cap=butt,line join=miter,dash pattern=](13pt,30.8106pt)--(21.3581pt,30.8106pt)--(21.3581pt,23.595pt)--(13pt,23.595pt)--cycle;
\path[-,line width=0.4pt,line cap=butt,line join=miter,dash pattern=](15pt,7.0678pt)--(19.4165pt,7.0678pt)--(19.4165pt,3.8522pt)--(15pt,3.8522pt)--cycle;
\node[anchor=north west,inner sep=0] at (15pt,7.0678pt){\savebox{\marxupbox}{{\({\scriptstyle o}\)}}\immediate\write\boxesfile{70}\immediate\write\boxesfile{\number\wd\marxupbox}\immediate\write\boxesfile{\number\ht\marxupbox}\immediate\write\boxesfile{\number\dp\marxupbox}\box\marxupbox};
\path[-,line width=0.4pt,line cap=butt,line join=miter,dash pattern=](13pt,9.0678pt)--(21.4165pt,9.0678pt)--(21.4165pt,1.8522pt)--(13pt,1.8522pt)--cycle;
\path[-,line width=0.4pt,line cap=butt,line join=miter,dash pattern=](15pt,1.8522pt)--(20.2231pt,1.8522pt)--(20.2231pt,-2.7723pt)--(15pt,-2.7723pt)--cycle;
\node[anchor=north west,inner sep=0] at (15pt,1.8522pt){\savebox{\marxupbox}{{\({\scriptstyle p}\)}}\immediate\write\boxesfile{71}\immediate\write\boxesfile{\number\wd\marxupbox}\immediate\write\boxesfile{\number\ht\marxupbox}\immediate\write\boxesfile{\number\dp\marxupbox}\box\marxupbox};
\path[-,line width=0.4pt,line cap=butt,line join=miter,dash pattern=](13pt,3.8522pt)--(22.2231pt,3.8522pt)--(22.2231pt,-4.7723pt)--(13pt,-4.7723pt)--cycle;
\end{tikzpicture}}.

This diagram notation can be generalised to
all morphisms in a {\sc{}smc}, and is known as \emph{string diagrams}. It is a
two-dimensional instance of the abstract categorical structure.  It is
also fully abstract, in the sense that every diagram can be mapped to
a unique morphism in the underlying {\sc{}smc}. \cref{63} shows several of the
atomic diagrams which make up {\sc{}smc}s.  The guide for this notation is
that each morphism is represented by a network of wires. Wires are
drawn in a way that makes it clear which inputs are connected to which
outputs. Because unit objects can be
added and dropped at will (using \(ρ\) and \(\bar{ρ}\)),
under some conventions the corresponding wires are not drawn at
all. Here we choose to draw them as gray lines.

The diagram notation is defined in such a way that morphisms that are
equal under the category laws have topologically equivalent diagram
representations \citep{selinger2011survey}.  That is, if we can deform one diagram to another without cutting wires, then they are equivalent.
We can illustrate this kind of topological reasoning with the following
simple example.  Assuming an abstract tensor \(\mathsf{u}\mskip 3.0mu\mathnormal{:}\mskip 3.0mu\mathsf{T}\mskip 1.0mu\overset{z}{\leadsto }\mskip 1.0mu\mathsf{T}\),
one can check that \(σ\mskip 3.0mu\allowbreak{}\mathnormal{∘}\allowbreak{}\mskip 3.0mu\allowbreak{}\mathnormal{(}\mskip 0.0mu\mathsf{id}\mskip 3.0mu{⊗}\mskip 3.0mu\mathsf{u}\mskip 0.0mu\mathnormal{)}\allowbreak{}\mskip 3.0mu\allowbreak{}\mathnormal{∘}\allowbreak{}\mskip 3.0muσ\mskip 3.0mu\allowbreak{}\mathnormal{∘}\allowbreak{}\mskip 3.0mu\allowbreak{}\mathnormal{(}\mskip 0.0mu\mathsf{id}\mskip 3.0mu{⊗}\mskip 3.0mu\mathsf{u}\mskip 0.0mu\mathnormal{)}\allowbreak{}\) is equivalent to
\(\mathsf{u}\mskip 3.0mu{⊗}\mskip 3.0mu\mathsf{u}\) by applying a number of algebraic
laws, but this is an error-prone process. If we first convert the morphisms to diagram form, we need to check
{\begin{tikzpicture}[baseline={([yshift=-0.8ex]current bounding box.center)}]\path[-,draw=black,line width=0.4pt,line cap=butt,line join=miter,dash pattern=](-25.31pt,7.6159pt)--(-18.31pt,7.6159pt);
\path[-,draw=black,line width=0.4pt,line cap=butt,line join=miter,dash pattern=](-25.31pt,-0.599pt)--(-18.31pt,-0.599pt);
\path[-,draw=black,line width=0.4pt,line cap=butt,line join=miter,dash pattern=](19.31pt,7.0169pt)--(20.31pt,7.0169pt);
\path[-,draw=black,line width=0.4pt,line cap=butt,line join=miter,dash pattern=](19.31pt,0pt)--(20.31pt,0pt);
\path[-,draw=black,line width=0.4pt,line cap=butt,line join=miter,dash pattern=](0pt,7.6159pt)--(1pt,7.6159pt);
\path[-,draw=black,line width=0.4pt,line cap=butt,line join=miter,dash pattern=](0pt,-0.599pt)--(1pt,-0.599pt);
\path[-,draw=black,line width=0.4pt,line cap=butt,line join=miter,dash pattern=](12.31pt,7.6159pt)--(13.31pt,7.6159pt);
\path[-,draw=black,line width=0.4pt,line cap=butt,line join=miter,dash pattern=](12.31pt,-0.599pt)--(13.31pt,-0.599pt);
\path[-,draw=black,line width=0.4pt,line cap=butt,line join=miter,dash pattern=](13.31pt,7.6159pt)..controls(17.31pt,7.6159pt)and(15.31pt,0pt)..(19.31pt,0pt);
\path[-,draw=black,line width=0.4pt,line cap=butt,line join=miter,dash pattern=](13.31pt,-0.599pt)..controls(17.31pt,-0.599pt)and(15.31pt,7.0169pt)..(19.31pt,7.0169pt);
\path[-,draw=black,line width=0.4pt,line cap=butt,line join=miter,dash pattern=](1pt,7.6159pt)--(12.31pt,7.6159pt);
\path[-,line width=0.4pt,line cap=butt,line join=miter,dash pattern=](4pt,1.6159pt)--(9.31pt,1.6159pt)--(9.31pt,-2.814pt)--(4pt,-2.814pt)--cycle;
\node[anchor=north west,inner sep=0] at (4pt,1.6159pt){\savebox{\marxupbox}{{\(u\)}}\immediate\write\boxesfile{72}\immediate\write\boxesfile{\number\wd\marxupbox}\immediate\write\boxesfile{\number\ht\marxupbox}\immediate\write\boxesfile{\number\dp\marxupbox}\box\marxupbox};
\path[-,draw=black,line width=0.4pt,line cap=butt,line join=miter,dash pattern=](1pt,2.1159pt)..controls(1pt,3.4967pt)and(2.1193pt,4.6159pt)..(3.5pt,4.6159pt)--(9.81pt,4.6159pt)..controls(11.1907pt,4.6159pt)and(12.31pt,3.4967pt)..(12.31pt,2.1159pt)--(12.31pt,-3.314pt)..controls(12.31pt,-4.6947pt)and(11.1907pt,-5.814pt)..(9.81pt,-5.814pt)--(3.5pt,-5.814pt)..controls(2.1193pt,-5.814pt)and(1pt,-4.6947pt)..(1pt,-3.314pt)--cycle;
\path[-,draw=black,line width=0.4pt,line cap=butt,line join=miter,dash pattern=](12.31pt,-0.599pt)--(12.31pt,-0.599pt);
\path[-,draw=black,line width=0.4pt,line cap=butt,line join=miter,dash pattern=](1pt,-0.599pt)--(1pt,-0.599pt);
\path[-,draw=black,line width=0.4pt,line cap=butt,line join=miter,dash pattern=](-7pt,7.6159pt)--(-6pt,7.6159pt);
\path[-,draw=black,line width=0.4pt,line cap=butt,line join=miter,dash pattern=](-7pt,-0.599pt)--(-6pt,-0.599pt);
\path[-,draw=black,line width=0.4pt,line cap=butt,line join=miter,dash pattern=](-6pt,7.6159pt)..controls(-2pt,7.6159pt)and(-4pt,-0.599pt)..(0pt,-0.599pt);
\path[-,draw=black,line width=0.4pt,line cap=butt,line join=miter,dash pattern=](-6pt,-0.599pt)..controls(-2pt,-0.599pt)and(-4pt,7.6159pt)..(0pt,7.6159pt);
\path[-,draw=black,line width=0.4pt,line cap=butt,line join=miter,dash pattern=](-18.31pt,7.6159pt)--(-7pt,7.6159pt);
\path[-,line width=0.4pt,line cap=butt,line join=miter,dash pattern=](-15.31pt,1.6159pt)--(-10pt,1.6159pt)--(-10pt,-2.814pt)--(-15.31pt,-2.814pt)--cycle;
\node[anchor=north west,inner sep=0] at (-15.31pt,1.6159pt){\savebox{\marxupbox}{{\(u\)}}\immediate\write\boxesfile{73}\immediate\write\boxesfile{\number\wd\marxupbox}\immediate\write\boxesfile{\number\ht\marxupbox}\immediate\write\boxesfile{\number\dp\marxupbox}\box\marxupbox};
\path[-,draw=black,line width=0.4pt,line cap=butt,line join=miter,dash pattern=](-18.31pt,2.1159pt)..controls(-18.31pt,3.4967pt)and(-17.1907pt,4.6159pt)..(-15.81pt,4.6159pt)--(-9.5pt,4.6159pt)..controls(-8.1193pt,4.6159pt)and(-7pt,3.4967pt)..(-7pt,2.1159pt)--(-7pt,-3.314pt)..controls(-7pt,-4.6947pt)and(-8.1193pt,-5.814pt)..(-9.5pt,-5.814pt)--(-15.81pt,-5.814pt)..controls(-17.1907pt,-5.814pt)and(-18.31pt,-4.6947pt)..(-18.31pt,-3.314pt)--cycle;
\path[-,draw=black,line width=0.4pt,line cap=butt,line join=miter,dash pattern=](-7pt,-0.599pt)--(-7pt,-0.599pt);
\path[-,draw=black,line width=0.4pt,line cap=butt,line join=miter,dash pattern=](-18.31pt,-0.599pt)--(-18.31pt,-0.599pt);
\path[-,line width=0.4pt,line cap=butt,line join=miter,dash pattern=](-29.9416pt,9.9556pt)--(-27.31pt,9.9556pt)--(-27.31pt,5.2763pt)--(-29.9416pt,5.2763pt)--cycle;
\node[anchor=north west,inner sep=0] at (-29.9416pt,9.9556pt){\savebox{\marxupbox}{{\({\scriptstyle i}\)}}\immediate\write\boxesfile{74}\immediate\write\boxesfile{\number\wd\marxupbox}\immediate\write\boxesfile{\number\ht\marxupbox}\immediate\write\boxesfile{\number\dp\marxupbox}\box\marxupbox};
\path[-,line width=0.4pt,line cap=butt,line join=miter,dash pattern=](-31.9416pt,11.9556pt)--(-25.31pt,11.9556pt)--(-25.31pt,3.2763pt)--(-31.9416pt,3.2763pt)--cycle;
\path[-,line width=0.4pt,line cap=butt,line join=miter,dash pattern=](-31.1717pt,2.4797pt)--(-27.31pt,2.4797pt)--(-27.31pt,-3.6778pt)--(-31.1717pt,-3.6778pt)--cycle;
\node[anchor=north west,inner sep=0] at (-31.1717pt,2.4797pt){\savebox{\marxupbox}{{\({\scriptstyle j}\)}}\immediate\write\boxesfile{75}\immediate\write\boxesfile{\number\wd\marxupbox}\immediate\write\boxesfile{\number\ht\marxupbox}\immediate\write\boxesfile{\number\dp\marxupbox}\box\marxupbox};
\path[-,line width=0.4pt,line cap=butt,line join=miter,dash pattern=](-33.1717pt,4.4797pt)--(-25.31pt,4.4797pt)--(-25.31pt,-5.6778pt)--(-33.1717pt,-5.6778pt)--cycle;
\path[-,line width=0.4pt,line cap=butt,line join=miter,dash pattern=](22.31pt,9.508pt)--(26.6498pt,9.508pt)--(26.6498pt,4.5258pt)--(22.31pt,4.5258pt)--cycle;
\node[anchor=north west,inner sep=0] at (22.31pt,9.508pt){\savebox{\marxupbox}{{\({\scriptstyle k}\)}}\immediate\write\boxesfile{76}\immediate\write\boxesfile{\number\wd\marxupbox}\immediate\write\boxesfile{\number\ht\marxupbox}\immediate\write\boxesfile{\number\dp\marxupbox}\box\marxupbox};
\path[-,line width=0.4pt,line cap=butt,line join=miter,dash pattern=](20.31pt,11.508pt)--(28.6498pt,11.508pt)--(28.6498pt,2.5258pt)--(20.31pt,2.5258pt)--cycle;
\path[-,line width=0.4pt,line cap=butt,line join=miter,dash pattern=](22.31pt,2.5258pt)--(24.8869pt,2.5258pt)--(24.8869pt,-2.5258pt)--(22.31pt,-2.5258pt)--cycle;
\node[anchor=north west,inner sep=0] at (22.31pt,2.5258pt){\savebox{\marxupbox}{{\({\scriptstyle l}\)}}\immediate\write\boxesfile{77}\immediate\write\boxesfile{\number\wd\marxupbox}\immediate\write\boxesfile{\number\ht\marxupbox}\immediate\write\boxesfile{\number\dp\marxupbox}\box\marxupbox};
\path[-,line width=0.4pt,line cap=butt,line join=miter,dash pattern=](20.31pt,4.5258pt)--(26.8869pt,4.5258pt)--(26.8869pt,-4.5258pt)--(20.31pt,-4.5258pt)--cycle;
\end{tikzpicture}} =
{\begin{tikzpicture}[baseline={([yshift=-0.8ex]current bounding box.center)}]\path[-,draw=black,line width=0.4pt,line cap=butt,line join=miter,dash pattern=](-18.31pt,5.215pt)--(-11.31pt,5.215pt);
\path[-,draw=black,line width=0.4pt,line cap=butt,line join=miter,dash pattern=](-18.31pt,-11.215pt)--(-11.31pt,-11.215pt);
\path[-,draw=black,line width=0.4pt,line cap=butt,line join=miter,dash pattern=](0pt,5.215pt)--(7pt,5.215pt);
\path[-,draw=black,line width=0.4pt,line cap=butt,line join=miter,dash pattern=](0pt,-11.215pt)--(7pt,-11.215pt);
\path[-,line width=0.4pt,line cap=butt,line join=miter,dash pattern=](-8.31pt,7.43pt)--(-3pt,7.43pt)--(-3pt,3pt)--(-8.31pt,3pt)--cycle;
\node[anchor=north west,inner sep=0] at (-8.31pt,7.43pt){\savebox{\marxupbox}{{\(u\)}}\immediate\write\boxesfile{78}\immediate\write\boxesfile{\number\wd\marxupbox}\immediate\write\boxesfile{\number\ht\marxupbox}\immediate\write\boxesfile{\number\dp\marxupbox}\box\marxupbox};
\path[-,draw=black,line width=0.4pt,line cap=butt,line join=miter,dash pattern=](-11.31pt,7.93pt)..controls(-11.31pt,9.3107pt)and(-10.1907pt,10.43pt)..(-8.81pt,10.43pt)--(-2.5pt,10.43pt)..controls(-1.1193pt,10.43pt)and(0pt,9.3107pt)..(0pt,7.93pt)--(0pt,2.5pt)..controls(0pt,1.1193pt)and(-1.1193pt,0pt)..(-2.5pt,0pt)--(-8.81pt,0pt)..controls(-10.1907pt,0pt)and(-11.31pt,1.1193pt)..(-11.31pt,2.5pt)--cycle;
\path[-,draw=black,line width=0.4pt,line cap=butt,line join=miter,dash pattern=](0pt,5.215pt)--(0pt,5.215pt);
\path[-,draw=black,line width=0.4pt,line cap=butt,line join=miter,dash pattern=](-11.31pt,5.215pt)--(-11.31pt,5.215pt);
\path[-,line width=0.4pt,line cap=butt,line join=miter,dash pattern=](-8.31pt,-9pt)--(-3pt,-9pt)--(-3pt,-13.43pt)--(-8.31pt,-13.43pt)--cycle;
\node[anchor=north west,inner sep=0] at (-8.31pt,-9pt){\savebox{\marxupbox}{{\(u\)}}\immediate\write\boxesfile{79}\immediate\write\boxesfile{\number\wd\marxupbox}\immediate\write\boxesfile{\number\ht\marxupbox}\immediate\write\boxesfile{\number\dp\marxupbox}\box\marxupbox};
\path[-,draw=black,line width=0.4pt,line cap=butt,line join=miter,dash pattern=](-11.31pt,-8.5pt)..controls(-11.31pt,-7.1193pt)and(-10.1907pt,-6pt)..(-8.81pt,-6pt)--(-2.5pt,-6pt)..controls(-1.1193pt,-6pt)and(0pt,-7.1193pt)..(0pt,-8.5pt)--(0pt,-13.93pt)..controls(0pt,-15.3107pt)and(-1.1193pt,-16.43pt)..(-2.5pt,-16.43pt)--(-8.81pt,-16.43pt)..controls(-10.1907pt,-16.43pt)and(-11.31pt,-15.3107pt)..(-11.31pt,-13.93pt)--cycle;
\path[-,draw=black,line width=0.4pt,line cap=butt,line join=miter,dash pattern=](0pt,-11.215pt)--(0pt,-11.215pt);
\path[-,draw=black,line width=0.4pt,line cap=butt,line join=miter,dash pattern=](-11.31pt,-11.215pt)--(-11.31pt,-11.215pt);
\path[-,line width=0.4pt,line cap=butt,line join=miter,dash pattern=](-22.9416pt,7.5546pt)--(-20.31pt,7.5546pt)--(-20.31pt,2.8754pt)--(-22.9416pt,2.8754pt)--cycle;
\node[anchor=north west,inner sep=0] at (-22.9416pt,7.5546pt){\savebox{\marxupbox}{{\({\scriptstyle i}\)}}\immediate\write\boxesfile{80}\immediate\write\boxesfile{\number\wd\marxupbox}\immediate\write\boxesfile{\number\ht\marxupbox}\immediate\write\boxesfile{\number\dp\marxupbox}\box\marxupbox};
\path[-,line width=0.4pt,line cap=butt,line join=miter,dash pattern=](-24.9416pt,9.5546pt)--(-18.31pt,9.5546pt)--(-18.31pt,0.8754pt)--(-24.9416pt,0.8754pt)--cycle;
\path[-,line width=0.4pt,line cap=butt,line join=miter,dash pattern=](-24.1717pt,-8.1362pt)--(-20.31pt,-8.1362pt)--(-20.31pt,-14.2938pt)--(-24.1717pt,-14.2938pt)--cycle;
\node[anchor=north west,inner sep=0] at (-24.1717pt,-8.1362pt){\savebox{\marxupbox}{{\({\scriptstyle j}\)}}\immediate\write\boxesfile{81}\immediate\write\boxesfile{\number\wd\marxupbox}\immediate\write\boxesfile{\number\ht\marxupbox}\immediate\write\boxesfile{\number\dp\marxupbox}\box\marxupbox};
\path[-,line width=0.4pt,line cap=butt,line join=miter,dash pattern=](-26.1717pt,-6.1362pt)--(-18.31pt,-6.1362pt)--(-18.31pt,-16.2938pt)--(-26.1717pt,-16.2938pt)--cycle;
\path[-,line width=0.4pt,line cap=butt,line join=miter,dash pattern=](9pt,7.7061pt)--(13.3398pt,7.7061pt)--(13.3398pt,2.7239pt)--(9pt,2.7239pt)--cycle;
\node[anchor=north west,inner sep=0] at (9pt,7.7061pt){\savebox{\marxupbox}{{\({\scriptstyle k}\)}}\immediate\write\boxesfile{82}\immediate\write\boxesfile{\number\wd\marxupbox}\immediate\write\boxesfile{\number\ht\marxupbox}\immediate\write\boxesfile{\number\dp\marxupbox}\box\marxupbox};
\path[-,line width=0.4pt,line cap=butt,line join=miter,dash pattern=](7pt,9.7061pt)--(15.3398pt,9.7061pt)--(15.3398pt,0.7239pt)--(7pt,0.7239pt)--cycle;
\path[-,line width=0.4pt,line cap=butt,line join=miter,dash pattern=](9pt,-8.6892pt)--(11.5769pt,-8.6892pt)--(11.5769pt,-13.7408pt)--(9pt,-13.7408pt)--cycle;
\node[anchor=north west,inner sep=0] at (9pt,-8.6892pt){\savebox{\marxupbox}{{\({\scriptstyle l}\)}}\immediate\write\boxesfile{83}\immediate\write\boxesfile{\number\wd\marxupbox}\immediate\write\boxesfile{\number\ht\marxupbox}\immediate\write\boxesfile{\number\dp\marxupbox}\box\marxupbox};
\path[-,line width=0.4pt,line cap=butt,line join=miter,dash pattern=](7pt,-6.6892pt)--(13.5769pt,-6.6892pt)--(13.5769pt,-15.7408pt)--(7pt,-15.7408pt)--cycle;
\end{tikzpicture}}, which is a matter of
repositioning the second box.\footnote{The property that
equivalent morphisms have the same representation is also satisfied by
the Einstein notation (up to α-renaming). Indeed, a connection in the
diagrammatic notation is represented by marking the two ends of the
connection by the same index name (in sub- or super-script position).
Besides, rearranging the inputs or outputs of a morphism is
implemented by a renaming of indices.
For instance, the right identity
law, \(\mathsf{t}\mskip 3.0mu\allowbreak{}\mathnormal{∘}\allowbreak{}\mskip 3.0mu\mathsf{id}\) \(\mathnormal{=}\mskip 3.0mu\mathsf{t}\) becomes \(δᵢᵏ tₖʲ =
tᵢʲ\). (That is, multiplication by \(\mathsf{δ}\) can act like a variable substitution
operator for indices.) The associativity law \(\mathsf{t}\mskip 3.0mu\allowbreak{}\mathnormal{∘}\allowbreak{}\mskip 3.0mu\allowbreak{}\mathnormal{(}\mskip 0.0mu\mathsf{u}\mskip 3.0mu\allowbreak{}\mathnormal{∘}\allowbreak{}\mskip 3.0mu\mathsf{v}\mskip 0.0mu\mathnormal{)}\allowbreak{}\mskip 3.0mu\mathnormal{=}\mskip 3.0mu\allowbreak{}\mathnormal{(}\mskip 0.0mu\mathsf{t}\mskip 3.0mu\allowbreak{}\mathnormal{∘}\allowbreak{}\mskip 3.0mu\mathsf{u}\mskip 0.0mu\mathnormal{)}\allowbreak{}\mskip 3.0mu\allowbreak{}\mathnormal{∘}\allowbreak{}\mskip 3.0mu\mathsf{v}\) becomes \((vᵢˡ uₗᵏ) tₖʲ = vᵢᵏ (uₖˡ tₗʲ)\): a combination
of α-equivalence and associativity of multiplication.
\label{84}} 

At this point, a reader familiar with programming languages might be
tempted to assume that \(\mathsf{V}\mskip 3.0mu\mathnormal{⊗}\mskip 3.0mu\mathsf{W}\) is like a pair of \(\mathsf{V}\) and
\(\mathsf{W}\). That is, that tensors would not only form a category, but even a Cartesian category. This is not
the case: tensors are not equipped with projections nor
duplications. This observation justifies the fact
that contraction in Einstein notation must involve exactly two
indices.  Indeed, contraction corresponds to connecting loose wires in
the diagram notation, and because we do not have a Cartesian category,
only two loose wires can be connected (to make a new continuous wire).
 
\paragraph*{Addition and scaling}\hspace{1.0ex}\label{85} 
As we saw, tensors of the same type (same domain and codomain) form
themselves a vector space, and as such can be scaled and added together.
The corresponding categorical structure is called an additive
category. Thus every tensor category \(z\) will satisfy the
\(\mathsf{Additive}\) constraint:
\begin{list}{}{\setlength\leftmargin{1.0em}}\item\relax
\ensuremath{\begin{parray}\column{B}{@{}>{}l<{}@{}}\column[0em]{1}{@{}>{}l<{}@{}}\column{2}{@{}>{}l<{}@{}}\column{3}{@{}>{}l<{}@{}}\column{E}{@{}>{}l<{}@{}}%
\>[1]{\mathbf{type}\mskip 3.0mu\mathsf{Additive}\mskip 3.0muz\mskip 3.0mu}\>[2]{\mathnormal{=}\mskip 3.0mu∀\mskip 3.0mu\mathsf{a}\mskip 3.0mu\mathsf{b}\mskip 1.0mu.\mskip 3.0mu\mathsf{VectorSpace}\mskip 3.0mu\allowbreak{}\mathnormal{(}\mskip 0.0mu}\>[3]{\mathsf{a}\mskip 1.0mu\overset{z}{\leadsto }\mskip 1.0mu\mathsf{b}\mskip 0.0mu\mathnormal{)}\allowbreak{}}\<[E]\end{parray}}\end{list} Recalling the definition of \(\mathsf{VectorSpace}\) from \cref{21},
\(\mathsf{Additive}\) implies that we have the following two operations
for every \(\mathsf{a}\) and \(\mathsf{b}\):
\begin{list}{}{\setlength\leftmargin{1.0em}}\item\relax
\ensuremath{\begin{parray}\column{B}{@{}>{}l<{}@{}}\column[0em]{1}{@{}>{}l<{}@{}}\column{2}{@{}>{}l<{}@{}}\column{3}{@{}>{}l<{}@{}}\column{E}{@{}>{}l<{}@{}}%
\>[1]{\allowbreak{}\mathnormal{(}\mskip 0.0mu\mathnormal{+}\mskip 0.0mu\mathnormal{)}\allowbreak{}\mskip 3.0mu}\>[2]{\mathnormal{::}\mskip 3.0mu}\>[3]{\allowbreak{}\mathnormal{(}\mskip 0.0mu\mathsf{a}\mskip 1.0mu\overset{z}{\leadsto }\mskip 1.0mu\mathsf{b}\mskip 0.0mu\mathnormal{)}\allowbreak{}\mskip 3.0mu\mathnormal{\rightarrow }\mskip 3.0mu\allowbreak{}\mathnormal{(}\mskip 0.0mu\mathsf{a}\mskip 1.0mu\overset{z}{\leadsto }\mskip 1.0mu\mathsf{b}\mskip 0.0mu\mathnormal{)}\allowbreak{}\mskip 3.0mu\mathnormal{\rightarrow }\mskip 3.0mu\allowbreak{}\mathnormal{(}\mskip 0.0mu\mathsf{a}\mskip 1.0mu\overset{z}{\leadsto }\mskip 1.0mu\mathsf{b}\mskip 0.0mu\mathnormal{)}\allowbreak{}}\<[E]\\
\>[1]{\allowbreak{}\mathnormal{(}\mskip 0.0mu\smalltriangleleft \mskip 0.0mu\mathnormal{)}\allowbreak{}\mskip 3.0mu}\>[2]{\mathnormal{::}\mskip 3.0mu}\>[3]{\mathsf{S}\mskip 3.0mu\mathnormal{\rightarrow }\mskip 3.0mu\allowbreak{}\mathnormal{(}\mskip 0.0mu\mathsf{a}\mskip 1.0mu\overset{z}{\leadsto }\mskip 1.0mu\mathsf{b}\mskip 0.0mu\mathnormal{)}\allowbreak{}\mskip 3.0mu\mathnormal{\rightarrow }\mskip 3.0mu\allowbreak{}\mathnormal{(}\mskip 0.0mu\mathsf{a}\mskip 1.0mu\overset{z}{\leadsto }\mskip 1.0mu\mathsf{b}\mskip 0.0mu\mathnormal{)}\allowbreak{}}\<[E]\end{parray}}\end{list} An additive category requires that composition \(\allowbreak{}\mathnormal{(}\mskip 0.0mu\allowbreak{}\mathnormal{∘}\allowbreak{}\mskip 0.0mu\mathnormal{)}\allowbreak{}\) and tensor
products \(\allowbreak{}\mathnormal{(}\mskip 0.0mu{⊗}\mskip 0.0mu\mathnormal{)}\allowbreak{}\) are bilinear. In full:

\begin{list}{}{\setlength\leftmargin{1.0em}}\item\relax
\begin{tabular}{c@{\quad\quad}c}\ensuremath{\begin{parray}\column{B}{@{}>{}l<{}@{}}\column[0em]{1}{@{}>{}l<{}@{}}\column{2}{@{}>{}l<{}@{}}\column{E}{@{}>{}l<{}@{}}%
\>[1]{\mathsf{t}\mskip 3.0mu\allowbreak{}\mathnormal{∘}\allowbreak{}\mskip 3.0mu\allowbreak{}\mathnormal{(}\mskip 0.0mu\mathsf{u}\mskip 3.0mu\mathnormal{+}\mskip 3.0mu\mathsf{v}\mskip 0.0mu\mathnormal{)}\allowbreak{}\mskip 3.0mu}\>[2]{\mathnormal{=}\mskip 3.0mu\allowbreak{}\mathnormal{(}\mskip 0.0mu\mathsf{t}\mskip 3.0mu\allowbreak{}\mathnormal{∘}\allowbreak{}\mskip 3.0mu\mathsf{u}\mskip 0.0mu\mathnormal{)}\allowbreak{}\mskip 3.0mu\mathnormal{+}\mskip 3.0mu\allowbreak{}\mathnormal{(}\mskip 0.0mu\mathsf{t}\mskip 3.0mu\allowbreak{}\mathnormal{∘}\allowbreak{}\mskip 3.0mu\mathsf{v}\mskip 0.0mu\mathnormal{)}\allowbreak{}}\<[E]\\
\>[1]{\allowbreak{}\mathnormal{(}\mskip 0.0mu\mathsf{t}\mskip 3.0mu\mathnormal{+}\mskip 3.0mu\mathsf{u}\mskip 0.0mu\mathnormal{)}\allowbreak{}\mskip 3.0mu\allowbreak{}\mathnormal{∘}\allowbreak{}\mskip 3.0mu\mathsf{v}\mskip 3.0mu}\>[2]{\mathnormal{=}\mskip 3.0mu\allowbreak{}\mathnormal{(}\mskip 0.0mu\mathsf{t}\mskip 3.0mu\allowbreak{}\mathnormal{∘}\allowbreak{}\mskip 3.0mu\mathsf{v}\mskip 0.0mu\mathnormal{)}\allowbreak{}\mskip 3.0mu\mathnormal{+}\mskip 3.0mu\allowbreak{}\mathnormal{(}\mskip 0.0mu\mathsf{u}\mskip 3.0mu\allowbreak{}\mathnormal{∘}\allowbreak{}\mskip 3.0mu\mathsf{v}\mskip 0.0mu\mathnormal{)}\allowbreak{}}\<[E]\\
\>[1]{\allowbreak{}\mathnormal{(}\mskip 0.0mu\mathsf{α}\mskip 3.0mu\smalltriangleleft \mskip 3.0mu\mathsf{t}\mskip 0.0mu\mathnormal{)}\allowbreak{}\mskip 3.0mu\allowbreak{}\mathnormal{∘}\allowbreak{}\mskip 3.0mu\mathsf{u}\mskip 3.0mu}\>[2]{\mathnormal{=}\mskip 3.0mu\mathsf{α}\mskip 3.0mu\smalltriangleleft \mskip 3.0mu\allowbreak{}\mathnormal{(}\mskip 0.0mu\mathsf{t}\mskip 3.0mu\allowbreak{}\mathnormal{∘}\allowbreak{}\mskip 3.0mu\mathsf{u}\mskip 0.0mu\mathnormal{)}\allowbreak{}}\<[E]\\
\>[1]{\mathsf{t}\mskip 3.0mu\allowbreak{}\mathnormal{∘}\allowbreak{}\mskip 3.0mu\allowbreak{}\mathnormal{(}\mskip 0.0mu\mathsf{α}\mskip 3.0mu\smalltriangleleft \mskip 3.0mu\mathsf{u}\mskip 0.0mu\mathnormal{)}\allowbreak{}\mskip 3.0mu}\>[2]{\mathnormal{=}\mskip 3.0mu\mathsf{α}\mskip 3.0mu\smalltriangleleft \mskip 3.0mu\allowbreak{}\mathnormal{(}\mskip 0.0mu\mathsf{t}\mskip 3.0mu\allowbreak{}\mathnormal{∘}\allowbreak{}\mskip 3.0mu\mathsf{u}\mskip 0.0mu\mathnormal{)}\allowbreak{}}\<[E]\end{parray}}&\ensuremath{\begin{parray}\column{B}{@{}>{}l<{}@{}}\column[0em]{1}{@{}>{}l<{}@{}}\column{2}{@{}>{}l<{}@{}}\column{E}{@{}>{}l<{}@{}}%
\>[1]{\mathsf{t}\mskip 3.0mu{⊗}\mskip 3.0mu\allowbreak{}\mathnormal{(}\mskip 0.0mu\mathsf{u}\mskip 3.0mu\mathnormal{+}\mskip 3.0mu\mathsf{v}\mskip 0.0mu\mathnormal{)}\allowbreak{}\mskip 3.0mu}\>[2]{\mathnormal{=}\mskip 3.0mu\allowbreak{}\mathnormal{(}\mskip 0.0mu\mathsf{t}\mskip 3.0mu{⊗}\mskip 3.0mu\mathsf{u}\mskip 0.0mu\mathnormal{)}\allowbreak{}\mskip 3.0mu\mathnormal{+}\mskip 3.0mu\allowbreak{}\mathnormal{(}\mskip 0.0mu\mathsf{t}\mskip 3.0mu{⊗}\mskip 3.0mu\mathsf{v}\mskip 0.0mu\mathnormal{)}\allowbreak{}}\<[E]\\
\>[1]{\allowbreak{}\mathnormal{(}\mskip 0.0mu\mathsf{t}\mskip 3.0mu\mathnormal{+}\mskip 3.0mu\mathsf{u}\mskip 0.0mu\mathnormal{)}\allowbreak{}\mskip 3.0mu{⊗}\mskip 3.0mu\mathsf{v}\mskip 3.0mu}\>[2]{\mathnormal{=}\mskip 3.0mu\allowbreak{}\mathnormal{(}\mskip 0.0mu\mathsf{t}\mskip 3.0mu{⊗}\mskip 3.0mu\mathsf{v}\mskip 0.0mu\mathnormal{)}\allowbreak{}\mskip 3.0mu\mathnormal{+}\mskip 3.0mu\allowbreak{}\mathnormal{(}\mskip 0.0mu\mathsf{u}\mskip 3.0mu{⊗}\mskip 3.0mu\mathsf{v}\mskip 0.0mu\mathnormal{)}\allowbreak{}}\<[E]\\
\>[1]{\allowbreak{}\mathnormal{(}\mskip 0.0mu\mathsf{α}\mskip 3.0mu\smalltriangleleft \mskip 3.0mu\mathsf{t}\mskip 0.0mu\mathnormal{)}\allowbreak{}\mskip 3.0mu{⊗}\mskip 3.0mu\mathsf{u}\mskip 3.0mu}\>[2]{\mathnormal{=}\mskip 3.0mu\mathsf{α}\mskip 3.0mu\smalltriangleleft \mskip 3.0mu\allowbreak{}\mathnormal{(}\mskip 0.0mu\mathsf{t}\mskip 3.0mu{⊗}\mskip 3.0mu\mathsf{u}\mskip 0.0mu\mathnormal{)}\allowbreak{}}\<[E]\\
\>[1]{\mathsf{t}\mskip 3.0mu{⊗}\mskip 3.0mu\allowbreak{}\mathnormal{(}\mskip 0.0mu\mathsf{α}\mskip 3.0mu\smalltriangleleft \mskip 3.0mu\mathsf{u}\mskip 0.0mu\mathnormal{)}\allowbreak{}\mskip 3.0mu}\>[2]{\mathnormal{=}\mskip 3.0mu\mathsf{α}\mskip 3.0mu\smalltriangleleft \mskip 3.0mu\allowbreak{}\mathnormal{(}\mskip 0.0mu\mathsf{t}\mskip 3.0mu{⊗}\mskip 3.0mu\mathsf{u}\mskip 0.0mu\mathnormal{)}\allowbreak{}}\<[E]\end{parray}}\end{tabular}\end{list} 
We note in passing that there is no obviously good way to represent
addition using diagrams.  If diagrams should be added together we
write them side by side with a plus sign in between.

\paragraph*{Compact Closed Category}\hspace{1.0ex}\label{86} 
\begin{figure*}[]\begin{subfigure}[t]{0.4\textwidth}\begin{tabular}{cc}{\begin{tikzpicture}\path[-,draw=lightgray,line width=0.4pt,line cap=butt,line join=miter,dash pattern=](-13pt,-3.7092pt)--(-6pt,-3.7092pt);
\path[-,draw=black,line width=0.4pt,line cap=butt,line join=miter,dash pattern=](0pt,0pt)--(1pt,0pt);
\path[-,draw=black,line width=0.4pt,line cap=butt,line join=miter,dash pattern=](0pt,-7.4184pt)--(1pt,-7.4184pt);
\path[to-,draw=black,line width=0.4pt,line cap=butt,line join=miter,dash pattern=](-3.06pt,-0.3287pt)..controls(-2.22pt,-0.1088pt)and(-1.2pt,0pt)..(0pt,0pt);
\path[-,draw=black,line width=0.4pt,line cap=butt,line join=miter,dash pattern=](-6pt,-3.7092pt)..controls(-6pt,-1.9592pt)and(-5.02pt,-0.8417pt)..(-3.06pt,-0.3287pt);
\path[-,draw=black,line width=0.4pt,line cap=butt,line join=miter,dash pattern=](-3.06pt,-7.0897pt)..controls(-2.22pt,-7.3096pt)and(-1.2pt,-7.4184pt)..(0pt,-7.4184pt);
\path[-,draw=black,line width=0.4pt,line cap=butt,line join=miter,dash pattern=](-6pt,-3.7092pt)..controls(-6pt,-5.4592pt)and(-5.02pt,-6.5767pt)..(-3.06pt,-7.0897pt);
\path[-,line width=0.4pt,line cap=butt,line join=miter,dash pattern=](3pt,2.3396pt)--(5.6316pt,2.3396pt)--(5.6316pt,-2.3396pt)--(3pt,-2.3396pt)--cycle;
\node[anchor=north west,inner sep=0] at (3pt,2.3396pt){\savebox{\marxupbox}{{\({\scriptstyle i}\)}}\immediate\write\boxesfile{87}\immediate\write\boxesfile{\number\wd\marxupbox}\immediate\write\boxesfile{\number\ht\marxupbox}\immediate\write\boxesfile{\number\dp\marxupbox}\box\marxupbox};
\path[-,line width=0.4pt,line cap=butt,line join=miter,dash pattern=](1pt,4.3396pt)--(7.6316pt,4.3396pt)--(7.6316pt,-4.3396pt)--(1pt,-4.3396pt)--cycle;
\path[-,line width=0.4pt,line cap=butt,line join=miter,dash pattern=](3pt,-4.3396pt)--(6.8617pt,-4.3396pt)--(6.8617pt,-10.4972pt)--(3pt,-10.4972pt)--cycle;
\node[anchor=north west,inner sep=0] at (3pt,-4.3396pt){\savebox{\marxupbox}{{\({\scriptstyle j}\)}}\immediate\write\boxesfile{88}\immediate\write\boxesfile{\number\wd\marxupbox}\immediate\write\boxesfile{\number\ht\marxupbox}\immediate\write\boxesfile{\number\dp\marxupbox}\box\marxupbox};
\path[-,line width=0.4pt,line cap=butt,line join=miter,dash pattern=](1pt,-2.3396pt)--(8.8617pt,-2.3396pt)--(8.8617pt,-12.4972pt)--(1pt,-12.4972pt)--cycle;
\end{tikzpicture}}&{\begin{tikzpicture}\path[-,draw=black,line width=0.4pt,line cap=butt,line join=miter,dash pattern=](-7pt,0pt)--(-6pt,0pt);
\path[-,draw=black,line width=0.4pt,line cap=butt,line join=miter,dash pattern=](-7pt,-7.4184pt)--(-6pt,-7.4184pt);
\path[-,draw=lightgray,line width=0.4pt,line cap=butt,line join=miter,dash pattern=](0pt,-3.7092pt)--(7pt,-3.7092pt);
\path[-,draw=black,line width=0.4pt,line cap=butt,line join=miter,dash pattern=](-2.94pt,-0.3287pt)..controls(-0.98pt,-0.8417pt)and(-0pt,-1.9592pt)..(0pt,-3.7092pt);
\path[-,draw=black,line width=0.4pt,line cap=butt,line join=miter,dash pattern=](-6pt,0pt)..controls(-4.8pt,0pt)and(-3.78pt,-0.1088pt)..(-2.94pt,-0.3287pt);
\path[to-,draw=black,line width=0.4pt,line cap=butt,line join=miter,dash pattern=](-2.94pt,-7.0897pt)..controls(-0.98pt,-6.5767pt)and(-0pt,-5.4592pt)..(0pt,-3.7092pt);
\path[-,draw=black,line width=0.4pt,line cap=butt,line join=miter,dash pattern=](-6pt,-7.4184pt)..controls(-4.8pt,-7.4184pt)and(-3.78pt,-7.3096pt)..(-2.94pt,-7.0897pt);
\path[-,line width=0.4pt,line cap=butt,line join=miter,dash pattern=](-11.6316pt,2.3396pt)--(-9pt,2.3396pt)--(-9pt,-2.3396pt)--(-11.6316pt,-2.3396pt)--cycle;
\node[anchor=north west,inner sep=0] at (-11.6316pt,2.3396pt){\savebox{\marxupbox}{{\({\scriptstyle i}\)}}\immediate\write\boxesfile{89}\immediate\write\boxesfile{\number\wd\marxupbox}\immediate\write\boxesfile{\number\ht\marxupbox}\immediate\write\boxesfile{\number\dp\marxupbox}\box\marxupbox};
\path[-,line width=0.4pt,line cap=butt,line join=miter,dash pattern=](-13.6316pt,4.3396pt)--(-7pt,4.3396pt)--(-7pt,-4.3396pt)--(-13.6316pt,-4.3396pt)--cycle;
\path[-,line width=0.4pt,line cap=butt,line join=miter,dash pattern=](-12.8617pt,-4.3396pt)--(-9pt,-4.3396pt)--(-9pt,-10.4972pt)--(-12.8617pt,-10.4972pt)--cycle;
\node[anchor=north west,inner sep=0] at (-12.8617pt,-4.3396pt){\savebox{\marxupbox}{{\({\scriptstyle j}\)}}\immediate\write\boxesfile{90}\immediate\write\boxesfile{\number\wd\marxupbox}\immediate\write\boxesfile{\number\ht\marxupbox}\immediate\write\boxesfile{\number\dp\marxupbox}\box\marxupbox};
\path[-,line width=0.4pt,line cap=butt,line join=miter,dash pattern=](-14.8617pt,-2.3396pt)--(-7pt,-2.3396pt)--(-7pt,-12.4972pt)--(-14.8617pt,-12.4972pt)--cycle;
\end{tikzpicture}}\\\(η\)&\(ϵ\)\\\(δ{_j}{^i}\)&\(δ{_i}{^j}\)\\\(\mathbf{1}\mskip 1.0mu\overset{z}{\leadsto }\mskip 1.0mu\dual{\mathsf{T}}\mskip 3.0mu\mathnormal{⊗}\mskip 3.0mu\mathsf{T}\)&\(\mathsf{T}\mskip 3.0mu\mathnormal{⊗}\mskip 3.0mu\dual{\mathsf{T}}\mskip 1.0mu\overset{z}{\leadsto }\mskip 1.0mu\mathbf{1}\)\end{tabular}\caption{Unit and counit.}\label{91}\end{subfigure}\quad{}\begin{subfigure}[t]{0.6\textwidth}\begin{tabular}{ccc}{\begin{tikzpicture}\path[-,draw=lightgray,line width=0.4pt,line cap=butt,line join=miter,dash pattern=](-23.32pt,10.1287pt)--(-16.32pt,10.1287pt);
\path[-,draw=black,line width=0.4pt,line cap=butt,line join=miter,dash pattern=](0pt,14.505pt)--(7pt,14.505pt);
\path[-,draw=black,line width=0.4pt,line cap=butt,line join=miter,dash pattern=](0pt,5.7525pt)--(7pt,5.7525pt);
\path[-,draw=black,line width=0.4pt,line cap=butt,line join=miter,dash pattern=](-10.32pt,14.505pt)--(-9.32pt,14.505pt);
\path[-,draw=black,line width=0.4pt,line cap=butt,line join=miter,dash pattern=](-10.32pt,5.7525pt)--(-9.32pt,5.7525pt);
\path[-,draw=black,line width=0.4pt,line cap=butt,line join=miter,dash pattern=](-9.32pt,14.505pt)--(0pt,14.505pt);
\path[-,line width=0.4pt,line cap=butt,line join=miter,dash pattern=](-6.32pt,8.505pt)--(-3pt,8.505pt)--(-3pt,3pt)--(-6.32pt,3pt)--cycle;
\node[anchor=north west,inner sep=0] at (-6.32pt,8.505pt){\savebox{\marxupbox}{{\(t\)}}\immediate\write\boxesfile{92}\immediate\write\boxesfile{\number\wd\marxupbox}\immediate\write\boxesfile{\number\ht\marxupbox}\immediate\write\boxesfile{\number\dp\marxupbox}\box\marxupbox};
\path[-,draw=black,line width=0.4pt,line cap=butt,line join=miter,dash pattern=](-9.32pt,9.005pt)..controls(-9.32pt,10.3857pt)and(-8.2007pt,11.505pt)..(-6.82pt,11.505pt)--(-2.5pt,11.505pt)..controls(-1.1193pt,11.505pt)and(0pt,10.3857pt)..(0pt,9.005pt)--(0pt,2.5pt)..controls(0pt,1.1193pt)and(-1.1193pt,0pt)..(-2.5pt,0pt)--(-6.82pt,0pt)..controls(-8.2007pt,0pt)and(-9.32pt,1.1193pt)..(-9.32pt,2.5pt)--cycle;
\path[-,draw=black,line width=0.4pt,line cap=butt,line join=miter,dash pattern=](0pt,5.7525pt)--(0pt,5.7525pt);
\path[-,draw=black,line width=0.4pt,line cap=butt,line join=miter,dash pattern=](-9.32pt,5.7525pt)--(-9.32pt,5.7525pt);
\path[to-,draw=black,line width=0.4pt,line cap=butt,line join=miter,dash pattern=](-13.38pt,14.0322pt)..controls(-12.54pt,14.3361pt)and(-11.52pt,14.505pt)..(-10.32pt,14.505pt);
\path[-,draw=black,line width=0.4pt,line cap=butt,line join=miter,dash pattern=](-16.32pt,10.1287pt)..controls(-16.32pt,11.8787pt)and(-15.34pt,13.3231pt)..(-13.38pt,14.0322pt);
\path[-,draw=black,line width=0.4pt,line cap=butt,line join=miter,dash pattern=](-13.38pt,6.2253pt)..controls(-12.54pt,5.9213pt)and(-11.52pt,5.7525pt)..(-10.32pt,5.7525pt);
\path[-,draw=black,line width=0.4pt,line cap=butt,line join=miter,dash pattern=](-16.32pt,10.1287pt)..controls(-16.32pt,8.3787pt)and(-15.34pt,6.9344pt)..(-13.38pt,6.2253pt);
\path[-,line width=0.4pt,line cap=butt,line join=miter,dash pattern=](9pt,16.8446pt)--(11.6316pt,16.8446pt)--(11.6316pt,12.1653pt)--(9pt,12.1653pt)--cycle;
\node[anchor=north west,inner sep=0] at (9pt,16.8446pt){\savebox{\marxupbox}{{\({\scriptstyle i}\)}}\immediate\write\boxesfile{93}\immediate\write\boxesfile{\number\wd\marxupbox}\immediate\write\boxesfile{\number\ht\marxupbox}\immediate\write\boxesfile{\number\dp\marxupbox}\box\marxupbox};
\path[-,line width=0.4pt,line cap=butt,line join=miter,dash pattern=](7pt,18.8446pt)--(13.6316pt,18.8446pt)--(13.6316pt,10.1653pt)--(7pt,10.1653pt)--cycle;
\path[-,line width=0.4pt,line cap=butt,line join=miter,dash pattern=](9pt,8.8313pt)--(12.8617pt,8.8313pt)--(12.8617pt,2.6737pt)--(9pt,2.6737pt)--cycle;
\node[anchor=north west,inner sep=0] at (9pt,8.8313pt){\savebox{\marxupbox}{{\({\scriptstyle j}\)}}\immediate\write\boxesfile{94}\immediate\write\boxesfile{\number\wd\marxupbox}\immediate\write\boxesfile{\number\ht\marxupbox}\immediate\write\boxesfile{\number\dp\marxupbox}\box\marxupbox};
\path[-,line width=0.4pt,line cap=butt,line join=miter,dash pattern=](7pt,10.8313pt)--(14.8617pt,10.8313pt)--(14.8617pt,0.6737pt)--(7pt,0.6737pt)--cycle;
\end{tikzpicture}}&{\begin{tikzpicture}\path[-,draw=black,line width=0.4pt,line cap=butt,line join=miter,dash pattern=](-16.32pt,5.7525pt)--(-9.32pt,5.7525pt);
\path[-,draw=black,line width=0.4pt,line cap=butt,line join=miter,dash pattern=](-16.32pt,-3pt)--(-9.32pt,-3pt);
\path[-,draw=lightgray,line width=0.4pt,line cap=butt,line join=miter,dash pattern=](7pt,1.3762pt)--(14pt,1.3762pt);
\path[-,draw=black,line width=0.4pt,line cap=butt,line join=miter,dash pattern=](0pt,5.7525pt)--(1pt,5.7525pt);
\path[-,draw=black,line width=0.4pt,line cap=butt,line join=miter,dash pattern=](0pt,-3pt)--(1pt,-3pt);
\path[-,draw=black,line width=0.4pt,line cap=butt,line join=miter,dash pattern=](4.06pt,5.2797pt)..controls(6.02pt,4.5706pt)and(7pt,3.1262pt)..(7pt,1.3762pt);
\path[-,draw=black,line width=0.4pt,line cap=butt,line join=miter,dash pattern=](1pt,5.7525pt)..controls(2.2pt,5.7525pt)and(3.22pt,5.5836pt)..(4.06pt,5.2797pt);
\path[to-,draw=black,line width=0.4pt,line cap=butt,line join=miter,dash pattern=](4.06pt,-2.5272pt)..controls(6.02pt,-1.8181pt)and(7pt,-0.3738pt)..(7pt,1.3762pt);
\path[-,draw=black,line width=0.4pt,line cap=butt,line join=miter,dash pattern=](1pt,-3pt)..controls(2.2pt,-3pt)and(3.22pt,-2.8311pt)..(4.06pt,-2.5272pt);
\path[-,line width=0.4pt,line cap=butt,line join=miter,dash pattern=](-6.32pt,8.505pt)--(-3pt,8.505pt)--(-3pt,3pt)--(-6.32pt,3pt)--cycle;
\node[anchor=north west,inner sep=0] at (-6.32pt,8.505pt){\savebox{\marxupbox}{{\(t\)}}\immediate\write\boxesfile{95}\immediate\write\boxesfile{\number\wd\marxupbox}\immediate\write\boxesfile{\number\ht\marxupbox}\immediate\write\boxesfile{\number\dp\marxupbox}\box\marxupbox};
\path[-,draw=black,line width=0.4pt,line cap=butt,line join=miter,dash pattern=](-9.32pt,9.005pt)..controls(-9.32pt,10.3857pt)and(-8.2007pt,11.505pt)..(-6.82pt,11.505pt)--(-2.5pt,11.505pt)..controls(-1.1193pt,11.505pt)and(0pt,10.3857pt)..(0pt,9.005pt)--(0pt,2.5pt)..controls(0pt,1.1193pt)and(-1.1193pt,0pt)..(-2.5pt,0pt)--(-6.82pt,0pt)..controls(-8.2007pt,0pt)and(-9.32pt,1.1193pt)..(-9.32pt,2.5pt)--cycle;
\path[-,draw=black,line width=0.4pt,line cap=butt,line join=miter,dash pattern=](0pt,5.7525pt)--(0pt,5.7525pt);
\path[-,draw=black,line width=0.4pt,line cap=butt,line join=miter,dash pattern=](-9.32pt,5.7525pt)--(-9.32pt,5.7525pt);
\path[-,draw=black,line width=0.4pt,line cap=butt,line join=miter,dash pattern=](-9.32pt,-3pt)--(0pt,-3pt);
\path[-,line width=0.4pt,line cap=butt,line join=miter,dash pattern=](-20.9516pt,8.0921pt)--(-18.32pt,8.0921pt)--(-18.32pt,3.4128pt)--(-20.9516pt,3.4128pt)--cycle;
\node[anchor=north west,inner sep=0] at (-20.9516pt,8.0921pt){\savebox{\marxupbox}{{\({\scriptstyle i}\)}}\immediate\write\boxesfile{96}\immediate\write\boxesfile{\number\wd\marxupbox}\immediate\write\boxesfile{\number\ht\marxupbox}\immediate\write\boxesfile{\number\dp\marxupbox}\box\marxupbox};
\path[-,line width=0.4pt,line cap=butt,line join=miter,dash pattern=](-22.9516pt,10.0921pt)--(-16.32pt,10.0921pt)--(-16.32pt,1.4128pt)--(-22.9516pt,1.4128pt)--cycle;
\path[-,line width=0.4pt,line cap=butt,line join=miter,dash pattern=](-22.1817pt,0.0788pt)--(-18.32pt,0.0788pt)--(-18.32pt,-6.0788pt)--(-22.1817pt,-6.0788pt)--cycle;
\node[anchor=north west,inner sep=0] at (-22.1817pt,0.0788pt){\savebox{\marxupbox}{{\({\scriptstyle j}\)}}\immediate\write\boxesfile{97}\immediate\write\boxesfile{\number\wd\marxupbox}\immediate\write\boxesfile{\number\ht\marxupbox}\immediate\write\boxesfile{\number\dp\marxupbox}\box\marxupbox};
\path[-,line width=0.4pt,line cap=butt,line join=miter,dash pattern=](-24.1817pt,2.0788pt)--(-16.32pt,2.0788pt)--(-16.32pt,-8.0788pt)--(-24.1817pt,-8.0788pt)--cycle;
\end{tikzpicture}}&{\begin{tikzpicture}\path[-,draw=lightgray,line width=0.4pt,line cap=butt,line join=miter,dash pattern=](-29.32pt,1.3762pt)--(-22.32pt,1.3762pt);
\path[-,draw=lightgray,line width=0.4pt,line cap=butt,line join=miter,dash pattern=](7pt,1.3762pt)--(14pt,1.3762pt);
\path[-,draw=black,line width=0.4pt,line cap=butt,line join=miter,dash pattern=](-16.32pt,3.8762pt)--(-16.32pt,3.8762pt);
\path[-,draw=black,line width=0.4pt,line cap=butt,line join=miter,dash pattern=](-16.32pt,-1.1238pt)--(-16.32pt,-1.1238pt);
\path[-,draw=black,line width=0.4pt,line cap=butt,line join=miter,dash pattern=](-10.32pt,5.7525pt)--(-9.32pt,5.7525pt);
\path[-,draw=black,line width=0.4pt,line cap=butt,line join=miter,dash pattern=](-10.32pt,-3pt)--(-9.32pt,-3pt);
\path[-,draw=black,line width=0.4pt,line cap=butt,line join=miter,dash pattern=](0pt,5.7525pt)--(1pt,5.7525pt);
\path[-,draw=black,line width=0.4pt,line cap=butt,line join=miter,dash pattern=](0pt,-3pt)--(1pt,-3pt);
\path[-,draw=black,line width=0.4pt,line cap=butt,line join=miter,dash pattern=](4.06pt,5.2797pt)..controls(6.02pt,4.5706pt)and(7pt,3.1262pt)..(7pt,1.3762pt);
\path[-,draw=black,line width=0.4pt,line cap=butt,line join=miter,dash pattern=](1pt,5.7525pt)..controls(2.2pt,5.7525pt)and(3.22pt,5.5836pt)..(4.06pt,5.2797pt);
\path[to-,draw=black,line width=0.4pt,line cap=butt,line join=miter,dash pattern=](4.06pt,-2.5272pt)..controls(6.02pt,-1.8181pt)and(7pt,-0.3738pt)..(7pt,1.3762pt);
\path[-,draw=black,line width=0.4pt,line cap=butt,line join=miter,dash pattern=](1pt,-3pt)..controls(2.2pt,-3pt)and(3.22pt,-2.8311pt)..(4.06pt,-2.5272pt);
\path[-,line width=0.4pt,line cap=butt,line join=miter,dash pattern=](-6.32pt,8.505pt)--(-3pt,8.505pt)--(-3pt,3pt)--(-6.32pt,3pt)--cycle;
\node[anchor=north west,inner sep=0] at (-6.32pt,8.505pt){\savebox{\marxupbox}{{\(t\)}}\immediate\write\boxesfile{98}\immediate\write\boxesfile{\number\wd\marxupbox}\immediate\write\boxesfile{\number\ht\marxupbox}\immediate\write\boxesfile{\number\dp\marxupbox}\box\marxupbox};
\path[-,draw=black,line width=0.4pt,line cap=butt,line join=miter,dash pattern=](-9.32pt,9.005pt)..controls(-9.32pt,10.3857pt)and(-8.2007pt,11.505pt)..(-6.82pt,11.505pt)--(-2.5pt,11.505pt)..controls(-1.1193pt,11.505pt)and(0pt,10.3857pt)..(0pt,9.005pt)--(0pt,2.5pt)..controls(0pt,1.1193pt)and(-1.1193pt,0pt)..(-2.5pt,0pt)--(-6.82pt,0pt)..controls(-8.2007pt,0pt)and(-9.32pt,1.1193pt)..(-9.32pt,2.5pt)--cycle;
\path[-,draw=black,line width=0.4pt,line cap=butt,line join=miter,dash pattern=](0pt,5.7525pt)--(0pt,5.7525pt);
\path[-,draw=black,line width=0.4pt,line cap=butt,line join=miter,dash pattern=](-9.32pt,5.7525pt)--(-9.32pt,5.7525pt);
\path[-,draw=black,line width=0.4pt,line cap=butt,line join=miter,dash pattern=](-9.32pt,-3pt)--(0pt,-3pt);
\path[-,draw=black,line width=0.4pt,line cap=butt,line join=miter,dash pattern=](-16.32pt,3.8762pt)..controls(-12.32pt,3.8762pt)and(-14.32pt,-3pt)..(-10.32pt,-3pt);
\path[-,draw=black,line width=0.4pt,line cap=butt,line join=miter,dash pattern=](-16.32pt,-1.1238pt)..controls(-12.32pt,-1.1238pt)and(-14.32pt,5.7525pt)..(-10.32pt,5.7525pt);
\path[to-,draw=black,line width=0.4pt,line cap=butt,line join=miter,dash pattern=](-19.38pt,3.8087pt)..controls(-18.54pt,3.8762pt)and(-17.52pt,3.8762pt)..(-16.32pt,3.8762pt);
\path[-,draw=black,line width=0.4pt,line cap=butt,line join=miter,dash pattern=](-22.32pt,1.3762pt)..controls(-22.32pt,3.1262pt)and(-21.34pt,3.6512pt)..(-19.38pt,3.8087pt);
\path[-,draw=black,line width=0.4pt,line cap=butt,line join=miter,dash pattern=](-19.38pt,-1.0563pt)..controls(-18.54pt,-1.1238pt)and(-17.52pt,-1.1238pt)..(-16.32pt,-1.1238pt);
\path[-,draw=black,line width=0.4pt,line cap=butt,line join=miter,dash pattern=](-22.32pt,1.3762pt)..controls(-22.32pt,-0.3738pt)and(-21.34pt,-0.8988pt)..(-19.38pt,-1.0563pt);
\end{tikzpicture}}\\\(\allowbreak{}\mathnormal{(}\mskip 0.0mu\mathsf{id}\mskip 3.0mu{⊗}\mskip 3.0mu\mathsf{t}\mskip 0.0mu\mathnormal{)}\allowbreak{}\mskip 3.0mu\allowbreak{}\mathnormal{∘}\allowbreak{}\mskip 3.0muη\)&\(ϵ\mskip 3.0mu\allowbreak{}\mathnormal{∘}\allowbreak{}\mskip 3.0mu\allowbreak{}\mathnormal{(}\mskip 0.0mu\mathsf{t}\mskip 3.0mu{⊗}\mskip 3.0mu\mathsf{id}\mskip 0.0mu\mathnormal{)}\allowbreak{}\)&\(ϵ\mskip 3.0mu\allowbreak{}\mathnormal{∘}\allowbreak{}\mskip 3.0mu\allowbreak{}\mathnormal{(}\mskip 0.0mu\mathsf{t}\mskip 3.0mu{⊗}\mskip 3.0mu\mathsf{id}\mskip 0.0mu\mathnormal{)}\allowbreak{}\mskip 3.0mu\allowbreak{}\mathnormal{∘}\allowbreak{}\mskip 3.0muσ\mskip 3.0mu\allowbreak{}\mathnormal{∘}\allowbreak{}\mskip 3.0muη\)\\\(t{_i}{^j}\)&\(t{_i}{^j}\)&\(t{_i}{^i}\)\\\(\mathbf{1}\mskip 1.0mu\overset{z}{\leadsto }\mskip 1.0mu\dual{\mathsf{T}}\mskip 3.0mu\mathnormal{⊗}\mskip 3.0mu\mathsf{T}\)&\(\mathsf{T}\mskip 3.0mu\mathnormal{⊗}\mskip 3.0mu\dual{\mathsf{T}}\mskip 1.0mu\overset{z}{\leadsto }\mskip 1.0mu\mathbf{1}\)&\(\mathbf{1}\mskip 1.0mu\overset{z}{\leadsto }\mskip 1.0mu\mathbf{1}\)\end{tabular}\caption{Bending input and output connections.}\label{99}\end{subfigure}\caption{Illustration of compact closed categories in various notations. Note that Einstein notation does not change when
bending connections using \(η\) or \(ϵ\), though in the
third example, the new connection is notated by repeated use of the index.}\label{100}\end{figure*} 
There remains to capture the
relationship between a vector space \(V\) and its associated
covector space \(\dual V\).  This is done abstractly using a compact
closed category structure \citep{selinger2011survey}.
In a compact closed category, every object has
a dual, and duals generalise the notion of co-vector space.
\begin{list}{}{\setlength\leftmargin{1.0em}}\item\relax
\ensuremath{\begin{parray}\column{B}{@{}>{}l<{}@{}}\column[0em]{1}{@{}>{}l<{}@{}}\column[1em]{2}{@{}>{}l<{}@{}}\column{3}{@{}>{}l<{}@{}}\column{4}{@{}>{}l<{}@{}}\column{5}{@{}>{}l<{}@{}}\column{E}{@{}>{}l<{}@{}}%
\>[1]{\mathbf{class}\mskip 3.0mu\allowbreak{}\mathnormal{(}\mskip 0.0mu}\>[5]{\mathsf{SymmetricMonoidal}\mskip 3.0muz\mskip 0.0mu\mathnormal{)}\allowbreak{}\mskip 3.0mu\mathnormal{\Rightarrow }\mskip 3.0mu\mathsf{CompactClosed}\mskip 3.0muz\mskip 3.0mu\mathbf{where}}\<[E]\\
\>[2]{η\mskip 3.0mu}\>[3]{\mathnormal{::}\mskip 3.0mu}\>[4]{\mathbf{1}\mskip 1.0mu\overset{z}{\leadsto }\mskip 1.0mu\allowbreak{}\mathnormal{(}\mskip 0.0mu\dual{\mathsf{a}}\mskip 3.0mu\mathnormal{⊗}\mskip 3.0mu\mathsf{a}\mskip 0.0mu\mathnormal{)}\allowbreak{}}\<[E]\\
\>[2]{ϵ\mskip 3.0mu}\>[3]{\mathnormal{::}\mskip 3.0mu}\>[4]{\allowbreak{}\mathnormal{(}\mskip 0.0mu\mathsf{a}\mskip 3.0mu\mathnormal{⊗}\mskip 3.0mu\dual{\mathsf{a}}\mskip 0.0mu\mathnormal{)}\allowbreak{}\mskip 1.0mu\overset{z}{\leadsto }\mskip 1.0mu\mathbf{1}}\<[E]\end{parray}}\end{list} 

 In the tensor instance, \(η\) and \(ϵ\) produce and consume
correlated products of vectors and covectors. While the algebraic
view is very abstract and can be hard to grasp (it is for the authors), the diagrams help.
Compact closed categories are required to satisfy the so-called
snake laws: one is \(\allowbreak{}\mathnormal{(}\mskip 0.0muϵ\mskip 3.0mu{⊗}\mskip 3.0mu\mathsf{id}\mskip 0.0mu\mathnormal{)}\allowbreak{}\mskip 3.0mu\allowbreak{}\mathnormal{∘}\allowbreak{}\mskip 3.0mu\bar{α}\mskip 3.0mu\allowbreak{}\mathnormal{∘}\allowbreak{}\mskip 3.0mu\allowbreak{}\mathnormal{(}\mskip 0.0mu\mathsf{id}\mskip 3.0mu{⊗}\mskip 3.0muη\mskip 0.0mu\mathnormal{)}\allowbreak{}\) =
\(σ\), or ({\begin{tikzpicture}[baseline={([yshift=-0.8ex]current bounding box.center)}]\path[-,draw=black,line width=0.4pt,line cap=butt,line join=miter,dash pattern=](-19pt,0pt)--(-12pt,0pt);
\path[-,draw=lightgray,line width=0.4pt,line cap=butt,line join=miter,dash pattern=](-19pt,-12.5pt)--(-12pt,-12.5pt);
\path[-,draw=lightgray,line width=0.4pt,line cap=butt,line join=miter,dash pattern=](6pt,-2.5pt)--(13pt,-2.5pt);
\path[-,draw=black,line width=0.4pt,line cap=butt,line join=miter,dash pattern=](6pt,-15pt)--(13pt,-15pt);
\path[-,draw=black,line width=0.4pt,line cap=butt,line join=miter,dash pattern=](-6pt,0pt)--(-6pt,0pt);
\path[-,draw=black,line width=0.4pt,line cap=butt,line join=miter,dash pattern=](-6pt,-10pt)--(-6pt,-10pt);
\path[-,draw=black,line width=0.4pt,line cap=butt,line join=miter,dash pattern=](-6pt,-15pt)--(-6pt,-15pt);
\path[-,draw=black,line width=0.4pt,line cap=butt,line join=miter,dash pattern=](0pt,0pt)--(0pt,0pt);
\path[-,draw=black,line width=0.4pt,line cap=butt,line join=miter,dash pattern=](0pt,-5pt)--(0pt,-5pt);
\path[-,draw=black,line width=0.4pt,line cap=butt,line join=miter,dash pattern=](0pt,-15pt)--(0pt,-15pt);
\path[-,draw=black,line width=0.4pt,line cap=butt,line join=miter,dash pattern=](3.06pt,-0.0675pt)..controls(5.02pt,-0.225pt)and(6pt,-0.75pt)..(6pt,-2.5pt);
\path[-,draw=black,line width=0.4pt,line cap=butt,line join=miter,dash pattern=](0pt,0pt)..controls(1.2pt,0pt)and(2.22pt,0pt)..(3.06pt,-0.0675pt);
\path[to-,draw=black,line width=0.4pt,line cap=butt,line join=miter,dash pattern=](3.06pt,-4.9325pt)..controls(5.02pt,-4.775pt)and(6pt,-4.25pt)..(6pt,-2.5pt);
\path[-,draw=black,line width=0.4pt,line cap=butt,line join=miter,dash pattern=](0pt,-5pt)..controls(1.2pt,-5pt)and(2.22pt,-5pt)..(3.06pt,-4.9325pt);
\path[-,draw=black,line width=0.4pt,line cap=butt,line join=miter,dash pattern=](0pt,-15pt)--(6pt,-15pt);
\path[-,draw=black,line width=0.4pt,line cap=butt,line join=miter,dash pattern=](-6pt,0pt)--(0pt,0pt);
\path[-,draw=black,line width=0.4pt,line cap=butt,line join=miter,dash pattern=](-6pt,-10pt)..controls(-2pt,-10pt)and(-4pt,-5pt)..(0pt,-5pt);
\path[-,draw=black,line width=0.4pt,line cap=butt,line join=miter,dash pattern=](-6pt,-15pt)--(0pt,-15pt);
\path[-,draw=black,line width=0.4pt,line cap=butt,line join=miter,dash pattern=](-12pt,0pt)--(-6pt,0pt);
\path[to-,draw=black,line width=0.4pt,line cap=butt,line join=miter,dash pattern=](-9.06pt,-10.0675pt)..controls(-8.22pt,-10pt)and(-7.2pt,-10pt)..(-6pt,-10pt);
\path[-,draw=black,line width=0.4pt,line cap=butt,line join=miter,dash pattern=](-12pt,-12.5pt)..controls(-12pt,-10.75pt)and(-11.02pt,-10.225pt)..(-9.06pt,-10.0675pt);
\path[-,draw=black,line width=0.4pt,line cap=butt,line join=miter,dash pattern=](-9.06pt,-14.9325pt)..controls(-8.22pt,-15pt)and(-7.2pt,-15pt)..(-6pt,-15pt);
\path[-,draw=black,line width=0.4pt,line cap=butt,line join=miter,dash pattern=](-12pt,-12.5pt)..controls(-12pt,-14.25pt)and(-11.02pt,-14.775pt)..(-9.06pt,-14.9325pt);
\path[-,line width=0.4pt,line cap=butt,line join=miter,dash pattern=](-23.6316pt,2.3396pt)--(-21pt,2.3396pt)--(-21pt,-2.3396pt)--(-23.6316pt,-2.3396pt)--cycle;
\node[anchor=north west,inner sep=0] at (-23.6316pt,2.3396pt){\savebox{\marxupbox}{{\({\scriptstyle i}\)}}\immediate\write\boxesfile{101}\immediate\write\boxesfile{\number\wd\marxupbox}\immediate\write\boxesfile{\number\ht\marxupbox}\immediate\write\boxesfile{\number\dp\marxupbox}\box\marxupbox};
\path[-,line width=0.4pt,line cap=butt,line join=miter,dash pattern=](-25.6316pt,4.3396pt)--(-19pt,4.3396pt)--(-19pt,-4.3396pt)--(-25.6316pt,-4.3396pt)--cycle;
\path[-,line width=0.4pt,line cap=butt,line join=miter,dash pattern=](15pt,-11.9212pt)--(18.8617pt,-11.9212pt)--(18.8617pt,-18.0788pt)--(15pt,-18.0788pt)--cycle;
\node[anchor=north west,inner sep=0] at (15pt,-11.9212pt){\savebox{\marxupbox}{{\({\scriptstyle j}\)}}\immediate\write\boxesfile{102}\immediate\write\boxesfile{\number\wd\marxupbox}\immediate\write\boxesfile{\number\ht\marxupbox}\immediate\write\boxesfile{\number\dp\marxupbox}\box\marxupbox};
\path[-,line width=0.4pt,line cap=butt,line join=miter,dash pattern=](13pt,-9.9212pt)--(20.8617pt,-9.9212pt)--(20.8617pt,-20.0788pt)--(13pt,-20.0788pt)--cycle;
\end{tikzpicture}} = {\begin{tikzpicture}[baseline={([yshift=-0.8ex]current bounding box.center)}]\path[-,draw=black,line width=0.4pt,line cap=butt,line join=miter,dash pattern=](-14pt,5pt)--(-7pt,5pt);
\path[-,draw=lightgray,line width=0.4pt,line cap=butt,line join=miter,dash pattern=](-14pt,0pt)--(-7pt,0pt);
\path[-,draw=lightgray,line width=0.4pt,line cap=butt,line join=miter,dash pattern=](7pt,5pt)--(8pt,5pt);
\path[-,draw=black,line width=0.4pt,line cap=butt,line join=miter,dash pattern=](7pt,0pt)--(8pt,0pt);
\path[-,draw=black,line width=0.4pt,line cap=butt,line join=miter,dash pattern=](-7pt,5pt)--(0pt,5pt);
\path[-,draw=black,line width=0.4pt,line cap=butt,line join=miter,dash pattern=](0pt,5pt)--(1pt,5pt);
\path[-,draw=lightgray,line width=0.4pt,line cap=butt,line join=miter,dash pattern=](0pt,0pt)--(1pt,0pt);
\path[-,draw=black,line width=0.4pt,line cap=butt,line join=miter,dash pattern=](1pt,5pt)..controls(5pt,5pt)and(3pt,0pt)..(7pt,0pt);
\path[-,draw=lightgray,line width=0.4pt,line cap=butt,line join=miter,dash pattern=](1pt,0pt)..controls(5pt,0pt)and(3pt,5pt)..(7pt,5pt);
\path[-,draw=black,line width=0.4pt,line cap=butt,line join=miter,dash pattern=](0pt,5pt)--(0pt,5pt);
\path[-,line width=0.4pt,line cap=butt,line join=miter,dash pattern=on 0.4pt off 1pt](0pt,5pt)--(0pt,5pt)--(0pt,0pt)--(0pt,0pt)--cycle;
\path[-,draw=black,line width=0.4pt,line cap=butt,line join=miter,dash pattern=on 0.4pt off 1pt](-3pt,8pt)--(3pt,8pt)--(3pt,-3pt)--(-3pt,-3pt)--cycle;
\path[-,fill=lightgray,line width=0.4pt,line cap=butt,line join=miter,dash pattern=](1pt,0pt)..controls(1pt,0.5523pt)and(0.5523pt,1pt)..(0pt,1pt)..controls(-0.5523pt,1pt)and(-1pt,0.5523pt)..(-1pt,0pt)..controls(-1pt,-0.5523pt)and(-0.5523pt,-1pt)..(0pt,-1pt)..controls(0.5523pt,-1pt)and(1pt,-0.5523pt)..(1pt,0pt)--cycle;
\path[-,draw=black,line width=0.4pt,line cap=butt,line join=miter,dash pattern=](-7pt,5pt)--(-7pt,5pt);
\path[-,line width=0.4pt,line cap=butt,line join=miter,dash pattern=on 0.4pt off 1pt](-7pt,5pt)--(-7pt,5pt)--(-7pt,0pt)--(-7pt,0pt)--cycle;
\path[-,draw=black,line width=0.4pt,line cap=butt,line join=miter,dash pattern=on 0.4pt off 1pt](-10pt,8pt)--(-4pt,8pt)--(-4pt,-3pt)--(-10pt,-3pt)--cycle;
\path[-,fill=lightgray,line width=0.4pt,line cap=butt,line join=miter,dash pattern=](-6pt,0pt)..controls(-6pt,0.5523pt)and(-6.4477pt,1pt)..(-7pt,1pt)..controls(-7.5523pt,1pt)and(-8pt,0.5523pt)..(-8pt,0pt)..controls(-8pt,-0.5523pt)and(-7.5523pt,-1pt)..(-7pt,-1pt)..controls(-6.4477pt,-1pt)and(-6pt,-0.5523pt)..(-6pt,0pt)--cycle;
\path[-,line width=0.4pt,line cap=butt,line join=miter,dash pattern=](-18.6316pt,7.3396pt)--(-16pt,7.3396pt)--(-16pt,2.6604pt)--(-18.6316pt,2.6604pt)--cycle;
\node[anchor=north west,inner sep=0] at (-18.6316pt,7.3396pt){\savebox{\marxupbox}{{\({\scriptstyle i}\)}}\immediate\write\boxesfile{103}\immediate\write\boxesfile{\number\wd\marxupbox}\immediate\write\boxesfile{\number\ht\marxupbox}\immediate\write\boxesfile{\number\dp\marxupbox}\box\marxupbox};
\path[-,line width=0.4pt,line cap=butt,line join=miter,dash pattern=](-20.6316pt,9.3396pt)--(-14pt,9.3396pt)--(-14pt,0.6604pt)--(-20.6316pt,0.6604pt)--cycle;
\path[-,line width=0.4pt,line cap=butt,line join=miter,dash pattern=](10pt,3.0788pt)--(13.8617pt,3.0788pt)--(13.8617pt,-3.0788pt)--(10pt,-3.0788pt)--cycle;
\node[anchor=north west,inner sep=0] at (10pt,3.0788pt){\savebox{\marxupbox}{{\({\scriptstyle j}\)}}\immediate\write\boxesfile{104}\immediate\write\boxesfile{\number\wd\marxupbox}\immediate\write\boxesfile{\number\ht\marxupbox}\immediate\write\boxesfile{\number\dp\marxupbox}\box\marxupbox};
\path[-,line width=0.4pt,line cap=butt,line join=miter,dash pattern=](8pt,5.0788pt)--(15.8617pt,5.0788pt)--(15.8617pt,-5.0788pt)--(8pt,-5.0788pt)--cycle;
\end{tikzpicture}}), and
the other is symmetrical. These laws ensure that the object \(\dual{\mathsf{a}}\) is just like the object \(\mathsf{a}\), but travelling backwards (input
and output roles are exchanged). To reflect this, in the diagrammatic
representation of \(η\) and \(ϵ\), we indicate the
\(\dual{\mathsf{a}}\) object with a left-pointing arrow, as shown in
\cref{91}.  Indeed, there is no difference between an input vector
space \(\mathsf{a}\) and an output vector space \(\dual{\mathsf{a}}\). Accordingly,
in the Einstein notation, no difference is being made between
inputs and outputs. Instead only co- or contra-variance is reflected
notationally. Consequently neither \(η\) nor \(ϵ\) are
visible in the Einstein notation, except perhaps as a Kronecker
\(δ\) (see \cref{91}).  For instance, the morphism \(ϵ\mskip 3.0mu\allowbreak{}\mathnormal{∘}\allowbreak{}\mskip 3.0mu\allowbreak{}\mathnormal{(}\mskip 0.0mu\bar{ρ}\mskip 3.0mu\allowbreak{}\mathnormal{∘}\allowbreak{}\mskip 3.0mu\allowbreak{}\mathnormal{(}\mskip 0.0mu\mathsf{id}\mskip 3.0mu{⊗}\mskip 3.0muϵ\mskip 0.0mu\mathnormal{)}\allowbreak{}\mskip 3.0mu{⊗}\mskip 3.0mu\mathsf{id}\mskip 0.0mu\mathnormal{)}\allowbreak{}\mskip 3.0mu\allowbreak{}\mathnormal{∘}\allowbreak{}\mskip 3.0mu\allowbreak{}\mathnormal{(}\mskip 0.0muα\mskip 3.0mu{⊗}\mskip 3.0mu\mathsf{id}\mskip 0.0mu\mathnormal{)}\allowbreak{}\mskip 3.0mu\allowbreak{}\mathnormal{∘}\allowbreak{}\mskip 3.0mu\allowbreak{}\mathnormal{(}\mskip 0.0mu\allowbreak{}\mathnormal{(}\mskip 0.0muσ\mskip 3.0mu{⊗}\mskip 3.0mu\mathsf{id}\mskip 0.0mu\mathnormal{)}\allowbreak{}\mskip 3.0mu{⊗}\mskip 3.0mu\mathsf{id}\mskip 0.0mu\mathnormal{)}\allowbreak{}\),
or {\begin{tikzpicture}[baseline={([yshift=-0.8ex]current bounding box.center)}]\path[-,draw=black,line width=0.4pt,line cap=butt,line join=miter,dash pattern=](-13pt,32.4184pt)--(-12pt,32.4184pt);
\path[-,draw=black,line width=0.4pt,line cap=butt,line join=miter,dash pattern=](-13pt,25pt)--(-12pt,25pt);
\path[-,draw=black,line width=0.4pt,line cap=butt,line join=miter,dash pattern=](-13pt,15pt)--(-12pt,15pt);
\path[-,draw=black,line width=0.4pt,line cap=butt,line join=miter,dash pattern=](-13pt,0pt)--(-12pt,0pt);
\path[-,draw=lightgray,line width=0.4pt,line cap=butt,line join=miter,dash pattern=](20pt,15.6046pt)--(27pt,15.6046pt);
\path[-,draw=black,line width=0.4pt,line cap=butt,line join=miter,dash pattern=](-6pt,31.2092pt)--(-6pt,31.2092pt);
\path[-,draw=black,line width=0.4pt,line cap=butt,line join=miter,dash pattern=](-6pt,26.2092pt)--(-6pt,26.2092pt);
\path[-,draw=black,line width=0.4pt,line cap=butt,line join=miter,dash pattern=](-6pt,15pt)--(-6pt,15pt);
\path[-,draw=black,line width=0.4pt,line cap=butt,line join=miter,dash pattern=](-6pt,0pt)--(-6pt,0pt);
\path[-,draw=black,line width=0.4pt,line cap=butt,line join=miter,dash pattern=](0pt,31.2092pt)--(0pt,31.2092pt);
\path[-,draw=black,line width=0.4pt,line cap=butt,line join=miter,dash pattern=](0pt,20pt)--(0pt,20pt);
\path[-,draw=black,line width=0.4pt,line cap=butt,line join=miter,dash pattern=](0pt,15pt)--(0pt,15pt);
\path[-,draw=black,line width=0.4pt,line cap=butt,line join=miter,dash pattern=](0pt,0pt)--(0pt,0pt);
\path[-,draw=black,line width=0.4pt,line cap=butt,line join=miter,dash pattern=](13pt,31.2092pt)--(14pt,31.2092pt);
\path[-,draw=black,line width=0.4pt,line cap=butt,line join=miter,dash pattern=](13pt,0pt)--(14pt,0pt);
\path[-,draw=black,line width=0.4pt,line cap=butt,line join=miter,dash pattern=](17.06pt,28.3111pt)..controls(19.02pt,24.3009pt)and(20pt,17.3546pt)..(20pt,15.6046pt);
\path[-,draw=black,line width=0.4pt,line cap=butt,line join=miter,dash pattern=](14pt,31.2092pt)..controls(15.2pt,31.2092pt)and(16.22pt,30.0298pt)..(17.06pt,28.3111pt);
\path[to-,draw=black,line width=0.4pt,line cap=butt,line join=miter,dash pattern=](17.06pt,2.8981pt)..controls(19.02pt,6.9083pt)and(20pt,13.8546pt)..(20pt,15.6046pt);
\path[-,draw=black,line width=0.4pt,line cap=butt,line join=miter,dash pattern=](14pt,0pt)..controls(15.2pt,0pt)and(16.22pt,1.1794pt)..(17.06pt,2.8981pt);
\path[-,draw=black,line width=0.4pt,line cap=butt,line join=miter,dash pattern=](6pt,31.2092pt)--(13pt,31.2092pt);
\path[-,draw=lightgray,line width=0.4pt,line cap=butt,line join=miter,dash pattern=](6pt,17.5pt)--(13pt,17.5pt);
\path[-,draw=black,line width=0.4pt,line cap=butt,line join=miter,dash pattern=](13pt,31.2092pt)--(13pt,31.2092pt);
\path[-,line width=0.4pt,line cap=butt,line join=miter,dash pattern=on 0.4pt off 1pt](13pt,31.2092pt)--(13pt,31.2092pt)--(13pt,17.5pt)--(13pt,17.5pt)--cycle;
\path[-,draw=black,line width=0.4pt,line cap=butt,line join=miter,dash pattern=on 0.4pt off 1pt](10pt,34.2092pt)--(16pt,34.2092pt)--(16pt,14.5pt)--(10pt,14.5pt)--cycle;
\path[-,fill=lightgray,line width=0.4pt,line cap=butt,line join=miter,dash pattern=](14pt,17.5pt)..controls(14pt,18.0523pt)and(13.5523pt,18.5pt)..(13pt,18.5pt)..controls(12.4477pt,18.5pt)and(12pt,18.0523pt)..(12pt,17.5pt)..controls(12pt,16.9477pt)and(12.4477pt,16.5pt)..(13pt,16.5pt)..controls(13.5523pt,16.5pt)and(14pt,16.9477pt)..(14pt,17.5pt)--cycle;
\path[-,draw=black,line width=0.4pt,line cap=butt,line join=miter,dash pattern=](0pt,31.2092pt)--(6pt,31.2092pt);
\path[-,draw=black,line width=0.4pt,line cap=butt,line join=miter,dash pattern=](3.06pt,19.9325pt)..controls(5.02pt,19.775pt)and(6pt,19.25pt)..(6pt,17.5pt);
\path[-,draw=black,line width=0.4pt,line cap=butt,line join=miter,dash pattern=](0pt,20pt)..controls(1.2pt,20pt)and(2.22pt,20pt)..(3.06pt,19.9325pt);
\path[to-,draw=black,line width=0.4pt,line cap=butt,line join=miter,dash pattern=](3.06pt,15.0675pt)..controls(5.02pt,15.225pt)and(6pt,15.75pt)..(6pt,17.5pt);
\path[-,draw=black,line width=0.4pt,line cap=butt,line join=miter,dash pattern=](0pt,15pt)..controls(1.2pt,15pt)and(2.22pt,15pt)..(3.06pt,15.0675pt);
\path[-,draw=black,line width=0.4pt,line cap=butt,line join=miter,dash pattern=](0pt,0pt)--(13pt,0pt);
\path[-,draw=black,line width=0.4pt,line cap=butt,line join=miter,dash pattern=](-6pt,31.2092pt)--(0pt,31.2092pt);
\path[-,draw=black,line width=0.4pt,line cap=butt,line join=miter,dash pattern=](-6pt,26.2092pt)..controls(-2pt,26.2092pt)and(-4pt,20pt)..(0pt,20pt);
\path[-,draw=black,line width=0.4pt,line cap=butt,line join=miter,dash pattern=](-6pt,15pt)--(0pt,15pt);
\path[-,draw=black,line width=0.4pt,line cap=butt,line join=miter,dash pattern=](-6pt,0pt)--(0pt,0pt);
\path[-,draw=black,line width=0.4pt,line cap=butt,line join=miter,dash pattern=](-12pt,32.4184pt)..controls(-8pt,32.4184pt)and(-10pt,26.2092pt)..(-6pt,26.2092pt);
\path[-,draw=black,line width=0.4pt,line cap=butt,line join=miter,dash pattern=](-12pt,25pt)..controls(-8pt,25pt)and(-10pt,31.2092pt)..(-6pt,31.2092pt);
\path[-,draw=black,line width=0.4pt,line cap=butt,line join=miter,dash pattern=](-12pt,15pt)--(-6pt,15pt);
\path[-,draw=black,line width=0.4pt,line cap=butt,line join=miter,dash pattern=](-12pt,0pt)--(-6pt,0pt);
\path[-,line width=0.4pt,line cap=butt,line join=miter,dash pattern=](-17.6316pt,34.758pt)--(-15pt,34.758pt)--(-15pt,30.0788pt)--(-17.6316pt,30.0788pt)--cycle;
\node[anchor=north west,inner sep=0] at (-17.6316pt,34.758pt){\savebox{\marxupbox}{{\({\scriptstyle i}\)}}\immediate\write\boxesfile{105}\immediate\write\boxesfile{\number\wd\marxupbox}\immediate\write\boxesfile{\number\ht\marxupbox}\immediate\write\boxesfile{\number\dp\marxupbox}\box\marxupbox};
\path[-,line width=0.4pt,line cap=butt,line join=miter,dash pattern=](-19.6316pt,36.758pt)--(-13pt,36.758pt)--(-13pt,28.0788pt)--(-19.6316pt,28.0788pt)--cycle;
\path[-,line width=0.4pt,line cap=butt,line join=miter,dash pattern=](-18.8617pt,28.0788pt)--(-15pt,28.0788pt)--(-15pt,21.9212pt)--(-18.8617pt,21.9212pt)--cycle;
\node[anchor=north west,inner sep=0] at (-18.8617pt,28.0788pt){\savebox{\marxupbox}{{\({\scriptstyle j}\)}}\immediate\write\boxesfile{106}\immediate\write\boxesfile{\number\wd\marxupbox}\immediate\write\boxesfile{\number\ht\marxupbox}\immediate\write\boxesfile{\number\dp\marxupbox}\box\marxupbox};
\path[-,line width=0.4pt,line cap=butt,line join=miter,dash pattern=](-20.8617pt,30.0788pt)--(-13pt,30.0788pt)--(-13pt,19.9212pt)--(-20.8617pt,19.9212pt)--cycle;
\path[-,line width=0.4pt,line cap=butt,line join=miter,dash pattern=](-19.3398pt,17.4911pt)--(-15pt,17.4911pt)--(-15pt,12.5089pt)--(-19.3398pt,12.5089pt)--cycle;
\node[anchor=north west,inner sep=0] at (-19.3398pt,17.4911pt){\savebox{\marxupbox}{{\({\scriptstyle k}\)}}\immediate\write\boxesfile{107}\immediate\write\boxesfile{\number\wd\marxupbox}\immediate\write\boxesfile{\number\ht\marxupbox}\immediate\write\boxesfile{\number\dp\marxupbox}\box\marxupbox};
\path[-,line width=0.4pt,line cap=butt,line join=miter,dash pattern=](-21.3398pt,19.4911pt)--(-13pt,19.4911pt)--(-13pt,10.5089pt)--(-21.3398pt,10.5089pt)--cycle;
\path[-,line width=0.4pt,line cap=butt,line join=miter,dash pattern=](-17.5769pt,2.5258pt)--(-15pt,2.5258pt)--(-15pt,-2.5258pt)--(-17.5769pt,-2.5258pt)--cycle;
\node[anchor=north west,inner sep=0] at (-17.5769pt,2.5258pt){\savebox{\marxupbox}{{\({\scriptstyle l}\)}}\immediate\write\boxesfile{108}\immediate\write\boxesfile{\number\wd\marxupbox}\immediate\write\boxesfile{\number\ht\marxupbox}\immediate\write\boxesfile{\number\dp\marxupbox}\box\marxupbox};
\path[-,line width=0.4pt,line cap=butt,line join=miter,dash pattern=](-19.5769pt,4.5258pt)--(-13pt,4.5258pt)--(-13pt,-4.5258pt)--(-19.5769pt,-4.5258pt)--cycle;
\end{tikzpicture}} in diagram notation, is written
\(δ{_i}{^k}δ{_j}{^l}\) in Einstein notation.
\cref{99} shows how an input object \(\mathsf{a}\) (of any morphism) can be converted to an output \(\dual{\mathsf{a}}\),
and \textit{vice versa}.  One can even combine both ideas and connect
the output of a morphism \(\mathsf{t}\) back to its input.  By doing so, one
constructs the \emph{trace} of \(\mathsf{t}\).\footnote{When the contracted index corresponds to a vector space of dimension \(\mathsf{n}\) and letting
\(\mathsf{t}\) be the identity, we find that \(ϵ\mskip 3.0mu\allowbreak{}\mathnormal{∘}\allowbreak{}\mskip 3.0muσ\mskip 3.0mu\allowbreak{}\mathnormal{∘}\allowbreak{}\mskip 3.0muη\mskip 3.0mu\mathnormal{=}\mskip 3.0mu\mathsf{trace}\mskip 3.0mu\mathsf{δ}\mskip 3.0mu\mathnormal{=}\mskip 3.0mu\mathsf{n}\).\label{109}} 

We now have a complete description of the tensor combinators --- the core of {\sc{}Roger}.
Unfortunately, in practice it is inconvenient to use as
such. Indeed, most of the tensor expressions encountered in practice
consists of building a network of connections between atomic
blocks. Unfortunately, using the categorical combinators for this
purpose is tedious. For instance, contracting two input indices is
particularly tedious in the point-free notation, because it is
realised as a composition with \(η\) or \(ϵ\) with a
large number of {\sc{}smc} combinators to select the appropriate
dimensions to contract.  It is akin to programming with SKI
combinators instead of using the lambda calculus. Using variable names for indices,
like in the Einstein notation, would be much more convenient.  We will get there in \cref{114}.

\subsection{Matrix Instances}\label{110} 
An important instance of the compact closed category structure is the
category of matrices of coefficients, which we encountered in
\cref{30}.  In our host functional language we define them as a
function from (both input and output) indices to coefficients (of type
\(\mathsf{S}\)):\footnote{This functional representation is for
expository purposes. It must be tabulated to avoid bad
runtime behaviour.\label{111}}  \begin{list}{}{\setlength\leftmargin{1.0em}}\item\relax
\ensuremath{\begin{parray}\column{B}{@{}>{}l<{}@{}}\column[0em]{1}{@{}>{}l<{}@{}}\column{E}{@{}>{}l<{}@{}}%
\>[1]{\mathbf{newtype}\mskip 3.0muM_{\mathsf{e}}\mskip 3.0mu\mathsf{a}\mskip 3.0mu\mathsf{b}\mskip 3.0mu\mathnormal{=}\mskip 3.0mu\mathsf{Tab}\mskip 3.0mu\allowbreak{}\mathnormal{(}\mskip 0.0mu\mathsf{a}\mskip 3.0mu\mathnormal{\rightarrow }\mskip 3.0mu\mathsf{b}\mskip 3.0mu\mathnormal{\rightarrow }\mskip 3.0mu\mathsf{S}\mskip 0.0mu\mathnormal{)}\allowbreak{}}\<[E]\end{parray}}\end{list} 
To emphasise the dependency on the basis, we use a subscript when
referring to a specific matrix category morphism, as in \(M_{\mathsf{b}}\) where \(\mathsf{b}\) is a reference to the choice of basis. In the Haskell implementation,
this basis is represented by a \emph{phantom} type parameter.

The identity morphism is the identity matrix, and composition is matrix multiplication:
\begin{list}{}{\setlength\leftmargin{1.0em}}\item\relax
\ensuremath{\begin{parray}\column{B}{@{}>{}l<{}@{}}\column[0em]{1}{@{}>{}l<{}@{}}\column[1em]{2}{@{}>{}l<{}@{}}\column{3}{@{}>{}l<{}@{}}\column{E}{@{}>{}l<{}@{}}%
\>[1]{\mathbf{instance}\mskip 3.0mu}\>[3]{\mathsf{Category}\mskip 3.0muM_{\mathsf{e}}\mskip 3.0mu\mathbf{where}}\<[E]\\
\>[2]{\mathsf{id}\mskip 3.0mu\mathnormal{=}\mskip 3.0mu\mathsf{Tab}\mskip 3.0muδ}\<[E]\\
\>[2]{\mathsf{Tab}\mskip 3.0mu\mathsf{g}\mskip 3.0mu\allowbreak{}\mathnormal{∘}\allowbreak{}\mskip 3.0mu\mathsf{Tab}\mskip 3.0mu\mathsf{f}\mskip 3.0mu\mathnormal{=}\mskip 3.0mu\mathsf{Tab}\mskip 3.0mu\allowbreak{}\mathnormal{(}\mskip 0.0muλ\mskip 3.0mu\mathsf{i}\mskip 3.0mu\mathsf{j}\mskip 3.0mu\mathnormal{\rightarrow }\mskip 3.0mu\mathsf{summation}\mskip 3.0mu\allowbreak{}\mathnormal{(}\mskip 0.0muλ\mskip 3.0mu\mathsf{k}\mskip 3.0mu\mathnormal{\rightarrow }\mskip 3.0mu\mathsf{f}\mskip 3.0mu\mathsf{i}\mskip 3.0mu\mathsf{k}\mskip 3.0mu\mathnormal{*}\mskip 3.0mu\mathsf{g}\mskip 3.0mu\mathsf{k}\mskip 3.0mu\mathsf{j}\mskip 0.0mu\mathnormal{)}\allowbreak{}\mskip 0.0mu\mathnormal{)}\allowbreak{}}\<[E]\\
\>[1]{δ\mskip 3.0mu\mathsf{i}\mskip 3.0mu\mathsf{j}\mskip 3.0mu\mathnormal{=}\mskip 3.0mu\mathbf{if}\mskip 3.0mu\mathsf{i}\mskip 3.0mu\allowbreak{}\doubleequals \allowbreak{}\mskip 3.0mu\mathsf{j}\mskip 3.0mu\mathbf{then}\mskip 3.0mu\mathrm{1}\mskip 3.0mu\mathbf{else}\mskip 3.0mu\mathrm{0}}\<[E]\end{parray}}\end{list} In this instance the
objects are identified with sets that
index the bases of the vector spaces that they stand for. These sets
are assumed to be enumerable and bounded (so we have access to their
\(\mathsf{inhabitants}\)) and we can compare indices for equality.\footnote{In Haskell, the constraints on indices \(\allowbreak{}\mathnormal{(}\mskip 0.0mu\mathsf{Enum}\mskip 0.0mu\mathnormal{,}\mskip 3.0mu\mathsf{Bounded}\mskip 0.0mu\mathnormal{,}\mskip 3.0mu\mathsf{Eq}\mskip 0.0mu\mathnormal{)}\allowbreak{}\) are tracked using an \emph{associated class constraint} on objects, which adds significant verbosity but is a well-understood
technique
\citep{orchard_haskell_2010,sculthorpe_constrained-monad_2013}.  For
concision we omit object constraints entirely in this presentation.\label{112}} 
The instance of the {\sc{}smc} structure for matrix
representations in coherent bases is then: \begin{list}{}{\setlength\leftmargin{1.0em}}\item\relax
\ensuremath{\begin{parray}\column{B}{@{}>{}l<{}@{}}\column[0em]{1}{@{}>{}l<{}@{}}\column[1em]{2}{@{}>{}l<{}@{}}\column{3}{@{}>{}l<{}@{}}\column{4}{@{}>{}l<{}@{}}\column{E}{@{}>{}l<{}@{}}%
\>[1]{\mathbf{instance}\mskip 3.0mu}\>[4]{\mathsf{SymmetricMonoidal}\mskip 3.0muM_{\mathsf{e}}\mskip 3.0mu\mathbf{where}}\<[E]\\
\>[2]{\allowbreak{}\mathnormal{(}\mskip 0.0mu{⊗}\mskip 0.0mu\mathnormal{)}\allowbreak{}\mskip 3.0mu\mathnormal{=}\mskip 3.0mu\mathsf{kroneckerProduct}}\<[E]\\
\>[2]{ρ\mskip 3.0mu}\>[3]{\mathnormal{=}\mskip 3.0mu\mathsf{Tab}\mskip 3.0mu\allowbreak{}\mathnormal{(}\mskip 0.0muλ\mskip 3.0mu\mathsf{x}\mskip 3.0mu\allowbreak{}\mathnormal{(}\mskip 0.0mu\mathsf{y}\mskip 0.0mu\mathnormal{,}\mskip 3.0mu\allowbreak{}\mathnormal{(}\mskip 0.0mu\mathnormal{)}\allowbreak{}\mskip 0.0mu\mathnormal{)}\allowbreak{}\mskip 3.0mu\mathnormal{\rightarrow }\mskip 3.0muδ\mskip 3.0mu\mathsf{x}\mskip 3.0mu\mathsf{y}\mskip 0.0mu\mathnormal{)}\allowbreak{}}\<[E]\\
\>[2]{\bar{ρ}\mskip 3.0mu}\>[3]{\mathnormal{=}\mskip 3.0mu\mathsf{Tab}\mskip 3.0mu\allowbreak{}\mathnormal{(}\mskip 0.0muλ\mskip 3.0mu\allowbreak{}\mathnormal{(}\mskip 0.0mu\mathsf{y}\mskip 0.0mu\mathnormal{,}\mskip 3.0mu\allowbreak{}\mathnormal{(}\mskip 0.0mu\mathnormal{)}\allowbreak{}\mskip 0.0mu\mathnormal{)}\allowbreak{}\mskip 3.0mu\mathsf{x}\mskip 3.0mu\mathnormal{\rightarrow }\mskip 3.0muδ\mskip 3.0mu\mathsf{x}\mskip 3.0mu\mathsf{y}\mskip 0.0mu\mathnormal{)}\allowbreak{}}\<[E]\\
\>[2]{α\mskip 3.0mu}\>[3]{\mathnormal{=}\mskip 3.0mu\mathsf{Tab}\mskip 3.0mu\allowbreak{}\mathnormal{(}\mskip 0.0muλ\mskip 3.0mu\allowbreak{}\mathnormal{(}\mskip 0.0mu\allowbreak{}\mathnormal{(}\mskip 0.0mu\mathsf{x}\mskip 0.0mu\mathnormal{,}\mskip 3.0mu\mathsf{y}\mskip 0.0mu\mathnormal{)}\allowbreak{}\mskip 0.0mu\mathnormal{,}\mskip 3.0mu\mathsf{z}\mskip 0.0mu\mathnormal{)}\allowbreak{}\mskip 3.0mu\allowbreak{}\mathnormal{(}\mskip 0.0mu\mathsf{x'}\mskip 0.0mu\mathnormal{,}\mskip 3.0mu\allowbreak{}\mathnormal{(}\mskip 0.0mu\mathsf{y'}\mskip 0.0mu\mathnormal{,}\mskip 3.0mu\mathsf{z'}\mskip 0.0mu\mathnormal{)}\allowbreak{}\mskip 0.0mu\mathnormal{)}\allowbreak{}\mskip 3.0mu\mathnormal{\rightarrow }\mskip 3.0muδ\mskip 3.0mu\allowbreak{}\mathnormal{(}\mskip 0.0mu\allowbreak{}\mathnormal{(}\mskip 0.0mu\mathsf{x}\mskip 0.0mu\mathnormal{,}\mskip 3.0mu\mathsf{y}\mskip 0.0mu\mathnormal{)}\allowbreak{}\mskip 0.0mu\mathnormal{,}\mskip 3.0mu\mathsf{z}\mskip 0.0mu\mathnormal{)}\allowbreak{}\mskip 3.0mu\allowbreak{}\mathnormal{(}\mskip 0.0mu\allowbreak{}\mathnormal{(}\mskip 0.0mu\mathsf{x'}\mskip 0.0mu\mathnormal{,}\mskip 3.0mu\mathsf{y'}\mskip 0.0mu\mathnormal{)}\allowbreak{}\mskip 0.0mu\mathnormal{,}\mskip 3.0mu\mathsf{z'}\mskip 0.0mu\mathnormal{)}\allowbreak{}\mskip 0.0mu\mathnormal{)}\allowbreak{}}\<[E]\\
\>[2]{\bar{α}\mskip 3.0mu}\>[3]{\mathnormal{=}\mskip 3.0mu\mathsf{Tab}\mskip 3.0mu\allowbreak{}\mathnormal{(}\mskip 0.0muλ\mskip 3.0mu\allowbreak{}\mathnormal{(}\mskip 0.0mu\mathsf{x'}\mskip 0.0mu\mathnormal{,}\mskip 3.0mu\allowbreak{}\mathnormal{(}\mskip 0.0mu\mathsf{y'}\mskip 0.0mu\mathnormal{,}\mskip 3.0mu\mathsf{z'}\mskip 0.0mu\mathnormal{)}\allowbreak{}\mskip 0.0mu\mathnormal{)}\allowbreak{}\mskip 3.0mu\allowbreak{}\mathnormal{(}\mskip 0.0mu\allowbreak{}\mathnormal{(}\mskip 0.0mu\mathsf{x}\mskip 0.0mu\mathnormal{,}\mskip 3.0mu\mathsf{y}\mskip 0.0mu\mathnormal{)}\allowbreak{}\mskip 0.0mu\mathnormal{,}\mskip 3.0mu\mathsf{z}\mskip 0.0mu\mathnormal{)}\allowbreak{}\mskip 3.0mu\mathnormal{\rightarrow }\mskip 3.0muδ\mskip 3.0mu\allowbreak{}\mathnormal{(}\mskip 0.0mu\allowbreak{}\mathnormal{(}\mskip 0.0mu\mathsf{x}\mskip 0.0mu\mathnormal{,}\mskip 3.0mu\mathsf{y}\mskip 0.0mu\mathnormal{)}\allowbreak{}\mskip 0.0mu\mathnormal{,}\mskip 3.0mu\mathsf{z}\mskip 0.0mu\mathnormal{)}\allowbreak{}\mskip 3.0mu\allowbreak{}\mathnormal{(}\mskip 0.0mu\allowbreak{}\mathnormal{(}\mskip 0.0mu\mathsf{x'}\mskip 0.0mu\mathnormal{,}\mskip 3.0mu\mathsf{y'}\mskip 0.0mu\mathnormal{)}\allowbreak{}\mskip 0.0mu\mathnormal{,}\mskip 3.0mu\mathsf{z'}\mskip 0.0mu\mathnormal{)}\allowbreak{}\mskip 0.0mu\mathnormal{)}\allowbreak{}}\<[E]\\
\>[2]{σ\mskip 3.0mu}\>[3]{\mathnormal{=}\mskip 3.0mu\mathsf{Tab}\mskip 3.0mu\allowbreak{}\mathnormal{(}\mskip 0.0muλ\mskip 3.0mu\allowbreak{}\mathnormal{(}\mskip 0.0mu\mathsf{x}\mskip 0.0mu\mathnormal{,}\mskip 3.0mu\mathsf{y}\mskip 0.0mu\mathnormal{)}\allowbreak{}\mskip 3.0mu\allowbreak{}\mathnormal{(}\mskip 0.0mu\mathsf{y'}\mskip 0.0mu\mathnormal{,}\mskip 3.0mu\mathsf{x'}\mskip 0.0mu\mathnormal{)}\allowbreak{}\mskip 3.0mu\mathnormal{\rightarrow }\mskip 3.0muδ\mskip 3.0mu\allowbreak{}\mathnormal{(}\mskip 0.0mu\mathsf{x}\mskip 0.0mu\mathnormal{,}\mskip 3.0mu\mathsf{y}\mskip 0.0mu\mathnormal{)}\allowbreak{}\mskip 3.0mu\allowbreak{}\mathnormal{(}\mskip 0.0mu\mathsf{x'}\mskip 0.0mu\mathnormal{,}\mskip 3.0mu\mathsf{y'}\mskip 0.0mu\mathnormal{)}\allowbreak{}\mskip 0.0mu\mathnormal{)}\allowbreak{}}\<[E]\\
\>[1]{\mathsf{kroneckerProduct}\mskip 3.0mu\allowbreak{}\mathnormal{(}\mskip 0.0mu\mathsf{Tab}\mskip 3.0mu\mathsf{f}\mskip 0.0mu\mathnormal{)}\allowbreak{}\mskip 3.0mu\allowbreak{}\mathnormal{(}\mskip 0.0mu\mathsf{Tab}\mskip 3.0mu\mathsf{g}\mskip 0.0mu\mathnormal{)}\allowbreak{}\mskip 3.0mu\mathnormal{=}\mskip 3.0mu\mathsf{Tab}\mskip 3.0mu\allowbreak{}\mathnormal{(}\mskip 0.0muλ\mskip 3.0mu\allowbreak{}\mathnormal{(}\mskip 0.0mu\mathsf{i}\mskip 0.0mu\mathnormal{,}\mskip 3.0mu\mathsf{k}\mskip 0.0mu\mathnormal{)}\allowbreak{}\mskip 3.0mu\allowbreak{}\mathnormal{(}\mskip 0.0mu\mathsf{j}\mskip 0.0mu\mathnormal{,}\mskip 3.0mu\mathsf{l}\mskip 0.0mu\mathnormal{)}\allowbreak{}\mskip 3.0mu\mathnormal{\rightarrow }\mskip 3.0mu\mathsf{f}\mskip 3.0mu\mathsf{i}\mskip 3.0mu\mathsf{j}\mskip 3.0mu\mathnormal{*}\mskip 3.0mu\mathsf{g}\mskip 3.0mu\mathsf{k}\mskip 3.0mu\mathsf{l}\mskip 0.0mu\mathnormal{)}\allowbreak{}}\<[E]\end{parray}}\end{list} Because objects index the bases of the corresponding vector
spaces, tensor products are represented as usual pairs.
In the above definition, the right-hand sides are Haskell code.  This
means that the asterisk operator \(\allowbreak{}\mathnormal{(}\mskip 0.0mu\mathnormal{*}\mskip 0.0mu\mathnormal{)}\allowbreak{}\) denotes multiplication
between scalars \emph{as components of matrices}.  In contrast, the
operator \(\allowbreak{}\mathnormal{(}\mskip 0.0mu{\tikzstar{0.11}{0.25}{5}{-18}{fill=black}}\mskip 0.0mu\mathnormal{)}\allowbreak{}\) defined in \cref{119} denotes multiplication between abstract scalar
(order-0) tensor expressions (independent of the chosen tensor
representation).

With the coherent choice of bases, \(η\) and \(ϵ\) are
simply realised as the identity.  
\begin{list}{}{\setlength\leftmargin{1.0em}}\item\relax
\ensuremath{\begin{parray}\column{B}{@{}>{}l<{}@{}}\column[0em]{1}{@{}>{}l<{}@{}}\column[1em]{2}{@{}>{}l<{}@{}}\column{3}{@{}>{}l<{}@{}}\column{E}{@{}>{}l<{}@{}}%
\>[1]{\mathbf{instance}\mskip 3.0mu}\>[3]{\mathsf{CompactClosed}\mskip 3.0muM_{\mathsf{e}}\mskip 3.0mu\mathbf{where}}\<[E]\\
\>[2]{η\mskip 3.0mu\mathnormal{=}\mskip 3.0mu\mathsf{Tab}\mskip 3.0mu\allowbreak{}\mathnormal{(}\mskip 0.0muλ\mskip 3.0mu\allowbreak{}\mathnormal{(}\mskip 0.0mu\mathnormal{)}\allowbreak{}\mskip 3.0mu\allowbreak{}\mathnormal{(}\mskip 0.0mu\mathsf{Dual}\mskip 3.0mu\mathsf{x}\mskip 0.0mu\mathnormal{,}\mskip 3.0mu\mathsf{y}\mskip 0.0mu\mathnormal{)}\allowbreak{}\mskip 3.0mu\mathnormal{\rightarrow }\mskip 3.0muδ\mskip 3.0mu\mathsf{x}\mskip 3.0mu\mathsf{y}\mskip 0.0mu\mathnormal{)}\allowbreak{}}\<[E]\\
\>[2]{ϵ\mskip 3.0mu\mathnormal{=}\mskip 3.0mu\mathsf{Tab}\mskip 3.0mu\allowbreak{}\mathnormal{(}\mskip 0.0muλ\mskip 3.0mu\allowbreak{}\mathnormal{(}\mskip 0.0mu\mathsf{x}\mskip 0.0mu\mathnormal{,}\mskip 3.0mu\mathsf{Dual}\mskip 3.0mu\mathsf{y}\mskip 0.0mu\mathnormal{)}\allowbreak{}\mskip 3.0mu\allowbreak{}\mathnormal{(}\mskip 0.0mu\mathnormal{)}\allowbreak{}\mskip 3.0mu\mathnormal{\rightarrow }\mskip 3.0muδ\mskip 3.0mu\mathsf{x}\mskip 3.0mu\mathsf{y}\mskip 0.0mu\mathnormal{)}\allowbreak{}}\<[E]\end{parray}}\end{list} The object \(\dual{\mathsf{a}}\) has the same dimensionality as
\(\mathsf{a}\), so in our Haskell encoding we use \(\mathbf{newtype}\) for it,
\ensuremath{\begin{parray}\column{B}{@{}>{}l<{}@{}}\column[0em]{1}{@{}>{}l<{}@{}}\column{E}{@{}>{}l<{}@{}}%
\>[1]{\mathbf{newtype}\mskip 3.0mu\mathsf{DualObject}\mskip 3.0mu\mathsf{a}\mskip 3.0mu\mathnormal{=}\mskip 3.0mu\mathsf{Dual}\mskip 3.0mu\mathsf{a}}\<[E]\end{parray}}. For concision (and following tradition) we use an asterisk as a shorthand,
so \(\dual{\mathsf{a}}\) stands for the type \(\mathsf{DualObject}\mskip 3.0mu\mathsf{a}\).

\paragraph*{Coordinate representation functors.}\hspace{1.0ex}\label{113} It is worth mentioning that the transformation functions between
linear maps \(\mathsf{L}\) and their representations \(M_{\mathsf{e}}\) in a given basis \(\mathsf{e}\) are a pair of functors which are the identity on objects and just change the morphisms:
\begin{list}{}{\setlength\leftmargin{1.0em}}\item\relax
\ensuremath{\begin{parray}\column{B}{@{}>{}l<{}@{}}\column[0em]{1}{@{}>{}l<{}@{}}\column{2}{@{}>{}l<{}@{}}\column{3}{@{}>{}l<{}@{}}\column{4}{@{}>{}l<{}@{}}\column{5}{@{}>{}l<{}@{}}\column{E}{@{}>{}l<{}@{}}%
\>[1]{\mathsf{toCoordinates}_{e}\mskip 3.0mu}\>[2]{\mathnormal{::}\mskip 3.0mu}\>[3]{\allowbreak{}\mathnormal{(}\mskip 0.0mu\mathsf{a}\mskip 1.0mu\overset{\mathsf{L}}{\leadsto }\mskip 1.0mu\mathsf{b}\mskip 0.0mu\mathnormal{)}\allowbreak{}\mskip 3.0mu}\>[4]{\mathnormal{\rightarrow }\mskip 3.0mu\allowbreak{}\mathnormal{(}\mskip 0.0mu\mathsf{a}\mskip 1.0mu\overset{M_{\mathsf{e}}}{\leadsto }\mskip 1.0mu\mathsf{b}\mskip 0.0mu\mathnormal{)}\allowbreak{}}\<[E]\\
\>[1]{\mathsf{fromCoordinates}_{d}\mskip 3.0mu}\>[2]{\mathnormal{::}\mskip 3.0mu}\>[3]{\allowbreak{}\mathnormal{(}\mskip 0.0mu\mathsf{a}\mskip 1.0mu\overset{M_{\mathsf{d}}}{\leadsto }\mskip 1.0mu\mathsf{b}\mskip 0.0mu\mathnormal{)}\allowbreak{}\mskip 3.0mu}\>[5]{\mathnormal{\rightarrow }\mskip 3.0mu\allowbreak{}\mathnormal{(}\mskip 0.0mu\mathsf{a}\mskip 1.0mu\overset{\mathsf{L}}{\leadsto }\mskip 1.0mu\mathsf{b}\mskip 0.0mu\mathnormal{)}\allowbreak{}}\<[E]\end{parray}}\end{list} Furthermore, this pair defines an isomorphism between the respective compact-closed categories.
Therefore, even though different representations form different categories, one can always transform one to another.
 The transformation between systems of coordinates, usually presented using
transformation matrices (see \cref{30}) can
be understood as the composition of \(\mathsf{fromCoordinates}\) in the source basis, and
\(\mathsf{toCoordinates}\) in the target basis:
\begin{list}{}{\setlength\leftmargin{1.0em}}\item\relax
\ensuremath{\begin{parray}\column{B}{@{}>{}l<{}@{}}\column[0em]{1}{@{}>{}l<{}@{}}\column{2}{@{}>{}l<{}@{}}\column{3}{@{}>{}l<{}@{}}\column{E}{@{}>{}l<{}@{}}%
\>[1]{\mathsf{transform}\mskip 3.0mu}\>[2]{\mathnormal{::}\mskip 3.0mu\allowbreak{}\mathnormal{(}\mskip 0.0mu\mathsf{a}\mskip 1.0mu\overset{M_{\mathsf{d}}}{\leadsto }\mskip 1.0mu\mathsf{b}\mskip 0.0mu\mathnormal{)}\allowbreak{}\mskip 3.0mu}\>[3]{\mathnormal{\rightarrow }\mskip 3.0mu\allowbreak{}\mathnormal{(}\mskip 0.0mu\mathsf{a}\mskip 1.0mu\overset{M_{\mathsf{e}}}{\leadsto }\mskip 1.0mu\mathsf{b}\mskip 0.0mu\mathnormal{)}\allowbreak{}}\<[E]\\
\>[1]{\mathsf{transform}\mskip 3.0mu\mathnormal{=}\mskip 3.0mu\mathsf{toCoordinates}_{e}\mskip 3.0mu\allowbreak{}\mathnormal{∘}\allowbreak{}\mskip 3.0mu\mathsf{fromCoordinates}_{d}}\<[E]\end{parray}}\end{list}  
\section{Design of {\sc{}Albert}}\label{114} 
In this section we provide the design of {\sc{}Albert}.
We intend our design to match Einstein notation as closely as possible. This
aim is achieved except for the following two differences:
\begin{enumerate}\item{}Indices can range over the dimensions of any space (not just atomic vector spaces\footnote{Einstein notation can be easily extended to make indices range over the dimensions of any space, even though this is rarely found in the literature.\label{115}})\item{}Indices are always explicitly bound\end{enumerate} The first difference is motivated by polymorphism considerations: we make many functions polymorphic over the vector space that they manipulate, and as a consequence the corresponding indices can range over the dimensions of tensor or scalar vector spaces.
For instance, when we write \(\mathsf{delta}\mskip 3.0mu_{\mathsf{i}}\mskip 3.0mu^{\mathsf{j}}\), indices (\(_{\mathsf{i}}\),
\(^{\mathsf{j}}\)) may range over order-2 tensor spaces, in which case the corresponding Einstein notation would be the product of two Kronecker deltas.

The second difference is motivated by the need to follow the functional programming conventions, which is required to
embed the {\sc{}dsl} in Haskell, or any functional language without macros.
Besides, to avoid confusion, in {\sc{}Albert} we choose names in all letters for combinators. For instance, when the conventional notation is the Greek letter \(δ\),
we write \(\mathsf{delta}\) in {\sc{}Albert}.

The principles and most of the interface of {\sc{}Albert} is presented in this
section. The tensor-field specific functions are discussed in
\cref{135}. The complete interface is summarised in \cref{263} and \cref{264}.

\paragraph*{Types}\hspace{1.0ex}\label{116} All types are parameterised by \(z\), the category which tensors inhabit.
The type of an index ranging over the dimensions of vector space
\(\mathsf{a}\) is \(\mathsf{P}\mskip 3.0muz\mskip 3.0mu\mathsf{r}\mskip 3.0mu\mathsf{a}\) (think of \(\mathsf{P}\) as ``port'' or ``end of a wire carrying \(\mathsf{a}\)'' in
the diagrams), where the variable \(\mathsf{r}\) is a technical (scoping) device
(made precise in \cref{394}). For the purpose of using {\sc{}Albert}, it
suffices to know that this variable \(\mathsf{r}\) should be consistent
throughout any given expression.

The type of expressions is \(\mathsf{R}\mskip 3.0muz\mskip 3.0mu\mathsf{r}\). Expressions of this type
closely match expressions in Einstein notation.  In
particular, expressions with several free index variables
correspond to higher-order tensors. For example, assuming two
free index variables
\(_{\mathsf{i}}\mskip 3.0mu\mathnormal{::}\mskip 3.0mu\mathsf{P}\mskip 3.0muz\mskip 3.0mu\mathsf{r}\mskip 3.0mu\mathsf{T}\) and
\(^{\mathsf{j}}\mskip 3.0mu\mathnormal{::}\mskip 3.0mu\mathsf{P}\mskip 3.0muz\mskip 3.0mu\mathsf{r}\mskip 3.0mu\dual{\mathsf{T}}\),
then
\(\mathsf{w}\mskip 3.0mu_{\mathsf{i}}\mskip 3.0mu\mathnormal{::}\mskip 3.0mu\mathsf{R}\mskip 3.0muz\mskip 3.0mu\mathsf{r}\) represents a covector over \(\mathsf{T}\);
\(\mathsf{v}\mskip 3.0mu^{\mathsf{j}}\mskip 3.0mu\mathnormal{::}\mskip 3.0mu\mathsf{R}\mskip 3.0muz\mskip 3.0mu\mathsf{r}\) represents a vector in \(\mathsf{T}\);
and
\(\mathsf{t}\mskip 3.0mu_{\mathsf{i}}\mskip 3.0mu^{\mathsf{j}}\mskip 3.0mu\mathnormal{::}\mskip 3.0mu\mathsf{R}\mskip 3.0muz\mskip 3.0mu\mathsf{r}\) represents a linear map from \(\mathsf{T}\) to \(\mathsf{T}\).
(Why this is so will become clear when we present the semantics of tensors, which we will do before the end of the section.)

In sum, exactly as in Einstein notation, our tensor expressions define
and manipulate tensors as (abstract) scalar-valued expressions depending on
indices. Likewise, the order of the underlying tensor is the sum
of the order of the free index variables occurring in it.
Even though we present index variables as either super- or subscripts,
they are just regular variable names.

Because the underlying category \(z\) is not Cartesian, every input must be connected to a single output, and \textit{vice-versa}.
Hence, index variables occur exactly once in each term. We
enforce this restriction by using (and binding)
index variables linearly.\footnote{There are two meanings to the word ``linear'', which we both use in this
paper. The first meaning, exclusively used above
refers to linear algebra (\cref{24}). The second meaning is evoked by the phrase ``linear type'',
and we use it to enforce usage restrictions of the indices appearing
in the tensor expressions. When any ambiguity might exist, we write
``type-linear'' to refer to this second meaning.\label{117}} Accordingly, the types of the variables \(\mathsf{v}\mskip 0.0mu\mathnormal{,}\mskip 3.0mu\mathsf{w}\mskip 0.0mu\mathnormal{,}\mskip 3.0mu\mathsf{t}\) mentioned above
involve (type-)linear functions. For instance
\(\mathsf{w}\mskip 3.0mu\mathnormal{::}\mskip 3.0mu\mathsf{P}\mskip 3.0muz\mskip 3.0mu\mathsf{r}\mskip 3.0mu\mathsf{T}\mskip 3.0mu\mathnormal{⊸}\mskip 3.0mu\mathsf{R}\mskip 3.0muz\mskip 3.0mu\mathsf{r}\) is a covector over \(\mathsf{T}\);
\(\mathsf{v}\mskip 3.0mu\mathnormal{::}\mskip 3.0mu\mathsf{P}\mskip 3.0muz\mskip 3.0mu\mathsf{r}\mskip 3.0mu\dual{\mathsf{T}}\mskip 3.0mu\mathnormal{⊸}\mskip 3.0mu\mathsf{R}\mskip 3.0muz\mskip 3.0mu\mathsf{r}\) is a vector in \(\mathsf{T}\);
and
\(\mathsf{t}\mskip 3.0mu\mathnormal{::}\mskip 3.0mu\mathsf{P}\mskip 3.0muz\mskip 3.0mu\mathsf{r}\mskip 3.0mu\mathsf{T}\mskip 3.0mu\mathnormal{⊸}\mskip 3.0mu\mathsf{P}\mskip 3.0muz\mskip 3.0mu\mathsf{r}\mskip 3.0mu\dual{\mathsf{T}}\mskip 3.0mu\mathnormal{⊸}\mskip 3.0mu\mathsf{R}\mskip 3.0muz\mskip 3.0mu\mathsf{r}\) is a linear map over \(\mathsf{T}\).
This means that {\sc{}Albert} uses a higher-order abstract syntax, that is, the abstraction
mechanism of the host language provides us with the means to abstract
over index variables.
The order of a tensor variable is given by taking the sum of the order of the index parameters in its type.
So, for instance, \(\mathsf{delta}\) has order \(2m\) if its type argument \(\mathsf{a}\) stands for a vector space of order \(m\):
\begin{list}{}{\setlength\leftmargin{1.0em}}\item\relax
\ensuremath{\begin{parray}\column{B}{@{}>{}l<{}@{}}\column[0em]{1}{@{}>{}l<{}@{}}\column{2}{@{}>{}l<{}@{}}\column{E}{@{}>{}l<{}@{}}%
\>[1]{\mathsf{delta}\mskip 3.0mu\mathnormal{::}\mskip 3.0mu}\>[2]{\mathsf{P}\mskip 3.0muz\mskip 3.0mu\mathsf{r}\mskip 3.0mu\mathsf{a}\mskip 3.0mu\mathnormal{⊸}\mskip 3.0mu\mathsf{P}\mskip 3.0muz\mskip 3.0mu\mathsf{r}\mskip 3.0mu\dual{\mathsf{a}}\mskip 3.0mu\mathnormal{⊸}\mskip 3.0mu\mathsf{R}\mskip 3.0muz\mskip 3.0mu\mathsf{r}}\<[E]\end{parray}}\end{list} 
\paragraph*{Tensor embedding, evaluation, and index manipulation}\hspace{1.0ex}\label{118} Next, we turn our
attention to embedding {\sc{}Roger} into {\sc{}Albert}. This is done by means of the following
combinators: \begin{list}{}{\setlength\leftmargin{1.0em}}\item\relax
\ensuremath{\begin{parray}\column{B}{@{}>{}l<{}@{}}\column[0em]{1}{@{}>{}l<{}@{}}\column{2}{@{}>{}l<{}@{}}\column{3}{@{}>{}l<{}@{}}\column{4}{@{}>{}l<{}@{}}\column{5}{@{}>{}l<{}@{}}\column{6}{@{}>{}l<{}@{}}\column{7}{@{}>{}l<{}@{}}\column{8}{@{}>{}l<{}@{}}\column{E}{@{}>{}l<{}@{}}%
\>[1]{\mathsf{tensorEmbed}\mskip 3.0mu}\>[2]{\mathnormal{::}\mskip 3.0mu}\>[4]{\mathsf{CompactClosed}\mskip 3.0muz\mskip 3.0mu}\>[5]{\mathnormal{\Rightarrow }\mskip 3.0mu\allowbreak{}\mathnormal{(}\mskip 0.0mu\mathsf{a}\mskip 1.0mu\overset{z}{\leadsto }\mskip 1.0mu\mathsf{b}\mskip 0.0mu\mathnormal{)}\allowbreak{}\mskip 3.0mu}\>[6]{\mathnormal{\rightarrow }\mskip 3.0mu\allowbreak{}\mathnormal{(}\mskip 0.0mu∀\mskip 3.0mu\mathsf{r}\mskip 1.0mu.\mskip 3.0mu}\>[7]{\mathsf{P}\mskip 3.0muz\mskip 3.0mu\mathsf{r}\mskip 3.0mu\mathsf{a}\mskip 3.0mu\mathnormal{⊸}\mskip 3.0mu\mathsf{P}\mskip 3.0muz\mskip 3.0mu\mathsf{r}\mskip 3.0mu\dual{\mathsf{b}}\mskip 3.0mu}\>[8]{\mathnormal{⊸}\mskip 3.0mu\mathsf{R}\mskip 3.0muz\mskip 3.0mu\mathsf{r}\mskip 0.0mu\mathnormal{)}\allowbreak{}}\<[E]\\
\>[1]{\mathsf{tensorEmbed}_{1}\mskip 3.0mu}\>[2]{\mathnormal{::}\mskip 3.0mu}\>[3]{\mathsf{SymmetricMonoidal}\mskip 3.0muz\mskip 3.0mu}\>[5]{\mathnormal{\Rightarrow }\mskip 3.0mu\allowbreak{}\mathnormal{(}\mskip 0.0mu\mathsf{a}\mskip 1.0mu\overset{z}{\leadsto }\mskip 1.0mu\mathbf{1}\mskip 0.0mu\mathnormal{)}\allowbreak{}\mskip 3.0mu}\>[6]{\mathnormal{\rightarrow }\mskip 3.0mu\allowbreak{}\mathnormal{(}\mskip 0.0mu∀\mskip 3.0mu\mathsf{r}\mskip 1.0mu.\mskip 3.0mu}\>[7]{\mathsf{P}\mskip 3.0muz\mskip 3.0mu\mathsf{r}\mskip 3.0mu\mathsf{a}\mskip 3.0mu}\>[8]{\mathnormal{⊸}\mskip 3.0mu\mathsf{R}\mskip 3.0muz\mskip 3.0mu\mathsf{r}\mskip 0.0mu\mathnormal{)}\allowbreak{}}\<[E]\end{parray}}\end{list} The special case of a vector, where the target object is the unit,
is common enough that it deserves a function of its own. In the general case, we take advantage of the
compact-closed structure, and turn the output object (\(\mathsf{b}\)) of the morphism into
an input of an index over the dual object (\(\dual{\mathsf{b}}\)).

The converse operation consists in evaluating a tensor expression into a morphism:
\begin{list}{}{\setlength\leftmargin{1.0em}}\item\relax
\ensuremath{\begin{parray}\column{B}{@{}>{}l<{}@{}}\column[0em]{1}{@{}>{}l<{}@{}}\column{2}{@{}>{}l<{}@{}}\column{3}{@{}>{}l<{}@{}}\column{4}{@{}>{}l<{}@{}}\column{5}{@{}>{}l<{}@{}}\column{6}{@{}>{}l<{}@{}}\column{7}{@{}>{}l<{}@{}}\column{8}{@{}>{}l<{}@{}}\column{9}{@{}>{}l<{}@{}}\column{10}{@{}>{}l<{}@{}}\column{E}{@{}>{}l<{}@{}}%
\>[1]{\mathsf{tensorEval}\mskip 3.0mu}\>[2]{\mathnormal{::}\mskip 3.0mu}\>[3]{\allowbreak{}\mathnormal{(}\mskip 0.0mu}\>[7]{\mathsf{CompactClosed}\mskip 3.0muz\mskip 0.0mu\mathnormal{,}\mskip 3.0mu\mathsf{Additive}\mskip 3.0mu}\>[9]{z\mskip 0.0mu\mathnormal{)}\allowbreak{}\mskip 3.0mu}\>[10]{\mathnormal{\Rightarrow }}\<[E]\\
\>[3]{\allowbreak{}\mathnormal{(}\mskip 0.0mu∀\mskip 3.0mu\mathsf{r}\mskip 1.0mu.\mskip 3.0mu}\>[4]{\mathsf{P}\mskip 3.0muz\mskip 3.0mu\mathsf{r}\mskip 3.0mu\mathsf{a}\mskip 3.0mu\mathnormal{⊸}\mskip 3.0mu\mathsf{P}\mskip 3.0muz\mskip 3.0mu\mathsf{r}\mskip 3.0mu\dual{\mathsf{b}}\mskip 3.0mu}\>[5]{\mathnormal{⊸}\mskip 3.0mu\mathsf{R}\mskip 3.0muz\mskip 3.0mu\mathsf{r}\mskip 0.0mu\mathnormal{)}\allowbreak{}\mskip 3.0mu\mathnormal{\rightarrow }\mskip 3.0mu\allowbreak{}\mathnormal{(}\mskip 0.0mu\mathsf{a}\mskip 1.0mu\overset{z}{\leadsto }\mskip 1.0mu\mathsf{b}\mskip 0.0mu\mathnormal{)}\allowbreak{}}\<[E]\\
\>[1]{\mathsf{tensorEval}_{1}\mskip 3.0mu}\>[2]{\mathnormal{::}\mskip 3.0mu}\>[3]{\allowbreak{}\mathnormal{(}\mskip 0.0mu}\>[6]{\mathsf{SymmetricMonoidal}\mskip 3.0muz\mskip 0.0mu\mathnormal{,}\mskip 3.0mu\mathsf{Additive}\mskip 3.0mu}\>[8]{z\mskip 0.0mu\mathnormal{)}\allowbreak{}\mskip 3.0mu}\>[10]{\mathnormal{\Rightarrow }}\<[E]\\
\>[3]{\allowbreak{}\mathnormal{(}\mskip 0.0mu∀\mskip 3.0mu\mathsf{r}\mskip 1.0mu.\mskip 3.0mu}\>[4]{\mathsf{P}\mskip 3.0muz\mskip 3.0mu\mathsf{r}\mskip 3.0mu\mathsf{a}\mskip 3.0mu}\>[5]{\mathnormal{⊸}\mskip 3.0mu\mathsf{R}\mskip 3.0muz\mskip 3.0mu\mathsf{r}\mskip 0.0mu\mathnormal{)}\allowbreak{}\mskip 3.0mu\mathnormal{\rightarrow }\mskip 3.0mu\allowbreak{}\mathnormal{(}\mskip 0.0mu\mathsf{a}\mskip 1.0mu\overset{z}{\leadsto }\mskip 1.0mu\mathbf{1}\mskip 0.0mu\mathnormal{)}\allowbreak{}}\<[E]\end{parray}}\end{list} 
The fact that we can move between these two {\sc{}dsl}s freely (using embedding and evaluation) means we can combine their
strengths. In both embedding and evaluation, neither \(\mathsf{a}\) nor \(\mathsf{b}\) need be atomic types.
To match the convention of Einstein notation, the user of {\sc{}Albert} can
break down the corresponding indices into their components after embedding, or conversely combine components before evaluation.
Likewise, unit objects might need to be introduced or discarded. The interface for performing such operations
is provided in the form of the following four combinators:
\begin{list}{}{\setlength\leftmargin{1.0em}}\item\relax
\ensuremath{\begin{parray}\column{B}{@{}>{}l<{}@{}}\column[0em]{1}{@{}>{}l<{}@{}}\column{2}{@{}>{}l<{}@{}}\column{3}{@{}>{}l<{}@{}}\column{4}{@{}>{}l<{}@{}}\column{5}{@{}>{}l<{}@{}}\column{E}{@{}>{}l<{}@{}}%
\>[1]{\mathsf{unit}\mskip 3.0mu}\>[2]{\mathnormal{::}\mskip 3.0mu}\>[3]{\mathsf{SymmetricMonoidal}\mskip 3.0muz\mskip 3.0mu}\>[4]{\mathnormal{\Rightarrow }\mskip 3.0mu\mathsf{P}\mskip 3.0muz\mskip 3.0mu\mathsf{r}\mskip 3.0mu\mathsf{a}\mskip 3.0mu}\>[5]{\mathnormal{⊸}\mskip 3.0mu\allowbreak{}\mathnormal{(}\mskip 0.0mu\mathsf{P}\mskip 3.0muz\mskip 3.0mu\mathsf{r}\mskip 3.0mu\mathsf{a}\mskip 0.0mu\mathnormal{,}\mskip 3.0mu\mathsf{P}\mskip 3.0muz\mskip 3.0mu\mathsf{r}\mskip 3.0mu\mathbf{1}\mskip 0.0mu\mathnormal{)}\allowbreak{}}\<[E]\\
\>[1]{\mathsf{unit'}\mskip 3.0mu}\>[2]{\mathnormal{::}\mskip 3.0mu}\>[3]{\mathsf{SymmetricMonoidal}\mskip 3.0muz\mskip 3.0mu}\>[4]{\mathnormal{\Rightarrow }\mskip 3.0mu\mathsf{P}\mskip 3.0muz\mskip 3.0mu\mathsf{r}\mskip 3.0mu\mathbf{1}\mskip 3.0mu\mathnormal{⊸}\mskip 3.0mu\mathsf{P}\mskip 3.0muz\mskip 3.0mu\mathsf{r}\mskip 3.0mu\mathsf{a}\mskip 3.0mu\mathnormal{⊸}\mskip 3.0mu\mathsf{P}\mskip 3.0muz\mskip 3.0mu\mathsf{r}\mskip 3.0mu\mathsf{a}}\<[E]\\
\>[1]{\mathsf{split}\mskip 3.0mu}\>[2]{\mathnormal{::}\mskip 3.0mu}\>[3]{\mathsf{SymmetricMonoidal}\mskip 3.0muz\mskip 3.0mu}\>[4]{\mathnormal{\Rightarrow }\mskip 3.0mu\mathsf{P}\mskip 3.0muz\mskip 3.0mu\mathsf{r}\mskip 3.0mu\allowbreak{}\mathnormal{(}\mskip 0.0mu\mathsf{a}\mskip 3.0mu\mathnormal{⊗}\mskip 3.0mu\mathsf{b}\mskip 0.0mu\mathnormal{)}\allowbreak{}\mskip 3.0mu\mathnormal{⊸}\mskip 3.0mu\allowbreak{}\mathnormal{(}\mskip 0.0mu\mathsf{P}\mskip 3.0muz\mskip 3.0mu\mathsf{r}\mskip 3.0mu\mathsf{a}\mskip 0.0mu\mathnormal{,}\mskip 3.0mu\mathsf{P}\mskip 3.0muz\mskip 3.0mu\mathsf{r}\mskip 3.0mu\mathsf{b}\mskip 0.0mu\mathnormal{)}\allowbreak{}}\<[E]\\
\>[1]{\mathsf{merge}\mskip 3.0mu}\>[2]{\mathnormal{::}\mskip 3.0mu}\>[3]{\mathsf{SymmetricMonoidal}\mskip 3.0muz\mskip 3.0mu}\>[4]{\mathnormal{\Rightarrow }\mskip 3.0mu\allowbreak{}\mathnormal{(}\mskip 0.0mu\mathsf{P}\mskip 3.0muz\mskip 3.0mu\mathsf{r}\mskip 3.0mu\mathsf{a}\mskip 0.0mu\mathnormal{,}\mskip 3.0mu\mathsf{P}\mskip 3.0muz\mskip 3.0mu\mathsf{r}\mskip 3.0mu\mathsf{b}\mskip 0.0mu\mathnormal{)}\allowbreak{}\mskip 3.0mu\mathnormal{⊸}\mskip 3.0mu\mathsf{P}\mskip 3.0muz\mskip 3.0mu\mathsf{r}\mskip 3.0mu\allowbreak{}\mathnormal{(}\mskip 0.0mu\mathsf{a}\mskip 3.0mu\mathnormal{⊗}\mskip 3.0mu\mathsf{b}\mskip 0.0mu\mathnormal{)}\allowbreak{}}\<[E]\end{parray}}\end{list} To sum up, when \(z\) is a {\sc{}smc}, the \(\mathsf{P}\mskip 3.0muz\mskip 3.0mu\mathsf{r}\) type transformer defines a homomorphism between the monoid of (linear)
Haskell pairs and that of tensor products of the category \(z\).
As an illustration, a function \(\mathsf{t}\) of type \(\mathsf{P}\mskip 3.0muz\mskip 3.0mu\mathsf{r}\mskip 3.0mu\allowbreak{}\mathnormal{(}\mskip 0.0mu\mathsf{a}\mskip 3.0mu\mathnormal{⊗}\mskip 3.0mu\dual{\mathsf{b}}\mskip 0.0mu\mathnormal{)}\allowbreak{}\mskip 3.0mu\mathnormal{⊸}\mskip 3.0mu\mathsf{R}\mskip 3.0muz\mskip 3.0mu\mathsf{r}\) can be curried to \(\mathsf{t'}\mskip 3.0mu\mathnormal{::}\mskip 3.0mu\mathsf{P}\mskip 3.0muz\mskip 3.0mu\mathsf{r}\mskip 3.0mu\mathsf{a}\mskip 3.0mu\mathnormal{⊸}\mskip 3.0mu\mathsf{P}\mskip 3.0muz\mskip 3.0mu\mathsf{r}\mskip 3.0mu\dual{\mathsf{b}}\mskip 3.0mu\mathnormal{⊸}\mskip 3.0mu\mathsf{R}\mskip 3.0muz\mskip 3.0mu\mathsf{r}\). When using \(\mathsf{t'}\), each index is its own variable,
closely matching Einstein notation.

\paragraph*{Multiplication and contraction}\hspace{1.0ex}\label{119} Another pervasive operation in Einstein notation is multiplication.
In {\sc{}Albert} we use a multiplication operator with a linear type:
\begin{list}{}{\setlength\leftmargin{1.0em}}\item\relax
\ensuremath{\begin{parray}\column{B}{@{}>{}l<{}@{}}\column[0em]{1}{@{}>{}l<{}@{}}\column{E}{@{}>{}l<{}@{}}%
\>[1]{\allowbreak{}\mathnormal{(}\mskip 0.0mu{\tikzstar{0.11}{0.25}{5}{-18}{fill=black}}\mskip 0.0mu\mathnormal{)}\allowbreak{}\mskip 3.0mu\mathnormal{::}\mskip 3.0mu\mathsf{R}\mskip 3.0muz\mskip 3.0mu\mathsf{r}\mskip 3.0mu\mathnormal{⊸}\mskip 3.0mu\mathsf{R}\mskip 3.0muz\mskip 3.0mu\mathsf{r}\mskip 3.0mu\mathnormal{⊸}\mskip 3.0mu\mathsf{R}\mskip 3.0muz\mskip 3.0mu\mathsf{r}}\<[E]\end{parray}}\end{list} According to the typing rules of the linear function types, the occurrences of variables are accumulated in a function call.
 This way, the order of the product \(\mathsf{t}\mskip 3.0mu{\tikzstar{0.11}{0.25}{5}{-18}{fill=black}}\mskip 3.0mu\mathsf{u}\) is the sum of the orders of \(\mathsf{t}\) and \(\mathsf{u}\).
Contraction is realised by the following
combinator.  \begin{list}{}{\setlength\leftmargin{1.0em}}\item\relax
\ensuremath{\begin{parray}\column{B}{@{}>{}l<{}@{}}\column[0em]{1}{@{}>{}l<{}@{}}\column{2}{@{}>{}l<{}@{}}\column{E}{@{}>{}l<{}@{}}%
\>[1]{\mathsf{contract}\mskip 3.0mu\mathnormal{::}\mskip 3.0mu\mathsf{CompactClosed}\mskip 3.0muz\mskip 3.0mu\mathnormal{\Rightarrow }\mskip 3.0mu}\>[2]{\allowbreak{}\mathnormal{(}\mskip 0.0mu\mathsf{P}\mskip 3.0muz\mskip 3.0mu\mathsf{r}\mskip 3.0mu\dual{\mathsf{a}}\mskip 3.0mu\mathnormal{⊸}\mskip 3.0mu\mathsf{P}\mskip 3.0muz\mskip 3.0mu\mathsf{r}\mskip 3.0mu\mathsf{a}\mskip 3.0mu\mathnormal{⊸}\mskip 3.0mu\mathsf{R}\mskip 3.0muz\mskip 3.0mu\mathsf{r}\mskip 0.0mu\mathnormal{)}\allowbreak{}\mskip 3.0mu\mathnormal{⊸}\mskip 3.0mu\mathsf{R}\mskip 3.0muz\mskip 3.0mu\mathsf{r}}\<[E]\end{parray}}\end{list} 
There are a couple of contrasting points when compared to the Einstein
notation. First, we \emph{bind index variables explicitly}, and thus
we use an explicit contraction combinator. Indeed, while in Einstein
notation indices are not explicitly bound, this liberty cannot be
taken in an {\sc{}edsl} based on a lambda calculus.  Second, we consider the high and low indices
involved in the contraction to be \emph{separate variables}. Indeed,
in Einstein notation each version of the index (high or low) must
occur exactly once, and thus making them separate linearly bound
variables is natural. One can think of it as the contraction creating a wire,
with each end of the wire bound to a separate name.
Nonetheless, the convention to use the same
variable name in different positions is a convenient one. We
recover it in this paper by a typographical trick: we use the same Latin letter
for both indices, and make the position as sub- or superscript
integral to variable names. (This is purely a matter of convention and users of
{\sc{}Albert} are free to use whichever variable names they prefer.)
Therefore, for instance, in {\sc{}Albert} the composition of two linear transformations
\(\mathsf{t}\) and \(\mathsf{u}\), as shown in \cref{44}, is realised as
\ensuremath{\begin{parray}\column{B}{@{}>{}l<{}@{}}\column[0em]{1}{@{}>{}l<{}@{}}\column{E}{@{}>{}l<{}@{}}%
\>[1]{λ\mskip 3.0mu_{\mathsf{i}}\mskip 3.0mu^{\mathsf{j}}\mskip 3.0mu\mathnormal{\rightarrow }\mskip 3.0mu\mathsf{contract}\mskip 3.0mu\allowbreak{}\mathnormal{(}\mskip 0.0muλ\mskip 3.0mu^{\mathsf{k}}\mskip 3.0mu_{\mathsf{k}}\mskip 3.0mu\mathnormal{\rightarrow }\mskip 3.0mu\mathsf{t}\mskip 3.0mu_{\mathsf{i}}\mskip 3.0mu^{\mathsf{k}}\mskip 3.0mu{\tikzstar{0.11}{0.25}{5}{-18}{fill=black}}\mskip 3.0mu\mathsf{u}\mskip 3.0mu_{\mathsf{k}}\mskip 3.0mu^{\mathsf{j}}\mskip 0.0mu\mathnormal{)}\allowbreak{}}\<[E]\end{parray}}.

\paragraph*{Addition and zero}\hspace{1.0ex}\label{120} 
In Einstein notation, one can use the addition operator as if it were the point-wise
addition of each of the components, for instance \(tᵢʲ + uᵢʲ\). Note that the
live indices are used in each of the operands of the sum, thus repeated in
the whole expression.
This means that the following linear type for the sum operator would be incorrect:
\begin{list}{}{\setlength\leftmargin{1.0em}}\item\relax
\ensuremath{\begin{parray}\column{B}{@{}>{}l<{}@{}}\column[0em]{1}{@{}>{}l<{}@{}}\column{E}{@{}>{}l<{}@{}}%
\>[1]{\mathsf{plus}_{wrong}\mskip 3.0mu\mathnormal{::}\mskip 3.0mu\mathsf{R}\mskip 3.0muz\mskip 3.0mu\mathsf{r}\mskip 3.0mu\mathnormal{⊸}\mskip 3.0mu\mathsf{R}\mskip 3.0muz\mskip 3.0mu\mathsf{r}\mskip 3.0mu\mathnormal{⊸}\mskip 3.0mu\mathsf{R}\mskip 3.0muz\mskip 3.0mu\mathsf{r}}\<[E]\end{parray}}\end{list} This is because in the expression \(\mathsf{plus}_{wrong}\mskip 3.0mu\allowbreak{}\mathnormal{(}\mskip 0.0mu\mathsf{t}\mskip 3.0mu_{\mathsf{i}}\mskip 3.0mu^{\mathsf{j}}\mskip 0.0mu\mathnormal{)}\allowbreak{}\mskip 3.0mu\allowbreak{}\mathnormal{(}\mskip 0.0mu\mathsf{u}\mskip 3.0mu_{\mathsf{i}}\mskip 3.0mu^{\mathsf{j}}\mskip 0.0mu\mathnormal{)}\allowbreak{}\), both
\(\mathsf{i}\) and \(\mathsf{j}\) occur twice, while the type would require indices to be split between the left and right operand.
Thus we must use another type. We settle on the following one:
\begin{list}{}{\setlength\leftmargin{1.0em}}\item\relax
\ensuremath{\begin{parray}\column{B}{@{}>{}l<{}@{}}\column[0em]{1}{@{}>{}l<{}@{}}\column{E}{@{}>{}l<{}@{}}%
\>[1]{\mathsf{plus}\mskip 3.0mu\mathnormal{::}\mskip 3.0mu\allowbreak{}\mathnormal{(}\mskip 0.0mu\mathsf{Bool}\mskip 3.0mu\mathnormal{\rightarrow }\mskip 3.0mu\mathsf{R}\mskip 3.0muz\mskip 3.0mu\mathsf{r}\mskip 0.0mu\mathnormal{)}\allowbreak{}\mskip 3.0mu\mathnormal{⊸}\mskip 3.0mu\mathsf{R}\mskip 3.0muz\mskip 3.0mu\mathsf{r}}\<[E]\end{parray}}\end{list} This type allows to code \(tᵢʲ + uᵢʲ\) as follows:\footnote{In what follows the \(\mathsf{False}\) branch corresponds to the left operand and \(\mathsf{True}\) to the right one. Any two-element type would do.\label{121}} \begin{list}{}{\setlength\leftmargin{1.0em}}\item\relax
\ensuremath{\begin{parray}\column{B}{@{}>{}l<{}@{}}\column[0em]{1}{@{}>{}l<{}@{}}\column[1em]{2}{@{}>{}l<{}@{}}\column{3}{@{}>{}l<{}@{}}\column{E}{@{}>{}l<{}@{}}%
\>[1]{\mathsf{plus}\mskip 3.0mu\allowbreak{}\mathnormal{(}\mskip 0.0muλ\mskip 3.0mu\mathsf{c}\mskip 3.0mu\mathnormal{\rightarrow }\mskip 3.0mu\mathbf{case}\mskip 3.0mu\mathsf{c}\mskip 3.0mu\mathbf{of}}\<[E]\\
\>[2]{\mathsf{False}\mskip 3.0mu}\>[3]{\mathnormal{\rightarrow }\mskip 3.0mu\mathsf{t}\mskip 3.0mu_{\mathsf{i}}\mskip 3.0mu^{\mathsf{j}}}\<[E]\\
\>[2]{\mathsf{True}\mskip 3.0mu}\>[3]{\mathnormal{\rightarrow }\mskip 3.0mu\mathsf{u}\mskip 3.0mu_{\mathsf{i}}\mskip 3.0mu^{\mathsf{j}}\mskip 0.0mu\mathnormal{)}\allowbreak{}}\<[E]\end{parray}}\end{list} 
The above is well typed. Indeed, 1. the argument of \(\mathsf{plus}\) is
type-linear, so any use of indices in its body is considered
type-linear; and 2. only one branch of a \(\mathbf{case}\) is considered to be run,
and therefore the same indices can (and must) be used in all the
branches. The fact that only one branch is run is counter-intuitive, because
the semantics depends on both of them. We explain our solution to this
apparent contradiction in \cref{396}.

Conversely, there is a zero tensor of every possible order. Thus
we have a zero combinator with an index argument ranging over an arbitrary
space: \begin{list}{}{\setlength\leftmargin{1.0em}}\item\relax
\ensuremath{\begin{parray}\column{B}{@{}>{}l<{}@{}}\column[0em]{1}{@{}>{}l<{}@{}}\column{2}{@{}>{}l<{}@{}}\column{3}{@{}>{}l<{}@{}}\column{4}{@{}>{}l<{}@{}}\column{E}{@{}>{}l<{}@{}}%
\>[1]{\mathsf{zeroTensor}\mskip 3.0mu}\>[2]{\mathnormal{::}\mskip 3.0mu\allowbreak{}\mathnormal{(}\mskip 0.0mu\mathsf{CompactClosed}\mskip 3.0muz\mskip 0.0mu\mathnormal{,}\mskip 3.0mu\mathsf{Additive}\mskip 3.0mu}\>[3]{z\mskip 0.0mu}\>[4]{\mathnormal{)}\allowbreak{}\mskip 3.0mu\mathnormal{\Rightarrow }\mskip 3.0mu\mathsf{P}\mskip 3.0muz\mskip 3.0mu\mathsf{r}\mskip 3.0mu\mathsf{a}\mskip 3.0mu\mathnormal{⊸}\mskip 3.0mu\mathsf{R}\mskip 3.0muz\mskip 3.0mu\mathsf{r}}\<[E]\end{parray}}\end{list} 
The scaling operator \(\allowbreak{}\mathnormal{(}\mskip 0.0mu\smalltriangleleft \mskip 0.0mu\mathnormal{)}\allowbreak{}\) underpins non-zero constants, with no
particular difficulty.

With the primitives of additive categories, one can construct the
tensor \(\mathsf{antisym}\mskip 3.0mu\mathnormal{=}\) \(\mathsf{id}\mskip 3.0mu\mathnormal{-}\mskip 3.0muσ\) :: \(\mathsf{T}\mskip 3.0mu\mathnormal{⊗}\mskip 3.0mu\mathsf{T}\mskip 1.0mu\overset{z}{\leadsto }\mskip 1.0mu\mathsf{T}\mskip 3.0mu\mathnormal{⊗}\mskip 3.0mu\mathsf{T}\).  This tensor can be rendered graphically as the
difference \({\begin{tikzpicture}[baseline=(current bounding box.center)]\path[-,draw=black,line width=0.4pt,line cap=butt,line join=miter,dash pattern=](0pt,0pt)--(1pt,0pt);
\path[-,draw=black,line width=0.4pt,line cap=butt,line join=miter,dash pattern=](0pt,-7.4184pt)--(1pt,-7.4184pt);
\path[-,draw=black,line width=0.4pt,line cap=butt,line join=miter,dash pattern=](7pt,0pt)--(8pt,0pt);
\path[-,draw=black,line width=0.4pt,line cap=butt,line join=miter,dash pattern=](7pt,-7.4184pt)--(8pt,-7.4184pt);
\path[-,draw=black,line width=0.4pt,line cap=butt,line join=miter,dash pattern=](1pt,0pt)--(7pt,0pt);
\path[-,draw=black,line width=0.4pt,line cap=butt,line join=miter,dash pattern=](1pt,-7.4184pt)--(7pt,-7.4184pt);
\path[-,line width=0.4pt,line cap=butt,line join=miter,dash pattern=](-4.6316pt,2.3396pt)--(-2pt,2.3396pt)--(-2pt,-2.3396pt)--(-4.6316pt,-2.3396pt)--cycle;
\node[anchor=north west,inner sep=0] at (-4.6316pt,2.3396pt){\savebox{\marxupbox}{{\({\scriptstyle i}\)}}\immediate\write\boxesfile{122}\immediate\write\boxesfile{\number\wd\marxupbox}\immediate\write\boxesfile{\number\ht\marxupbox}\immediate\write\boxesfile{\number\dp\marxupbox}\box\marxupbox};
\path[-,line width=0.4pt,line cap=butt,line join=miter,dash pattern=](-6.6316pt,4.3396pt)--(0pt,4.3396pt)--(0pt,-4.3396pt)--(-6.6316pt,-4.3396pt)--cycle;
\path[-,line width=0.4pt,line cap=butt,line join=miter,dash pattern=](-5.8617pt,-4.3396pt)--(-2pt,-4.3396pt)--(-2pt,-10.4972pt)--(-5.8617pt,-10.4972pt)--cycle;
\node[anchor=north west,inner sep=0] at (-5.8617pt,-4.3396pt){\savebox{\marxupbox}{{\({\scriptstyle j}\)}}\immediate\write\boxesfile{123}\immediate\write\boxesfile{\number\wd\marxupbox}\immediate\write\boxesfile{\number\ht\marxupbox}\immediate\write\boxesfile{\number\dp\marxupbox}\box\marxupbox};
\path[-,line width=0.4pt,line cap=butt,line join=miter,dash pattern=](-7.8617pt,-2.3396pt)--(0pt,-2.3396pt)--(0pt,-12.4972pt)--(-7.8617pt,-12.4972pt)--cycle;
\path[-,line width=0.4pt,line cap=butt,line join=miter,dash pattern=](10pt,2.4911pt)--(14.3398pt,2.4911pt)--(14.3398pt,-2.4911pt)--(10pt,-2.4911pt)--cycle;
\node[anchor=north west,inner sep=0] at (10pt,2.4911pt){\savebox{\marxupbox}{{\({\scriptstyle k}\)}}\immediate\write\boxesfile{124}\immediate\write\boxesfile{\number\wd\marxupbox}\immediate\write\boxesfile{\number\ht\marxupbox}\immediate\write\boxesfile{\number\dp\marxupbox}\box\marxupbox};
\path[-,line width=0.4pt,line cap=butt,line join=miter,dash pattern=](8pt,4.4911pt)--(16.3398pt,4.4911pt)--(16.3398pt,-4.4911pt)--(8pt,-4.4911pt)--cycle;
\path[-,line width=0.4pt,line cap=butt,line join=miter,dash pattern=](10pt,-4.8926pt)--(12.5769pt,-4.8926pt)--(12.5769pt,-9.9442pt)--(10pt,-9.9442pt)--cycle;
\node[anchor=north west,inner sep=0] at (10pt,-4.8926pt){\savebox{\marxupbox}{{\({\scriptstyle l}\)}}\immediate\write\boxesfile{125}\immediate\write\boxesfile{\number\wd\marxupbox}\immediate\write\boxesfile{\number\ht\marxupbox}\immediate\write\boxesfile{\number\dp\marxupbox}\box\marxupbox};
\path[-,line width=0.4pt,line cap=butt,line join=miter,dash pattern=](8pt,-2.8926pt)--(14.5769pt,-2.8926pt)--(14.5769pt,-11.9442pt)--(8pt,-11.9442pt)--cycle;
\end{tikzpicture}}-{\begin{tikzpicture}[baseline=(current bounding box.center)]\path[-,draw=black,line width=0.4pt,line cap=butt,line join=miter,dash pattern=](-1pt,7.2177pt)--(0pt,7.2177pt);
\path[-,draw=black,line width=0.4pt,line cap=butt,line join=miter,dash pattern=](-1pt,-0.2008pt)--(0pt,-0.2008pt);
\path[-,draw=black,line width=0.4pt,line cap=butt,line join=miter,dash pattern=](6pt,7.0169pt)--(7pt,7.0169pt);
\path[-,draw=black,line width=0.4pt,line cap=butt,line join=miter,dash pattern=](6pt,0pt)--(7pt,0pt);
\path[-,draw=black,line width=0.4pt,line cap=butt,line join=miter,dash pattern=](0pt,7.2177pt)..controls(4pt,7.2177pt)and(2pt,0pt)..(6pt,0pt);
\path[-,draw=black,line width=0.4pt,line cap=butt,line join=miter,dash pattern=](0pt,-0.2008pt)..controls(4pt,-0.2008pt)and(2pt,7.0169pt)..(6pt,7.0169pt);
\path[-,line width=0.4pt,line cap=butt,line join=miter,dash pattern=](-5.6316pt,9.5573pt)--(-3pt,9.5573pt)--(-3pt,4.878pt)--(-5.6316pt,4.878pt)--cycle;
\node[anchor=north west,inner sep=0] at (-5.6316pt,9.5573pt){\savebox{\marxupbox}{{\({\scriptstyle i}\)}}\immediate\write\boxesfile{126}\immediate\write\boxesfile{\number\wd\marxupbox}\immediate\write\boxesfile{\number\ht\marxupbox}\immediate\write\boxesfile{\number\dp\marxupbox}\box\marxupbox};
\path[-,line width=0.4pt,line cap=butt,line join=miter,dash pattern=](-7.6316pt,11.5573pt)--(-1pt,11.5573pt)--(-1pt,2.878pt)--(-7.6316pt,2.878pt)--cycle;
\path[-,line width=0.4pt,line cap=butt,line join=miter,dash pattern=](-6.8617pt,2.878pt)--(-3pt,2.878pt)--(-3pt,-3.2795pt)--(-6.8617pt,-3.2795pt)--cycle;
\node[anchor=north west,inner sep=0] at (-6.8617pt,2.878pt){\savebox{\marxupbox}{{\({\scriptstyle j}\)}}\immediate\write\boxesfile{127}\immediate\write\boxesfile{\number\wd\marxupbox}\immediate\write\boxesfile{\number\ht\marxupbox}\immediate\write\boxesfile{\number\dp\marxupbox}\box\marxupbox};
\path[-,line width=0.4pt,line cap=butt,line join=miter,dash pattern=](-8.8617pt,4.878pt)--(-1pt,4.878pt)--(-1pt,-5.2795pt)--(-8.8617pt,-5.2795pt)--cycle;
\path[-,line width=0.4pt,line cap=butt,line join=miter,dash pattern=](9pt,9.508pt)--(13.3398pt,9.508pt)--(13.3398pt,4.5258pt)--(9pt,4.5258pt)--cycle;
\node[anchor=north west,inner sep=0] at (9pt,9.508pt){\savebox{\marxupbox}{{\({\scriptstyle k}\)}}\immediate\write\boxesfile{128}\immediate\write\boxesfile{\number\wd\marxupbox}\immediate\write\boxesfile{\number\ht\marxupbox}\immediate\write\boxesfile{\number\dp\marxupbox}\box\marxupbox};
\path[-,line width=0.4pt,line cap=butt,line join=miter,dash pattern=](7pt,11.508pt)--(15.3398pt,11.508pt)--(15.3398pt,2.5258pt)--(7pt,2.5258pt)--cycle;
\path[-,line width=0.4pt,line cap=butt,line join=miter,dash pattern=](9pt,2.5258pt)--(11.5769pt,2.5258pt)--(11.5769pt,-2.5258pt)--(9pt,-2.5258pt)--cycle;
\node[anchor=north west,inner sep=0] at (9pt,2.5258pt){\savebox{\marxupbox}{{\({\scriptstyle l}\)}}\immediate\write\boxesfile{129}\immediate\write\boxesfile{\number\wd\marxupbox}\immediate\write\boxesfile{\number\ht\marxupbox}\immediate\write\boxesfile{\number\dp\marxupbox}\box\marxupbox};
\path[-,line width=0.4pt,line cap=butt,line join=miter,dash pattern=](7pt,4.5258pt)--(13.5769pt,4.5258pt)--(13.5769pt,-4.5258pt)--(7pt,-4.5258pt)--cycle;
\end{tikzpicture}}\), but it is useful enough to
receive the special notation {\begin{tikzpicture}[baseline={([yshift=-0.8ex]current bounding box.center)}]\path[-,draw=black,line width=0.4pt,line cap=butt,line join=miter,dash pattern=](0pt,0pt)--(1pt,0pt);
\path[-,draw=black,line width=0.4pt,line cap=butt,line join=miter,dash pattern=](0pt,-7.4184pt)--(1pt,-7.4184pt);
\path[-,draw=black,line width=0.4pt,line cap=butt,line join=miter,dash pattern=](7pt,0pt)--(8pt,0pt);
\path[-,draw=black,line width=0.4pt,line cap=butt,line join=miter,dash pattern=](7pt,-7.4184pt)--(8pt,-7.4184pt);
\path[-,draw=black,line width=0.4pt,line cap=butt,line join=miter,dash pattern=](1pt,0pt)--(7pt,0pt);
\path[-,draw=black,line width=0.4pt,line cap=butt,line join=miter,dash pattern=](1pt,-7.4184pt)--(7pt,-7.4184pt);
\path[-,draw=black,line width=1.2pt,line cap=butt,line join=miter,dash pattern=](4pt,1pt)--(4pt,-8.4184pt)--cycle;
\path[-,line width=0.4pt,line cap=butt,line join=miter,dash pattern=](-4.6316pt,2.3396pt)--(-2pt,2.3396pt)--(-2pt,-2.3396pt)--(-4.6316pt,-2.3396pt)--cycle;
\node[anchor=north west,inner sep=0] at (-4.6316pt,2.3396pt){\savebox{\marxupbox}{{\({\scriptstyle i}\)}}\immediate\write\boxesfile{130}\immediate\write\boxesfile{\number\wd\marxupbox}\immediate\write\boxesfile{\number\ht\marxupbox}\immediate\write\boxesfile{\number\dp\marxupbox}\box\marxupbox};
\path[-,line width=0.4pt,line cap=butt,line join=miter,dash pattern=](-6.6316pt,4.3396pt)--(0pt,4.3396pt)--(0pt,-4.3396pt)--(-6.6316pt,-4.3396pt)--cycle;
\path[-,line width=0.4pt,line cap=butt,line join=miter,dash pattern=](-5.8617pt,-4.3396pt)--(-2pt,-4.3396pt)--(-2pt,-10.4972pt)--(-5.8617pt,-10.4972pt)--cycle;
\node[anchor=north west,inner sep=0] at (-5.8617pt,-4.3396pt){\savebox{\marxupbox}{{\({\scriptstyle j}\)}}\immediate\write\boxesfile{131}\immediate\write\boxesfile{\number\wd\marxupbox}\immediate\write\boxesfile{\number\ht\marxupbox}\immediate\write\boxesfile{\number\dp\marxupbox}\box\marxupbox};
\path[-,line width=0.4pt,line cap=butt,line join=miter,dash pattern=](-7.8617pt,-2.3396pt)--(0pt,-2.3396pt)--(0pt,-12.4972pt)--(-7.8617pt,-12.4972pt)--cycle;
\path[-,line width=0.4pt,line cap=butt,line join=miter,dash pattern=](10pt,2.4911pt)--(14.3398pt,2.4911pt)--(14.3398pt,-2.4911pt)--(10pt,-2.4911pt)--cycle;
\node[anchor=north west,inner sep=0] at (10pt,2.4911pt){\savebox{\marxupbox}{{\({\scriptstyle k}\)}}\immediate\write\boxesfile{132}\immediate\write\boxesfile{\number\wd\marxupbox}\immediate\write\boxesfile{\number\ht\marxupbox}\immediate\write\boxesfile{\number\dp\marxupbox}\box\marxupbox};
\path[-,line width=0.4pt,line cap=butt,line join=miter,dash pattern=](8pt,4.4911pt)--(16.3398pt,4.4911pt)--(16.3398pt,-4.4911pt)--(8pt,-4.4911pt)--cycle;
\path[-,line width=0.4pt,line cap=butt,line join=miter,dash pattern=](10pt,-4.8926pt)--(12.5769pt,-4.8926pt)--(12.5769pt,-9.9442pt)--(10pt,-9.9442pt)--cycle;
\node[anchor=north west,inner sep=0] at (10pt,-4.8926pt){\savebox{\marxupbox}{{\({\scriptstyle l}\)}}\immediate\write\boxesfile{133}\immediate\write\boxesfile{\number\wd\marxupbox}\immediate\write\boxesfile{\number\ht\marxupbox}\immediate\write\boxesfile{\number\dp\marxupbox}\box\marxupbox};
\path[-,line width=0.4pt,line cap=butt,line join=miter,dash pattern=](8pt,-2.8926pt)--(14.5769pt,-2.8926pt)--(14.5769pt,-11.9442pt)--(8pt,-11.9442pt)--cycle;
\end{tikzpicture}}.
Indeed, composing it with an arbitrary tensor gives its antisymmetric
part with respect to the two connected indices.\footnote{Consider the
case of a matrix \(M\). Let \(A_{ij} = M_{ij} - M_{ji}\) and
\(S_{ij} = M_{ij} + M_{ji}\).  It is easy to see that \(S\) is
symmetric and that \(A\) is antisymmetric, and \(A\) is the
composition of \(\mathsf{antisym}\) with \(M\). Furthermore, \(M\) is
the average of \(A\) and \(S\): \(M = (A + S) / 2\). This is
why \(A\) is called the antisymmetric part of \(M\).\label{134}} 

\section{Tensor Calculus: Fields and Their Derivatives}\label{135} 
We have up to now worked with tensors as elements of certain vector
spaces, but to further illustrate the capabilities of {\sc{}Albert}, we apply
it to tensor \emph{calculus}, starting with the notion of fields.

\subsection{Tensor Fields}\label{136} 
In this context a field means that a different value is associated
with every position on a given manifold.\footnote{A manifold is a
topological space that looks like Euclidean space near each point.\label{137}}  We
denote the position parameter by \(\mathbf X\). Hence, the scalars
(from \cref{22}) are no longer just real numbers, but rather
real-valued expressions depending on \(\mathbf X\).\footnote{We
support such expressions in our implementation, but this
generalisation being orthogonal to the rest of the development, we
won't discuss it further.\label{138}}  For instance, \(\mathbf X\) could be a
position on the surface of the earth, and a scalar field could be the
temperature at each such point.

A vector field associates a vector with every
position; for instance, the wind
direction.  The perhaps surprising aspect is that each of these vectors
may inhabit a different, \emph{local}, vector space, which can be
thought of as tangent to the manifold at the considered point.
So in our example, we assumed that the wind is parallel to the earth surface.
Hereafter we
assume such a local space for each category \(z\), and call it \(T_{z}\),
leaving the dependency on position \(\mathbf X\) implicit.
Even though in the typical case the
local vector space is different at each position, it keeps the same
dimensionality. Therefore, as an object, it is independent of \(\mathbf X\).
In Haskell terms, \(T_{z}\) is an associated type; see \cref{144}.

When we deal with matrix representations and want to perform
computations with them, we need a way to identify the position
\(\mathbf X\).  For a general manifold, this is difficult to do, but we
will restrict our scope to the case where a single coordinate system
is sufficient.  We also need a basis at each position, which gives a
meaning to the entries in a tensor matrix representation (the meaning
of these coordinates change with position).  Furthermore, different
choices of coordinate system are possible for the same manifold.  The
coordinates used to identify the position will be referred to as the
global coordinates, while the coordinates of a tensor will be referred
to as local coordinates. This terminology is not usual in mathematical
praxis. However, we find that making this distinction is useful to
lift ambiguities.\footnote{In general, the global view is an
\emph{atlas} (a collection of charts) and \emph{transition maps} (between overlapping charts). The coordinates within each chart are
called local coordinates in the usual praxis. What is called “local
basis” here is usually called a “frame” (as part of a “frame bundle”).\label{139}} 
\begin{figure*}[]\begin{subfigure}[t]{0.5\textwidth}\begin{center}{
}\end{center}\caption{Cartesian}\label{140}\end{subfigure}\quad{}\begin{subfigure}[t]{0.5\textwidth}\begin{center}{%
}\end{center}\caption{Polar}\label{141}\end{subfigure}\caption{Possible fields of bases for a local space
field covering the Euclidean plane. In both examples, basis vectors
are tangent to coordinate lines; either Cartesian or
polar coordinates.  In the second instance the basis vectors are
undefined at the origin.}\label{142}\end{figure*} 

While the choice of basis field is arbitrary from an algebraic
perspective, some choices of basis will make certain computations
easier than others. Given a system of coordinates to identify
positions in the manifold, there is a canonical way to define the
local basis field. It is to let the base vectors be the partial
derivatives of the position \(\mathbf X\) with respect to each
global coordinate.  This yields base vectors which are tangent to coordinate
lines in the manifold.  In \cref{142}, one example follows Cartesian
coordinate lines, and the other polar coordinate lines. In the polar
case, we have the base vector fields \((\mathbf e_ρ, \mathbf
e_θ)\)\footnote{When used as indices \(ρ\) and \(θ\) act as
symbols which the index variables range over. Essentially, \(ρ=1\) and \(θ=2\). They are not index variables themselves.\label{143}} as basis for
\(T_{M_{\mathsf{p}}}\), with \(\mathbf e_ρ = \partial \mathbf X / \partial ρ\) and \(\mathbf e_θ = \partial \mathbf X / \partial θ\).
 
\subsection{Metrics and Index Juggling}\label{144} 
An important additional structure that one can add to vector spaces are associated
\emph{metrics}.  The (covariant) metric, noted \(g\), is what defines
the inner product of (local) vectors.
It is a tensor field that, when given two vectors as input, returns a scalar, their inner product.
One defines the contravariant metric \(g'\) as the inverse of the
covariant metric.\footnote{In the literature, both metrics are written using the same symbol (typically \(g\)), relying on the context to disambiguate.\label{145}} This can be specified as
\(g{_i}{_k}g'{^k}{^j}\) = \(δᵢʲ\).

To capture this algebraic structure, we distinguish the vector space for which we define the metric as an associated type, \(T_{z}\), of a new category class \(\mathsf{MetricCategory}\).

The various notations for metrics are shown in \cref{170}.
 \begin{list}{}{\setlength\leftmargin{1.0em}}\item\relax
\ensuremath{\begin{parray}\column{B}{@{}>{}l<{}@{}}\column[0em]{1}{@{}>{}l<{}@{}}\column[1em]{2}{@{}>{}l<{}@{}}\column{3}{@{}>{}l<{}@{}}\column{4}{@{}>{}l<{}@{}}\column{5}{@{}>{}l<{}@{}}\column{E}{@{}>{}l<{}@{}}%
\>[1]{\mathbf{class}\mskip 3.0mu\allowbreak{}\mathnormal{(}\mskip 0.0mu}\>[4]{\mathsf{CompactClosed}\mskip 3.0muz\mskip 0.0mu\mathnormal{)}\allowbreak{}\mskip 3.0mu\mathnormal{\Rightarrow }\mskip 3.0mu\mathsf{MetricCategory}\mskip 3.0muz\mskip 3.0mu\mathbf{where}}\<[E]\\
\>[2]{\mathbf{type}\mskip 3.0muT_{z}}\<[E]\\
\>[2]{g\mskip 3.0mu}\>[3]{\mathnormal{::}\mskip 3.0mu}\>[5]{\allowbreak{}\mathnormal{(}\mskip 0.0muT_{z}\mskip 3.0mu\mathnormal{⊗}\mskip 3.0muT_{z}\mskip 0.0mu\mathnormal{)}\allowbreak{}\mskip 1.0mu\overset{z}{\leadsto }\mskip 1.0mu\mathbf{1}}\<[E]\\
\>[2]{g'\mskip 3.0mu}\>[3]{\mathnormal{::}\mskip 3.0mu}\>[5]{\mathbf{1}\mskip 1.0mu\overset{z}{\leadsto }\mskip 1.0mu\allowbreak{}\mathnormal{(}\mskip 0.0muT_{z}\mskip 3.0mu\mathnormal{⊗}\mskip 3.0muT_{z}\mskip 0.0mu\mathnormal{)}\allowbreak{}}\<[E]\end{parray}}\end{list} Then we can embed the metric morphism in {\sc{}Albert} as follows:
\begin{list}{}{\setlength\leftmargin{1.0em}}\item\relax
 \ensuremath{\begin{parray}\column{B}{@{}>{}l<{}@{}}\column[0em]{1}{@{}>{}l<{}@{}}\column{2}{@{}>{}l<{}@{}}\column{E}{@{}>{}l<{}@{}}%
\>[1]{\mathsf{metric}\mskip 3.0mu\mathnormal{::}\mskip 3.0mu}\>[2]{\mathsf{MetricCategory}\mskip 3.0muz\mskip 3.0mu\mathnormal{\Rightarrow }\mskip 3.0mu\mathsf{P}\mskip 3.0muz\mskip 3.0mu\mathsf{r}\mskip 3.0muT_{z}\mskip 3.0mu\mathnormal{⊸}\mskip 3.0mu\mathsf{P}\mskip 3.0muz\mskip 3.0mu\mathsf{r}\mskip 3.0muT_{z}\mskip 3.0mu\mathnormal{⊸}\mskip 3.0mu\mathsf{R}\mskip 3.0muz\mskip 3.0mu\mathsf{r}}\<[E]\end{parray}} 

\ensuremath{\begin{parray}\column{B}{@{}>{}l<{}@{}}\column[0em]{1}{@{}>{}l<{}@{}}\column{E}{@{}>{}l<{}@{}}%
\>[1]{\mathsf{metric}\mskip 3.0mu_{\mathsf{i}}\mskip 3.0mu_{\mathsf{j}}\mskip 3.0mu\mathnormal{=}\mskip 3.0mu\mathsf{tensorEmbed}_{1}\mskip 3.0mug\mskip 3.0mu\allowbreak{}\mathnormal{(}\mskip 0.0mu\mathsf{merge}\mskip 3.0mu\allowbreak{}\mathnormal{(}\mskip 0.0mu_{\mathsf{i}}\mskip 0.0mu\mathnormal{,}\mskip 3.0mu_{\mathsf{j}}\mskip 0.0mu\mathnormal{)}\allowbreak{}\mskip 0.0mu\mathnormal{)}\allowbreak{}}\<[E]\end{parray}} \end{list} 
The metric tensor is symmetric--- a fact which can be expressed as its antisymmetric part ({\begin{tikzpicture}[baseline={([yshift=-0.8ex]current bounding box.center)}]\path[-,draw=black,line width=0.4pt,line cap=butt,line join=miter,dash pattern=](-7pt,7.4184pt)--(-6pt,7.4184pt);
\path[-,draw=black,line width=0.4pt,line cap=butt,line join=miter,dash pattern=](-7pt,0pt)--(-6pt,0pt);
\path[-,draw=lightgray,line width=0.4pt,line cap=butt,line join=miter,dash pattern=](6pt,3.7092pt)--(13pt,3.7092pt);
\path[-,draw=black,line width=0.4pt,line cap=butt,line join=miter,dash pattern=](0pt,7.4184pt)--(0pt,7.4184pt);
\path[-,draw=black,line width=0.4pt,line cap=butt,line join=miter,dash pattern=](0pt,0pt)--(0pt,0pt);
\path[-,fill=black,line width=0.4pt,line cap=butt,line join=miter,dash pattern=](7pt,3.7092pt)..controls(7pt,4.2615pt)and(6.5523pt,4.7092pt)..(6pt,4.7092pt)..controls(5.4477pt,4.7092pt)and(5pt,4.2615pt)..(5pt,3.7092pt)..controls(5pt,3.1569pt)and(5.4477pt,2.7092pt)..(6pt,2.7092pt)..controls(6.5523pt,2.7092pt)and(7pt,3.1569pt)..(7pt,3.7092pt)--cycle;
\path[-,draw=black,line width=0.4pt,line cap=butt,line join=miter,dash pattern=](3.06pt,7.0897pt)..controls(5.02pt,6.5767pt)and(6pt,5.4592pt)..(6pt,3.7092pt);
\path[-,draw=black,line width=0.4pt,line cap=butt,line join=miter,dash pattern=](0pt,7.4184pt)..controls(1.2pt,7.4184pt)and(2.22pt,7.3096pt)..(3.06pt,7.0897pt);
\path[-,draw=black,line width=0.4pt,line cap=butt,line join=miter,dash pattern=](3.06pt,0.3287pt)..controls(5.02pt,0.8417pt)and(6pt,1.9592pt)..(6pt,3.7092pt);
\path[-,draw=black,line width=0.4pt,line cap=butt,line join=miter,dash pattern=](0pt,0pt)..controls(1.2pt,0pt)and(2.22pt,0.1088pt)..(3.06pt,0.3287pt);
\path[-,draw=black,line width=0.4pt,line cap=butt,line join=miter,dash pattern=](-6pt,7.4184pt)--(0pt,7.4184pt);
\path[-,draw=black,line width=0.4pt,line cap=butt,line join=miter,dash pattern=](-6pt,0pt)--(0pt,0pt);
\path[-,draw=black,line width=1.2pt,line cap=butt,line join=miter,dash pattern=](-3pt,8.4184pt)--(-3pt,-1pt)--cycle;
\path[-,line width=0.4pt,line cap=butt,line join=miter,dash pattern=](-11.6316pt,9.758pt)--(-9pt,9.758pt)--(-9pt,5.0788pt)--(-11.6316pt,5.0788pt)--cycle;
\node[anchor=north west,inner sep=0] at (-11.6316pt,9.758pt){\savebox{\marxupbox}{{\({\scriptstyle i}\)}}\immediate\write\boxesfile{146}\immediate\write\boxesfile{\number\wd\marxupbox}\immediate\write\boxesfile{\number\ht\marxupbox}\immediate\write\boxesfile{\number\dp\marxupbox}\box\marxupbox};
\path[-,line width=0.4pt,line cap=butt,line join=miter,dash pattern=](-13.6316pt,11.758pt)--(-7pt,11.758pt)--(-7pt,3.0788pt)--(-13.6316pt,3.0788pt)--cycle;
\path[-,line width=0.4pt,line cap=butt,line join=miter,dash pattern=](-12.8617pt,3.0788pt)--(-9pt,3.0788pt)--(-9pt,-3.0788pt)--(-12.8617pt,-3.0788pt)--cycle;
\node[anchor=north west,inner sep=0] at (-12.8617pt,3.0788pt){\savebox{\marxupbox}{{\({\scriptstyle j}\)}}\immediate\write\boxesfile{147}\immediate\write\boxesfile{\number\wd\marxupbox}\immediate\write\boxesfile{\number\ht\marxupbox}\immediate\write\boxesfile{\number\dp\marxupbox}\box\marxupbox};
\path[-,line width=0.4pt,line cap=butt,line join=miter,dash pattern=](-14.8617pt,5.0788pt)--(-7pt,5.0788pt)--(-7pt,-5.0788pt)--(-14.8617pt,-5.0788pt)--cycle;
\end{tikzpicture}}) being zero.
The coefficients of the matrix representation of the metric is given by the
inner product of the basis vectors.
For the usual Cartesian basis, the metric is represented by the identity matrix.
For the basis \((\mathbf e_ρ, \mathbf e_θ)\) defined above, the matrix representation of the metric is:
\begin{displaymath}g_\mathsf{Polar} =\left[\begin{array}{cc}ρ^{2}&0\\0&1\end{array}\right]\end{displaymath} With this knowledge, we can define an instance for coordinate representations of the \(\mathsf{MetricCategory}\) class:
\begin{list}{}{\setlength\leftmargin{1.0em}}\item\relax
\ensuremath{\begin{parray}\column{B}{@{}>{}l<{}@{}}\column[0em]{1}{@{}>{}l<{}@{}}\column[1em]{2}{@{}>{}l<{}@{}}\column[2em]{3}{@{}>{}l<{}@{}}\column{4}{@{}>{}l<{}@{}}\column{5}{@{}>{}l<{}@{}}\column{E}{@{}>{}l<{}@{}}%
\>[1]{\mathbf{instance}\mskip 3.0mu\mathsf{MetricCategory}\mskip 3.0muM_{\mathsf{Polar}}\mskip 3.0mu\mathbf{where}}\<[E]\\
\>[2]{\mathbf{type}\mskip 3.0muT_{M_{\mathsf{Polar}}}\mskip 3.0mu\mathnormal{=}\mskip 3.0mu\mathsf{Atom}\mskip 3.0mu\mathsf{Polar}}\<[E]\\
\>[2]{g\mskip 3.0mu\mathnormal{=}\mskip 3.0mu\mathsf{Tab}\mskip 3.0mu\mathnormal{\$}\mskip 3.0muλ\mskip 3.0mu\allowbreak{}\mathnormal{(}\mskip 0.0mu\mathsf{Atom}\mskip 3.0mu\mathsf{i}\mskip 0.0mu\mathnormal{,}\mskip 3.0mu\mathsf{Atom}\mskip 3.0mu\mathsf{j}\mskip 0.0mu\mathnormal{)}\allowbreak{}\mskip 3.0mu\mathnormal{\_}\mskip 3.0mu\mathnormal{\rightarrow }\mskip 3.0mu\mathbf{case}\mskip 3.0mu\allowbreak{}\mathnormal{(}\mskip 0.0mu\mathsf{i}\mskip 0.0mu\mathnormal{,}\mskip 3.0mu\mathsf{j}\mskip 0.0mu\mathnormal{)}\allowbreak{}\mskip 3.0mu\mathbf{of}}\<[E]\\
\>[3]{\allowbreak{}\mathnormal{(}\mskip 0.0mu\mathsf{Rho}\mskip 0.0mu\mathnormal{,}\mskip 3.0mu}\>[4]{\mathsf{Rho}\mskip 0.0mu\mathnormal{)}\allowbreak{}\mskip 3.0mu}\>[5]{\mathnormal{\rightarrow }\mskip 3.0mu\mathsf{variable}\mskip 3.0mu\mathsf{Rho}\mskip 3.0mu\string^\mskip 3.0mu\mathrm{2}}\<[E]\\
\>[3]{\allowbreak{}\mathnormal{(}\mskip 0.0mu\mathsf{Theta}\mskip 0.0mu\mathnormal{,}\mskip 3.0mu}\>[4]{\mathsf{Theta}\mskip 0.0mu\mathnormal{)}\allowbreak{}\mskip 3.0mu}\>[5]{\mathnormal{\rightarrow }\mskip 3.0mu\mathrm{1}}\<[E]\\
\>[3]{\mathnormal{\_}\mskip 3.0mu}\>[5]{\mathnormal{\rightarrow }\mskip 3.0mu0}\<[E]\end{parray}}\end{list} The scalars associated with this additive category are expressions
where the variables ρ and θ occur, and the \(\mathsf{variable}\) function embeds a coordinate name into such expression types:
\begin{list}{}{\setlength\leftmargin{1.0em}}\item\relax
\ensuremath{\begin{parray}\column{B}{@{}>{}l<{}@{}}\column[0em]{1}{@{}>{}l<{}@{}}\column{E}{@{}>{}l<{}@{}}%
\>[1]{\mathsf{variable}\mskip 3.0mu\mathnormal{::}\mskip 3.0mu\mathsf{e}\mskip 3.0mu\mathnormal{\rightarrow }\mskip 3.0mu\mathsf{S}_{\mathsf{e}}}\<[E]\end{parray}}\end{list} Thus the \(\mathsf{Polar}\) data type
serves triple duty. First, it is the set of variables
used in such expressions. Second it serves to identify the meaning of
coordinates as a parameter in the \ensuremath{\begin{parray}\column{B}{@{}>{}l<{}@{}}\column[0em]{1}{@{}>{}l<{}@{}}\column{E}{@{}>{}l<{}@{}}%
\>[1]{M_{\mathsf{Polar}}}\<[E]\end{parray}} type. Third, it is used to construct
a representation of the atomic vector space. For the matrix category, an
object is a set indexing the base vectors, so \(\mathsf{Atom}\) is a simple wrapper around
the coordinate type:
\begin{list}{}{\setlength\leftmargin{1.0em}}\item\relax
\ensuremath{\begin{parray}\column{B}{@{}>{}l<{}@{}}\column[0em]{1}{@{}>{}l<{}@{}}\column{E}{@{}>{}l<{}@{}}%
\>[1]{\mathbf{newtype}\mskip 3.0mu\mathsf{Atom}\mskip 3.0mu\mathsf{s}\mskip 3.0mu\mathnormal{=}\mskip 3.0mu\mathsf{Atom}\mskip 3.0mu\mathsf{s}}\<[E]\end{parray}}\end{list} 
One can contrast the type of metrics with that of \(η\) and
\(ϵ\). First, only the tangent space \(T_{z}\) has a metric, while
every object has a dual.  Indeed, \(η\) and \(ϵ\) are
both realised as \(δ\) in the tensor instance, and are an identity
operation in Einstein notation. They have no geometric significance
and thus can be defined generically. In contrast, the metric depends on
the geometric properties of the space \(T_{z}\) that they
operate upon. Second, the metrics do not dualise objects, whereas
\(η\) and \(ϵ\) do.  Accordingly, our diagrammatic
notation does not make any special mark on the input/output wires of
metrics.
Yet, metrics satisfy a variant of the snake laws: {\begin{tikzpicture}[baseline={([yshift=-0.8ex]current bounding box.center)}]\path[-,draw=black,line width=0.4pt,line cap=butt,line join=miter,dash pattern=](-19pt,0pt)--(-12pt,0pt);
\path[-,draw=lightgray,line width=0.4pt,line cap=butt,line join=miter,dash pattern=](-19pt,-12.5pt)--(-12pt,-12.5pt);
\path[-,draw=lightgray,line width=0.4pt,line cap=butt,line join=miter,dash pattern=](6pt,-2.5pt)--(13pt,-2.5pt);
\path[-,draw=black,line width=0.4pt,line cap=butt,line join=miter,dash pattern=](6pt,-15pt)--(13pt,-15pt);
\path[-,draw=black,line width=0.4pt,line cap=butt,line join=miter,dash pattern=](-6pt,0pt)--(-6pt,0pt);
\path[-,draw=black,line width=0.4pt,line cap=butt,line join=miter,dash pattern=](-6pt,-10pt)--(-6pt,-10pt);
\path[-,draw=black,line width=0.4pt,line cap=butt,line join=miter,dash pattern=](-6pt,-15pt)--(-6pt,-15pt);
\path[-,draw=black,line width=0.4pt,line cap=butt,line join=miter,dash pattern=](0pt,0pt)--(0pt,0pt);
\path[-,draw=black,line width=0.4pt,line cap=butt,line join=miter,dash pattern=](0pt,-5pt)--(0pt,-5pt);
\path[-,draw=black,line width=0.4pt,line cap=butt,line join=miter,dash pattern=](0pt,-15pt)--(0pt,-15pt);
\path[-,fill=black,line width=0.4pt,line cap=butt,line join=miter,dash pattern=](7pt,-2.5pt)..controls(7pt,-1.9477pt)and(6.5523pt,-1.5pt)..(6pt,-1.5pt)..controls(5.4477pt,-1.5pt)and(5pt,-1.9477pt)..(5pt,-2.5pt)..controls(5pt,-3.0523pt)and(5.4477pt,-3.5pt)..(6pt,-3.5pt)..controls(6.5523pt,-3.5pt)and(7pt,-3.0523pt)..(7pt,-2.5pt)--cycle;
\path[-,draw=black,line width=0.4pt,line cap=butt,line join=miter,dash pattern=](3.06pt,-0.0675pt)..controls(5.02pt,-0.225pt)and(6pt,-0.75pt)..(6pt,-2.5pt);
\path[-,draw=black,line width=0.4pt,line cap=butt,line join=miter,dash pattern=](0pt,0pt)..controls(1.2pt,0pt)and(2.22pt,0pt)..(3.06pt,-0.0675pt);
\path[-,draw=black,line width=0.4pt,line cap=butt,line join=miter,dash pattern=](3.06pt,-4.9325pt)..controls(5.02pt,-4.775pt)and(6pt,-4.25pt)..(6pt,-2.5pt);
\path[-,draw=black,line width=0.4pt,line cap=butt,line join=miter,dash pattern=](0pt,-5pt)..controls(1.2pt,-5pt)and(2.22pt,-5pt)..(3.06pt,-4.9325pt);
\path[-,draw=black,line width=0.4pt,line cap=butt,line join=miter,dash pattern=](0pt,-15pt)--(6pt,-15pt);
\path[-,draw=black,line width=0.4pt,line cap=butt,line join=miter,dash pattern=](-6pt,0pt)--(0pt,0pt);
\path[-,draw=black,line width=0.4pt,line cap=butt,line join=miter,dash pattern=](-6pt,-10pt)..controls(-2pt,-10pt)and(-4pt,-5pt)..(0pt,-5pt);
\path[-,draw=black,line width=0.4pt,line cap=butt,line join=miter,dash pattern=](-6pt,-15pt)--(0pt,-15pt);
\path[-,draw=black,line width=0.4pt,line cap=butt,line join=miter,dash pattern=](-12pt,0pt)--(-6pt,0pt);
\path[-,fill=black,line width=0.4pt,line cap=butt,line join=miter,dash pattern=](-11pt,-12.5pt)..controls(-11pt,-11.9477pt)and(-11.4477pt,-11.5pt)..(-12pt,-11.5pt)..controls(-12.5523pt,-11.5pt)and(-13pt,-11.9477pt)..(-13pt,-12.5pt)..controls(-13pt,-13.0523pt)and(-12.5523pt,-13.5pt)..(-12pt,-13.5pt)..controls(-11.4477pt,-13.5pt)and(-11pt,-13.0523pt)..(-11pt,-12.5pt)--cycle;
\path[-,draw=black,line width=0.4pt,line cap=butt,line join=miter,dash pattern=](-9.06pt,-10.0675pt)..controls(-8.22pt,-10pt)and(-7.2pt,-10pt)..(-6pt,-10pt);
\path[-,draw=black,line width=0.4pt,line cap=butt,line join=miter,dash pattern=](-12pt,-12.5pt)..controls(-12pt,-10.75pt)and(-11.02pt,-10.225pt)..(-9.06pt,-10.0675pt);
\path[-,draw=black,line width=0.4pt,line cap=butt,line join=miter,dash pattern=](-9.06pt,-14.9325pt)..controls(-8.22pt,-15pt)and(-7.2pt,-15pt)..(-6pt,-15pt);
\path[-,draw=black,line width=0.4pt,line cap=butt,line join=miter,dash pattern=](-12pt,-12.5pt)..controls(-12pt,-14.25pt)and(-11.02pt,-14.775pt)..(-9.06pt,-14.9325pt);
\path[-,line width=0.4pt,line cap=butt,line join=miter,dash pattern=](-23.6316pt,2.3396pt)--(-21pt,2.3396pt)--(-21pt,-2.3396pt)--(-23.6316pt,-2.3396pt)--cycle;
\node[anchor=north west,inner sep=0] at (-23.6316pt,2.3396pt){\savebox{\marxupbox}{{\({\scriptstyle i}\)}}\immediate\write\boxesfile{148}\immediate\write\boxesfile{\number\wd\marxupbox}\immediate\write\boxesfile{\number\ht\marxupbox}\immediate\write\boxesfile{\number\dp\marxupbox}\box\marxupbox};
\path[-,line width=0.4pt,line cap=butt,line join=miter,dash pattern=](-25.6316pt,4.3396pt)--(-19pt,4.3396pt)--(-19pt,-4.3396pt)--(-25.6316pt,-4.3396pt)--cycle;
\path[-,line width=0.4pt,line cap=butt,line join=miter,dash pattern=](15pt,-11.9212pt)--(18.8617pt,-11.9212pt)--(18.8617pt,-18.0788pt)--(15pt,-18.0788pt)--cycle;
\node[anchor=north west,inner sep=0] at (15pt,-11.9212pt){\savebox{\marxupbox}{{\({\scriptstyle j}\)}}\immediate\write\boxesfile{149}\immediate\write\boxesfile{\number\wd\marxupbox}\immediate\write\boxesfile{\number\ht\marxupbox}\immediate\write\boxesfile{\number\dp\marxupbox}\box\marxupbox};
\path[-,line width=0.4pt,line cap=butt,line join=miter,dash pattern=](13pt,-9.9212pt)--(20.8617pt,-9.9212pt)--(20.8617pt,-20.0788pt)--(13pt,-20.0788pt)--cycle;
\end{tikzpicture}} is equal to {\begin{tikzpicture}[baseline={([yshift=-0.8ex]current bounding box.center)}]\path[-,draw=black,line width=0.4pt,line cap=butt,line join=miter,dash pattern=](-14pt,5pt)--(-7pt,5pt);
\path[-,draw=lightgray,line width=0.4pt,line cap=butt,line join=miter,dash pattern=](-14pt,0pt)--(-7pt,0pt);
\path[-,draw=lightgray,line width=0.4pt,line cap=butt,line join=miter,dash pattern=](7pt,5pt)--(8pt,5pt);
\path[-,draw=black,line width=0.4pt,line cap=butt,line join=miter,dash pattern=](7pt,0pt)--(8pt,0pt);
\path[-,draw=black,line width=0.4pt,line cap=butt,line join=miter,dash pattern=](-7pt,5pt)--(0pt,5pt);
\path[-,draw=black,line width=0.4pt,line cap=butt,line join=miter,dash pattern=](0pt,5pt)--(1pt,5pt);
\path[-,draw=lightgray,line width=0.4pt,line cap=butt,line join=miter,dash pattern=](0pt,0pt)--(1pt,0pt);
\path[-,draw=black,line width=0.4pt,line cap=butt,line join=miter,dash pattern=](1pt,5pt)..controls(5pt,5pt)and(3pt,0pt)..(7pt,0pt);
\path[-,draw=lightgray,line width=0.4pt,line cap=butt,line join=miter,dash pattern=](1pt,0pt)..controls(5pt,0pt)and(3pt,5pt)..(7pt,5pt);
\path[-,draw=black,line width=0.4pt,line cap=butt,line join=miter,dash pattern=](0pt,5pt)--(0pt,5pt);
\path[-,line width=0.4pt,line cap=butt,line join=miter,dash pattern=on 0.4pt off 1pt](0pt,5pt)--(0pt,5pt)--(0pt,0pt)--(0pt,0pt)--cycle;
\path[-,draw=black,line width=0.4pt,line cap=butt,line join=miter,dash pattern=on 0.4pt off 1pt](-3pt,8pt)--(3pt,8pt)--(3pt,-3pt)--(-3pt,-3pt)--cycle;
\path[-,fill=lightgray,line width=0.4pt,line cap=butt,line join=miter,dash pattern=](1pt,0pt)..controls(1pt,0.5523pt)and(0.5523pt,1pt)..(0pt,1pt)..controls(-0.5523pt,1pt)and(-1pt,0.5523pt)..(-1pt,0pt)..controls(-1pt,-0.5523pt)and(-0.5523pt,-1pt)..(0pt,-1pt)..controls(0.5523pt,-1pt)and(1pt,-0.5523pt)..(1pt,0pt)--cycle;
\path[-,draw=black,line width=0.4pt,line cap=butt,line join=miter,dash pattern=](-7pt,5pt)--(-7pt,5pt);
\path[-,line width=0.4pt,line cap=butt,line join=miter,dash pattern=on 0.4pt off 1pt](-7pt,5pt)--(-7pt,5pt)--(-7pt,0pt)--(-7pt,0pt)--cycle;
\path[-,draw=black,line width=0.4pt,line cap=butt,line join=miter,dash pattern=on 0.4pt off 1pt](-10pt,8pt)--(-4pt,8pt)--(-4pt,-3pt)--(-10pt,-3pt)--cycle;
\path[-,fill=lightgray,line width=0.4pt,line cap=butt,line join=miter,dash pattern=](-6pt,0pt)..controls(-6pt,0.5523pt)and(-6.4477pt,1pt)..(-7pt,1pt)..controls(-7.5523pt,1pt)and(-8pt,0.5523pt)..(-8pt,0pt)..controls(-8pt,-0.5523pt)and(-7.5523pt,-1pt)..(-7pt,-1pt)..controls(-6.4477pt,-1pt)and(-6pt,-0.5523pt)..(-6pt,0pt)--cycle;
\path[-,line width=0.4pt,line cap=butt,line join=miter,dash pattern=](-18.6316pt,7.3396pt)--(-16pt,7.3396pt)--(-16pt,2.6604pt)--(-18.6316pt,2.6604pt)--cycle;
\node[anchor=north west,inner sep=0] at (-18.6316pt,7.3396pt){\savebox{\marxupbox}{{\({\scriptstyle i}\)}}\immediate\write\boxesfile{150}\immediate\write\boxesfile{\number\wd\marxupbox}\immediate\write\boxesfile{\number\ht\marxupbox}\immediate\write\boxesfile{\number\dp\marxupbox}\box\marxupbox};
\path[-,line width=0.4pt,line cap=butt,line join=miter,dash pattern=](-20.6316pt,9.3396pt)--(-14pt,9.3396pt)--(-14pt,0.6604pt)--(-20.6316pt,0.6604pt)--cycle;
\path[-,line width=0.4pt,line cap=butt,line join=miter,dash pattern=](10pt,3.0788pt)--(13.8617pt,3.0788pt)--(13.8617pt,-3.0788pt)--(10pt,-3.0788pt)--cycle;
\node[anchor=north west,inner sep=0] at (10pt,3.0788pt){\savebox{\marxupbox}{{\({\scriptstyle j}\)}}\immediate\write\boxesfile{151}\immediate\write\boxesfile{\number\wd\marxupbox}\immediate\write\boxesfile{\number\ht\marxupbox}\immediate\write\boxesfile{\number\dp\marxupbox}\box\marxupbox};
\path[-,line width=0.4pt,line cap=butt,line join=miter,dash pattern=](8pt,5.0788pt)--(15.8617pt,5.0788pt)--(15.8617pt,-5.0788pt)--(8pt,-5.0788pt)--cycle;
\end{tikzpicture}}.

By combining the compact closed categorical structure with metrics,
one can construct the following two morphisms (in {\sc{}Roger}):
\begin{list}{}{\setlength\leftmargin{1.0em}}\item\relax
\ensuremath{\begin{parray}\column{B}{@{}>{}l<{}@{}}\column[0em]{1}{@{}>{}l<{}@{}}\column{2}{@{}>{}l<{}@{}}\column{3}{@{}>{}l<{}@{}}\column{E}{@{}>{}l<{}@{}}%
\>[1]{\mathsf{juggleDown}\mskip 3.0mu}\>[2]{\mathnormal{::}\mskip 3.0mu}\>[3]{\mathsf{MetricCategory}\mskip 3.0muz\mskip 3.0mu\mathnormal{\Rightarrow }\mskip 3.0mu\dual{T_{z}}\mskip 1.0mu\overset{z}{\leadsto }\mskip 1.0muT_{z}}\<[E]\\
\>[1]{\mathsf{juggleUp}\mskip 3.0mu}\>[2]{\mathnormal{::}\mskip 3.0mu}\>[3]{\mathsf{MetricCategory}\mskip 3.0muz\mskip 3.0mu\mathnormal{\Rightarrow }\mskip 3.0muT_{z}\mskip 1.0mu\overset{z}{\leadsto }\mskip 1.0mu\dual{T_{z}}}\<[E]\\
\>[1]{\mathsf{juggleDown}\mskip 3.0mu\mathnormal{=}\mskip 3.0mu\bar{ρ}\mskip 3.0mu\allowbreak{}\mathnormal{∘}\allowbreak{}\mskip 3.0muσ\mskip 3.0mu\allowbreak{}\mathnormal{∘}\allowbreak{}\mskip 3.0mu\allowbreak{}\mathnormal{(}\mskip 0.0mu\allowbreak{}\mathnormal{(}\mskip 0.0muϵ\mskip 3.0mu\allowbreak{}\mathnormal{∘}\allowbreak{}\mskip 3.0muσ\mskip 0.0mu\mathnormal{)}\allowbreak{}\mskip 3.0mu{⊗}\mskip 3.0mu\mathsf{id}\mskip 0.0mu\mathnormal{)}\allowbreak{}\mskip 3.0mu\allowbreak{}\mathnormal{∘}\allowbreak{}\mskip 3.0mu\bar{α}\mskip 3.0mu\allowbreak{}\mathnormal{∘}\allowbreak{}\mskip 3.0mu\allowbreak{}\mathnormal{(}\mskip 0.0mu\mathsf{id}\mskip 3.0mu{⊗}\mskip 3.0mug'\mskip 0.0mu\mathnormal{)}\allowbreak{}\mskip 3.0mu\allowbreak{}\mathnormal{∘}\allowbreak{}\mskip 3.0muρ}\<[E]\\
\>[1]{\mathsf{juggleUp}\mskip 3.0mu\mathnormal{=}\mskip 3.0mu\bar{ρ}\mskip 3.0mu\allowbreak{}\mathnormal{∘}\allowbreak{}\mskip 3.0mu\allowbreak{}\mathnormal{(}\mskip 0.0mu\mathsf{id}\mskip 3.0mu{⊗}\mskip 3.0mug\mskip 0.0mu\mathnormal{)}\allowbreak{}\mskip 3.0mu\allowbreak{}\mathnormal{∘}\allowbreak{}\mskip 3.0muα\mskip 3.0mu\allowbreak{}\mathnormal{∘}\allowbreak{}\mskip 3.0mu\allowbreak{}\mathnormal{(}\mskip 0.0muσ\mskip 3.0mu{⊗}\mskip 3.0mu\mathsf{id}\mskip 0.0mu\mathnormal{)}\allowbreak{}\mskip 3.0mu\allowbreak{}\mathnormal{∘}\allowbreak{}\mskip 3.0mu\bar{α}\mskip 3.0mu\allowbreak{}\mathnormal{∘}\allowbreak{}\mskip 3.0mu\allowbreak{}\mathnormal{(}\mskip 0.0mu\mathsf{id}\mskip 3.0mu{⊗}\mskip 3.0muη\mskip 0.0mu\mathnormal{)}\allowbreak{}\mskip 3.0mu\allowbreak{}\mathnormal{∘}\allowbreak{}\mskip 3.0muρ}\<[E]\end{parray}}\end{list} In diagram form \(\mathsf{juggleUp}\) is {\begin{tikzpicture}[baseline={([yshift=-0.8ex]current bounding box.center)}]\path[-,draw=black,line width=0.4pt,line cap=butt,line join=miter,dash pattern=](-51pt,0pt)--(-44pt,0pt);
\path[-,draw=black,line width=0.4pt,line cap=butt,line join=miter,dash pattern=](0pt,0pt)--(7pt,0pt);
\path[-,draw=black,line width=0.4pt,line cap=butt,line join=miter,dash pattern=](-44pt,0pt)--(-37pt,0pt);
\path[-,draw=lightgray,line width=0.4pt,line cap=butt,line join=miter,dash pattern=](-44pt,-12.5pt)--(-37pt,-12.5pt);
\path[-,draw=black,line width=0.4pt,line cap=butt,line join=miter,dash pattern=](-31pt,0pt)--(-31pt,0pt);
\path[-,draw=black,line width=0.4pt,line cap=butt,line join=miter,dash pattern=](-31pt,-10pt)--(-31pt,-10pt);
\path[-,draw=black,line width=0.4pt,line cap=butt,line join=miter,dash pattern=](-31pt,-15pt)--(-31pt,-15pt);
\path[-,draw=black,line width=0.4pt,line cap=butt,line join=miter,dash pattern=](-25pt,0pt)--(-25pt,0pt);
\path[-,draw=black,line width=0.4pt,line cap=butt,line join=miter,dash pattern=](-25pt,-5pt)--(-25pt,-5pt);
\path[-,draw=black,line width=0.4pt,line cap=butt,line join=miter,dash pattern=](-25pt,-15pt)--(-25pt,-15pt);
\path[-,draw=black,line width=0.4pt,line cap=butt,line join=miter,dash pattern=](-19pt,0pt)--(-19pt,0pt);
\path[-,draw=black,line width=0.4pt,line cap=butt,line join=miter,dash pattern=](-19pt,-5pt)--(-19pt,-5pt);
\path[-,draw=black,line width=0.4pt,line cap=butt,line join=miter,dash pattern=](-19pt,-15pt)--(-19pt,-15pt);
\path[-,draw=black,line width=0.4pt,line cap=butt,line join=miter,dash pattern=](-13pt,0pt)--(-13pt,0pt);
\path[-,draw=black,line width=0.4pt,line cap=butt,line join=miter,dash pattern=](-13pt,-10pt)--(-13pt,-10pt);
\path[-,draw=black,line width=0.4pt,line cap=butt,line join=miter,dash pattern=](-13pt,-15pt)--(-13pt,-15pt);
\path[-,draw=black,line width=0.4pt,line cap=butt,line join=miter,dash pattern=](-7pt,0pt)--(0pt,0pt);
\path[-,draw=lightgray,line width=0.4pt,line cap=butt,line join=miter,dash pattern=](-7pt,-12.5pt)--(0pt,-12.5pt);
\path[-,draw=black,line width=0.4pt,line cap=butt,line join=miter,dash pattern=](0pt,0pt)--(0pt,0pt);
\path[-,line width=0.4pt,line cap=butt,line join=miter,dash pattern=on 0.4pt off 1pt](0pt,0pt)--(0pt,0pt)--(0pt,-12.5pt)--(0pt,-12.5pt)--cycle;
\path[-,draw=black,line width=0.4pt,line cap=butt,line join=miter,dash pattern=on 0.4pt off 1pt](-3pt,3pt)--(3pt,3pt)--(3pt,-15.5pt)--(-3pt,-15.5pt)--cycle;
\path[-,fill=lightgray,line width=0.4pt,line cap=butt,line join=miter,dash pattern=](1pt,-12.5pt)..controls(1pt,-11.9477pt)and(0.5523pt,-11.5pt)..(0pt,-11.5pt)..controls(-0.5523pt,-11.5pt)and(-1pt,-11.9477pt)..(-1pt,-12.5pt)..controls(-1pt,-13.0523pt)and(-0.5523pt,-13.5pt)..(0pt,-13.5pt)..controls(0.5523pt,-13.5pt)and(1pt,-13.0523pt)..(1pt,-12.5pt)--cycle;
\path[-,draw=black,line width=0.4pt,line cap=butt,line join=miter,dash pattern=](-13pt,0pt)--(-7pt,0pt);
\path[-,fill=black,line width=0.4pt,line cap=butt,line join=miter,dash pattern=](-6pt,-12.5pt)..controls(-6pt,-11.9477pt)and(-6.4477pt,-11.5pt)..(-7pt,-11.5pt)..controls(-7.5523pt,-11.5pt)and(-8pt,-11.9477pt)..(-8pt,-12.5pt)..controls(-8pt,-13.0523pt)and(-7.5523pt,-13.5pt)..(-7pt,-13.5pt)..controls(-6.4477pt,-13.5pt)and(-6pt,-13.0523pt)..(-6pt,-12.5pt)--cycle;
\path[-,draw=black,line width=0.4pt,line cap=butt,line join=miter,dash pattern=](-9.94pt,-10.0675pt)..controls(-7.98pt,-10.225pt)and(-7pt,-10.75pt)..(-7pt,-12.5pt);
\path[-,draw=black,line width=0.4pt,line cap=butt,line join=miter,dash pattern=](-13pt,-10pt)..controls(-11.8pt,-10pt)and(-10.78pt,-10pt)..(-9.94pt,-10.0675pt);
\path[-,draw=black,line width=0.4pt,line cap=butt,line join=miter,dash pattern=](-9.94pt,-14.9325pt)..controls(-7.98pt,-14.775pt)and(-7pt,-14.25pt)..(-7pt,-12.5pt);
\path[-,draw=black,line width=0.4pt,line cap=butt,line join=miter,dash pattern=](-13pt,-15pt)..controls(-11.8pt,-15pt)and(-10.78pt,-15pt)..(-9.94pt,-14.9325pt);
\path[-,draw=black,line width=0.4pt,line cap=butt,line join=miter,dash pattern=](-19pt,0pt)--(-13pt,0pt);
\path[-,draw=black,line width=0.4pt,line cap=butt,line join=miter,dash pattern=](-19pt,-5pt)..controls(-15pt,-5pt)and(-17pt,-10pt)..(-13pt,-10pt);
\path[-,draw=black,line width=0.4pt,line cap=butt,line join=miter,dash pattern=](-19pt,-15pt)--(-13pt,-15pt);
\path[-,draw=black,line width=0.4pt,line cap=butt,line join=miter,dash pattern=](-25pt,0pt)..controls(-21pt,0pt)and(-23pt,-5pt)..(-19pt,-5pt);
\path[-,draw=black,line width=0.4pt,line cap=butt,line join=miter,dash pattern=](-25pt,-5pt)..controls(-21pt,-5pt)and(-23pt,0pt)..(-19pt,0pt);
\path[-,draw=black,line width=0.4pt,line cap=butt,line join=miter,dash pattern=](-25pt,-15pt)--(-19pt,-15pt);
\path[-,draw=black,line width=0.4pt,line cap=butt,line join=miter,dash pattern=](-31pt,0pt)--(-25pt,0pt);
\path[-,draw=black,line width=0.4pt,line cap=butt,line join=miter,dash pattern=](-31pt,-10pt)..controls(-27pt,-10pt)and(-29pt,-5pt)..(-25pt,-5pt);
\path[-,draw=black,line width=0.4pt,line cap=butt,line join=miter,dash pattern=](-31pt,-15pt)--(-25pt,-15pt);
\path[-,draw=black,line width=0.4pt,line cap=butt,line join=miter,dash pattern=](-37pt,0pt)--(-31pt,0pt);
\path[to-,draw=black,line width=0.4pt,line cap=butt,line join=miter,dash pattern=](-34.06pt,-10.0675pt)..controls(-33.22pt,-10pt)and(-32.2pt,-10pt)..(-31pt,-10pt);
\path[-,draw=black,line width=0.4pt,line cap=butt,line join=miter,dash pattern=](-37pt,-12.5pt)..controls(-37pt,-10.75pt)and(-36.02pt,-10.225pt)..(-34.06pt,-10.0675pt);
\path[-,draw=black,line width=0.4pt,line cap=butt,line join=miter,dash pattern=](-34.06pt,-14.9325pt)..controls(-33.22pt,-15pt)and(-32.2pt,-15pt)..(-31pt,-15pt);
\path[-,draw=black,line width=0.4pt,line cap=butt,line join=miter,dash pattern=](-37pt,-12.5pt)..controls(-37pt,-14.25pt)and(-36.02pt,-14.775pt)..(-34.06pt,-14.9325pt);
\path[-,draw=black,line width=0.4pt,line cap=butt,line join=miter,dash pattern=](-44pt,0pt)--(-44pt,0pt);
\path[-,line width=0.4pt,line cap=butt,line join=miter,dash pattern=on 0.4pt off 1pt](-44pt,0pt)--(-44pt,0pt)--(-44pt,-12.5pt)--(-44pt,-12.5pt)--cycle;
\path[-,draw=black,line width=0.4pt,line cap=butt,line join=miter,dash pattern=on 0.4pt off 1pt](-47pt,3pt)--(-41pt,3pt)--(-41pt,-15.5pt)--(-47pt,-15.5pt)--cycle;
\path[-,fill=lightgray,line width=0.4pt,line cap=butt,line join=miter,dash pattern=](-43pt,-12.5pt)..controls(-43pt,-11.9477pt)and(-43.4477pt,-11.5pt)..(-44pt,-11.5pt)..controls(-44.5523pt,-11.5pt)and(-45pt,-11.9477pt)..(-45pt,-12.5pt)..controls(-45pt,-13.0523pt)and(-44.5523pt,-13.5pt)..(-44pt,-13.5pt)..controls(-43.4477pt,-13.5pt)and(-43pt,-13.0523pt)..(-43pt,-12.5pt)--cycle;
\path[-,line width=0.4pt,line cap=butt,line join=miter,dash pattern=](-55.6316pt,2.3396pt)--(-53pt,2.3396pt)--(-53pt,-2.3396pt)--(-55.6316pt,-2.3396pt)--cycle;
\node[anchor=north west,inner sep=0] at (-55.6316pt,2.3396pt){\savebox{\marxupbox}{{\({\scriptstyle i}\)}}\immediate\write\boxesfile{152}\immediate\write\boxesfile{\number\wd\marxupbox}\immediate\write\boxesfile{\number\ht\marxupbox}\immediate\write\boxesfile{\number\dp\marxupbox}\box\marxupbox};
\path[-,line width=0.4pt,line cap=butt,line join=miter,dash pattern=](-57.6316pt,4.3396pt)--(-51pt,4.3396pt)--(-51pt,-4.3396pt)--(-57.6316pt,-4.3396pt)--cycle;
\path[-,line width=0.4pt,line cap=butt,line join=miter,dash pattern=](9pt,3.0788pt)--(12.8617pt,3.0788pt)--(12.8617pt,-3.0788pt)--(9pt,-3.0788pt)--cycle;
\node[anchor=north west,inner sep=0] at (9pt,3.0788pt){\savebox{\marxupbox}{{\({\scriptstyle j}\)}}\immediate\write\boxesfile{153}\immediate\write\boxesfile{\number\wd\marxupbox}\immediate\write\boxesfile{\number\ht\marxupbox}\immediate\write\boxesfile{\number\dp\marxupbox}\box\marxupbox};
\path[-,line width=0.4pt,line cap=butt,line join=miter,dash pattern=](7pt,5.0788pt)--(14.8617pt,5.0788pt)--(14.8617pt,-5.0788pt)--(7pt,-5.0788pt)--cycle;
\end{tikzpicture}}.
The existence of such morphisms have a direct consequence
on Einstein notation: any subscript index \(\mathsf{P}\mskip 3.0muz\mskip 3.0mu\mathsf{r}\mskip 3.0mu\mathsf{a}\) can be
raised into \(\mathsf{P}\mskip 3.0muz\mskip 3.0mu\mathsf{r}\mskip 3.0mu\dual{\mathsf{a}}\), and \textit{vice versa}. In other
words, any index can be used as super or sub-script as needed. As a
first example, given a vector \(v^{i}\), the
covector \(v_i\) is a shorthand for \(v^k
g_{ki}\). It is important to note that the position of the indices
changes the value of the expression: the actual numbers (the components) of a
vector \(v^{i}\) are different from those of the corresponding
covector \(v_i\) unless the metric representation is the identity (as in global Cartesian
coordinates). As another example, assuming a tensor \(\mathsf{t}\mskip 3.0mu\mathnormal{:}\mskip 3.0muT_{z}\mskip 3.0mu\mathnormal{⊗}\mskip 3.0muT_{z}\mskip 1.0mu\overset{z}{\leadsto }\mskip 1.0muT_{z}\), the expression \(t_{ijl}\) is a shorthand for
\(t_{ij}{}^{k} g_{kl}\).

In the mathematical praxis, raising and lowering indices is also
referred to as index juggling.  In {\sc{}Albert}, index juggling is
realised explicitly: our recent example can be written
\(\mathsf{t}\mskip 3.0mu_{\mathsf{i}}\mskip 3.0mu_{\mathsf{j}}\mskip 3.0mu\allowbreak{}\mathnormal{(}\mskip 0.0mu\mathsf{raise}\mskip 3.0mu_{\mathsf{l}}\mskip 0.0mu\mathnormal{)}\allowbreak{}\).  The raising and lowering
functions have the following types, and are implemented by embedding \(\mathsf{juggleUp}\) and \(\mathsf{juggleDown}\):
\begin{list}{}{\setlength\leftmargin{1.0em}}\item\relax
\ensuremath{\begin{parray}\column{B}{@{}>{}l<{}@{}}\column[0em]{1}{@{}>{}l<{}@{}}\column{2}{@{}>{}l<{}@{}}\column{3}{@{}>{}l<{}@{}}\column{E}{@{}>{}l<{}@{}}%
\>[1]{\mathsf{raise}\mskip 3.0mu}\>[2]{\mathnormal{::}\mskip 3.0mu\mathsf{MetricCategory}\mskip 3.0muz\mskip 3.0mu\mathnormal{\Rightarrow }\mskip 3.0mu}\>[3]{\mathsf{P}\mskip 3.0muz\mskip 3.0mu\mathsf{r}\mskip 3.0muT_{z}\mskip 3.0mu\mathnormal{⊸}\mskip 3.0mu\mathsf{P}\mskip 3.0muz\mskip 3.0mu\mathsf{r}\mskip 3.0mu\dual{T_{z}}}\<[E]\\
\>[1]{\mathsf{lower}\mskip 3.0mu}\>[2]{\mathnormal{::}\mskip 3.0mu\mathsf{MetricCategory}\mskip 3.0muz\mskip 3.0mu\mathnormal{\Rightarrow }\mskip 3.0mu}\>[3]{\mathsf{P}\mskip 3.0muz\mskip 3.0mu\mathsf{r}\mskip 3.0mu\dual{T_{z}}\mskip 3.0mu\mathnormal{⊸}\mskip 3.0mu\mathsf{P}\mskip 3.0muz\mskip 3.0mu\mathsf{r}\mskip 3.0muT_{z}}\<[E]\end{parray}}\end{list} 
\subsection{Change of global coordinate system}\label{154} As we have seen, Einstein
notation is carefully set up to work without reference to the system
of coordinates used to identify positions in the manifold. In
practice, this means that one can reason about tensors, their
derivatives (\cref{171}), etc. without worrying about the choice
of coordinates. It is only at the very last stage that one chooses a
system of global coordinates where the data of the problem is easy to express
and calculate with. We have seen that one can convert between bases by
applying the appropriate transformations on the representation matrices (see \cref{32}).
For canonical bases (defined as partial derivatives of the position), the
\(\mathsf{transform}\) function is obtained by composing with
Jacobian.
In our running example, we have at our disposal the Jacobian \(J(T)\) and its inverse \(J(\dual T)\) to convert between polar and Cartesian
global coordinates with \(J(\dual T) = \) \(\left[\begin{array}{cc}\cos{θ}&\sin{θ}\\ρ^{-1}\sin{θ}&-ρ^{-1}\cos{θ}\end{array}\right]\).

As an illustration, let \(\tilde v_{\mathsf{Polar}}\) =
\(\left[\begin{array}{cc}-ρ^{-1}&0\end{array}\right]\) be the local coordinates of a co-vector field in the polar tangent basis
\(\tilde v_{\mathsf{Polar}}\)\(\mathnormal{:}\mskip 3.0muT_{M_{\mathsf{Polar}}}\mskip 1.0mu\overset{M_{\mathsf{Polar}}}{\leadsto }\mskip 1.0mu\mathbf{1}\).
 Then the local coordinates of the co-vector field in the Cartesian tangent
basis are given by \({v_{\mathsf{Cartesian}}}_i = J(T)_i^j {v_{\mathsf{Polar}}}_j\), and we have: \(\tilde v_{\mathsf{Cartesian}}\) = \(\left[\begin{array}{cc}-ρ^{-1}\cos{θ}&-ρ^{-1}\sin{θ}\end{array}\right]\).

\begin{figure}[]\begin{center}\begin{tabular}{ccccc}{\begin{tikzpicture}\path[-,draw=black,line width=0.4pt,line cap=butt,line join=miter,dash pattern=](-22.32pt,13.1709pt)--(-19.32pt,13.1709pt);
\path[-,draw=black,line width=0.4pt,line cap=butt,line join=miter,dash pattern=](-22.32pt,5.7525pt)--(-19.32pt,5.7525pt);
\path[-,draw=black,line width=0.4pt,line cap=butt,line join=miter,dash pattern=](4pt,5.7525pt)--(7pt,5.7525pt);
\path[-,draw=black,line width=0.8pt,line cap=butt,line join=miter,dash pattern=](-15.32pt,14.1709pt)..controls(-15.32pt,15.8277pt)and(-13.9768pt,17.1709pt)..(-12.32pt,17.1709pt)--(-3pt,17.1709pt)..controls(-1.3431pt,17.1709pt)and(0pt,15.8277pt)..(0pt,14.1709pt)--(0pt,0pt)..controls(0pt,-1.6569pt)and(-1.3431pt,-3pt)..(-3pt,-3pt)--(-12.32pt,-3pt)..controls(-13.9768pt,-3pt)and(-15.32pt,-1.6569pt)..(-15.32pt,0pt)--cycle;
\path[-,draw=black,line width=0.4pt,line cap=butt,line join=miter,dash pattern=](-15.32pt,5.7525pt)--(-12.32pt,5.7525pt);
\path[-,draw=black,line width=0.4pt,line cap=butt,line join=miter,dash pattern=](-15.32pt,13.1709pt)--(-15.32pt,13.1709pt);
\path[-,draw=black,line width=0.4pt,line cap=butt,line join=miter,dash pattern=](-3pt,5.7525pt)--(0pt,5.7525pt);
\path[-,draw=black,line width=0.4pt,line cap=butt,line join=miter,dash pattern=](-19.32pt,13.1709pt)--(-15.32pt,13.1709pt);
\path[-,draw=black,line width=0.4pt,line cap=butt,line join=miter,dash pattern=](-19.32pt,5.7525pt)--(-15.32pt,5.7525pt);
\path[-,draw=black,line width=0.4pt,line cap=butt,line join=miter,dash pattern=](0pt,5.7525pt)--(4pt,5.7525pt);
\path[-,line width=0.4pt,line cap=butt,line join=miter,dash pattern=](-9.32pt,8.505pt)--(-6pt,8.505pt)--(-6pt,3pt)--(-9.32pt,3pt)--cycle;
\node[anchor=north west,inner sep=0] at (-9.32pt,8.505pt){\savebox{\marxupbox}{{\(t\)}}\immediate\write\boxesfile{155}\immediate\write\boxesfile{\number\wd\marxupbox}\immediate\write\boxesfile{\number\ht\marxupbox}\immediate\write\boxesfile{\number\dp\marxupbox}\box\marxupbox};
\path[-,draw=black,line width=0.4pt,line cap=butt,line join=miter,dash pattern=](-12.32pt,9.005pt)..controls(-12.32pt,10.3857pt)and(-11.2007pt,11.505pt)..(-9.82pt,11.505pt)--(-5.5pt,11.505pt)..controls(-4.1193pt,11.505pt)and(-3pt,10.3857pt)..(-3pt,9.005pt)--(-3pt,2.5pt)..controls(-3pt,1.1193pt)and(-4.1193pt,0pt)..(-5.5pt,0pt)--(-9.82pt,0pt)..controls(-11.2007pt,0pt)and(-12.32pt,1.1193pt)..(-12.32pt,2.5pt)--cycle;
\path[-,draw=black,line width=0.4pt,line cap=butt,line join=miter,dash pattern=](-3pt,5.7525pt)--(-3pt,5.7525pt);
\path[-,draw=black,line width=0.4pt,line cap=butt,line join=miter,dash pattern=](-12.32pt,5.7525pt)--(-12.32pt,5.7525pt);
\path[-,line width=0.4pt,line cap=butt,line join=miter,dash pattern=](-26.9516pt,15.5105pt)--(-24.32pt,15.5105pt)--(-24.32pt,10.8313pt)--(-26.9516pt,10.8313pt)--cycle;
\node[anchor=north west,inner sep=0] at (-26.9516pt,15.5105pt){\savebox{\marxupbox}{{\({\scriptstyle i}\)}}\immediate\write\boxesfile{156}\immediate\write\boxesfile{\number\wd\marxupbox}\immediate\write\boxesfile{\number\ht\marxupbox}\immediate\write\boxesfile{\number\dp\marxupbox}\box\marxupbox};
\path[-,line width=0.4pt,line cap=butt,line join=miter,dash pattern=](-28.9516pt,17.5105pt)--(-22.32pt,17.5105pt)--(-22.32pt,8.8313pt)--(-28.9516pt,8.8313pt)--cycle;
\path[-,line width=0.4pt,line cap=butt,line join=miter,dash pattern=](-28.1817pt,8.8313pt)--(-24.32pt,8.8313pt)--(-24.32pt,2.6737pt)--(-28.1817pt,2.6737pt)--cycle;
\node[anchor=north west,inner sep=0] at (-28.1817pt,8.8313pt){\savebox{\marxupbox}{{\({\scriptstyle j}\)}}\immediate\write\boxesfile{157}\immediate\write\boxesfile{\number\wd\marxupbox}\immediate\write\boxesfile{\number\ht\marxupbox}\immediate\write\boxesfile{\number\dp\marxupbox}\box\marxupbox};
\path[-,line width=0.4pt,line cap=butt,line join=miter,dash pattern=](-30.1817pt,10.8313pt)--(-22.32pt,10.8313pt)--(-22.32pt,0.6737pt)--(-30.1817pt,0.6737pt)--cycle;
\path[-,line width=0.4pt,line cap=butt,line join=miter,dash pattern=](9pt,8.2436pt)--(13.3398pt,8.2436pt)--(13.3398pt,3.2614pt)--(9pt,3.2614pt)--cycle;
\node[anchor=north west,inner sep=0] at (9pt,8.2436pt){\savebox{\marxupbox}{{\({\scriptstyle k}\)}}\immediate\write\boxesfile{158}\immediate\write\boxesfile{\number\wd\marxupbox}\immediate\write\boxesfile{\number\ht\marxupbox}\immediate\write\boxesfile{\number\dp\marxupbox}\box\marxupbox};
\path[-,line width=0.4pt,line cap=butt,line join=miter,dash pattern=](7pt,10.2436pt)--(15.3398pt,10.2436pt)--(15.3398pt,1.2614pt)--(7pt,1.2614pt)--cycle;
\end{tikzpicture}}&{\begin{tikzpicture}\path[-,draw=black,line width=0.4pt,line cap=butt,line join=miter,dash pattern=](-22.32pt,13.1709pt)--(-19.32pt,13.1709pt);
\path[-,draw=black,line width=0.4pt,line cap=butt,line join=miter,dash pattern=](-22.32pt,5.7525pt)--(-19.32pt,5.7525pt);
\path[-,draw=black,line width=0.4pt,line cap=butt,line join=miter,dash pattern=](4pt,5.7525pt)--(7pt,5.7525pt);
\path[-,draw=black,line width=0.4pt,line cap=butt,line join=miter,dash pattern=](-15.32pt,17.1709pt)--(0pt,17.1709pt)--(0pt,-3pt)--(-15.32pt,-3pt)--cycle;
\path[-,draw=black,line width=0.4pt,line cap=butt,line join=miter,dash pattern=](-15.32pt,5.7525pt)--(-12.32pt,5.7525pt);
\path[-,draw=black,line width=0.4pt,line cap=butt,line join=miter,dash pattern=](-15.32pt,13.1709pt)--(-15.32pt,13.1709pt);
\path[-,draw=black,line width=0.4pt,line cap=butt,line join=miter,dash pattern=](-3pt,5.7525pt)--(0pt,5.7525pt);
\path[-,draw=black,line width=0.4pt,line cap=butt,line join=miter,dash pattern=](-19.32pt,13.1709pt)--(-15.32pt,13.1709pt);
\path[-,draw=black,line width=0.4pt,line cap=butt,line join=miter,dash pattern=](-19.32pt,5.7525pt)--(-15.32pt,5.7525pt);
\path[-,draw=black,line width=0.4pt,line cap=butt,line join=miter,dash pattern=](0pt,5.7525pt)--(4pt,5.7525pt);
\path[-,line width=0.4pt,line cap=butt,line join=miter,dash pattern=](-9.32pt,8.505pt)--(-6pt,8.505pt)--(-6pt,3pt)--(-9.32pt,3pt)--cycle;
\node[anchor=north west,inner sep=0] at (-9.32pt,8.505pt){\savebox{\marxupbox}{{\(t\)}}\immediate\write\boxesfile{159}\immediate\write\boxesfile{\number\wd\marxupbox}\immediate\write\boxesfile{\number\ht\marxupbox}\immediate\write\boxesfile{\number\dp\marxupbox}\box\marxupbox};
\path[-,draw=black,line width=0.4pt,line cap=butt,line join=miter,dash pattern=](-12.32pt,9.005pt)..controls(-12.32pt,10.3857pt)and(-11.2007pt,11.505pt)..(-9.82pt,11.505pt)--(-5.5pt,11.505pt)..controls(-4.1193pt,11.505pt)and(-3pt,10.3857pt)..(-3pt,9.005pt)--(-3pt,2.5pt)..controls(-3pt,1.1193pt)and(-4.1193pt,0pt)..(-5.5pt,0pt)--(-9.82pt,0pt)..controls(-11.2007pt,0pt)and(-12.32pt,1.1193pt)..(-12.32pt,2.5pt)--cycle;
\path[-,draw=black,line width=0.4pt,line cap=butt,line join=miter,dash pattern=](-3pt,5.7525pt)--(-3pt,5.7525pt);
\path[-,draw=black,line width=0.4pt,line cap=butt,line join=miter,dash pattern=](-12.32pt,5.7525pt)--(-12.32pt,5.7525pt);
\path[-,line width=0.4pt,line cap=butt,line join=miter,dash pattern=](-26.9516pt,15.5105pt)--(-24.32pt,15.5105pt)--(-24.32pt,10.8313pt)--(-26.9516pt,10.8313pt)--cycle;
\node[anchor=north west,inner sep=0] at (-26.9516pt,15.5105pt){\savebox{\marxupbox}{{\({\scriptstyle i}\)}}\immediate\write\boxesfile{160}\immediate\write\boxesfile{\number\wd\marxupbox}\immediate\write\boxesfile{\number\ht\marxupbox}\immediate\write\boxesfile{\number\dp\marxupbox}\box\marxupbox};
\path[-,line width=0.4pt,line cap=butt,line join=miter,dash pattern=](-28.9516pt,17.5105pt)--(-22.32pt,17.5105pt)--(-22.32pt,8.8313pt)--(-28.9516pt,8.8313pt)--cycle;
\path[-,line width=0.4pt,line cap=butt,line join=miter,dash pattern=](-28.1817pt,8.8313pt)--(-24.32pt,8.8313pt)--(-24.32pt,2.6737pt)--(-28.1817pt,2.6737pt)--cycle;
\node[anchor=north west,inner sep=0] at (-28.1817pt,8.8313pt){\savebox{\marxupbox}{{\({\scriptstyle j}\)}}\immediate\write\boxesfile{161}\immediate\write\boxesfile{\number\wd\marxupbox}\immediate\write\boxesfile{\number\ht\marxupbox}\immediate\write\boxesfile{\number\dp\marxupbox}\box\marxupbox};
\path[-,line width=0.4pt,line cap=butt,line join=miter,dash pattern=](-30.1817pt,10.8313pt)--(-22.32pt,10.8313pt)--(-22.32pt,0.6737pt)--(-30.1817pt,0.6737pt)--cycle;
\path[-,line width=0.4pt,line cap=butt,line join=miter,dash pattern=](9pt,8.2436pt)--(13.3398pt,8.2436pt)--(13.3398pt,3.2614pt)--(9pt,3.2614pt)--cycle;
\node[anchor=north west,inner sep=0] at (9pt,8.2436pt){\savebox{\marxupbox}{{\({\scriptstyle k}\)}}\immediate\write\boxesfile{162}\immediate\write\boxesfile{\number\wd\marxupbox}\immediate\write\boxesfile{\number\ht\marxupbox}\immediate\write\boxesfile{\number\dp\marxupbox}\box\marxupbox};
\path[-,line width=0.4pt,line cap=butt,line join=miter,dash pattern=](7pt,10.2436pt)--(15.3398pt,10.2436pt)--(15.3398pt,1.2614pt)--(7pt,1.2614pt)--cycle;
\end{tikzpicture}}&{\begin{tikzpicture}\path[-,draw=black,line width=0.4pt,line cap=butt,line join=miter,dash pattern=](-13pt,0pt)--(-6pt,0pt);
\path[-,draw=black,line width=0.4pt,line cap=butt,line join=miter,dash pattern=](-13pt,-7.4184pt)--(-6pt,-7.4184pt);
\path[-,draw=black,line width=0.4pt,line cap=butt,line join=miter,dash pattern=](0pt,-3.7092pt)--(7pt,-3.7092pt);
\path[-,draw=black,fill= gray,line width=0.4pt,line cap=butt,line join=miter,dash pattern=](-6pt,1pt)..controls(-2pt,1pt)and(0pt,-1.7092pt)..(0pt,-3.7092pt)..controls(0pt,-5.7092pt)and(-2pt,-8.4184pt)..(-6pt,-8.4184pt)--cycle;
\path[-,line width=0.4pt,line cap=butt,line join=miter,dash pattern=](-17.6316pt,2.3396pt)--(-15pt,2.3396pt)--(-15pt,-2.3396pt)--(-17.6316pt,-2.3396pt)--cycle;
\node[anchor=north west,inner sep=0] at (-17.6316pt,2.3396pt){\savebox{\marxupbox}{{\({\scriptstyle i}\)}}\immediate\write\boxesfile{163}\immediate\write\boxesfile{\number\wd\marxupbox}\immediate\write\boxesfile{\number\ht\marxupbox}\immediate\write\boxesfile{\number\dp\marxupbox}\box\marxupbox};
\path[-,line width=0.4pt,line cap=butt,line join=miter,dash pattern=](-19.6316pt,4.3396pt)--(-13pt,4.3396pt)--(-13pt,-4.3396pt)--(-19.6316pt,-4.3396pt)--cycle;
\path[-,line width=0.4pt,line cap=butt,line join=miter,dash pattern=](-18.8617pt,-4.3396pt)--(-15pt,-4.3396pt)--(-15pt,-10.4972pt)--(-18.8617pt,-10.4972pt)--cycle;
\node[anchor=north west,inner sep=0] at (-18.8617pt,-4.3396pt){\savebox{\marxupbox}{{\({\scriptstyle j}\)}}\immediate\write\boxesfile{164}\immediate\write\boxesfile{\number\wd\marxupbox}\immediate\write\boxesfile{\number\ht\marxupbox}\immediate\write\boxesfile{\number\dp\marxupbox}\box\marxupbox};
\path[-,line width=0.4pt,line cap=butt,line join=miter,dash pattern=](-20.8617pt,-2.3396pt)--(-13pt,-2.3396pt)--(-13pt,-12.4972pt)--(-20.8617pt,-12.4972pt)--cycle;
\path[-,line width=0.4pt,line cap=butt,line join=miter,dash pattern=](9pt,-1.2181pt)--(13.3398pt,-1.2181pt)--(13.3398pt,-6.2003pt)--(9pt,-6.2003pt)--cycle;
\node[anchor=north west,inner sep=0] at (9pt,-1.2181pt){\savebox{\marxupbox}{{\({\scriptstyle k}\)}}\immediate\write\boxesfile{165}\immediate\write\boxesfile{\number\wd\marxupbox}\immediate\write\boxesfile{\number\ht\marxupbox}\immediate\write\boxesfile{\number\dp\marxupbox}\box\marxupbox};
\path[-,line width=0.4pt,line cap=butt,line join=miter,dash pattern=](7pt,0.7819pt)--(15.3398pt,0.7819pt)--(15.3398pt,-8.2003pt)--(7pt,-8.2003pt)--cycle;
\end{tikzpicture}}&{\begin{tikzpicture}\path[-,draw=lightgray,line width=0.4pt,line cap=butt,line join=miter,dash pattern=](-13pt,-3.7092pt)--(-6pt,-3.7092pt);
\path[-,draw=black,line width=0.4pt,line cap=butt,line join=miter,dash pattern=](0pt,0pt)--(1pt,0pt);
\path[-,draw=black,line width=0.4pt,line cap=butt,line join=miter,dash pattern=](0pt,-7.4184pt)--(1pt,-7.4184pt);
\path[-,fill=black,line width=0.4pt,line cap=butt,line join=miter,dash pattern=](-5pt,-3.7092pt)..controls(-5pt,-3.1569pt)and(-5.4477pt,-2.7092pt)..(-6pt,-2.7092pt)..controls(-6.5523pt,-2.7092pt)and(-7pt,-3.1569pt)..(-7pt,-3.7092pt)..controls(-7pt,-4.2615pt)and(-6.5523pt,-4.7092pt)..(-6pt,-4.7092pt)..controls(-5.4477pt,-4.7092pt)and(-5pt,-4.2615pt)..(-5pt,-3.7092pt)--cycle;
\path[-,draw=black,line width=0.4pt,line cap=butt,line join=miter,dash pattern=](-3.06pt,-0.3287pt)..controls(-2.22pt,-0.1088pt)and(-1.2pt,0pt)..(0pt,0pt);
\path[-,draw=black,line width=0.4pt,line cap=butt,line join=miter,dash pattern=](-6pt,-3.7092pt)..controls(-6pt,-1.9592pt)and(-5.02pt,-0.8417pt)..(-3.06pt,-0.3287pt);
\path[-,draw=black,line width=0.4pt,line cap=butt,line join=miter,dash pattern=](-3.06pt,-7.0897pt)..controls(-2.22pt,-7.3096pt)and(-1.2pt,-7.4184pt)..(0pt,-7.4184pt);
\path[-,draw=black,line width=0.4pt,line cap=butt,line join=miter,dash pattern=](-6pt,-3.7092pt)..controls(-6pt,-5.4592pt)and(-5.02pt,-6.5767pt)..(-3.06pt,-7.0897pt);
\path[-,line width=0.4pt,line cap=butt,line join=miter,dash pattern=](3pt,2.3396pt)--(5.6316pt,2.3396pt)--(5.6316pt,-2.3396pt)--(3pt,-2.3396pt)--cycle;
\node[anchor=north west,inner sep=0] at (3pt,2.3396pt){\savebox{\marxupbox}{{\({\scriptstyle i}\)}}\immediate\write\boxesfile{166}\immediate\write\boxesfile{\number\wd\marxupbox}\immediate\write\boxesfile{\number\ht\marxupbox}\immediate\write\boxesfile{\number\dp\marxupbox}\box\marxupbox};
\path[-,line width=0.4pt,line cap=butt,line join=miter,dash pattern=](1pt,4.3396pt)--(7.6316pt,4.3396pt)--(7.6316pt,-4.3396pt)--(1pt,-4.3396pt)--cycle;
\path[-,line width=0.4pt,line cap=butt,line join=miter,dash pattern=](3pt,-4.3396pt)--(6.8617pt,-4.3396pt)--(6.8617pt,-10.4972pt)--(3pt,-10.4972pt)--cycle;
\node[anchor=north west,inner sep=0] at (3pt,-4.3396pt){\savebox{\marxupbox}{{\({\scriptstyle j}\)}}\immediate\write\boxesfile{167}\immediate\write\boxesfile{\number\wd\marxupbox}\immediate\write\boxesfile{\number\ht\marxupbox}\immediate\write\boxesfile{\number\dp\marxupbox}\box\marxupbox};
\path[-,line width=0.4pt,line cap=butt,line join=miter,dash pattern=](1pt,-2.3396pt)--(8.8617pt,-2.3396pt)--(8.8617pt,-12.4972pt)--(1pt,-12.4972pt)--cycle;
\end{tikzpicture}}&{\begin{tikzpicture}\path[-,draw=black,line width=0.4pt,line cap=butt,line join=miter,dash pattern=](-7pt,0pt)--(-6pt,0pt);
\path[-,draw=black,line width=0.4pt,line cap=butt,line join=miter,dash pattern=](-7pt,-7.4184pt)--(-6pt,-7.4184pt);
\path[-,draw=lightgray,line width=0.4pt,line cap=butt,line join=miter,dash pattern=](0pt,-3.7092pt)--(7pt,-3.7092pt);
\path[-,fill=black,line width=0.4pt,line cap=butt,line join=miter,dash pattern=](1pt,-3.7092pt)..controls(1pt,-3.1569pt)and(0.5523pt,-2.7092pt)..(0pt,-2.7092pt)..controls(-0.5523pt,-2.7092pt)and(-1pt,-3.1569pt)..(-1pt,-3.7092pt)..controls(-1pt,-4.2615pt)and(-0.5523pt,-4.7092pt)..(0pt,-4.7092pt)..controls(0.5523pt,-4.7092pt)and(1pt,-4.2615pt)..(1pt,-3.7092pt)--cycle;
\path[-,draw=black,line width=0.4pt,line cap=butt,line join=miter,dash pattern=](-2.94pt,-0.3287pt)..controls(-0.98pt,-0.8417pt)and(-0pt,-1.9592pt)..(0pt,-3.7092pt);
\path[-,draw=black,line width=0.4pt,line cap=butt,line join=miter,dash pattern=](-6pt,0pt)..controls(-4.8pt,0pt)and(-3.78pt,-0.1088pt)..(-2.94pt,-0.3287pt);
\path[-,draw=black,line width=0.4pt,line cap=butt,line join=miter,dash pattern=](-2.94pt,-7.0897pt)..controls(-0.98pt,-6.5767pt)and(-0pt,-5.4592pt)..(0pt,-3.7092pt);
\path[-,draw=black,line width=0.4pt,line cap=butt,line join=miter,dash pattern=](-6pt,-7.4184pt)..controls(-4.8pt,-7.4184pt)and(-3.78pt,-7.3096pt)..(-2.94pt,-7.0897pt);
\path[-,line width=0.4pt,line cap=butt,line join=miter,dash pattern=](-11.6316pt,2.3396pt)--(-9pt,2.3396pt)--(-9pt,-2.3396pt)--(-11.6316pt,-2.3396pt)--cycle;
\node[anchor=north west,inner sep=0] at (-11.6316pt,2.3396pt){\savebox{\marxupbox}{{\({\scriptstyle i}\)}}\immediate\write\boxesfile{168}\immediate\write\boxesfile{\number\wd\marxupbox}\immediate\write\boxesfile{\number\ht\marxupbox}\immediate\write\boxesfile{\number\dp\marxupbox}\box\marxupbox};
\path[-,line width=0.4pt,line cap=butt,line join=miter,dash pattern=](-13.6316pt,4.3396pt)--(-7pt,4.3396pt)--(-7pt,-4.3396pt)--(-13.6316pt,-4.3396pt)--cycle;
\path[-,line width=0.4pt,line cap=butt,line join=miter,dash pattern=](-12.8617pt,-4.3396pt)--(-9pt,-4.3396pt)--(-9pt,-10.4972pt)--(-12.8617pt,-10.4972pt)--cycle;
\node[anchor=north west,inner sep=0] at (-12.8617pt,-4.3396pt){\savebox{\marxupbox}{{\({\scriptstyle j}\)}}\immediate\write\boxesfile{169}\immediate\write\boxesfile{\number\wd\marxupbox}\immediate\write\boxesfile{\number\ht\marxupbox}\immediate\write\boxesfile{\number\dp\marxupbox}\box\marxupbox};
\path[-,line width=0.4pt,line cap=butt,line join=miter,dash pattern=](-14.8617pt,-2.3396pt)--(-7pt,-2.3396pt)--(-7pt,-12.4972pt)--(-14.8617pt,-12.4972pt)--cycle;
\end{tikzpicture}}\\\(\mathnormal{∇}\mskip 3.0mu\mathsf{t}\)&\(\mathnormal{∂}\mskip 3.0mu\mathsf{t}\)&\(Γ\)&\(\mathsf{g'}\)&\(\mathsf{g}\)\\\(∇{_i}t{_j}{^k}\)&\(∂{_i}t{_j}{^k}\)&\(Γ{_i}{_j}{^k}\)&\(g'{^i}{^j}\)&\(g{_i}{_j}\)\\\(T_{z}\mskip 3.0mu\mathnormal{⊗}\mskip 3.0muT_{z}\mskip 1.0mu\overset{z}{\leadsto }\mskip 1.0muT_{z}\)&\(T_{z}\mskip 3.0mu\mathnormal{⊗}\mskip 3.0muT_{z}\mskip 1.0mu\overset{z}{\leadsto }\mskip 1.0muT_{z}\)&\(T_{z}\mskip 3.0mu\mathnormal{⊗}\mskip 3.0muT_{z}\mskip 1.0mu\overset{z}{\leadsto }\mskip 1.0muT_{z}\)&\(\mathbf{1}\mskip 1.0mu\overset{z}{\leadsto }\mskip 1.0muT_{z}\mskip 3.0mu\mathnormal{⊗}\mskip 3.0muT_{z}\)&\(T_{z}\mskip 3.0mu\mathnormal{⊗}\mskip 3.0muT_{z}\mskip 1.0mu\overset{z}{\leadsto }\mskip 1.0mu\mathbf{1}\)\end{tabular}\end{center}\caption{Tensor field primitives in various notations, and their types.}\label{170}\end{figure} 
\subsection{Spatial Derivative: Levi-Civita connection}\label{171} 
The local vector space can change from position to position, but it is assumed to vary smoothly.
One says that a \emph{connection} is defined between neighbouring spaces.
This means that we can take the derivative of tensor fields with respect to position.
Two cautionary remarks are in order regarding our use of the phrase \emph{spatial derivative}.
First, the terminology is meant to include hyperbolic geometries, thus space-time (Minkowski space) is covered.
Second, there are other notions of spatial derivatives, but here we will only consider the case of the Levi-Civita connection, more precisely, which has additional properties: see footnote \ref{224}.

Considering the simplest case, the derivative of a scalar field \(\mathsf{s}\) is a covector field: its
gradient. That is, given a direction vector \(\vec v\), the gradient will return the slope of
\(\mathsf{s}\) in the \(\vec v\) direction.
In general, the spatial derivative of a tensor takes a
vector argument and returns the variation in this direction.
So the derivative of a tensor field is itself a tensor field, whose
(covariant) order is one more than its
argument, and for this reason the spatial derivative of tensors is
called the covariant derivative.
We capture
this in the following class, which signals the presence of a connection (and spatial derivative) in a category.
\begin{list}{}{\setlength\leftmargin{1.0em}}\item\relax
\ensuremath{\begin{parray}\column{B}{@{}>{}l<{}@{}}\column[0em]{1}{@{}>{}l<{}@{}}\column[1em]{2}{@{}>{}l<{}@{}}\column{3}{@{}>{}l<{}@{}}\column{4}{@{}>{}l<{}@{}}\column{E}{@{}>{}l<{}@{}}%
\>[1]{\mathbf{class}\mskip 3.0mu\mathsf{MetricCategory}\mskip 3.0muz\mskip 3.0mu\mathnormal{\Rightarrow }\mskip 3.0mu\mathsf{ConnectionCategory}\mskip 3.0muz\mskip 3.0mu\mathbf{where}}\<[E]\\
\>[2]{∇\mskip 3.0mu}\>[3]{\mathnormal{::}\mskip 3.0mu}\>[4]{\allowbreak{}\mathnormal{(}\mskip 0.0mu\mathsf{a}\mskip 1.0mu\overset{z}{\leadsto }\mskip 1.0mu\mathsf{b}\mskip 0.0mu\mathnormal{)}\allowbreak{}\mskip 3.0mu\mathnormal{\rightarrow }\mskip 3.0mu\allowbreak{}\mathnormal{(}\mskip 0.0mu\allowbreak{}\mathnormal{(}\mskip 0.0muT_{z}\mskip 3.0mu\mathnormal{⊗}\mskip 3.0mu\mathsf{a}\mskip 0.0mu\mathnormal{)}\allowbreak{}\mskip 1.0mu\overset{z}{\leadsto }\mskip 1.0mu\mathsf{b}\mskip 0.0mu\mathnormal{)}\allowbreak{}}\<[E]\end{parray}}\end{list} Accordingly, in Einstein notation, the covariant
derivative uses an additional (lower) index.
\begin{list}{}{\setlength\leftmargin{1.0em}}\item\relax
\ensuremath{\begin{parray}\column{B}{@{}>{}l<{}@{}}\column[0em]{1}{@{}>{}l<{}@{}}\column{2}{@{}>{}l<{}@{}}\column{3}{@{}>{}l<{}@{}}\column{E}{@{}>{}l<{}@{}}%
\>[1]{\mathsf{deriv}\mskip 3.0mu}\>[2]{\mathnormal{::}\mskip 3.0mu}\>[3]{\mathsf{ConnectionCategory}\mskip 3.0muz\mskip 3.0mu\mathnormal{\Rightarrow }\mskip 3.0mu\mathsf{P}\mskip 3.0muz\mskip 3.0mu\mathsf{r}\mskip 3.0mu\mathsf{v}\mskip 3.0mu\mathnormal{⊸}\mskip 3.0mu\mathsf{R}\mskip 3.0muz\mskip 3.0mu\mathsf{r}\mskip 3.0mu\mathnormal{⊸}\mskip 3.0mu\mathsf{R}\mskip 3.0muz\mskip 3.0mu\mathsf{r}}\<[E]\end{parray}}\end{list} The actual implementation is too involved to present just yet, and deferred to \cref{394}.
In the diagrammatic notation we represent the covariant derivative as a thick
box around the tensor whose derivative is taken.
This box adds an input wire for the additional input vector, and propagates the wires of the tensor
which it encloses, reflecting the propagation of the types \(\mathsf{a}\) and \(\mathsf{b}\) in the type of \(∇\).
In \cref{170} we illustrate with a derivative
applied to a tensor of order (1,1), but it can be applied to
any morphism, with arbitrary domain and codomain---
and hence to tensors of arbitrary order.
For instance, the covariant derivative of a tensor \(\mathsf{t}\) of order (2,2)
is written \(∇{_i}t{_j}{_k}{^l}{^m}\) or
{\begin{tikzpicture}[baseline={([yshift=-0.8ex]current bounding box.center)}]\path[-,draw=black,line width=0.4pt,line cap=butt,line join=miter,dash pattern=](-22.32pt,16.8517pt)--(-19.32pt,16.8517pt);
\path[-,draw=black,line width=0.4pt,line cap=butt,line join=miter,dash pattern=](-22.32pt,6.8517pt)--(-19.32pt,6.8517pt);
\path[-,draw=black,line width=0.4pt,line cap=butt,line join=miter,dash pattern=](-22.32pt,-0.7181pt)--(-19.32pt,-0.7181pt);
\path[-,draw=black,line width=0.4pt,line cap=butt,line join=miter,dash pattern=](4pt,6.1336pt)--(7pt,6.1336pt);
\path[-,draw=black,line width=0.4pt,line cap=butt,line join=miter,dash pattern=](4pt,0pt)--(7pt,0pt);
\path[-,draw=black,line width=0.8pt,line cap=butt,line join=miter,dash pattern=](-15.32pt,17.8517pt)..controls(-15.32pt,19.5086pt)and(-13.9768pt,20.8517pt)..(-12.32pt,20.8517pt)--(-3pt,20.8517pt)..controls(-1.3431pt,20.8517pt)and(0pt,19.5086pt)..(0pt,17.8517pt)--(0pt,-4.7181pt)..controls(0pt,-6.375pt)and(-1.3431pt,-7.7181pt)..(-3pt,-7.7181pt)--(-12.32pt,-7.7181pt)..controls(-13.9768pt,-7.7181pt)and(-15.32pt,-6.375pt)..(-15.32pt,-4.7181pt)--cycle;
\path[-,draw=black,line width=0.4pt,line cap=butt,line join=miter,dash pattern=](-15.32pt,6.8517pt)--(-12.32pt,6.8517pt);
\path[-,draw=black,line width=0.4pt,line cap=butt,line join=miter,dash pattern=](-15.32pt,-0.7181pt)--(-12.32pt,-0.7181pt);
\path[-,draw=black,line width=0.4pt,line cap=butt,line join=miter,dash pattern=](-15.32pt,16.8517pt)--(-15.32pt,16.8517pt);
\path[-,draw=black,line width=0.4pt,line cap=butt,line join=miter,dash pattern=](-3pt,6.1336pt)--(0pt,6.1336pt);
\path[-,draw=black,line width=0.4pt,line cap=butt,line join=miter,dash pattern=](-3pt,0pt)--(0pt,0pt);
\path[-,draw=black,line width=0.4pt,line cap=butt,line join=miter,dash pattern=](-19.32pt,16.8517pt)--(-15.32pt,16.8517pt);
\path[-,draw=black,line width=0.4pt,line cap=butt,line join=miter,dash pattern=](-19.32pt,6.8517pt)--(-15.32pt,6.8517pt);
\path[-,draw=black,line width=0.4pt,line cap=butt,line join=miter,dash pattern=](-19.32pt,-0.7181pt)--(-15.32pt,-0.7181pt);
\path[-,draw=black,line width=0.4pt,line cap=butt,line join=miter,dash pattern=](0pt,6.1336pt)--(4pt,6.1336pt);
\path[-,draw=black,line width=0.4pt,line cap=butt,line join=miter,dash pattern=](0pt,0pt)--(4pt,0pt);
\path[-,line width=0.4pt,line cap=butt,line join=miter,dash pattern=](-9.32pt,5.8193pt)--(-6pt,5.8193pt)--(-6pt,0.3143pt)--(-9.32pt,0.3143pt)--cycle;
\node[anchor=north west,inner sep=0] at (-9.32pt,5.8193pt){\savebox{\marxupbox}{{\(t\)}}\immediate\write\boxesfile{172}\immediate\write\boxesfile{\number\wd\marxupbox}\immediate\write\boxesfile{\number\ht\marxupbox}\immediate\write\boxesfile{\number\dp\marxupbox}\box\marxupbox};
\path[-,draw=black,line width=0.4pt,line cap=butt,line join=miter,dash pattern=](-12.32pt,8.3517pt)..controls(-12.32pt,9.7324pt)and(-11.2007pt,10.8517pt)..(-9.82pt,10.8517pt)--(-5.5pt,10.8517pt)..controls(-4.1193pt,10.8517pt)and(-3pt,9.7324pt)..(-3pt,8.3517pt)--(-3pt,-2.2181pt)..controls(-3pt,-3.5989pt)and(-4.1193pt,-4.7181pt)..(-5.5pt,-4.7181pt)--(-9.82pt,-4.7181pt)..controls(-11.2007pt,-4.7181pt)and(-12.32pt,-3.5989pt)..(-12.32pt,-2.2181pt)--cycle;
\path[-,draw=black,line width=0.4pt,line cap=butt,line join=miter,dash pattern=](-3pt,6.1336pt)--(-3pt,6.1336pt);
\path[-,draw=black,line width=0.4pt,line cap=butt,line join=miter,dash pattern=](-3pt,0pt)--(-3pt,0pt);
\path[-,draw=black,line width=0.4pt,line cap=butt,line join=miter,dash pattern=](-12.32pt,6.8517pt)--(-12.32pt,6.8517pt);
\path[-,draw=black,line width=0.4pt,line cap=butt,line join=miter,dash pattern=](-12.32pt,-0.7181pt)--(-12.32pt,-0.7181pt);
\path[-,line width=0.4pt,line cap=butt,line join=miter,dash pattern=](-26.9516pt,19.1914pt)--(-24.32pt,19.1914pt)--(-24.32pt,14.5121pt)--(-26.9516pt,14.5121pt)--cycle;
\node[anchor=north west,inner sep=0] at (-26.9516pt,19.1914pt){\savebox{\marxupbox}{{\({\scriptstyle i}\)}}\immediate\write\boxesfile{173}\immediate\write\boxesfile{\number\wd\marxupbox}\immediate\write\boxesfile{\number\ht\marxupbox}\immediate\write\boxesfile{\number\dp\marxupbox}\box\marxupbox};
\path[-,line width=0.4pt,line cap=butt,line join=miter,dash pattern=](-28.9516pt,21.1914pt)--(-22.32pt,21.1914pt)--(-22.32pt,12.5121pt)--(-28.9516pt,12.5121pt)--cycle;
\path[-,line width=0.4pt,line cap=butt,line join=miter,dash pattern=](-28.1817pt,9.9305pt)--(-24.32pt,9.9305pt)--(-24.32pt,3.773pt)--(-28.1817pt,3.773pt)--cycle;
\node[anchor=north west,inner sep=0] at (-28.1817pt,9.9305pt){\savebox{\marxupbox}{{\({\scriptstyle j}\)}}\immediate\write\boxesfile{174}\immediate\write\boxesfile{\number\wd\marxupbox}\immediate\write\boxesfile{\number\ht\marxupbox}\immediate\write\boxesfile{\number\dp\marxupbox}\box\marxupbox};
\path[-,line width=0.4pt,line cap=butt,line join=miter,dash pattern=](-30.1817pt,11.9305pt)--(-22.32pt,11.9305pt)--(-22.32pt,1.773pt)--(-30.1817pt,1.773pt)--cycle;
\path[-,line width=0.4pt,line cap=butt,line join=miter,dash pattern=](-28.6598pt,1.773pt)--(-24.32pt,1.773pt)--(-24.32pt,-3.2093pt)--(-28.6598pt,-3.2093pt)--cycle;
\node[anchor=north west,inner sep=0] at (-28.6598pt,1.773pt){\savebox{\marxupbox}{{\({\scriptstyle k}\)}}\immediate\write\boxesfile{175}\immediate\write\boxesfile{\number\wd\marxupbox}\immediate\write\boxesfile{\number\ht\marxupbox}\immediate\write\boxesfile{\number\dp\marxupbox}\box\marxupbox};
\path[-,line width=0.4pt,line cap=butt,line join=miter,dash pattern=](-30.6598pt,3.773pt)--(-22.32pt,3.773pt)--(-22.32pt,-5.2093pt)--(-30.6598pt,-5.2093pt)--cycle;
\path[-,line width=0.4pt,line cap=butt,line join=miter,dash pattern=](9pt,8.6594pt)--(11.5769pt,8.6594pt)--(11.5769pt,3.6078pt)--(9pt,3.6078pt)--cycle;
\node[anchor=north west,inner sep=0] at (9pt,8.6594pt){\savebox{\marxupbox}{{\({\scriptstyle l}\)}}\immediate\write\boxesfile{176}\immediate\write\boxesfile{\number\wd\marxupbox}\immediate\write\boxesfile{\number\ht\marxupbox}\immediate\write\boxesfile{\number\dp\marxupbox}\box\marxupbox};
\path[-,line width=0.4pt,line cap=butt,line join=miter,dash pattern=](7pt,10.6594pt)--(13.5769pt,10.6594pt)--(13.5769pt,1.6078pt)--(7pt,1.6078pt)--cycle;
\path[-,line width=0.4pt,line cap=butt,line join=miter,dash pattern=](9pt,1.6078pt)--(15.1685pt,1.6078pt)--(15.1685pt,-1.6078pt)--(9pt,-1.6078pt)--cycle;
\node[anchor=north west,inner sep=0] at (9pt,1.6078pt){\savebox{\marxupbox}{{\({\scriptstyle m}\)}}\immediate\write\boxesfile{177}\immediate\write\boxesfile{\number\wd\marxupbox}\immediate\write\boxesfile{\number\ht\marxupbox}\immediate\write\boxesfile{\number\dp\marxupbox}\box\marxupbox};
\path[-,line width=0.4pt,line cap=butt,line join=miter,dash pattern=](7pt,3.6078pt)--(17.1685pt,3.6078pt)--(17.1685pt,-3.6078pt)--(7pt,-3.6078pt)--cycle;
\end{tikzpicture}}, but (still) \(\mathnormal{∇}\mskip 3.0mu\mathsf{t}\) in
{\sc{}Roger}.

\paragraph*{Example: divergence.}\hspace{1.0ex}\label{178} As a simple example we express the
divergence of a given scalar field \(P\).  Its physical meaning has
no bearing on our development; but if \(P\) is an  energy potential, its gradient is the corresponding force field, and
its divergence the density of an associated charge or mass. We write \(P\) for the potential field, or {\begin{tikzpicture}[baseline={([yshift=-0.8ex]current bounding box.center)}]\path[-,draw=lightgray,line width=0.4pt,line cap=butt,line join=miter,dash pattern=](-19.66pt,6.2675pt)--(-12.66pt,6.2675pt);
\path[-,draw=lightgray,line width=0.4pt,line cap=butt,line join=miter,dash pattern=](0pt,6.2675pt)--(7pt,6.2675pt);
\path[-,line width=0.4pt,line cap=butt,line join=miter,dash pattern=](-9.66pt,9.535pt)--(-3pt,9.535pt)--(-3pt,3pt)--(-9.66pt,3pt)--cycle;
\node[anchor=north west,inner sep=0] at (-9.66pt,9.535pt){\savebox{\marxupbox}{{\(P\)}}\immediate\write\boxesfile{179}\immediate\write\boxesfile{\number\wd\marxupbox}\immediate\write\boxesfile{\number\ht\marxupbox}\immediate\write\boxesfile{\number\dp\marxupbox}\box\marxupbox};
\path[-,draw=black,line width=0.4pt,line cap=butt,line join=miter,dash pattern=](-12.66pt,10.035pt)..controls(-12.66pt,11.4157pt)and(-11.5407pt,12.535pt)..(-10.16pt,12.535pt)--(-2.5pt,12.535pt)..controls(-1.1193pt,12.535pt)and(0pt,11.4157pt)..(0pt,10.035pt)--(0pt,2.5pt)..controls(0pt,1.1193pt)and(-1.1193pt,0pt)..(-2.5pt,0pt)--(-10.16pt,0pt)..controls(-11.5407pt,0pt)and(-12.66pt,1.1193pt)..(-12.66pt,2.5pt)--cycle;
\path[-,draw=lightgray,line width=0.4pt,line cap=butt,line join=miter,dash pattern=](0pt,6.2675pt)--(0pt,6.2675pt);
\path[-,draw=lightgray,line width=0.4pt,line cap=butt,line join=miter,dash pattern=](-12.66pt,6.2675pt)--(-12.66pt,6.2675pt);
\end{tikzpicture}} as a diagram. The
scalar character of this field is represented by the lack of indices,
or the use of the unit object for its domain and codomain.  Its
gradient is denoted \(∇{_i}P\), or {\begin{tikzpicture}[baseline={([yshift=-0.8ex]current bounding box.center)}]\path[-,draw=black,line width=0.4pt,line cap=butt,line join=miter,dash pattern=](-7pt,11.2675pt)--(-4pt,11.2675pt);
\path[-,draw=lightgray,line width=0.4pt,line cap=butt,line join=miter,dash pattern=](-7pt,6.2675pt)--(-4pt,6.2675pt);
\path[-,draw=lightgray,line width=0.4pt,line cap=butt,line join=miter,dash pattern=](22.66pt,6.2675pt)--(25.66pt,6.2675pt);
\path[-,draw=black,line width=0.8pt,line cap=butt,line join=miter,dash pattern=](0pt,12.535pt)..controls(0pt,14.1918pt)and(1.3431pt,15.535pt)..(3pt,15.535pt)--(15.66pt,15.535pt)..controls(17.3168pt,15.535pt)and(18.66pt,14.1918pt)..(18.66pt,12.535pt)--(18.66pt,0pt)..controls(18.66pt,-1.6569pt)and(17.3168pt,-3pt)..(15.66pt,-3pt)--(3pt,-3pt)..controls(1.3431pt,-3pt)and(0pt,-1.6569pt)..(0pt,0pt)--cycle;
\path[-,draw=lightgray,line width=0.4pt,line cap=butt,line join=miter,dash pattern=](0pt,6.2675pt)--(3pt,6.2675pt);
\path[-,draw=black,line width=0.4pt,line cap=butt,line join=miter,dash pattern=](0pt,11.2675pt)--(0pt,11.2675pt);
\path[-,draw=lightgray,line width=0.4pt,line cap=butt,line join=miter,dash pattern=](15.66pt,6.2675pt)--(18.66pt,6.2675pt);
\path[-,draw=black,line width=0.4pt,line cap=butt,line join=miter,dash pattern=](-4pt,11.2675pt)--(0pt,11.2675pt);
\path[-,draw=lightgray,line width=0.4pt,line cap=butt,line join=miter,dash pattern=](-4pt,6.2675pt)--(0pt,6.2675pt);
\path[-,draw=lightgray,line width=0.4pt,line cap=butt,line join=miter,dash pattern=](18.66pt,6.2675pt)--(22.66pt,6.2675pt);
\path[-,line width=0.4pt,line cap=butt,line join=miter,dash pattern=](6pt,9.535pt)--(12.66pt,9.535pt)--(12.66pt,3pt)--(6pt,3pt)--cycle;
\node[anchor=north west,inner sep=0] at (6pt,9.535pt){\savebox{\marxupbox}{{\(P\)}}\immediate\write\boxesfile{180}\immediate\write\boxesfile{\number\wd\marxupbox}\immediate\write\boxesfile{\number\ht\marxupbox}\immediate\write\boxesfile{\number\dp\marxupbox}\box\marxupbox};
\path[-,draw=black,line width=0.4pt,line cap=butt,line join=miter,dash pattern=](3pt,10.035pt)..controls(3pt,11.4157pt)and(4.1193pt,12.535pt)..(5.5pt,12.535pt)--(13.16pt,12.535pt)..controls(14.5407pt,12.535pt)and(15.66pt,11.4157pt)..(15.66pt,10.035pt)--(15.66pt,2.5pt)..controls(15.66pt,1.1193pt)and(14.5407pt,0pt)..(13.16pt,0pt)--(5.5pt,0pt)..controls(4.1193pt,0pt)and(3pt,1.1193pt)..(3pt,2.5pt)--cycle;
\path[-,draw=lightgray,line width=0.4pt,line cap=butt,line join=miter,dash pattern=](15.66pt,6.2675pt)--(15.66pt,6.2675pt);
\path[-,draw=lightgray,line width=0.4pt,line cap=butt,line join=miter,dash pattern=](3pt,6.2675pt)--(3pt,6.2675pt);
\path[-,line width=0.4pt,line cap=butt,line join=miter,dash pattern=](-11.6316pt,13.6071pt)--(-9pt,13.6071pt)--(-9pt,8.9279pt)--(-11.6316pt,8.9279pt)--cycle;
\node[anchor=north west,inner sep=0] at (-11.6316pt,13.6071pt){\savebox{\marxupbox}{{\({\scriptstyle i}\)}}\immediate\write\boxesfile{181}\immediate\write\boxesfile{\number\wd\marxupbox}\immediate\write\boxesfile{\number\ht\marxupbox}\immediate\write\boxesfile{\number\dp\marxupbox}\box\marxupbox};
\path[-,line width=0.4pt,line cap=butt,line join=miter,dash pattern=](-13.6316pt,15.6071pt)--(-7pt,15.6071pt)--(-7pt,6.9279pt)--(-13.6316pt,6.9279pt)--cycle;
\end{tikzpicture}}. Again, indices or objects indicate that we have an order (1,0)
(covector) field. At each point the local covector is a linear
function that takes a direction (vector) into the slope in that
direction (a scalar).

The divergence is a linear combination of second order derivatives. To
compute the linear combination we need to apply contraction, which
always expects an upper and a lower index, but we have two covariant (lower)
indices, so we must raise one index by multiplying with the
contravariant metric.  The Einstein notation is \(g'{^i}{^j}∇{_i}∇{_j}P\), the corresponding diagram is {\begin{tikzpicture}[baseline={([yshift=-0.8ex]current bounding box.center)}]\path[-,draw=lightgray,line width=0.4pt,line cap=butt,line join=miter,dash pattern=](-20pt,13.7675pt)--(-13pt,13.7675pt);
\path[-,draw=lightgray,line width=0.4pt,line cap=butt,line join=miter,dash pattern=](33.66pt,6.2675pt)--(36.66pt,6.2675pt);
\path[-,draw=black,line width=0.4pt,line cap=butt,line join=miter,dash pattern=](-7pt,16.2675pt)--(-7pt,16.2675pt);
\path[-,draw=black,line width=0.4pt,line cap=butt,line join=miter,dash pattern=](-7pt,11.2675pt)--(-7pt,11.2675pt);
\path[-,draw=black,line width=0.8pt,line cap=butt,line join=miter,dash pattern=](-3pt,17.2675pt)..controls(-3pt,18.9243pt)and(-1.6569pt,20.2675pt)..(0pt,20.2675pt)--(26.66pt,20.2675pt)..controls(28.3168pt,20.2675pt)and(29.66pt,18.9243pt)..(29.66pt,17.2675pt)--(29.66pt,-3pt)..controls(29.66pt,-4.6569pt)and(28.3168pt,-6pt)..(26.66pt,-6pt)--(0pt,-6pt)..controls(-1.6569pt,-6pt)and(-3pt,-4.6569pt)..(-3pt,-3pt)--cycle;
\path[-,draw=black,line width=0.4pt,line cap=butt,line join=miter,dash pattern=](-3pt,11.2675pt)--(0pt,11.2675pt);
\path[-,draw=black,line width=0.4pt,line cap=butt,line join=miter,dash pattern=](-3pt,16.2675pt)--(-3pt,16.2675pt);
\path[-,draw=lightgray,line width=0.4pt,line cap=butt,line join=miter,dash pattern=](29.66pt,6.2675pt)--(29.66pt,6.2675pt);
\path[-,draw=black,line width=0.4pt,line cap=butt,line join=miter,dash pattern=](-7pt,16.2675pt)--(-3pt,16.2675pt);
\path[-,draw=black,line width=0.4pt,line cap=butt,line join=miter,dash pattern=](-7pt,11.2675pt)--(-3pt,11.2675pt);
\path[-,draw=lightgray,line width=0.4pt,line cap=butt,line join=miter,dash pattern=](29.66pt,6.2675pt)--(33.66pt,6.2675pt);
\path[-,draw=black,line width=0.4pt,line cap=butt,line join=miter,dash pattern=](0pt,11.2675pt)--(3pt,11.2675pt);
\path[-,draw=lightgray,line width=0.4pt,line cap=butt,line join=miter,dash pattern=](0pt,6.2675pt)--(3pt,6.2675pt);
\path[-,draw=black,line width=0.8pt,line cap=butt,line join=miter,dash pattern=](7pt,12.535pt)..controls(7pt,14.1918pt)and(8.3431pt,15.535pt)..(10pt,15.535pt)--(22.66pt,15.535pt)..controls(24.3168pt,15.535pt)and(25.66pt,14.1918pt)..(25.66pt,12.535pt)--(25.66pt,0pt)..controls(25.66pt,-1.6569pt)and(24.3168pt,-3pt)..(22.66pt,-3pt)--(10pt,-3pt)..controls(8.3431pt,-3pt)and(7pt,-1.6569pt)..(7pt,0pt)--cycle;
\path[-,draw=lightgray,line width=0.4pt,line cap=butt,line join=miter,dash pattern=](7pt,6.2675pt)--(10pt,6.2675pt);
\path[-,draw=black,line width=0.4pt,line cap=butt,line join=miter,dash pattern=](7pt,11.2675pt)--(7pt,11.2675pt);
\path[-,draw=lightgray,line width=0.4pt,line cap=butt,line join=miter,dash pattern=](22.66pt,6.2675pt)--(25.66pt,6.2675pt);
\path[-,draw=black,line width=0.4pt,line cap=butt,line join=miter,dash pattern=](3pt,11.2675pt)--(7pt,11.2675pt);
\path[-,draw=lightgray,line width=0.4pt,line cap=butt,line join=miter,dash pattern=](3pt,6.2675pt)--(7pt,6.2675pt);
\path[-,draw=lightgray,line width=0.4pt,line cap=butt,line join=miter,dash pattern=](25.66pt,6.2675pt)--(29.66pt,6.2675pt);
\path[-,line width=0.4pt,line cap=butt,line join=miter,dash pattern=](13pt,9.535pt)--(19.66pt,9.535pt)--(19.66pt,3pt)--(13pt,3pt)--cycle;
\node[anchor=north west,inner sep=0] at (13pt,9.535pt){\savebox{\marxupbox}{{\(P\)}}\immediate\write\boxesfile{182}\immediate\write\boxesfile{\number\wd\marxupbox}\immediate\write\boxesfile{\number\ht\marxupbox}\immediate\write\boxesfile{\number\dp\marxupbox}\box\marxupbox};
\path[-,draw=black,line width=0.4pt,line cap=butt,line join=miter,dash pattern=](10pt,10.035pt)..controls(10pt,11.4157pt)and(11.1193pt,12.535pt)..(12.5pt,12.535pt)--(20.16pt,12.535pt)..controls(21.5407pt,12.535pt)and(22.66pt,11.4157pt)..(22.66pt,10.035pt)--(22.66pt,2.5pt)..controls(22.66pt,1.1193pt)and(21.5407pt,0pt)..(20.16pt,0pt)--(12.5pt,0pt)..controls(11.1193pt,0pt)and(10pt,1.1193pt)..(10pt,2.5pt)--cycle;
\path[-,draw=lightgray,line width=0.4pt,line cap=butt,line join=miter,dash pattern=](22.66pt,6.2675pt)--(22.66pt,6.2675pt);
\path[-,draw=lightgray,line width=0.4pt,line cap=butt,line join=miter,dash pattern=](10pt,6.2675pt)--(10pt,6.2675pt);
\path[-,draw=black,line width=0.4pt,line cap=butt,line join=miter,dash pattern=](0pt,11.2675pt)--(0pt,11.2675pt);
\path[-,line width=0.4pt,line cap=butt,line join=miter,dash pattern=on 0.4pt off 1pt](0pt,11.2675pt)--(0pt,11.2675pt)--(0pt,6.2675pt)--(0pt,6.2675pt)--cycle;
\path[-,draw=black,line width=0.4pt,line cap=butt,line join=miter,dash pattern=on 0.4pt off 1pt](-3pt,14.2675pt)--(3pt,14.2675pt)--(3pt,3.2675pt)--(-3pt,3.2675pt)--cycle;
\path[-,fill=lightgray,line width=0.4pt,line cap=butt,line join=miter,dash pattern=](1pt,6.2675pt)..controls(1pt,6.8198pt)and(0.5523pt,7.2675pt)..(0pt,7.2675pt)..controls(-0.5523pt,7.2675pt)and(-1pt,6.8198pt)..(-1pt,6.2675pt)..controls(-1pt,5.7152pt)and(-0.5523pt,5.2675pt)..(0pt,5.2675pt)..controls(0.5523pt,5.2675pt)and(1pt,5.7152pt)..(1pt,6.2675pt)--cycle;
\path[-,fill=black,line width=0.4pt,line cap=butt,line join=miter,dash pattern=](-12pt,13.7675pt)..controls(-12pt,14.3198pt)and(-12.4477pt,14.7675pt)..(-13pt,14.7675pt)..controls(-13.5523pt,14.7675pt)and(-14pt,14.3198pt)..(-14pt,13.7675pt)..controls(-14pt,13.2152pt)and(-13.5523pt,12.7675pt)..(-13pt,12.7675pt)..controls(-12.4477pt,12.7675pt)and(-12pt,13.2152pt)..(-12pt,13.7675pt)--cycle;
\path[-,draw=black,line width=0.4pt,line cap=butt,line join=miter,dash pattern=](-10.06pt,16.2pt)..controls(-9.22pt,16.2675pt)and(-8.2pt,16.2675pt)..(-7pt,16.2675pt);
\path[-,draw=black,line width=0.4pt,line cap=butt,line join=miter,dash pattern=](-13pt,13.7675pt)..controls(-13pt,15.5175pt)and(-12.02pt,16.0425pt)..(-10.06pt,16.2pt);
\path[-,draw=black,line width=0.4pt,line cap=butt,line join=miter,dash pattern=](-10.06pt,11.335pt)..controls(-9.22pt,11.2675pt)and(-8.2pt,11.2675pt)..(-7pt,11.2675pt);
\path[-,draw=black,line width=0.4pt,line cap=butt,line join=miter,dash pattern=](-13pt,13.7675pt)..controls(-13pt,12.0175pt)and(-12.02pt,11.4925pt)..(-10.06pt,11.335pt);
\end{tikzpicture}}, and in {\sc{}Roger} it is
\(\allowbreak{}\mathnormal{(}\mskip 0.0mu\mathnormal{∇}\mskip 3.0mu\allowbreak{}\mathnormal{(}\mskip 0.0mu\allowbreak{}\mathnormal{(}\mskip 0.0mu\mathnormal{∇}\mskip 3.0mu\mathsf{P}\mskip 0.0mu\mathnormal{)}\allowbreak{}\mskip 3.0mu\allowbreak{}\mathnormal{∘}\allowbreak{}\mskip 3.0muρ\mskip 0.0mu\mathnormal{)}\allowbreak{}\mskip 0.0mu\mathnormal{)}\allowbreak{}\mskip 3.0mu\allowbreak{}\mathnormal{∘}\allowbreak{}\mskip 3.0mu\mathsf{g'}\).  The Einstein notation, while already
economical, can be made even more concise by using the index juggling
convention: \(∇ᵢ∇ⁱP\).  In {\sc{}Albert}, the same thing would
be written \begin{list}{}{\setlength\leftmargin{1.0em}}\item\relax
\(\mathsf{contract}\mskip 3.0mu\allowbreak{}\mathnormal{(}\mskip 0.0muλ\mskip 3.0mu^{\mathsf{i}}\mskip 3.0mu_{\mathsf{i}}\mskip 3.0mu\mathnormal{\rightarrow }\mskip 3.0mu\mathsf{deriv}\mskip 3.0mu_{\mathsf{i}}\mskip 3.0mu\allowbreak{}\mathnormal{(}\mskip 0.0mu\mathsf{deriv}\mskip 3.0mu\allowbreak{}\mathnormal{(}\mskip 0.0mu\mathsf{lower}\mskip 3.0mu^{\mathsf{i}}\mskip 0.0mu\mathnormal{)}\allowbreak{}\mskip 3.0mu\mathsf{potential}\mskip 0.0mu\mathnormal{)}\allowbreak{}\mskip 0.0mu\mathnormal{)}\allowbreak{}\)\end{list} Here the high index from the contraction must be lowered
because \(\mathsf{deriv}\) can only take a low index as its first
argument.

\subsubsection{Laws of covariant derivatives}\label{183} 
As one might expect, the covariant derivative satisfies the product law for derivatives:
\begin{equation}∇ᵢ (t u) = (∇ᵢ t)  u + t  (∇ᵢ u)\label{184}\end{equation}  
The above formulation is concise, but using it as basis for implementation
can be tedious, because one needs to track free and bound
variables. A pitfall is that \(t\) and \(u\) stand for arbitrary expressions,
and index variables may occur free in them. Therefore a specific implementation difficulty is that one needs to preserve the linearity of index
variables at the level of the host language.\footnote{In a draft
version of this work, we attempted doing this and found the result
inscrutable.\label{185}}  Thus we find that the morphism {\sc{}edsl} {\sc{}Roger} is a better
implementation vehicle in this case. The laws become noticeably more
verbose, but not horribly so, and dispense of tracking free
variables. In this notation, the product law is expressed as two cases, one for each of the
composition \(\allowbreak{}\mathnormal{(}\mskip 0.0mu\allowbreak{}\mathnormal{∘}\allowbreak{}\mskip 0.0mu\mathnormal{)}\allowbreak{}\) and tensor \(\allowbreak{}\mathnormal{(}\mskip 0.0mu{⊗}\mskip 0.0mu\mathnormal{)}\allowbreak{}\) operators:
\begin{align*}\mathnormal{∇}\mskip 3.0mu\allowbreak{}\mathnormal{(}\mskip 0.0mu\mathsf{t}\mskip 3.0mu\allowbreak{}\mathnormal{∘}\allowbreak{}\mskip 3.0mu\mathsf{u}\mskip 0.0mu\mathnormal{)}\allowbreak{}&=\mathsf{t}\mskip 3.0mu\allowbreak{}\mathnormal{∘}\allowbreak{}\mskip 3.0mu\allowbreak{}\mathnormal{(}\mskip 0.0mu\mathnormal{∇}\mskip 3.0mu\mathsf{u}\mskip 0.0mu\mathnormal{)}\allowbreak{}\mskip 3.0mu\mathnormal{+}\mskip 3.0mu\allowbreak{}\mathnormal{(}\mskip 0.0mu\mathnormal{∇}\mskip 3.0mu\mathsf{t}\mskip 0.0mu\mathnormal{)}\allowbreak{}\mskip 3.0mu\allowbreak{}\mathnormal{∘}\allowbreak{}\mskip 3.0mu\allowbreak{}\mathnormal{(}\mskip 0.0mu\mathsf{id}\mskip 3.0mu{⊗}\mskip 3.0mu\mathsf{u}\mskip 0.0mu\mathnormal{)}\allowbreak{}\\\mathnormal{∇}\mskip 3.0mu\allowbreak{}\mathnormal{(}\mskip 0.0mu\mathsf{t}\mskip 3.0mu{⊗}\mskip 3.0mu\mathsf{u}\mskip 0.0mu\mathnormal{)}\allowbreak{}&=\allowbreak{}\mathnormal{(}\mskip 0.0mu\mathnormal{∇}\mskip 3.0mu\mathsf{t}\mskip 3.0mu{⊗}\mskip 3.0mu\mathsf{u}\mskip 0.0mu\mathnormal{)}\allowbreak{}\mskip 3.0mu\allowbreak{}\mathnormal{∘}\allowbreak{}\mskip 3.0mu\bar{α}\mskip 3.0mu\mathnormal{+}\mskip 3.0mu\allowbreak{}\mathnormal{(}\mskip 0.0mu\mathsf{t}\mskip 3.0mu{⊗}\mskip 3.0mu\mathnormal{∇}\mskip 3.0mu\mathsf{u}\mskip 0.0mu\mathnormal{)}\allowbreak{}\mskip 3.0mu\allowbreak{}\mathnormal{∘}\allowbreak{}\mskip 3.0muα\mskip 3.0mu\allowbreak{}\mathnormal{∘}\allowbreak{}\mskip 3.0mu\allowbreak{}\mathnormal{(}\mskip 0.0muσ\mskip 3.0mu{⊗}\mskip 3.0mu\mathsf{id}\mskip 0.0mu\mathnormal{)}\allowbreak{}\mskip 3.0mu\allowbreak{}\mathnormal{∘}\allowbreak{}\mskip 3.0mu\bar{α}\end{align*} As before, we find the corresponding diagrams more readable:
\begin{equation}{
}\label{223}\end{equation} 
The derivative of all constant morphisms is
zero: \(∇\mskip 3.0mu\mathsf{id}\mskip 3.0mu\mathnormal{=}\mskip 3.0mu\mathrm{0}\), \(∇\mskip 3.0muρ\mskip 3.0mu\mathnormal{=}\mskip 3.0mu\mathrm{0}\), \(∇\mskip 3.0muα\mskip 3.0mu\mathnormal{=}\mskip 3.0mu\mathrm{0}\), etc.  This property also holds for the (co)metric tensors:\footnote{The condition \(∇\mskip 3.0mug\mskip 3.0mu\mathnormal{=}\mskip 3.0mu\mathrm{0}\mskip 3.0mu\mathnormal{=}\mskip 3.0mu∇\mskip 3.0mug'\) holds only for Levi-Civita connections. Even though our general framework can be generalised to support other derivatives, all our examples fit this case.\label{224}} \(∇\mskip 3.0mug\mskip 3.0mu\mathnormal{=}\mskip 3.0mu\mathrm{0}\mskip 3.0mu\mathnormal{=}\mskip 3.0mu∇\mskip 3.0mug'\).  Together,
the above laws fully specify the structural behaviour of the derivative and the implementation of the
covariant derivative falls out from them with no additional difficulty.
However, the important case of the derivative of a tensor in a coordinate category remains to be addressed.
This question leads us to the concept of affinity.

\subsubsection{Partial Derivatives, Christoffel Symbols and Affinities}\label{225} 
One may be tempted to think that the coefficient representation of the covariant
derivative is the index-wise derivative of the coefficient
representations.  While this is true if the metric is the identity everywhere on the manifold, it is not the case in general.

Conventionally, one speaks of the ``partial derivative'' for
the index-wise derivative of coefficients and retains the term ``covariant
derivative'' for the spatial derivative. (The notations
are shown in \cref{170}, but note that
\(∂ᵢt\) stands for the partial derivative of the expression \(\mathsf{t}\) with respect to the \(i^\text{th}\) coordinate, sometimes also written \(∂t/∂xⁱ\).)
As usual, we express the availability of this new operation by means of a class:
\begin{list}{}{\setlength\leftmargin{1.0em}}\item\relax
\ensuremath{\begin{parray}\column{B}{@{}>{}l<{}@{}}\column[0em]{1}{@{}>{}l<{}@{}}\column[1em]{2}{@{}>{}l<{}@{}}\column{3}{@{}>{}l<{}@{}}\column{4}{@{}>{}l<{}@{}}\column{E}{@{}>{}l<{}@{}}%
\>[1]{\mathbf{class}\mskip 3.0mu\mathsf{MetricCategory}\mskip 3.0muz\mskip 3.0mu\mathnormal{\Rightarrow }\mskip 3.0mu\mathsf{CoordinateCategory}\mskip 3.0muz\mskip 3.0mu}\>[4]{\mathbf{where}}\<[E]\\
\>[2]{∂\mskip 3.0mu\mathnormal{::}\mskip 3.0mu}\>[3]{\allowbreak{}\mathnormal{(}\mskip 0.0mu\mathsf{a}\mskip 1.0mu\overset{z}{\leadsto }\mskip 1.0mu\mathsf{b}\mskip 0.0mu\mathnormal{)}\allowbreak{}\mskip 3.0mu\mathnormal{\rightarrow }\mskip 3.0mu\allowbreak{}\mathnormal{(}\mskip 0.0mu\allowbreak{}\mathnormal{(}\mskip 0.0muT_{z}\mskip 3.0mu\mathnormal{⊗}\mskip 3.0mu\mathsf{a}\mskip 0.0mu\mathnormal{)}\allowbreak{}\mskip 1.0mu\overset{z}{\leadsto }\mskip 1.0mu\mathsf{b}\mskip 0.0mu\mathnormal{)}\allowbreak{}}\<[E]\end{parray}}\end{list} The Levi-Civita connection is unique for a given metric, but we still
require the user to provide an implementation. For categories with
coordinates, discussed in the next subsection, a canonical
implementation of \(∇\) is provided.

We make the concept available in {\sc{}Albert} like this:
\begin{list}{}{\setlength\leftmargin{1.0em}}\item\relax
\ensuremath{\begin{parray}\column{B}{@{}>{}l<{}@{}}\column[0em]{1}{@{}>{}l<{}@{}}\column{2}{@{}>{}l<{}@{}}\column{3}{@{}>{}l<{}@{}}\column{E}{@{}>{}l<{}@{}}%
\>[1]{\mathsf{partial}\mskip 3.0mu}\>[2]{\mathnormal{::}\mskip 3.0mu}\>[3]{\mathsf{CoordinateCategory}\mskip 3.0muz\mskip 3.0mu\mathnormal{\Rightarrow }\mskip 3.0mu\mathsf{P}\mskip 3.0muz\mskip 3.0mu\mathsf{r}\mskip 3.0mu\mathsf{v}\mskip 3.0mu\mathnormal{⊸}\mskip 3.0mu\mathsf{R}\mskip 3.0muz\mskip 3.0mu\mathsf{r}\mskip 3.0mu\mathnormal{⊸}\mskip 3.0mu\mathsf{R}\mskip 3.0muz\mskip 3.0mu\mathsf{r}}\<[E]\end{parray}}\end{list} 
In the general case, to compute the covariant derivative, one must account for
the variation of the basis.
Therefore the partial derivative must be corrected by a so called affinity term.
The partial derivative accounts for the variation of (the representation of) the tensor field itself
as the position varies, while the affinity term accounts for the variation
of the basis.
The variation of the basis is measured by the \emph{Christoffel symbol},
denoted \(Γ\) and of type \(\allowbreak{}\mathnormal{(}\mskip 0.0muT_{z}\mskip 3.0mu\mathnormal{⊗}\mskip 3.0muT_{z}\mskip 0.0mu\mathnormal{)}\allowbreak{}\mskip 1.0mu\overset{z}{\leadsto }\mskip 1.0muT_{z}\) (\cref{170}).
Different choices of local basis field for the
\emph{same} manifold will yield different values for it.
(Therefore, even though Γ is a morphism in a matrix category, and even though it can be transformed
to another basis by multiplication with Jacobians, this transformed version will not
be the Christoffel symbol for the new basis.  This fact is sometimes expressed in textbooks as ``Γ is not a tensor'').

The Christoffel symbol is often treated abstractly, but it is determined by the (coefficient representation of)
the metric:
\begin{equation}Γ{_i}{_j}{^k} = \frac{1}{2}g'{^l}{^k}∂{_j}g{_i}{_l} + \frac{1}{2}g'{^m}{^k}∂{_i}g{_j}{_m} - \frac{1}{2}g'{^n}{^k}∂{_n}g{_j}{_i}\label{226}\end{equation} The Christoffel symbol is always symmetric in its first two indices, which is equivalent to
asserting that the following diagram is zero: \({\begin{tikzpicture}[baseline=(current bounding box.center)]\path[-,draw=black,line width=0.4pt,line cap=butt,line join=miter,dash pattern=](-8pt,7.4184pt)--(-7pt,7.4184pt);
\path[-,draw=black,line width=0.4pt,line cap=butt,line join=miter,dash pattern=](-8pt,0pt)--(-7pt,0pt);
\path[-,draw=black,line width=0.4pt,line cap=butt,line join=miter,dash pattern=](6pt,3.7092pt)--(13pt,3.7092pt);
\path[-,draw=black,line width=0.4pt,line cap=butt,line join=miter,dash pattern=](-1pt,7.4184pt)--(0pt,7.4184pt);
\path[-,draw=black,line width=0.4pt,line cap=butt,line join=miter,dash pattern=](-1pt,0pt)--(0pt,0pt);
\path[-,draw=black,fill= gray,line width=0.4pt,line cap=butt,line join=miter,dash pattern=](0pt,8.4184pt)..controls(4pt,8.4184pt)and(6pt,5.7092pt)..(6pt,3.7092pt)..controls(6pt,1.7092pt)and(4pt,-1pt)..(0pt,-1pt)--cycle;
\path[-,draw=black,line width=0.4pt,line cap=butt,line join=miter,dash pattern=](-7pt,7.4184pt)--(-1pt,7.4184pt);
\path[-,draw=black,line width=0.4pt,line cap=butt,line join=miter,dash pattern=](-7pt,0pt)--(-1pt,0pt);
\path[-,draw=black,line width=1.2pt,line cap=butt,line join=miter,dash pattern=](-4pt,8.4184pt)--(-4pt,-1pt)--cycle;
\path[-,line width=0.4pt,line cap=butt,line join=miter,dash pattern=](-12.6316pt,9.758pt)--(-10pt,9.758pt)--(-10pt,5.0788pt)--(-12.6316pt,5.0788pt)--cycle;
\node[anchor=north west,inner sep=0] at (-12.6316pt,9.758pt){\savebox{\marxupbox}{{\({\scriptstyle i}\)}}\immediate\write\boxesfile{227}\immediate\write\boxesfile{\number\wd\marxupbox}\immediate\write\boxesfile{\number\ht\marxupbox}\immediate\write\boxesfile{\number\dp\marxupbox}\box\marxupbox};
\path[-,line width=0.4pt,line cap=butt,line join=miter,dash pattern=](-14.6316pt,11.758pt)--(-8pt,11.758pt)--(-8pt,3.0788pt)--(-14.6316pt,3.0788pt)--cycle;
\path[-,line width=0.4pt,line cap=butt,line join=miter,dash pattern=](-13.8617pt,3.0788pt)--(-10pt,3.0788pt)--(-10pt,-3.0788pt)--(-13.8617pt,-3.0788pt)--cycle;
\node[anchor=north west,inner sep=0] at (-13.8617pt,3.0788pt){\savebox{\marxupbox}{{\({\scriptstyle j}\)}}\immediate\write\boxesfile{228}\immediate\write\boxesfile{\number\wd\marxupbox}\immediate\write\boxesfile{\number\ht\marxupbox}\immediate\write\boxesfile{\number\dp\marxupbox}\box\marxupbox};
\path[-,line width=0.4pt,line cap=butt,line join=miter,dash pattern=](-15.8617pt,5.0788pt)--(-8pt,5.0788pt)--(-8pt,-5.0788pt)--(-15.8617pt,-5.0788pt)--cycle;
\path[-,line width=0.4pt,line cap=butt,line join=miter,dash pattern=](15pt,6.2003pt)--(19.3398pt,6.2003pt)--(19.3398pt,1.2181pt)--(15pt,1.2181pt)--cycle;
\node[anchor=north west,inner sep=0] at (15pt,6.2003pt){\savebox{\marxupbox}{{\({\scriptstyle k}\)}}\immediate\write\boxesfile{229}\immediate\write\boxesfile{\number\wd\marxupbox}\immediate\write\boxesfile{\number\ht\marxupbox}\immediate\write\boxesfile{\number\dp\marxupbox}\box\marxupbox};
\path[-,line width=0.4pt,line cap=butt,line join=miter,dash pattern=](13pt,8.2003pt)--(21.3398pt,8.2003pt)--(21.3398pt,-0.7819pt)--(13pt,-0.7819pt)--cycle;
\end{tikzpicture}}\).
In the implementation, we make \(\Gamma\) a method of the class \(\mathsf{CoordinateCategory}\), but with a default definition in terms of \cref{226}.
We make it available in {\sc{}Albert} by embedding the morphism \(Γ\) as follows:
\begin{list}{}{\setlength\leftmargin{1.0em}}\item\relax
 \ensuremath{\begin{parray}\column{B}{@{}>{}l<{}@{}}\column[0em]{1}{@{}>{}l<{}@{}}\column{2}{@{}>{}l<{}@{}}\column{3}{@{}>{}l<{}@{}}\column{E}{@{}>{}l<{}@{}}%
\>[1]{\mathsf{christoffel}\mskip 3.0mu}\>[2]{\mathnormal{::}\mskip 3.0mu}\>[3]{\mathsf{CoordinateCategory}\mskip 3.0muz\mskip 3.0mu\mathnormal{\Rightarrow }\mskip 3.0mu\mathsf{P}\mskip 3.0muz\mskip 3.0mu\mathsf{r}\mskip 3.0muT_{z}\mskip 3.0mu\mathnormal{⊸}\mskip 3.0mu\mathsf{P}\mskip 3.0muz\mskip 3.0mu\mathsf{r}\mskip 3.0muT_{z}\mskip 3.0mu\mathnormal{⊸}\mskip 3.0mu\mathsf{P}\mskip 3.0muz\mskip 3.0mu\mathsf{r}\mskip 3.0mu\dual{T_{z}}\mskip 3.0mu\mathnormal{⊸}\mskip 3.0mu\mathsf{R}\mskip 3.0muz\mskip 3.0mu\mathsf{r}}\<[E]\end{parray}} \ensuremath{\begin{parray}\column{B}{@{}>{}l<{}@{}}\column[0em]{1}{@{}>{}l<{}@{}}\column{2}{@{}>{}l<{}@{}}\column{E}{@{}>{}l<{}@{}}%
\>[1]{\mathsf{christoffel}\mskip 3.0mu_{\mathsf{i}}\mskip 3.0mu_{\mathsf{j}}\mskip 3.0mu^{\mathsf{k}}\mskip 3.0mu\mathnormal{=}\mskip 3.0mu\mathsf{tensorEmbed}\mskip 3.0mu\Gamma\mskip 3.0mu}\>[2]{\allowbreak{}\mathnormal{(}\mskip 0.0mu\mathsf{merge}\mskip 3.0mu\allowbreak{}\mathnormal{(}\mskip 0.0mu_{\mathsf{i}}\mskip 0.0mu\mathnormal{,}\mskip 3.0mu_{\mathsf{j}}\mskip 0.0mu\mathnormal{)}\allowbreak{}\mskip 0.0mu\mathnormal{)}\allowbreak{}\mskip 3.0mu^{\mathsf{k}}}\<[E]\end{parray}} \end{list} 
For a 2-dimensional atomic vector space, the Christoffel symbol has \(2^3=8\) components. For our running example of the polar coordinate system and associated canonical
tangent space, we can write \(Γ\) as two 2×2 symmetric matrices, as follows:
\begin{mathpar}Γ^{ρ} =\left[\begin{array}{cc}0&0\\0&-ρ\end{array}\right]\and{}Γ^{θ} =\left[\begin{array}{cc}0&1/ρ\\1/ρ&0\end{array}\right]\end{mathpar} 

In textbooks on tensors, for instance that of
\citet{lovelock_tensors_1989}, one often sees the relation between
covariant and partial derivatives expressed as a family of equations,
depending on the order of the tensor whose derivative is taken:
 \begin{align*}∇ᵢ T& = ∂ᵢ T\\∇ᵢ Tⱼ& = ∂ᵢ Tⱼ - Γ_{ij}{ᵏ} Tₖ\\∇ᵢ Tʲ& = ∂ᵢ Tʲ + Γ_{ik}{ʲ} Tᵏ\\∇ᵢ Tⱼ{ᵏ}& = ∂ᵢ Tⱼ{ᵏ} - Γ_{il}{ᵏ} Tⱼ{ˡ} + Γ_{ij}{ˡ} Tₗ{ᵏ}\end{align*} etc. Using this sort of definition is particularly
error prone (can you spot quickly whether there is a mistake in the
last line?). In contrast, {\sc{}Roger} captures all cases in one go:
\begin{equation}\text{\ensuremath{\begin{parray}\column{B}{@{}>{}l<{}@{}}\column[0em]{1}{@{}>{}l<{}@{}}\column{E}{@{}>{}l<{}@{}}%
\>[1]{∇\mskip 3.0mu\mathsf{t}\mskip 3.0mu\mathnormal{=}\mskip 3.0mu∂\mskip 3.0mu\mathsf{t}\mskip 3.0mu\mathnormal{-}\mskip 3.0mu\allowbreak{}\mathnormal{(}\mskip 0.0mu\mathsf{t}\mskip 3.0mu\allowbreak{}\mathnormal{∘}\allowbreak{}\mskip 3.0mu\mathsf{affinity}\mskip 0.0mu\mathnormal{)}\allowbreak{}\mskip 3.0mu\mathnormal{+}\mskip 3.0mu\allowbreak{}\mathnormal{(}\mskip 0.0mu\mathsf{affinity}\mskip 3.0mu\allowbreak{}\mathnormal{∘}\allowbreak{}\mskip 3.0mu\allowbreak{}\mathnormal{(}\mskip 0.0mu\mathsf{id}\mskip 3.0mu{⊗}\mskip 3.0mu\mathsf{t}\mskip 0.0mu\mathnormal{)}\allowbreak{}\mskip 0.0mu\mathnormal{)}\allowbreak{}}\<[E]\end{parray}}}\label{230}\end{equation} The complexity is pushed
down into \(\mathsf{affinity}\), which is invoked once for the domain and once for the
codomain of \(\mathsf{t}\).
The affinity is a family of morphisms
\(\mathsf{affinity}\mskip 3.0mu\mathnormal{::}\mskip 3.0muT_{M_{\mathsf{b}}}\mskip 3.0mu\mathnormal{⊗}\mskip 3.0mu\mathsf{a}\mskip 1.0mu\overset{M_{\mathsf{b}}}{\leadsto }\mskip 1.0mu\mathsf{a}\) for any object \(\mathsf{a}\) constructed
from the local vector space \(T_{M_{\mathsf{b}}}\) and refers to a specific basis for it, using some coherent canonical choice of bases (with \(\mathsf{b}\) as a basis for \(T_{M_{\mathsf{b}}}\)).
  The affinity for arbitrary vector spaces
is defined by induction on the structure of the corresponding object. The affinity for a product
object is the sum of affinities for each of the components, leaving the
other component untouched.  The affinity for a dual object is the
negative affinity of the underlying object, with input and output
suitably swapped. This can be coded by a type-dependent set of
equations:\footnote{For readability, we take some liberties in the presentation of \(\mathsf{aff}\):
1. we pattern match on types, but Haskell requires using singleton
types for this purpose. 2. The \(\allowbreak{}\mathnormal{(}\mskip 0.0mu\mathnormal{\fatsemi }\mskip 0.0mu\mathnormal{)}\allowbreak{}\) operator is used in place
of \(\mathsf{merge}\) in expressions and in place of \(\mathsf{split}\) in
patterns. 3. We use the operators \(\allowbreak{}\mathnormal{(}\mskip 0.0mu\mathnormal{+}\mskip 0.0mu\mathnormal{)}\allowbreak{}\) and \(\allowbreak{}\mathnormal{(}\mskip 0.0mu\mathnormal{-}\mskip 0.0mu\mathnormal{)}\allowbreak{}\) for the addition and
scaling by \(\allowbreak{}\mathnormal{(}\mskip 0.0mu\mathnormal{-}\mskip 3.0mu\mathrm{1}\mskip 0.0mu\mathnormal{)}\allowbreak{}\) instead of the syntax suggested in \cref{85}.\label{231}} 
\begin{list}{}{\setlength\leftmargin{1.0em}}\item\relax
\ensuremath{\begin{parray}\column{B}{@{}>{}l<{}@{}}\column[0em]{1}{@{}>{}l<{}@{}}\column{2}{@{}>{}l<{}@{}}\column{3}{@{}>{}l<{}@{}}\column{4}{@{}>{}l<{}@{}}\column{5}{@{}>{}l<{}@{}}\column{6}{@{}>{}l<{}@{}}\column{7}{@{}>{}l<{}@{}}\column{8}{@{}>{}l<{}@{}}\column{9}{@{}>{}l<{}@{}}\column{10}{@{}>{}l<{}@{}}\column{11}{@{}>{}l<{}@{}}\column{12}{@{}>{}l<{}@{}}\column{13}{@{}>{}l<{}@{}}\column{14}{@{}>{}l<{}@{}}\column{15}{@{}>{}l<{}@{}}\column{16}{@{}>{}l<{}@{}}\column{17}{@{}>{}l<{}@{}}\column{18}{@{}>{}l<{}@{}}\column{19}{@{}>{}l<{}@{}}\column{20}{@{}>{}l<{}@{}}\column{E}{@{}>{}l<{}@{}}%
\>[1]{\mathsf{aff}\mskip 3.0mu\mathsf{a}\mskip 3.0mu\mathnormal{::}\mskip 3.0mu\mathsf{CoordinateCategory}\mskip 3.0muz\mskip 3.0mu\mathnormal{\Rightarrow }\mskip 3.0mu\mathsf{P}\mskip 3.0muz\mskip 3.0mu\mathsf{r}\mskip 3.0muT_{z}\mskip 3.0mu\mathnormal{⊸}\mskip 3.0mu\mathsf{P}\mskip 3.0muz\mskip 3.0mu\mathsf{r}\mskip 3.0mu\mathsf{a}\mskip 3.0mu\mathnormal{⊸}\mskip 3.0mu\mathsf{P}\mskip 3.0muz\mskip 3.0mu\mathsf{r}\mskip 3.0mu\dual{\mathsf{a}}\mskip 3.0mu\mathnormal{⊸}\mskip 3.0mu\mathsf{R}\mskip 3.0muz\mskip 3.0mu\mathsf{r}}\<[E]\\
\>[1]{\mathsf{aff}\mskip 3.0muT_{z}\mskip 3.0mu}\>[2]{\mathsf{ᵢ}\mskip 3.0mu}\>[3]{\mathsf{ⱼ}\mskip 3.0mu}\>[4]{\mathsf{ᵏ}\mskip 3.0mu}\>[5]{\mathnormal{=}\mskip 3.0mu}\>[6]{\Gamma\mskip 3.0mu\mathsf{ᵢ}\mskip 3.0mu}\>[11]{\mathsf{ⱼ}\mskip 3.0mu}\>[12]{\mathsf{ᵏ}}\<[E]\\
\>[1]{\mathsf{aff}\mskip 3.0mu\allowbreak{}\mathnormal{(}\mskip 0.0mu\mathsf{a}\mskip 3.0mu\mathnormal{⊗}\mskip 3.0mu\mathsf{b}\mskip 0.0mu\mathnormal{)}\allowbreak{}\mskip 3.0mu}\>[2]{\mathsf{ᵢ}\mskip 3.0mu}\>[3]{\allowbreak{}\mathnormal{(}\mskip 0.0mu\mathsf{ⱼ}\mskip 2.0mu\mathnormal{\fatsemi }\mskip 3.0mu\mathsf{ₖ}\mskip 0.0mu\mathnormal{)}\allowbreak{}\mskip 3.0mu}\>[4]{\allowbreak{}\mathnormal{(}\mskip 0.0mu\mathsf{ˡ}\mskip 2.0mu\mathnormal{\fatsemi }\mskip 3.0mu\mathsf{ᵐ}\mskip 0.0mu\mathnormal{)}\allowbreak{}\mskip 3.0mu}\>[5]{\mathnormal{=}\mskip 3.0mu}\>[6]{\mathsf{delta}\mskip 3.0mu\mathsf{ₖ}\mskip 3.0mu\mathsf{ᵐ}\mskip 3.0mu}\>[8]{{\tikzstar{0.11}{0.25}{5}{-18}{fill=black}}\mskip 3.0mu}\>[10]{\mathsf{aff}\mskip 3.0mu\mathsf{a}\mskip 3.0mu}\>[13]{\mathsf{ᵢ}\mskip 3.0mu}\>[14]{\mathsf{ⱼ}\mskip 3.0mu}\>[15]{\mathsf{ˡ}\mskip 3.0mu\mathnormal{+}\mskip 3.0mu}\>[16]{\mathsf{delta}\mskip 3.0mu\mathsf{ⱼ}\mskip 3.0mu\mathsf{ˡ}\mskip 3.0mu}\>[17]{{\tikzstar{0.11}{0.25}{5}{-18}{fill=black}}\mskip 3.0mu\mathsf{aff}\mskip 3.0mu\mathsf{a}\mskip 3.0mu}\>[18]{\mathsf{ᵢ}\mskip 3.0mu}\>[19]{\mathsf{ₖ}\mskip 3.0mu}\>[20]{\mathsf{ᵐ}}\<[E]\\
\>[1]{\mathsf{aff}\mskip 3.0mu\dual{\mathsf{a}}\mskip 3.0mu}\>[2]{\mathsf{ᵢ}\mskip 3.0mu}\>[3]{\mathsf{ʲ}\mskip 3.0mu}\>[4]{\mathsf{ₖ}\mskip 3.0mu}\>[5]{\mathnormal{=}\mskip 3.0mu\mathnormal{-}\mskip 3.0mu}\>[6]{\mathsf{aff}\mskip 3.0mu\mathsf{a}\mskip 3.0mu\mathsf{ᵢ}\mskip 3.0mu}\>[7]{\mathsf{ₖ}\mskip 3.0mu}\>[9]{\mathsf{ʲ}}\<[E]\\
\>[1]{\mathsf{aff}\mskip 3.0mu\mathbf{1}\mskip 3.0mu}\>[2]{\mathsf{ᵢ}\mskip 3.0mu}\>[3]{\mathsf{ⱼ}\mskip 3.0mu}\>[4]{\mathsf{ᵏ}\mskip 3.0mu}\>[5]{\mathnormal{=}\mskip 3.0mu}\>[6]{\mathsf{zeroTensor}\mskip 3.0mu\allowbreak{}\mathnormal{(}\mskip 0.0mu\mathsf{ᵢ}\mskip 2.0mu\mathnormal{\fatsemi }\mskip 3.0mu\mathsf{ⱼ}\mskip 2.0mu\mathnormal{\fatsemi }\mskip 3.0mu\mathsf{ᵏ}\mskip 0.0mu\mathnormal{)}\allowbreak{}}\<[E]\end{parray}}\end{list} 
Here, {\sc{}Albert} is much more concise than {\sc{}Roger} (we
omit the corresponding expressions entirely). The corresponding graphical notation is shown in \cref{248} for each case (but using atomic types in place of a proper induction).

\begin{figure}[]\begin{center}
\end{center}\caption{Affinities \(\mathsf{aff}\mskip 3.0mu\mathsf{a}\) at various types \(\mathsf{a}\) (shown in the third row).}\label{248}\end{figure} 

\subsection{Example: computing divergence in an arbitrary coordinate system}\label{249} 
Returning to the example of the potential field, \cref{230} and \(\mathsf{aff}\) tell us that its
covariant derivative is equal to its partial derivative, regardless of
the value of \(Γ\): \(∇{_i}P\) = \(∂{_i}P\).
Indeed, its domain and codomain are both \(\mathbf{1}\), and therefore the
affinities are both zero.  (To compute the value of \(∇{_i}P\) in a given coordinate
system, one would still need to multiply by the Jacobian as indicated
earlier.)

However, when computing the second derivative, a non-zero affinity
arises (because the 1st derivative has a non-unit domain.) One has
therefore: \begin{equation}∇ᵢ∇ⁱP =g'{^i}{^j}∇{_i}∇{_j}P=g'{^i}{^j}∂{_i}∂{_j}P - g'{^k}{^l}Γ{_k}{_l}{^m}∂{_m}P\label{250}\end{equation} In diagram notation:
{\begin{tikzpicture}[baseline={([yshift=-0.8ex]current bounding box.center)}]\path[-,draw=lightgray,line width=0.4pt,line cap=butt,line join=miter,dash pattern=](-20pt,13.7675pt)--(-13pt,13.7675pt);
\path[-,draw=lightgray,line width=0.4pt,line cap=butt,line join=miter,dash pattern=](33.66pt,6.2675pt)--(36.66pt,6.2675pt);
\path[-,draw=black,line width=0.4pt,line cap=butt,line join=miter,dash pattern=](-7pt,16.2675pt)--(-7pt,16.2675pt);
\path[-,draw=black,line width=0.4pt,line cap=butt,line join=miter,dash pattern=](-7pt,11.2675pt)--(-7pt,11.2675pt);
\path[-,draw=black,line width=0.8pt,line cap=butt,line join=miter,dash pattern=](-3pt,17.2675pt)..controls(-3pt,18.9243pt)and(-1.6569pt,20.2675pt)..(0pt,20.2675pt)--(26.66pt,20.2675pt)..controls(28.3168pt,20.2675pt)and(29.66pt,18.9243pt)..(29.66pt,17.2675pt)--(29.66pt,-3pt)..controls(29.66pt,-4.6569pt)and(28.3168pt,-6pt)..(26.66pt,-6pt)--(0pt,-6pt)..controls(-1.6569pt,-6pt)and(-3pt,-4.6569pt)..(-3pt,-3pt)--cycle;
\path[-,draw=black,line width=0.4pt,line cap=butt,line join=miter,dash pattern=](-3pt,11.2675pt)--(0pt,11.2675pt);
\path[-,draw=black,line width=0.4pt,line cap=butt,line join=miter,dash pattern=](-3pt,16.2675pt)--(-3pt,16.2675pt);
\path[-,draw=lightgray,line width=0.4pt,line cap=butt,line join=miter,dash pattern=](29.66pt,6.2675pt)--(29.66pt,6.2675pt);
\path[-,draw=black,line width=0.4pt,line cap=butt,line join=miter,dash pattern=](-7pt,16.2675pt)--(-3pt,16.2675pt);
\path[-,draw=black,line width=0.4pt,line cap=butt,line join=miter,dash pattern=](-7pt,11.2675pt)--(-3pt,11.2675pt);
\path[-,draw=lightgray,line width=0.4pt,line cap=butt,line join=miter,dash pattern=](29.66pt,6.2675pt)--(33.66pt,6.2675pt);
\path[-,draw=black,line width=0.4pt,line cap=butt,line join=miter,dash pattern=](0pt,11.2675pt)--(3pt,11.2675pt);
\path[-,draw=lightgray,line width=0.4pt,line cap=butt,line join=miter,dash pattern=](0pt,6.2675pt)--(3pt,6.2675pt);
\path[-,draw=black,line width=0.8pt,line cap=butt,line join=miter,dash pattern=](7pt,12.535pt)..controls(7pt,14.1918pt)and(8.3431pt,15.535pt)..(10pt,15.535pt)--(22.66pt,15.535pt)..controls(24.3168pt,15.535pt)and(25.66pt,14.1918pt)..(25.66pt,12.535pt)--(25.66pt,0pt)..controls(25.66pt,-1.6569pt)and(24.3168pt,-3pt)..(22.66pt,-3pt)--(10pt,-3pt)..controls(8.3431pt,-3pt)and(7pt,-1.6569pt)..(7pt,0pt)--cycle;
\path[-,draw=lightgray,line width=0.4pt,line cap=butt,line join=miter,dash pattern=](7pt,6.2675pt)--(10pt,6.2675pt);
\path[-,draw=black,line width=0.4pt,line cap=butt,line join=miter,dash pattern=](7pt,11.2675pt)--(7pt,11.2675pt);
\path[-,draw=lightgray,line width=0.4pt,line cap=butt,line join=miter,dash pattern=](22.66pt,6.2675pt)--(25.66pt,6.2675pt);
\path[-,draw=black,line width=0.4pt,line cap=butt,line join=miter,dash pattern=](3pt,11.2675pt)--(7pt,11.2675pt);
\path[-,draw=lightgray,line width=0.4pt,line cap=butt,line join=miter,dash pattern=](3pt,6.2675pt)--(7pt,6.2675pt);
\path[-,draw=lightgray,line width=0.4pt,line cap=butt,line join=miter,dash pattern=](25.66pt,6.2675pt)--(29.66pt,6.2675pt);
\path[-,line width=0.4pt,line cap=butt,line join=miter,dash pattern=](13pt,9.535pt)--(19.66pt,9.535pt)--(19.66pt,3pt)--(13pt,3pt)--cycle;
\node[anchor=north west,inner sep=0] at (13pt,9.535pt){\savebox{\marxupbox}{{\(P\)}}\immediate\write\boxesfile{251}\immediate\write\boxesfile{\number\wd\marxupbox}\immediate\write\boxesfile{\number\ht\marxupbox}\immediate\write\boxesfile{\number\dp\marxupbox}\box\marxupbox};
\path[-,draw=black,line width=0.4pt,line cap=butt,line join=miter,dash pattern=](10pt,10.035pt)..controls(10pt,11.4157pt)and(11.1193pt,12.535pt)..(12.5pt,12.535pt)--(20.16pt,12.535pt)..controls(21.5407pt,12.535pt)and(22.66pt,11.4157pt)..(22.66pt,10.035pt)--(22.66pt,2.5pt)..controls(22.66pt,1.1193pt)and(21.5407pt,0pt)..(20.16pt,0pt)--(12.5pt,0pt)..controls(11.1193pt,0pt)and(10pt,1.1193pt)..(10pt,2.5pt)--cycle;
\path[-,draw=lightgray,line width=0.4pt,line cap=butt,line join=miter,dash pattern=](22.66pt,6.2675pt)--(22.66pt,6.2675pt);
\path[-,draw=lightgray,line width=0.4pt,line cap=butt,line join=miter,dash pattern=](10pt,6.2675pt)--(10pt,6.2675pt);
\path[-,draw=black,line width=0.4pt,line cap=butt,line join=miter,dash pattern=](0pt,11.2675pt)--(0pt,11.2675pt);
\path[-,line width=0.4pt,line cap=butt,line join=miter,dash pattern=on 0.4pt off 1pt](0pt,11.2675pt)--(0pt,11.2675pt)--(0pt,6.2675pt)--(0pt,6.2675pt)--cycle;
\path[-,draw=black,line width=0.4pt,line cap=butt,line join=miter,dash pattern=on 0.4pt off 1pt](-3pt,14.2675pt)--(3pt,14.2675pt)--(3pt,3.2675pt)--(-3pt,3.2675pt)--cycle;
\path[-,fill=lightgray,line width=0.4pt,line cap=butt,line join=miter,dash pattern=](1pt,6.2675pt)..controls(1pt,6.8198pt)and(0.5523pt,7.2675pt)..(0pt,7.2675pt)..controls(-0.5523pt,7.2675pt)and(-1pt,6.8198pt)..(-1pt,6.2675pt)..controls(-1pt,5.7152pt)and(-0.5523pt,5.2675pt)..(0pt,5.2675pt)..controls(0.5523pt,5.2675pt)and(1pt,5.7152pt)..(1pt,6.2675pt)--cycle;
\path[-,fill=black,line width=0.4pt,line cap=butt,line join=miter,dash pattern=](-12pt,13.7675pt)..controls(-12pt,14.3198pt)and(-12.4477pt,14.7675pt)..(-13pt,14.7675pt)..controls(-13.5523pt,14.7675pt)and(-14pt,14.3198pt)..(-14pt,13.7675pt)..controls(-14pt,13.2152pt)and(-13.5523pt,12.7675pt)..(-13pt,12.7675pt)..controls(-12.4477pt,12.7675pt)and(-12pt,13.2152pt)..(-12pt,13.7675pt)--cycle;
\path[-,draw=black,line width=0.4pt,line cap=butt,line join=miter,dash pattern=](-10.06pt,16.2pt)..controls(-9.22pt,16.2675pt)and(-8.2pt,16.2675pt)..(-7pt,16.2675pt);
\path[-,draw=black,line width=0.4pt,line cap=butt,line join=miter,dash pattern=](-13pt,13.7675pt)..controls(-13pt,15.5175pt)and(-12.02pt,16.0425pt)..(-10.06pt,16.2pt);
\path[-,draw=black,line width=0.4pt,line cap=butt,line join=miter,dash pattern=](-10.06pt,11.335pt)..controls(-9.22pt,11.2675pt)and(-8.2pt,11.2675pt)..(-7pt,11.2675pt);
\path[-,draw=black,line width=0.4pt,line cap=butt,line join=miter,dash pattern=](-13pt,13.7675pt)..controls(-13pt,12.0175pt)and(-12.02pt,11.4925pt)..(-10.06pt,11.335pt);
\end{tikzpicture}} = \({\begin{tikzpicture}[baseline=(current bounding box.center)]\path[-,draw=lightgray,line width=0.4pt,line cap=butt,line join=miter,dash pattern=](-20pt,16.2675pt)--(-13pt,16.2675pt);
\path[-,draw=lightgray,line width=0.4pt,line cap=butt,line join=miter,dash pattern=](37.66pt,6.2675pt)--(40.66pt,6.2675pt);
\path[-,draw=black,line width=0.4pt,line cap=butt,line join=miter,dash pattern=](-7pt,21.2675pt)--(-6pt,21.2675pt);
\path[-,draw=black,line width=0.4pt,line cap=butt,line join=miter,dash pattern=](-7pt,11.2675pt)--(-6pt,11.2675pt);
\path[-,draw=black,line width=0.4pt,line cap=butt,line join=miter,dash pattern=](0pt,21.2675pt)--(3pt,21.2675pt);
\path[-,draw=black,line width=0.4pt,line cap=butt,line join=miter,dash pattern=](0pt,11.2675pt)--(3pt,11.2675pt);
\path[-,draw=lightgray,line width=0.4pt,line cap=butt,line join=miter,dash pattern=](0pt,6.2675pt)--(3pt,6.2675pt);
\path[-,draw=black,line width=0.4pt,line cap=butt,line join=miter,dash pattern=](7pt,25.2675pt)--(33.66pt,25.2675pt)--(33.66pt,-6pt)--(7pt,-6pt)--cycle;
\path[-,draw=black,line width=0.4pt,line cap=butt,line join=miter,dash pattern=](7pt,11.2675pt)--(7pt,11.2675pt);
\path[-,draw=lightgray,line width=0.4pt,line cap=butt,line join=miter,dash pattern=](7pt,6.2675pt)--(7pt,6.2675pt);
\path[-,draw=black,line width=0.4pt,line cap=butt,line join=miter,dash pattern=](7pt,21.2675pt)--(7pt,21.2675pt);
\path[-,draw=lightgray,line width=0.4pt,line cap=butt,line join=miter,dash pattern=](33.66pt,6.2675pt)--(33.66pt,6.2675pt);
\path[-,draw=black,line width=0.4pt,line cap=butt,line join=miter,dash pattern=](3pt,21.2675pt)--(7pt,21.2675pt);
\path[-,draw=black,line width=0.4pt,line cap=butt,line join=miter,dash pattern=](3pt,11.2675pt)--(7pt,11.2675pt);
\path[-,draw=lightgray,line width=0.4pt,line cap=butt,line join=miter,dash pattern=](3pt,6.2675pt)--(7pt,6.2675pt);
\path[-,draw=lightgray,line width=0.4pt,line cap=butt,line join=miter,dash pattern=](33.66pt,6.2675pt)--(37.66pt,6.2675pt);
\path[-,draw=black,line width=0.4pt,line cap=butt,line join=miter,dash pattern=](11pt,15.535pt)--(29.66pt,15.535pt)--(29.66pt,-3pt)--(11pt,-3pt)--cycle;
\path[-,draw=lightgray,line width=0.4pt,line cap=butt,line join=miter,dash pattern=](11pt,6.2675pt)--(14pt,6.2675pt);
\path[-,draw=black,line width=0.4pt,line cap=butt,line join=miter,dash pattern=](11pt,11.2675pt)--(11pt,11.2675pt);
\path[-,draw=lightgray,line width=0.4pt,line cap=butt,line join=miter,dash pattern=](26.66pt,6.2675pt)--(29.66pt,6.2675pt);
\path[-,draw=black,line width=0.4pt,line cap=butt,line join=miter,dash pattern=](7pt,11.2675pt)--(11pt,11.2675pt);
\path[-,draw=lightgray,line width=0.4pt,line cap=butt,line join=miter,dash pattern=](7pt,6.2675pt)--(11pt,6.2675pt);
\path[-,draw=lightgray,line width=0.4pt,line cap=butt,line join=miter,dash pattern=](29.66pt,6.2675pt)--(33.66pt,6.2675pt);
\path[-,line width=0.4pt,line cap=butt,line join=miter,dash pattern=](17pt,9.535pt)--(23.66pt,9.535pt)--(23.66pt,3pt)--(17pt,3pt)--cycle;
\node[anchor=north west,inner sep=0] at (17pt,9.535pt){\savebox{\marxupbox}{{\(P\)}}\immediate\write\boxesfile{252}\immediate\write\boxesfile{\number\wd\marxupbox}\immediate\write\boxesfile{\number\ht\marxupbox}\immediate\write\boxesfile{\number\dp\marxupbox}\box\marxupbox};
\path[-,draw=black,line width=0.4pt,line cap=butt,line join=miter,dash pattern=](14pt,10.035pt)..controls(14pt,11.4157pt)and(15.1193pt,12.535pt)..(16.5pt,12.535pt)--(24.16pt,12.535pt)..controls(25.5407pt,12.535pt)and(26.66pt,11.4157pt)..(26.66pt,10.035pt)--(26.66pt,2.5pt)..controls(26.66pt,1.1193pt)and(25.5407pt,0pt)..(24.16pt,0pt)--(16.5pt,0pt)..controls(15.1193pt,0pt)and(14pt,1.1193pt)..(14pt,2.5pt)--cycle;
\path[-,draw=lightgray,line width=0.4pt,line cap=butt,line join=miter,dash pattern=](26.66pt,6.2675pt)--(26.66pt,6.2675pt);
\path[-,draw=lightgray,line width=0.4pt,line cap=butt,line join=miter,dash pattern=](14pt,6.2675pt)--(14pt,6.2675pt);
\path[-,draw=black,line width=0.4pt,line cap=butt,line join=miter,dash pattern=](-6pt,21.2675pt)--(0pt,21.2675pt);
\path[-,draw=black,line width=0.4pt,line cap=butt,line join=miter,dash pattern=](-6pt,11.2675pt)--(0pt,11.2675pt);
\path[-,line width=0.4pt,line cap=butt,line join=miter,dash pattern=on 0.4pt off 1pt](0pt,11.2675pt)--(0pt,11.2675pt)--(0pt,6.2675pt)--(0pt,6.2675pt)--cycle;
\path[-,draw=black,line width=0.4pt,line cap=butt,line join=miter,dash pattern=on 0.4pt off 1pt](-3pt,14.2675pt)--(3pt,14.2675pt)--(3pt,3.2675pt)--(-3pt,3.2675pt)--cycle;
\path[-,fill=lightgray,line width=0.4pt,line cap=butt,line join=miter,dash pattern=](1pt,6.2675pt)..controls(1pt,6.8198pt)and(0.5523pt,7.2675pt)..(0pt,7.2675pt)..controls(-0.5523pt,7.2675pt)and(-1pt,6.8198pt)..(-1pt,6.2675pt)..controls(-1pt,5.7152pt)and(-0.5523pt,5.2675pt)..(0pt,5.2675pt)..controls(0.5523pt,5.2675pt)and(1pt,5.7152pt)..(1pt,6.2675pt)--cycle;
\path[-,fill=black,line width=0.4pt,line cap=butt,line join=miter,dash pattern=](-12pt,16.2675pt)..controls(-12pt,16.8198pt)and(-12.4477pt,17.2675pt)..(-13pt,17.2675pt)..controls(-13.5523pt,17.2675pt)and(-14pt,16.8198pt)..(-14pt,16.2675pt)..controls(-14pt,15.7152pt)and(-13.5523pt,15.2675pt)..(-13pt,15.2675pt)..controls(-12.4477pt,15.2675pt)and(-12pt,15.7152pt)..(-12pt,16.2675pt)--cycle;
\path[-,draw=black,line width=0.4pt,line cap=butt,line join=miter,dash pattern=](-10.06pt,20.66pt)..controls(-9.22pt,21.0425pt)and(-8.2pt,21.2675pt)..(-7pt,21.2675pt);
\path[-,draw=black,line width=0.4pt,line cap=butt,line join=miter,dash pattern=](-13pt,16.2675pt)..controls(-13pt,18.0175pt)and(-12.02pt,19.7675pt)..(-10.06pt,20.66pt);
\path[-,draw=black,line width=0.4pt,line cap=butt,line join=miter,dash pattern=](-10.06pt,11.875pt)..controls(-9.22pt,11.4925pt)and(-8.2pt,11.2675pt)..(-7pt,11.2675pt);
\path[-,draw=black,line width=0.4pt,line cap=butt,line join=miter,dash pattern=](-13pt,16.2675pt)..controls(-13pt,14.5175pt)and(-12.02pt,12.7675pt)..(-10.06pt,11.875pt);
\end{tikzpicture}}-{\begin{tikzpicture}[baseline=(current bounding box.center)]\path[-,draw=lightgray,line width=0.4pt,line cap=butt,line join=miter,dash pattern=](-20pt,10pt)--(-13pt,10pt);
\path[-,draw=lightgray,line width=0.4pt,line cap=butt,line join=miter,dash pattern=](43.66pt,0pt)--(46.66pt,0pt);
\path[-,draw=black,line width=0.4pt,line cap=butt,line join=miter,dash pattern=](-7pt,15pt)--(-6pt,15pt);
\path[-,draw=black,line width=0.4pt,line cap=butt,line join=miter,dash pattern=](-7pt,5pt)--(-6pt,5pt);
\path[-,draw=black,line width=0.4pt,line cap=butt,line join=miter,dash pattern=](0pt,15pt)--(1pt,15pt);
\path[-,draw=black,line width=0.4pt,line cap=butt,line join=miter,dash pattern=](0pt,5pt)--(1pt,5pt);
\path[-,draw=lightgray,line width=0.4pt,line cap=butt,line join=miter,dash pattern=](0pt,0pt)--(1pt,0pt);
\path[-,draw=black,line width=0.4pt,line cap=butt,line join=miter,dash pattern=](14pt,12.5pt)--(17pt,12.5pt);
\path[-,draw=lightgray,line width=0.4pt,line cap=butt,line join=miter,dash pattern=](14pt,0pt)--(17pt,0pt);
\path[-,draw=black,line width=0.4pt,line cap=butt,line join=miter,dash pattern=](21pt,16.5pt)--(39.66pt,16.5pt)--(39.66pt,-9.2675pt)--(21pt,-9.2675pt)--cycle;
\path[-,draw=lightgray,line width=0.4pt,line cap=butt,line join=miter,dash pattern=](21pt,0pt)--(24pt,0pt);
\path[-,draw=black,line width=0.4pt,line cap=butt,line join=miter,dash pattern=](21pt,12.5pt)--(21pt,12.5pt);
\path[-,draw=lightgray,line width=0.4pt,line cap=butt,line join=miter,dash pattern=](36.66pt,0pt)--(39.66pt,0pt);
\path[-,draw=black,line width=0.4pt,line cap=butt,line join=miter,dash pattern=](17pt,12.5pt)--(21pt,12.5pt);
\path[-,draw=lightgray,line width=0.4pt,line cap=butt,line join=miter,dash pattern=](17pt,0pt)--(21pt,0pt);
\path[-,draw=lightgray,line width=0.4pt,line cap=butt,line join=miter,dash pattern=](39.66pt,0pt)--(43.66pt,0pt);
\path[-,line width=0.4pt,line cap=butt,line join=miter,dash pattern=](27pt,3.2675pt)--(33.66pt,3.2675pt)--(33.66pt,-3.2675pt)--(27pt,-3.2675pt)--cycle;
\node[anchor=north west,inner sep=0] at (27pt,3.2675pt){\savebox{\marxupbox}{{\(P\)}}\immediate\write\boxesfile{253}\immediate\write\boxesfile{\number\wd\marxupbox}\immediate\write\boxesfile{\number\ht\marxupbox}\immediate\write\boxesfile{\number\dp\marxupbox}\box\marxupbox};
\path[-,draw=black,line width=0.4pt,line cap=butt,line join=miter,dash pattern=](24pt,3.7675pt)..controls(24pt,5.1482pt)and(25.1193pt,6.2675pt)..(26.5pt,6.2675pt)--(34.16pt,6.2675pt)..controls(35.5407pt,6.2675pt)and(36.66pt,5.1482pt)..(36.66pt,3.7675pt)--(36.66pt,-3.7675pt)..controls(36.66pt,-5.1482pt)and(35.5407pt,-6.2675pt)..(34.16pt,-6.2675pt)--(26.5pt,-6.2675pt)..controls(25.1193pt,-6.2675pt)and(24pt,-5.1482pt)..(24pt,-3.7675pt)--cycle;
\path[-,draw=lightgray,line width=0.4pt,line cap=butt,line join=miter,dash pattern=](36.66pt,0pt)--(36.66pt,0pt);
\path[-,draw=lightgray,line width=0.4pt,line cap=butt,line join=miter,dash pattern=](24pt,0pt)--(24pt,0pt);
\path[-,draw=black,line width=0.4pt,line cap=butt,line join=miter,dash pattern=](7pt,15pt)--(8pt,15pt);
\path[-,draw=black,line width=0.4pt,line cap=butt,line join=miter,dash pattern=](7pt,10pt)--(8pt,10pt);
\path[-,draw=lightgray,line width=0.4pt,line cap=butt,line join=miter,dash pattern=](7pt,0pt)--(8pt,0pt);
\path[-,draw=black,fill= gray,line width=0.4pt,line cap=butt,line join=miter,dash pattern=](8pt,16pt)..controls(12pt,16pt)and(14pt,14.5pt)..(14pt,12.5pt)..controls(14pt,10.5pt)and(12pt,9pt)..(8pt,9pt)--cycle;
\path[-,draw=lightgray,line width=0.4pt,line cap=butt,line join=miter,dash pattern=](8pt,0pt)--(14pt,0pt);
\path[-,draw=black,line width=0.4pt,line cap=butt,line join=miter,dash pattern=](1pt,15pt)--(7pt,15pt);
\path[-,draw=black,line width=0.4pt,line cap=butt,line join=miter,dash pattern=](1pt,5pt)..controls(5pt,5pt)and(3pt,10pt)..(7pt,10pt);
\path[-,draw=lightgray,line width=0.4pt,line cap=butt,line join=miter,dash pattern=](1pt,0pt)--(7pt,0pt);
\path[-,draw=black,line width=0.4pt,line cap=butt,line join=miter,dash pattern=](-6pt,15pt)--(0pt,15pt);
\path[-,draw=black,line width=0.4pt,line cap=butt,line join=miter,dash pattern=](-6pt,5pt)--(0pt,5pt);
\path[-,line width=0.4pt,line cap=butt,line join=miter,dash pattern=on 0.4pt off 1pt](0pt,5pt)--(0pt,5pt)--(0pt,0pt)--(0pt,0pt)--cycle;
\path[-,draw=black,line width=0.4pt,line cap=butt,line join=miter,dash pattern=on 0.4pt off 1pt](-3pt,8pt)--(3pt,8pt)--(3pt,-3pt)--(-3pt,-3pt)--cycle;
\path[-,fill=lightgray,line width=0.4pt,line cap=butt,line join=miter,dash pattern=](1pt,0pt)..controls(1pt,0.5523pt)and(0.5523pt,1pt)..(0pt,1pt)..controls(-0.5523pt,1pt)and(-1pt,0.5523pt)..(-1pt,0pt)..controls(-1pt,-0.5523pt)and(-0.5523pt,-1pt)..(0pt,-1pt)..controls(0.5523pt,-1pt)and(1pt,-0.5523pt)..(1pt,0pt)--cycle;
\path[-,fill=black,line width=0.4pt,line cap=butt,line join=miter,dash pattern=](-12pt,10pt)..controls(-12pt,10.5523pt)and(-12.4477pt,11pt)..(-13pt,11pt)..controls(-13.5523pt,11pt)and(-14pt,10.5523pt)..(-14pt,10pt)..controls(-14pt,9.4477pt)and(-13.5523pt,9pt)..(-13pt,9pt)..controls(-12.4477pt,9pt)and(-12pt,9.4477pt)..(-12pt,10pt)--cycle;
\path[-,draw=black,line width=0.4pt,line cap=butt,line join=miter,dash pattern=](-10.06pt,14.3925pt)..controls(-9.22pt,14.775pt)and(-8.2pt,15pt)..(-7pt,15pt);
\path[-,draw=black,line width=0.4pt,line cap=butt,line join=miter,dash pattern=](-13pt,10pt)..controls(-13pt,11.75pt)and(-12.02pt,13.5pt)..(-10.06pt,14.3925pt);
\path[-,draw=black,line width=0.4pt,line cap=butt,line join=miter,dash pattern=](-10.06pt,5.6075pt)..controls(-9.22pt,5.225pt)and(-8.2pt,5pt)..(-7pt,5pt);
\path[-,draw=black,line width=0.4pt,line cap=butt,line join=miter,dash pattern=](-13pt,10pt)..controls(-13pt,8.25pt)and(-12.02pt,6.5pt)..(-10.06pt,5.6075pt);
\end{tikzpicture}}\) 
 
As an illustration, let us compute the divergence of the scalar field defined as growing
with the negated logarithm of the distance to the origin: \(P = -\log(ρ)\).

Its covariant derivative (or gradient) is given by the partial
derivatives in polar coordinates (because \(\mathsf{aff}\mskip 3.0mu\mathbf{1}\mskip 3.0mu\mathnormal{=}\mskip 3.0mu\mathrm{0}\), as
already mentioned). The components in the polar tangent basis are:
\begin{displaymath}∇P = ∂P - 0 + 0 = \left[\begin{array}{cc}∂P/∂ρ&∂P/∂θ\end{array}\right]=\left[\begin{array}{cc}-ρ^{-1}&0\end{array}\right]\end{displaymath} 
We have seen above that, by composing with the appropriate Jacobian,
the components of the gradient in the Cartesian tangent basis are
\(\left[\begin{array}{cc}-ρ^{-1}\cos{θ}&-ρ^{-1}\sin{θ}\end{array}\right]\).

The second derivative (\(∇∇P\)) is a second order tensor, and
computing it manually in the Cartesian basis is error-prone. In contrast,
because \(P\) has radial symmetry, in the polar tangent basis its
partial derivative \(∂P\) has a simple expression (the co-vector \(\left[\begin{array}{cc}-ρ^{-1}&0\end{array}\right]\)).
Thus only the ρ-ρ-component of the second partial derivative is non-zero:
\(∂²P/∂ρ² = ρ^{-2}\). To compute the covariant derivative, we also need
the affinity term, which is obtained by multiplying the gradient
\(∇P\) by \(Γ\), then multiply by the contravariant metric
(recall \cref{250}).  Because only the \(\mathbf e_ρ\) coefficient of
\(∇P\) is non-zero, is suffices to multiply this component (\(-ρ^{-1}\)) by
\(Γ^\rho\) to obtain the affinity term \(\left[\begin{array}{cc}0&0\\0&1\end{array}\right]\).
Finally we get \(∇∇P\) (in polar tangent coordinates) from
\cref{230}: the partial derivative minus the affinity we computed above
plus the second affinity which is zero (because the output space is
\(\mathbf{1}\)) which gives us \(\left[\begin{array}{cc}ρ^{-2}&0\\0&-1\end{array}\right]\).

The divergence is obtained by 2-way contraction with the contravariant
metric. One way to do this is to multiply the above by the
contravariant metric, to get \(\left[\begin{array}{cc}ρ^{-2}&0\\0&-ρ^{-2}\end{array}\right]\),
and take the trace, which is zero for \(ρ>0\). (Note that even if
it were nonzero there was no need to multiply by any Jacobian because
the divergence field is scalar.) The fact that the divergence of this
scalar field is zero means that the charge density is zero away from
the origin: in a two-dimensional space the potential of a point charge at the origin is
proportional to this scalar field \(P\).

Beyond its illustrative benefits, a takeaway from this example is that
all computations were free of trigonometry, and involved many zeros,
thanks to the choice of a global coordinate system which had a zero gradient
in one of the axes.  Even though each step in the computations was presented for illustrative purposes, the
\(\mathsf{MetricCategory}\mskip 3.0muM_{\mathsf{Polar}}\) instance meant that we could run all
the above computations as Haskell programs.

\subsection{Tensor calculus summary}\label{254} 
At this point we have presented all the classes whose morphisms
constitute the combinators of {\sc{}Roger}. Their relationships are
summarised in \cref{262}.

\begin{figure}[]\begin{center}{\begin{tikzpicture}\path[-,line width=0.4pt,line cap=butt,line join=miter,dash pattern=](27.7724pt,76.8463pt)--(66.1898pt,76.8463pt)--(66.1898pt,67.827pt)--(27.7724pt,67.827pt)--cycle;
\node[anchor=north west,inner sep=0] at (27.7724pt,76.8463pt){\savebox{\marxupbox}{{\(\mathsf{Category}\)}}\immediate\write\boxesfile{255}\immediate\write\boxesfile{\number\wd\marxupbox}\immediate\write\boxesfile{\number\ht\marxupbox}\immediate\write\boxesfile{\number\dp\marxupbox}\box\marxupbox};
\path[-,line width=0.4pt,line cap=butt,line join=miter,dash pattern=](23.7724pt,80.8463pt)--(70.1898pt,80.8463pt)--(70.1898pt,63.827pt)--(23.7724pt,63.827pt)--cycle;
\path[-,draw=black,line width=0.4pt,line cap=butt,line join=miter,dash pattern=](23.7724pt,77.8463pt)..controls(23.7724pt,79.5031pt)and(25.1155pt,80.8463pt)..(26.7724pt,80.8463pt)--(67.1898pt,80.8463pt)..controls(68.8467pt,80.8463pt)and(70.1898pt,79.5031pt)..(70.1898pt,77.8463pt)--(70.1898pt,66.827pt)..controls(70.1898pt,65.1701pt)and(68.8467pt,63.827pt)..(67.1898pt,63.827pt)--(26.7724pt,63.827pt)..controls(25.1155pt,63.827pt)and(23.7724pt,65.1701pt)..(23.7724pt,66.827pt)--cycle;
\path[-,line width=0.4pt,line cap=butt,line join=miter,dash pattern=](5.6496pt,49.827pt)--(88.3126pt,49.827pt)--(88.3126pt,40.8877pt)--(5.6496pt,40.8877pt)--cycle;
\node[anchor=north west,inner sep=0] at (5.6496pt,49.827pt){\savebox{\marxupbox}{{\(\mathsf{SymmetricMonoidal}\)}}\immediate\write\boxesfile{256}\immediate\write\boxesfile{\number\wd\marxupbox}\immediate\write\boxesfile{\number\ht\marxupbox}\immediate\write\boxesfile{\number\dp\marxupbox}\box\marxupbox};
\path[-,line width=0.4pt,line cap=butt,line join=miter,dash pattern=](1.6496pt,53.827pt)--(92.3126pt,53.827pt)--(92.3126pt,36.8877pt)--(1.6496pt,36.8877pt)--cycle;
\path[-,draw=black,line width=0.4pt,line cap=butt,line join=miter,dash pattern=](1.6496pt,50.827pt)..controls(1.6496pt,52.4838pt)and(2.9928pt,53.827pt)..(4.6496pt,53.827pt)--(89.3126pt,53.827pt)..controls(90.9694pt,53.827pt)and(92.3126pt,52.4838pt)..(92.3126pt,50.827pt)--(92.3126pt,39.8877pt)..controls(92.3126pt,38.2308pt)and(90.9694pt,36.8877pt)..(89.3126pt,36.8877pt)--(4.6496pt,36.8877pt)..controls(2.9928pt,36.8877pt)and(1.6496pt,38.2308pt)..(1.6496pt,39.8877pt)--cycle;
\path[-,line width=0.4pt,line cap=butt,line join=miter,dash pattern=](106.3126pt,-4pt)--(170.4575pt,-4pt)--(170.4575pt,-13.0193pt)--(106.3126pt,-13.0193pt)--cycle;
\node[anchor=north west,inner sep=0] at (106.3126pt,-4pt){\savebox{\marxupbox}{{\(\mathsf{MetricCategory}\)}}\immediate\write\boxesfile{257}\immediate\write\boxesfile{\number\wd\marxupbox}\immediate\write\boxesfile{\number\ht\marxupbox}\immediate\write\boxesfile{\number\dp\marxupbox}\box\marxupbox};
\path[-,line width=0.4pt,line cap=butt,line join=miter,dash pattern=](102.3126pt,0pt)--(174.4575pt,0pt)--(174.4575pt,-17.0193pt)--(102.3126pt,-17.0193pt)--cycle;
\path[-,draw=black,line width=0.4pt,line cap=butt,line join=miter,dash pattern=](102.3126pt,-3pt)..controls(102.3126pt,-1.3431pt)and(103.6557pt,0pt)..(105.3126pt,0pt)--(171.4575pt,0pt)..controls(173.1143pt,0pt)and(174.4575pt,-1.3431pt)..(174.4575pt,-3pt)--(174.4575pt,-14.0193pt)..controls(174.4575pt,-15.6761pt)and(173.1143pt,-17.0193pt)..(171.4575pt,-17.0193pt)--(105.3126pt,-17.0193pt)..controls(103.6557pt,-17.0193pt)and(102.3126pt,-15.6761pt)..(102.3126pt,-14.0193pt)--cycle;
\path[-,line width=0.4pt,line cap=butt,line join=miter,dash pattern=](13.5502pt,22.8877pt)--(80.412pt,22.8877pt)--(80.412pt,14pt)--(13.5502pt,14pt)--cycle;
\node[anchor=north west,inner sep=0] at (13.5502pt,22.8877pt){\savebox{\marxupbox}{{\(\mathsf{CompactClosed}\)}}\immediate\write\boxesfile{258}\immediate\write\boxesfile{\number\wd\marxupbox}\immediate\write\boxesfile{\number\ht\marxupbox}\immediate\write\boxesfile{\number\dp\marxupbox}\box\marxupbox};
\path[-,line width=0.4pt,line cap=butt,line join=miter,dash pattern=](9.5502pt,26.8877pt)--(84.412pt,26.8877pt)--(84.412pt,10pt)--(9.5502pt,10pt)--cycle;
\path[-,draw=black,line width=0.4pt,line cap=butt,line join=miter,dash pattern=](9.5502pt,23.8877pt)..controls(9.5502pt,25.5445pt)and(10.8934pt,26.8877pt)..(12.5502pt,26.8877pt)--(81.412pt,26.8877pt)..controls(83.0688pt,26.8877pt)and(84.412pt,25.5445pt)..(84.412pt,23.8877pt)--(84.412pt,13pt)..controls(84.412pt,11.3431pt)and(83.0688pt,10pt)..(81.412pt,10pt)--(12.5502pt,10pt)..controls(10.8934pt,10pt)and(9.5502pt,11.3431pt)..(9.5502pt,13pt)--cycle;
\path[-,line width=0.4pt,line cap=butt,line join=miter,dash pattern=](188.4575pt,20.9464pt)--(221.6579pt,20.9464pt)--(221.6579pt,14pt)--(188.4575pt,14pt)--cycle;
\node[anchor=north west,inner sep=0] at (188.4575pt,20.9464pt){\savebox{\marxupbox}{{\(\mathsf{Additive}\)}}\immediate\write\boxesfile{259}\immediate\write\boxesfile{\number\wd\marxupbox}\immediate\write\boxesfile{\number\ht\marxupbox}\immediate\write\boxesfile{\number\dp\marxupbox}\box\marxupbox};
\path[-,line width=0.4pt,line cap=butt,line join=miter,dash pattern=](184.4575pt,24.9464pt)--(225.6579pt,24.9464pt)--(225.6579pt,10pt)--(184.4575pt,10pt)--cycle;
\path[-,draw=black,line width=0.4pt,line cap=butt,line join=miter,dash pattern=](184.4575pt,21.9464pt)..controls(184.4575pt,23.6033pt)and(185.8006pt,24.9464pt)..(187.4575pt,24.9464pt)--(222.6579pt,24.9464pt)..controls(224.3148pt,24.9464pt)and(225.6579pt,23.6033pt)..(225.6579pt,21.9464pt)--(225.6579pt,13pt)..controls(225.6579pt,11.3431pt)and(224.3148pt,10pt)..(222.6579pt,10pt)--(187.4575pt,10pt)..controls(185.8006pt,10pt)and(184.4575pt,11.3431pt)..(184.4575pt,13pt)--cycle;
\path[-,line width=0.4pt,line cap=butt,line join=miter,dash pattern=](4pt,-31.0193pt)--(89.9622pt,-31.0193pt)--(89.9622pt,-40.0386pt)--(4pt,-40.0386pt)--cycle;
\node[anchor=north west,inner sep=0] at (4pt,-31.0193pt){\savebox{\marxupbox}{{\(\mathsf{ConnectionCategory}\)}}\immediate\write\boxesfile{260}\immediate\write\boxesfile{\number\wd\marxupbox}\immediate\write\boxesfile{\number\ht\marxupbox}\immediate\write\boxesfile{\number\dp\marxupbox}\box\marxupbox};
\path[-,line width=0.4pt,line cap=butt,line join=miter,dash pattern=](0pt,-27.0193pt)--(93.9622pt,-27.0193pt)--(93.9622pt,-44.0386pt)--(0pt,-44.0386pt)--cycle;
\path[-,draw=black,line width=0.4pt,line cap=butt,line join=miter,dash pattern=](0pt,-30.0193pt)..controls(0pt,-28.3624pt)and(1.3431pt,-27.0193pt)..(3pt,-27.0193pt)--(90.9622pt,-27.0193pt)..controls(92.619pt,-27.0193pt)and(93.9622pt,-28.3624pt)..(93.9622pt,-30.0193pt)--(93.9622pt,-41.0386pt)..controls(93.9622pt,-42.6954pt)and(92.619pt,-44.0386pt)..(90.9622pt,-44.0386pt)--(3pt,-44.0386pt)..controls(1.3431pt,-44.0386pt)and(0pt,-42.6954pt)..(0pt,-41.0386pt)--cycle;
\path[-,line width=0.4pt,line cap=butt,line join=miter,dash pattern=](162.8615pt,-31.0193pt)--(247.2539pt,-31.0193pt)--(247.2539pt,-40.0386pt)--(162.8615pt,-40.0386pt)--cycle;
\node[anchor=north west,inner sep=0] at (162.8615pt,-31.0193pt){\savebox{\marxupbox}{{\(\mathsf{CoordinateCategory}\)}}\immediate\write\boxesfile{261}\immediate\write\boxesfile{\number\wd\marxupbox}\immediate\write\boxesfile{\number\ht\marxupbox}\immediate\write\boxesfile{\number\dp\marxupbox}\box\marxupbox};
\path[-,line width=0.4pt,line cap=butt,line join=miter,dash pattern=](158.8615pt,-27.0193pt)--(251.2539pt,-27.0193pt)--(251.2539pt,-44.0386pt)--(158.8615pt,-44.0386pt)--cycle;
\path[-,draw=black,line width=0.4pt,line cap=butt,line join=miter,dash pattern=](158.8615pt,-30.0193pt)..controls(158.8615pt,-28.3624pt)and(160.2047pt,-27.0193pt)..(161.8615pt,-27.0193pt)--(248.2539pt,-27.0193pt)..controls(249.9107pt,-27.0193pt)and(251.2539pt,-28.3624pt)..(251.2539pt,-30.0193pt)--(251.2539pt,-41.0386pt)..controls(251.2539pt,-42.6954pt)and(249.9107pt,-44.0386pt)..(248.2539pt,-44.0386pt)--(161.8615pt,-44.0386pt)..controls(160.2047pt,-44.0386pt)and(158.8615pt,-42.6954pt)..(158.8615pt,-41.0386pt)--cycle;
\path[-to,draw=black,line width=0.4pt,line cap=butt,line join=miter,dash pattern=](75.7685pt,-27.0193pt)..controls(88.5359pt,-23.2452pt)and(96.8302pt,-20.7934pt)..(109.5976pt,-17.0193pt);
\path[-to,draw=black,line width=0.4pt,line cap=butt,line join=miter,dash pattern=](184.0593pt,-27.0193pt)..controls(174.7464pt,-23.2452pt)and(168.6963pt,-20.7934pt)..(159.3834pt,-17.0193pt);
\path[-to,draw=black,line width=0.4pt,line cap=butt,line join=miter,dash pattern=](109.5273pt,0pt)..controls(96.7489pt,3.7681pt)and(88.4463pt,6.2164pt)..(75.6156pt,10pt);
\path[-to,draw=black,line width=0.4pt,line cap=butt,line join=miter,dash pattern=on 0.4pt off 1pt](160.221pt,0pt)..controls(169.8034pt,3.7344pt)and(176.0029pt,6.1503pt)..(186.3859pt,10.1967pt);
\path[-to,draw=black,line width=0.4pt,line cap=butt,line join=miter,dash pattern=](46.9811pt,26.8877pt)..controls(46.9811pt,30.6676pt)and(46.9811pt,33.1136pt)..(46.9811pt,36.8873pt);
\path[-to,draw=black,line width=0.4pt,line cap=butt,line join=miter,dash pattern=](46.9811pt,53.827pt)..controls(46.9811pt,57.6067pt)and(46.9811pt,60.0564pt)..(46.9811pt,63.8266pt);
\end{tikzpicture}}\end{center}\caption{Inheritance relationships between tensor structures. In the praxis, every instance of a category with a metric \(\mathsf{Additive}\) is also additive.
However, diagrams don't support a good representation for addition, so the dotted line is implemented as a subclass relationship in our library.}\label{262}\end{figure} 
When it comes to {\sc{}Albert}, we have two separate sub-languages.  First,
we have a number of combinators which only manipulate indices, shown
in \cref{263}. Indices can be split, merged, raised and
lowered. Indices for unit vector-spaces are unimportant and can be
created or discarded at will. Second, we have a number of combinators
which nominally manipulate scalar-valued expressions (addition,
multiplication, embedding of constant tensors, Kronecker delta, and
contraction, etc; see \cref{264}). Various combinators require
various amount of structure in the underlying category \(z\).

The semantics in terms of morphisms is provided by the
\(\mathsf{tensorEval}\) function, and \(\mathsf{tensorEval}_{1}\) for the special
case of closed tensor expressions.
For tensor fields, primitives for metrics
and derivatives are also available.
\begin{figure}[] \begin{list}{}{\setlength\leftmargin{1.0em}}\item\relax
\ensuremath{\begin{parray}\column{B}{@{}>{}l<{}@{}}\column[0em]{1}{@{}>{}l<{}@{}}\column{2}{@{}>{}l<{}@{}}\column{E}{@{}>{}l<{}@{}}%
\>[1]{\mathbf{type}\mskip 3.0mu\mathsf{P}\mskip 3.0mu}\>[2]{\mathnormal{::}\mskip 3.0mu\allowbreak{}\mathnormal{(}\mskip 0.0mu\mathsf{Type}\mskip 3.0mu\mathnormal{\rightarrow }\mskip 3.0mu\mathsf{Type}\mskip 3.0mu\mathnormal{\rightarrow }\mskip 3.0mu\mathsf{Type}\mskip 0.0mu\mathnormal{)}\allowbreak{}\mskip 3.0mu\mathnormal{\rightarrow }\mskip 3.0mu\mathsf{Type}\mskip 3.0mu\mathnormal{\rightarrow }\mskip 3.0mu\mathsf{Type}\mskip 3.0mu\mathnormal{\rightarrow }\mskip 3.0mu\mathsf{Type}}\<[E]\end{parray}}\end{list} \begin{list}{}{\setlength\leftmargin{1.0em}}\item\relax
\ensuremath{\begin{parray}\column{B}{@{}>{}l<{}@{}}\column[0em]{1}{@{}>{}l<{}@{}}\column{2}{@{}>{}l<{}@{}}\column{3}{@{}>{}l<{}@{}}\column{4}{@{}>{}l<{}@{}}\column{5}{@{}>{}l<{}@{}}\column{E}{@{}>{}l<{}@{}}%
\>[1]{\mathsf{unit}\mskip 3.0mu}\>[2]{\mathnormal{::}\mskip 3.0mu}\>[3]{\mathsf{SymmetricMonoidal}\mskip 3.0muz\mskip 3.0mu}\>[4]{\mathnormal{\Rightarrow }\mskip 3.0mu\mathsf{P}\mskip 3.0muz\mskip 3.0mu\mathsf{r}\mskip 3.0mu\mathsf{a}\mskip 3.0mu}\>[5]{\mathnormal{⊸}\mskip 3.0mu\allowbreak{}\mathnormal{(}\mskip 0.0mu\mathsf{P}\mskip 3.0muz\mskip 3.0mu\mathsf{r}\mskip 3.0mu\mathsf{a}\mskip 0.0mu\mathnormal{,}\mskip 3.0mu\mathsf{P}\mskip 3.0muz\mskip 3.0mu\mathsf{r}\mskip 3.0mu\mathbf{1}\mskip 0.0mu\mathnormal{)}\allowbreak{}}\<[E]\\
\>[1]{\mathsf{unit'}\mskip 3.0mu}\>[2]{\mathnormal{::}\mskip 3.0mu}\>[3]{\mathsf{SymmetricMonoidal}\mskip 3.0muz\mskip 3.0mu}\>[4]{\mathnormal{\Rightarrow }\mskip 3.0mu\mathsf{P}\mskip 3.0muz\mskip 3.0mu\mathsf{r}\mskip 3.0mu\mathbf{1}\mskip 3.0mu\mathnormal{⊸}\mskip 3.0mu\mathsf{P}\mskip 3.0muz\mskip 3.0mu\mathsf{r}\mskip 3.0mu\mathsf{a}\mskip 3.0mu\mathnormal{⊸}\mskip 3.0mu\mathsf{P}\mskip 3.0muz\mskip 3.0mu\mathsf{r}\mskip 3.0mu\mathsf{a}}\<[E]\\
\>[1]{\mathsf{split}\mskip 3.0mu}\>[2]{\mathnormal{::}\mskip 3.0mu}\>[3]{\mathsf{SymmetricMonoidal}\mskip 3.0muz\mskip 3.0mu}\>[4]{\mathnormal{\Rightarrow }\mskip 3.0mu\mathsf{P}\mskip 3.0muz\mskip 3.0mu\mathsf{r}\mskip 3.0mu\allowbreak{}\mathnormal{(}\mskip 0.0mu\mathsf{a}\mskip 3.0mu\mathnormal{⊗}\mskip 3.0mu\mathsf{b}\mskip 0.0mu\mathnormal{)}\allowbreak{}\mskip 3.0mu\mathnormal{⊸}\mskip 3.0mu\allowbreak{}\mathnormal{(}\mskip 0.0mu\mathsf{P}\mskip 3.0muz\mskip 3.0mu\mathsf{r}\mskip 3.0mu\mathsf{a}\mskip 0.0mu\mathnormal{,}\mskip 3.0mu\mathsf{P}\mskip 3.0muz\mskip 3.0mu\mathsf{r}\mskip 3.0mu\mathsf{b}\mskip 0.0mu\mathnormal{)}\allowbreak{}}\<[E]\\
\>[1]{\mathsf{merge}\mskip 3.0mu}\>[2]{\mathnormal{::}\mskip 3.0mu}\>[3]{\mathsf{SymmetricMonoidal}\mskip 3.0muz\mskip 3.0mu}\>[4]{\mathnormal{\Rightarrow }\mskip 3.0mu\allowbreak{}\mathnormal{(}\mskip 0.0mu\mathsf{P}\mskip 3.0muz\mskip 3.0mu\mathsf{r}\mskip 3.0mu\mathsf{a}\mskip 0.0mu\mathnormal{,}\mskip 3.0mu\mathsf{P}\mskip 3.0muz\mskip 3.0mu\mathsf{r}\mskip 3.0mu\mathsf{b}\mskip 0.0mu\mathnormal{)}\allowbreak{}\mskip 3.0mu\mathnormal{⊸}\mskip 3.0mu\mathsf{P}\mskip 3.0muz\mskip 3.0mu\mathsf{r}\mskip 3.0mu\allowbreak{}\mathnormal{(}\mskip 0.0mu\mathsf{a}\mskip 3.0mu\mathnormal{⊗}\mskip 3.0mu\mathsf{b}\mskip 0.0mu\mathnormal{)}\allowbreak{}}\<[E]\end{parray}}\end{list} \begin{list}{}{\setlength\leftmargin{1.0em}}\item\relax
\ensuremath{\begin{parray}\column{B}{@{}>{}l<{}@{}}\column[0em]{1}{@{}>{}l<{}@{}}\column{2}{@{}>{}l<{}@{}}\column{3}{@{}>{}l<{}@{}}\column{E}{@{}>{}l<{}@{}}%
\>[1]{\mathsf{raise}\mskip 3.0mu}\>[2]{\mathnormal{::}\mskip 3.0mu\mathsf{MetricCategory}\mskip 3.0muz\mskip 3.0mu\mathnormal{\Rightarrow }\mskip 3.0mu}\>[3]{\mathsf{P}\mskip 3.0muz\mskip 3.0mu\mathsf{r}\mskip 3.0muT_{z}\mskip 3.0mu\mathnormal{⊸}\mskip 3.0mu\mathsf{P}\mskip 3.0muz\mskip 3.0mu\mathsf{r}\mskip 3.0mu\dual{T_{z}}}\<[E]\\
\>[1]{\mathsf{lower}\mskip 3.0mu}\>[2]{\mathnormal{::}\mskip 3.0mu\mathsf{MetricCategory}\mskip 3.0muz\mskip 3.0mu\mathnormal{\Rightarrow }\mskip 3.0mu}\>[3]{\mathsf{P}\mskip 3.0muz\mskip 3.0mu\mathsf{r}\mskip 3.0mu\dual{T_{z}}\mskip 3.0mu\mathnormal{⊸}\mskip 3.0mu\mathsf{P}\mskip 3.0muz\mskip 3.0mu\mathsf{r}\mskip 3.0muT_{z}}\<[E]\end{parray}}\end{list} \caption{Syntax of the index language of {\sc{}Albert}, as types of combinators. }\label{263}\end{figure} 
\begin{figure}[] \begin{list}{}{\setlength\leftmargin{1.0em}}\item\relax
\ensuremath{\begin{parray}\column{B}{@{}>{}l<{}@{}}\column[0em]{1}{@{}>{}l<{}@{}}\column{E}{@{}>{}l<{}@{}}%
\>[1]{\mathbf{type}\mskip 3.0mu\mathsf{R}\mskip 3.0mu\mathnormal{::}\mskip 3.0mu\allowbreak{}\mathnormal{(}\mskip 0.0mu\mathsf{Type}\mskip 3.0mu\mathnormal{\rightarrow }\mskip 3.0mu\mathsf{Type}\mskip 3.0mu\mathnormal{\rightarrow }\mskip 3.0mu\mathsf{Type}\mskip 0.0mu\mathnormal{)}\allowbreak{}\mskip 3.0mu\mathnormal{\rightarrow }\mskip 3.0mu\mathsf{Type}\mskip 3.0mu\mathnormal{\rightarrow }\mskip 3.0mu\mathsf{Type}}\<[E]\end{parray}}\end{list} \begin{list}{}{\setlength\leftmargin{1.0em}}\item\relax
\ensuremath{\begin{parray}\column{B}{@{}>{}l<{}@{}}\column[0em]{1}{@{}>{}l<{}@{}}\column{2}{@{}>{}l<{}@{}}\column{3}{@{}>{}l<{}@{}}\column{4}{@{}>{}l<{}@{}}\column{5}{@{}>{}l<{}@{}}\column{6}{@{}>{}l<{}@{}}\column{7}{@{}>{}l<{}@{}}\column{8}{@{}>{}l<{}@{}}\column{9}{@{}>{}l<{}@{}}\column{10}{@{}>{}l<{}@{}}\column{11}{@{}>{}l<{}@{}}\column{E}{@{}>{}l<{}@{}}%
\>[1]{\mathsf{plus}\mskip 3.0mu}\>[2]{\mathnormal{::}\mskip 3.0mu\allowbreak{}\mathnormal{(}\mskip 0.0mu\mathsf{Bool}\mskip 3.0mu\mathnormal{\rightarrow }\mskip 3.0mu\mathsf{R}\mskip 3.0muz\mskip 3.0mu\mathsf{r}\mskip 0.0mu\mathnormal{)}\allowbreak{}\mskip 3.0mu\mathnormal{⊸}\mskip 3.0mu\mathsf{R}\mskip 3.0muz\mskip 3.0mu\mathsf{r}}\<[E]\\
\>[1]{\allowbreak{}\mathnormal{(}\mskip 0.0mu{\tikzstar{0.11}{0.25}{5}{-18}{fill=black}}\mskip 0.0mu\mathnormal{)}\allowbreak{}\mskip 3.0mu}\>[2]{\mathnormal{::}\mskip 3.0mu\mathsf{SymmetricMonoidal}\mskip 3.0muz\mskip 3.0mu\mathnormal{\Rightarrow }\mskip 3.0mu}\>[6]{\mathsf{R}\mskip 3.0muz\mskip 3.0mu\mathsf{r}\mskip 3.0mu\mathnormal{⊸}\mskip 3.0mu\mathsf{R}\mskip 3.0muz\mskip 3.0mu\mathsf{r}\mskip 3.0mu\mathnormal{⊸}\mskip 3.0mu\mathsf{R}\mskip 3.0muz\mskip 3.0mu\mathsf{r}}\<[E]\\
\>[1]{\mathsf{zeroTensor}\mskip 3.0mu}\>[2]{\mathnormal{::}\mskip 3.0mu\allowbreak{}\mathnormal{(}\mskip 0.0mu\mathsf{CompactClosed}\mskip 3.0muz\mskip 0.0mu\mathnormal{,}\mskip 3.0mu\mathsf{Additive}\mskip 3.0mu}\>[5]{z\mskip 0.0mu}\>[9]{\mathnormal{)}\allowbreak{}\mskip 3.0mu\mathnormal{\Rightarrow }\mskip 3.0mu\mathsf{P}\mskip 3.0muz\mskip 3.0mu\mathsf{r}\mskip 3.0mu\mathsf{a}\mskip 3.0mu\mathnormal{⊸}\mskip 3.0mu\mathsf{R}\mskip 3.0muz\mskip 3.0mu\mathsf{r}}\<[E]\\
\>[1]{\mathsf{constant}\mskip 3.0mu}\>[2]{\mathnormal{::}\mskip 3.0mu\allowbreak{}\mathnormal{(}\mskip 0.0mu\mathsf{SymmetricMonoidal}\mskip 3.0muz\mskip 0.0mu\mathnormal{,}\mskip 3.0mu}\>[3]{\mathsf{VectorSpace}\mskip 3.0mu\allowbreak{}\mathnormal{(}\mskip 0.0mu}\>[4]{\mathbf{1}\mskip 1.0mu\overset{z}{\leadsto }\mskip 1.0mu\mathbf{1}\mskip 0.0mu\mathnormal{)}\allowbreak{}\mskip 0.0mu}\>[10]{\mathnormal{)}\allowbreak{}\mskip 3.0mu\mathnormal{\Rightarrow }\mskip 3.0mu\mathsf{S}\mskip 3.0mu\mathnormal{\rightarrow }\mskip 3.0mu}\>[11]{\mathsf{R}\mskip 3.0muz\mskip 3.0mu\mathsf{r}}\<[E]\\
\>[1]{\mathsf{delta}\mskip 3.0mu}\>[2]{\mathnormal{::}\mskip 3.0mu\mathsf{CompactClosed}\mskip 3.0muz\mskip 3.0mu\mathnormal{\Rightarrow }\mskip 3.0mu}\>[7]{\mathsf{P}\mskip 3.0muz\mskip 3.0mu\mathsf{r}\mskip 3.0mu\mathsf{a}\mskip 3.0mu\mathnormal{⊸}\mskip 3.0mu\mathsf{P}\mskip 3.0muz\mskip 3.0mu\mathsf{r}\mskip 3.0mu\dual{\mathsf{a}}\mskip 3.0mu\mathnormal{⊸}\mskip 3.0mu\mathsf{R}\mskip 3.0muz\mskip 3.0mu\mathsf{r}}\<[E]\\
\>[1]{\mathsf{contract}\mskip 3.0mu}\>[2]{\mathnormal{::}\mskip 3.0mu\mathsf{CompactClosed}\mskip 3.0muz\mskip 3.0mu\mathnormal{\Rightarrow }\mskip 3.0mu}\>[8]{\allowbreak{}\mathnormal{(}\mskip 0.0mu\mathsf{P}\mskip 3.0muz\mskip 3.0mu\mathsf{r}\mskip 3.0mu\dual{\mathsf{a}}\mskip 3.0mu\mathnormal{⊸}\mskip 3.0mu\mathsf{P}\mskip 3.0muz\mskip 3.0mu\mathsf{r}\mskip 3.0mu\mathsf{a}\mskip 3.0mu\mathnormal{⊸}\mskip 3.0mu\mathsf{R}\mskip 3.0muz\mskip 3.0mu\mathsf{r}\mskip 0.0mu\mathnormal{)}\allowbreak{}\mskip 3.0mu\mathnormal{⊸}\mskip 3.0mu\mathsf{R}\mskip 3.0muz\mskip 3.0mu\mathsf{r}}\<[E]\end{parray}}\end{list} \begin{list}{}{\setlength\leftmargin{1.0em}}\item\relax
\ensuremath{\begin{parray}\column{B}{@{}>{}l<{}@{}}\column[0em]{1}{@{}>{}l<{}@{}}\column{2}{@{}>{}l<{}@{}}\column{3}{@{}>{}l<{}@{}}\column{4}{@{}>{}l<{}@{}}\column{5}{@{}>{}l<{}@{}}\column{6}{@{}>{}l<{}@{}}\column{7}{@{}>{}l<{}@{}}\column{8}{@{}>{}l<{}@{}}\column{E}{@{}>{}l<{}@{}}%
\>[1]{\mathsf{tensorEmbed}\mskip 3.0mu}\>[2]{\mathnormal{::}\mskip 3.0mu}\>[4]{\mathsf{CompactClosed}\mskip 3.0muz\mskip 3.0mu}\>[5]{\mathnormal{\Rightarrow }\mskip 3.0mu\allowbreak{}\mathnormal{(}\mskip 0.0mu\mathsf{a}\mskip 1.0mu\overset{z}{\leadsto }\mskip 1.0mu\mathsf{b}\mskip 0.0mu\mathnormal{)}\allowbreak{}\mskip 3.0mu}\>[6]{\mathnormal{\rightarrow }\mskip 3.0mu\allowbreak{}\mathnormal{(}\mskip 0.0mu∀\mskip 3.0mu\mathsf{r}\mskip 1.0mu.\mskip 3.0mu}\>[7]{\mathsf{P}\mskip 3.0muz\mskip 3.0mu\mathsf{r}\mskip 3.0mu\mathsf{a}\mskip 3.0mu\mathnormal{⊸}\mskip 3.0mu\mathsf{P}\mskip 3.0muz\mskip 3.0mu\mathsf{r}\mskip 3.0mu\dual{\mathsf{b}}\mskip 3.0mu}\>[8]{\mathnormal{⊸}\mskip 3.0mu\mathsf{R}\mskip 3.0muz\mskip 3.0mu\mathsf{r}\mskip 0.0mu\mathnormal{)}\allowbreak{}}\<[E]\\
\>[1]{\mathsf{tensorEmbed}_{1}\mskip 3.0mu}\>[2]{\mathnormal{::}\mskip 3.0mu}\>[3]{\mathsf{SymmetricMonoidal}\mskip 3.0muz\mskip 3.0mu}\>[5]{\mathnormal{\Rightarrow }\mskip 3.0mu\allowbreak{}\mathnormal{(}\mskip 0.0mu\mathsf{a}\mskip 1.0mu\overset{z}{\leadsto }\mskip 1.0mu\mathbf{1}\mskip 0.0mu\mathnormal{)}\allowbreak{}\mskip 3.0mu}\>[6]{\mathnormal{\rightarrow }\mskip 3.0mu\allowbreak{}\mathnormal{(}\mskip 0.0mu∀\mskip 3.0mu\mathsf{r}\mskip 1.0mu.\mskip 3.0mu}\>[7]{\mathsf{P}\mskip 3.0muz\mskip 3.0mu\mathsf{r}\mskip 3.0mu\mathsf{a}\mskip 3.0mu}\>[8]{\mathnormal{⊸}\mskip 3.0mu\mathsf{R}\mskip 3.0muz\mskip 3.0mu\mathsf{r}\mskip 0.0mu\mathnormal{)}\allowbreak{}}\<[E]\end{parray}}\end{list} \begin{list}{}{\setlength\leftmargin{1.0em}}\item\relax
\ensuremath{\begin{parray}\column{B}{@{}>{}l<{}@{}}\column[0em]{1}{@{}>{}l<{}@{}}\column{2}{@{}>{}l<{}@{}}\column{3}{@{}>{}l<{}@{}}\column{4}{@{}>{}l<{}@{}}\column{5}{@{}>{}l<{}@{}}\column{6}{@{}>{}l<{}@{}}\column{7}{@{}>{}l<{}@{}}\column{8}{@{}>{}l<{}@{}}\column{9}{@{}>{}l<{}@{}}\column{10}{@{}>{}l<{}@{}}\column{E}{@{}>{}l<{}@{}}%
\>[1]{\mathsf{tensorEval}\mskip 3.0mu}\>[2]{\mathnormal{::}\mskip 3.0mu}\>[3]{\allowbreak{}\mathnormal{(}\mskip 0.0mu}\>[7]{\mathsf{CompactClosed}\mskip 3.0muz\mskip 0.0mu\mathnormal{,}\mskip 3.0mu\mathsf{Additive}\mskip 3.0mu}\>[9]{z\mskip 0.0mu\mathnormal{)}\allowbreak{}\mskip 3.0mu}\>[10]{\mathnormal{\Rightarrow }}\<[E]\\
\>[3]{\allowbreak{}\mathnormal{(}\mskip 0.0mu∀\mskip 3.0mu\mathsf{r}\mskip 1.0mu.\mskip 3.0mu}\>[4]{\mathsf{P}\mskip 3.0muz\mskip 3.0mu\mathsf{r}\mskip 3.0mu\mathsf{a}\mskip 3.0mu\mathnormal{⊸}\mskip 3.0mu\mathsf{P}\mskip 3.0muz\mskip 3.0mu\mathsf{r}\mskip 3.0mu\dual{\mathsf{b}}\mskip 3.0mu}\>[5]{\mathnormal{⊸}\mskip 3.0mu\mathsf{R}\mskip 3.0muz\mskip 3.0mu\mathsf{r}\mskip 0.0mu\mathnormal{)}\allowbreak{}\mskip 3.0mu\mathnormal{\rightarrow }\mskip 3.0mu\allowbreak{}\mathnormal{(}\mskip 0.0mu\mathsf{a}\mskip 1.0mu\overset{z}{\leadsto }\mskip 1.0mu\mathsf{b}\mskip 0.0mu\mathnormal{)}\allowbreak{}}\<[E]\\
\>[1]{\mathsf{tensorEval}_{1}\mskip 3.0mu}\>[2]{\mathnormal{::}\mskip 3.0mu}\>[3]{\allowbreak{}\mathnormal{(}\mskip 0.0mu}\>[6]{\mathsf{SymmetricMonoidal}\mskip 3.0muz\mskip 0.0mu\mathnormal{,}\mskip 3.0mu\mathsf{Additive}\mskip 3.0mu}\>[8]{z\mskip 0.0mu\mathnormal{)}\allowbreak{}\mskip 3.0mu}\>[10]{\mathnormal{\Rightarrow }}\<[E]\\
\>[3]{\allowbreak{}\mathnormal{(}\mskip 0.0mu∀\mskip 3.0mu\mathsf{r}\mskip 1.0mu.\mskip 3.0mu}\>[4]{\mathsf{P}\mskip 3.0muz\mskip 3.0mu\mathsf{r}\mskip 3.0mu\mathsf{a}\mskip 3.0mu}\>[5]{\mathnormal{⊸}\mskip 3.0mu\mathsf{R}\mskip 3.0muz\mskip 3.0mu\mathsf{r}\mskip 0.0mu\mathnormal{)}\allowbreak{}\mskip 3.0mu\mathnormal{\rightarrow }\mskip 3.0mu\allowbreak{}\mathnormal{(}\mskip 0.0mu\mathsf{a}\mskip 1.0mu\overset{z}{\leadsto }\mskip 1.0mu\mathbf{1}\mskip 0.0mu\mathnormal{)}\allowbreak{}}\<[E]\end{parray}}\end{list} \begin{list}{}{\setlength\leftmargin{1.0em}}\item\relax
 \ensuremath{\begin{parray}\column{B}{@{}>{}l<{}@{}}\column[0em]{1}{@{}>{}l<{}@{}}\column{2}{@{}>{}l<{}@{}}\column{3}{@{}>{}l<{}@{}}\column{E}{@{}>{}l<{}@{}}%
\>[1]{\mathsf{deriv}\mskip 3.0mu}\>[2]{\mathnormal{::}\mskip 3.0mu}\>[3]{\mathsf{ConnectionCategory}\mskip 3.0muz\mskip 3.0mu\mathnormal{\Rightarrow }\mskip 3.0mu\mathsf{P}\mskip 3.0muz\mskip 3.0mu\mathsf{r}\mskip 3.0mu\mathsf{v}\mskip 3.0mu\mathnormal{⊸}\mskip 3.0mu\mathsf{R}\mskip 3.0muz\mskip 3.0mu\mathsf{r}\mskip 3.0mu\mathnormal{⊸}\mskip 3.0mu\mathsf{R}\mskip 3.0muz\mskip 3.0mu\mathsf{r}}\<[E]\end{parray}} \ensuremath{\begin{parray}\column{B}{@{}>{}l<{}@{}}\column[0em]{1}{@{}>{}l<{}@{}}\column{2}{@{}>{}l<{}@{}}\column{3}{@{}>{}l<{}@{}}\column{E}{@{}>{}l<{}@{}}%
\>[1]{\mathsf{partial}\mskip 3.0mu}\>[2]{\mathnormal{::}\mskip 3.0mu}\>[3]{\mathsf{CoordinateCategory}\mskip 3.0muz\mskip 3.0mu\mathnormal{\Rightarrow }\mskip 3.0mu\mathsf{P}\mskip 3.0muz\mskip 3.0mu\mathsf{r}\mskip 3.0mu\mathsf{v}\mskip 3.0mu\mathnormal{⊸}\mskip 3.0mu\mathsf{R}\mskip 3.0muz\mskip 3.0mu\mathsf{r}\mskip 3.0mu\mathnormal{⊸}\mskip 3.0mu\mathsf{R}\mskip 3.0muz\mskip 3.0mu\mathsf{r}}\<[E]\end{parray}} \ensuremath{\begin{parray}\column{B}{@{}>{}l<{}@{}}\column[0em]{1}{@{}>{}l<{}@{}}\column{2}{@{}>{}l<{}@{}}\column{E}{@{}>{}l<{}@{}}%
\>[1]{\mathsf{metric}\mskip 3.0mu\mathnormal{::}\mskip 3.0mu}\>[2]{\mathsf{MetricCategory}\mskip 3.0muz\mskip 3.0mu\mathnormal{\Rightarrow }\mskip 3.0mu\mathsf{P}\mskip 3.0muz\mskip 3.0mu\mathsf{r}\mskip 3.0muT_{z}\mskip 3.0mu\mathnormal{⊸}\mskip 3.0mu\mathsf{P}\mskip 3.0muz\mskip 3.0mu\mathsf{r}\mskip 3.0muT_{z}\mskip 3.0mu\mathnormal{⊸}\mskip 3.0mu\mathsf{R}\mskip 3.0muz\mskip 3.0mu\mathsf{r}}\<[E]\end{parray}} 

\ensuremath{\begin{parray}\column{B}{@{}>{}l<{}@{}}\column[0em]{1}{@{}>{}l<{}@{}}\column{2}{@{}>{}l<{}@{}}\column{3}{@{}>{}l<{}@{}}\column{E}{@{}>{}l<{}@{}}%
\>[1]{\mathsf{christoffel}\mskip 3.0mu}\>[2]{\mathnormal{::}\mskip 3.0mu}\>[3]{\mathsf{CoordinateCategory}\mskip 3.0muz\mskip 3.0mu\mathnormal{\Rightarrow }\mskip 3.0mu\mathsf{P}\mskip 3.0muz\mskip 3.0mu\mathsf{r}\mskip 3.0muT_{z}\mskip 3.0mu\mathnormal{⊸}\mskip 3.0mu\mathsf{P}\mskip 3.0muz\mskip 3.0mu\mathsf{r}\mskip 3.0muT_{z}\mskip 3.0mu\mathnormal{⊸}\mskip 3.0mu\mathsf{P}\mskip 3.0muz\mskip 3.0mu\mathsf{r}\mskip 3.0mu\dual{T_{z}}\mskip 3.0mu\mathnormal{⊸}\mskip 3.0mu\mathsf{R}\mskip 3.0muz\mskip 3.0mu\mathsf{r}}\<[E]\end{parray}} \end{list} \caption{Syntax of the expression sub-language of {\sc{}Albert}, as types of combinators. We repeat christoffel symbol and metric here even though they can be defined by the user as the embedding of the corresponding {\sc{}Roger} primitives.}\label{264}\end{figure}

\section{Application: Curvature and General Relativity}\label{265} 
To further demonstrate the applicability of {\sc{}Albert}, in this section
we present some concepts of general relativity, with particular focus
on the notion of curvature.
General relativity can be summarised as ``matter curves space-time''. This
informal statement can be expressed as a tensor equation as follows:
 
\begin{equation}R{^k}{_k}{_i}{_j} + \frac{1}{2}g{_i}{_j}g'{^l}{^m}R{^n}{_n}{_l}{_m} = κT{_i}{_j}\label{266}\end{equation} 
In the above, \(T{_i}{_j}\) represents the contents of
space-time in terms of energy, momentum, pressure, etc. depending on
the components of the tensor.
The gravitational constant is \(κ\).\footnote{We omit the cosmological term from the equation because it does not bring any more insight for our purposes.\label{267}} The tensor \(R{^l}{_i}{_j}{_k}\) captures the curvature properties of
space-time, and its value solely depends on the metric, as we will see below.
Thus, solving the equation for a given \(T{_i}{_j}\) amounts to finding a suitable metric.  Given such a
solution, we can compute the expression for the left-hand-side of the
equation and verify that it is equal to the right-hand-side.  We do so
for the example of a point mass in \cref{392}. Before that, we
discuss in more detail \(R{^l}{_i}{_j}{_k}\), the Riemann
curvature tensor.

\begin{definition}[Riemann curvature] The Riemann curvature is a 4-tensor, given by the following identity:
\(R{^l}{_i}{_j}{_k}\) = \(∂{_i}Γ{_j}{_k}{^l} - ∂{_j}Γ{_i}{_k}{^l} + Γ{_i}{_m}{^l}Γ{_j}{_k}{^m} - Γ{_j}{_n}{^l}Γ{_i}{_k}{^n}\).\label{268}\end{definition} Each pair of terms is the antisymmetric part of a 4-tensor. Taking advantage of this property,
we can make the diagram notation a sum of two terms: \({\begin{tikzpicture}[baseline=(current bounding box.center)]\path[-,draw=black,line width=0.4pt,line cap=butt,line join=miter,dash pattern=](-17pt,17.4184pt)--(-16pt,17.4184pt);
\path[-,draw=black,line width=0.4pt,line cap=butt,line join=miter,dash pattern=](-17pt,10pt)--(-16pt,10pt);
\path[-,draw=black,line width=0.4pt,line cap=butt,line join=miter,dash pattern=](-17pt,0pt)--(-16pt,0pt);
\path[-,draw=black,line width=0.4pt,line cap=butt,line join=miter,dash pattern=](16pt,2.5pt)--(19pt,2.5pt);
\path[-,draw=black,line width=0.4pt,line cap=butt,line join=miter,dash pattern=](-10pt,17.4184pt)--(-10pt,17.4184pt);
\path[-,draw=black,line width=0.4pt,line cap=butt,line join=miter,dash pattern=](-10pt,10pt)--(-10pt,10pt);
\path[-,draw=black,line width=0.4pt,line cap=butt,line join=miter,dash pattern=](-10pt,0pt)--(-10pt,0pt);
\path[-,draw=black,line width=0.4pt,line cap=butt,line join=miter,dash pattern=](-4pt,17.4184pt)--(-4pt,17.4184pt);
\path[-,draw=black,line width=0.4pt,line cap=butt,line join=miter,dash pattern=](-4pt,5pt)--(-4pt,5pt);
\path[-,draw=black,line width=0.4pt,line cap=butt,line join=miter,dash pattern=](-4pt,0pt)--(-4pt,0pt);
\path[-,draw=black,line width=0.4pt,line cap=butt,line join=miter,dash pattern=](0pt,21.4184pt)--(12pt,21.4184pt)--(12pt,-4pt)--(0pt,-4pt)--cycle;
\path[-,draw=black,line width=0.4pt,line cap=butt,line join=miter,dash pattern=](0pt,5pt)--(3pt,5pt);
\path[-,draw=black,line width=0.4pt,line cap=butt,line join=miter,dash pattern=](0pt,0pt)--(3pt,0pt);
\path[-,draw=black,line width=0.4pt,line cap=butt,line join=miter,dash pattern=](0pt,17.4184pt)--(0pt,17.4184pt);
\path[-,draw=black,line width=0.4pt,line cap=butt,line join=miter,dash pattern=](9pt,2.5pt)--(12pt,2.5pt);
\path[-,draw=black,line width=0.4pt,line cap=butt,line join=miter,dash pattern=](-4pt,17.4184pt)--(0pt,17.4184pt);
\path[-,draw=black,line width=0.4pt,line cap=butt,line join=miter,dash pattern=](-4pt,5pt)--(0pt,5pt);
\path[-,draw=black,line width=0.4pt,line cap=butt,line join=miter,dash pattern=](-4pt,0pt)--(0pt,0pt);
\path[-,draw=black,line width=0.4pt,line cap=butt,line join=miter,dash pattern=](12pt,2.5pt)--(16pt,2.5pt);
\path[-,draw=black,fill= gray,line width=0.4pt,line cap=butt,line join=miter,dash pattern=](3pt,6pt)..controls(7pt,6pt)and(9pt,4.5pt)..(9pt,2.5pt)..controls(9pt,0.5pt)and(7pt,-1pt)..(3pt,-1pt)--cycle;
\path[-,draw=black,line width=0.4pt,line cap=butt,line join=miter,dash pattern=](-10pt,17.4184pt)--(-4pt,17.4184pt);
\path[-,draw=black,line width=0.4pt,line cap=butt,line join=miter,dash pattern=](-10pt,10pt)..controls(-6pt,10pt)and(-8pt,5pt)..(-4pt,5pt);
\path[-,draw=black,line width=0.4pt,line cap=butt,line join=miter,dash pattern=](-10pt,0pt)--(-4pt,0pt);
\path[-,draw=black,line width=0.4pt,line cap=butt,line join=miter,dash pattern=](-16pt,17.4184pt)--(-10pt,17.4184pt);
\path[-,draw=black,line width=0.4pt,line cap=butt,line join=miter,dash pattern=](-16pt,10pt)--(-10pt,10pt);
\path[-,draw=black,line width=1.2pt,line cap=butt,line join=miter,dash pattern=](-13pt,18.4184pt)--(-13pt,9pt)--cycle;
\path[-,draw=black,line width=0.4pt,line cap=butt,line join=miter,dash pattern=](-16pt,0pt)--(-10pt,0pt);
\path[-,line width=0.4pt,line cap=butt,line join=miter,dash pattern=](-21.6316pt,19.758pt)--(-19pt,19.758pt)--(-19pt,15.0788pt)--(-21.6316pt,15.0788pt)--cycle;
\node[anchor=north west,inner sep=0] at (-21.6316pt,19.758pt){\savebox{\marxupbox}{{\({\scriptstyle i}\)}}\immediate\write\boxesfile{269}\immediate\write\boxesfile{\number\wd\marxupbox}\immediate\write\boxesfile{\number\ht\marxupbox}\immediate\write\boxesfile{\number\dp\marxupbox}\box\marxupbox};
\path[-,line width=0.4pt,line cap=butt,line join=miter,dash pattern=](-23.6316pt,21.758pt)--(-17pt,21.758pt)--(-17pt,13.0788pt)--(-23.6316pt,13.0788pt)--cycle;
\path[-,line width=0.4pt,line cap=butt,line join=miter,dash pattern=](-22.8617pt,13.0788pt)--(-19pt,13.0788pt)--(-19pt,6.9212pt)--(-22.8617pt,6.9212pt)--cycle;
\node[anchor=north west,inner sep=0] at (-22.8617pt,13.0788pt){\savebox{\marxupbox}{{\({\scriptstyle j}\)}}\immediate\write\boxesfile{270}\immediate\write\boxesfile{\number\wd\marxupbox}\immediate\write\boxesfile{\number\ht\marxupbox}\immediate\write\boxesfile{\number\dp\marxupbox}\box\marxupbox};
\path[-,line width=0.4pt,line cap=butt,line join=miter,dash pattern=](-24.8617pt,15.0788pt)--(-17pt,15.0788pt)--(-17pt,4.9212pt)--(-24.8617pt,4.9212pt)--cycle;
\path[-,line width=0.4pt,line cap=butt,line join=miter,dash pattern=](-23.3398pt,2.4911pt)--(-19pt,2.4911pt)--(-19pt,-2.4911pt)--(-23.3398pt,-2.4911pt)--cycle;
\node[anchor=north west,inner sep=0] at (-23.3398pt,2.4911pt){\savebox{\marxupbox}{{\({\scriptstyle k}\)}}\immediate\write\boxesfile{271}\immediate\write\boxesfile{\number\wd\marxupbox}\immediate\write\boxesfile{\number\ht\marxupbox}\immediate\write\boxesfile{\number\dp\marxupbox}\box\marxupbox};
\path[-,line width=0.4pt,line cap=butt,line join=miter,dash pattern=](-25.3398pt,4.4911pt)--(-17pt,4.4911pt)--(-17pt,-4.4911pt)--(-25.3398pt,-4.4911pt)--cycle;
\path[-,line width=0.4pt,line cap=butt,line join=miter,dash pattern=](21pt,5.0258pt)--(23.5769pt,5.0258pt)--(23.5769pt,-0.0258pt)--(21pt,-0.0258pt)--cycle;
\node[anchor=north west,inner sep=0] at (21pt,5.0258pt){\savebox{\marxupbox}{{\({\scriptstyle l}\)}}\immediate\write\boxesfile{272}\immediate\write\boxesfile{\number\wd\marxupbox}\immediate\write\boxesfile{\number\ht\marxupbox}\immediate\write\boxesfile{\number\dp\marxupbox}\box\marxupbox};
\path[-,line width=0.4pt,line cap=butt,line join=miter,dash pattern=](19pt,7.0258pt)--(25.5769pt,7.0258pt)--(25.5769pt,-2.0258pt)--(19pt,-2.0258pt)--cycle;
\end{tikzpicture}}+{\begin{tikzpicture}[baseline=(current bounding box.center)]\path[-,draw=black,line width=0.4pt,line cap=butt,line join=miter,dash pattern=](-20pt,17.4184pt)--(-19pt,17.4184pt);
\path[-,draw=black,line width=0.4pt,line cap=butt,line join=miter,dash pattern=](-20pt,10pt)--(-19pt,10pt);
\path[-,draw=black,line width=0.4pt,line cap=butt,line join=miter,dash pattern=](-20pt,0pt)--(-19pt,0pt);
\path[-,draw=black,line width=0.4pt,line cap=butt,line join=miter,dash pattern=](13pt,9.9592pt)--(20pt,9.9592pt);
\path[-,draw=black,line width=0.4pt,line cap=butt,line join=miter,dash pattern=](-13pt,17.4184pt)--(-13pt,17.4184pt);
\path[-,draw=black,line width=0.4pt,line cap=butt,line join=miter,dash pattern=](-13pt,10pt)--(-13pt,10pt);
\path[-,draw=black,line width=0.4pt,line cap=butt,line join=miter,dash pattern=](-13pt,0pt)--(-13pt,0pt);
\path[-,draw=black,line width=0.4pt,line cap=butt,line join=miter,dash pattern=](-7pt,17.4184pt)--(-6pt,17.4184pt);
\path[-,draw=black,line width=0.4pt,line cap=butt,line join=miter,dash pattern=](-7pt,5pt)--(-6pt,5pt);
\path[-,draw=black,line width=0.4pt,line cap=butt,line join=miter,dash pattern=](-7pt,0pt)--(-6pt,0pt);
\path[-,draw=black,line width=0.4pt,line cap=butt,line join=miter,dash pattern=](0pt,17.4184pt)--(7pt,17.4184pt);
\path[-,draw=black,line width=0.4pt,line cap=butt,line join=miter,dash pattern=](0pt,2.5pt)--(7pt,2.5pt);
\path[-,draw=black,fill= gray,line width=0.4pt,line cap=butt,line join=miter,dash pattern=](7pt,18.4184pt)..controls(11pt,18.4184pt)and(13pt,11.9592pt)..(13pt,9.9592pt)..controls(13pt,7.9592pt)and(11pt,1.5pt)..(7pt,1.5pt)--cycle;
\path[-,draw=black,line width=0.4pt,line cap=butt,line join=miter,dash pattern=](-6pt,17.4184pt)--(0pt,17.4184pt);
\path[-,draw=black,fill= gray,line width=0.4pt,line cap=butt,line join=miter,dash pattern=](-6pt,6pt)..controls(-2pt,6pt)and(0pt,4.5pt)..(0pt,2.5pt)..controls(0pt,0.5pt)and(-2pt,-1pt)..(-6pt,-1pt)--cycle;
\path[-,draw=black,line width=0.4pt,line cap=butt,line join=miter,dash pattern=](-13pt,17.4184pt)--(-7pt,17.4184pt);
\path[-,draw=black,line width=0.4pt,line cap=butt,line join=miter,dash pattern=](-13pt,10pt)..controls(-9pt,10pt)and(-11pt,5pt)..(-7pt,5pt);
\path[-,draw=black,line width=0.4pt,line cap=butt,line join=miter,dash pattern=](-13pt,0pt)--(-7pt,0pt);
\path[-,draw=black,line width=0.4pt,line cap=butt,line join=miter,dash pattern=](-19pt,17.4184pt)--(-13pt,17.4184pt);
\path[-,draw=black,line width=0.4pt,line cap=butt,line join=miter,dash pattern=](-19pt,10pt)--(-13pt,10pt);
\path[-,draw=black,line width=1.2pt,line cap=butt,line join=miter,dash pattern=](-16pt,18.4184pt)--(-16pt,9pt)--cycle;
\path[-,draw=black,line width=0.4pt,line cap=butt,line join=miter,dash pattern=](-19pt,0pt)--(-13pt,0pt);
\path[-,line width=0.4pt,line cap=butt,line join=miter,dash pattern=](-24.6316pt,19.758pt)--(-22pt,19.758pt)--(-22pt,15.0788pt)--(-24.6316pt,15.0788pt)--cycle;
\node[anchor=north west,inner sep=0] at (-24.6316pt,19.758pt){\savebox{\marxupbox}{{\({\scriptstyle i}\)}}\immediate\write\boxesfile{273}\immediate\write\boxesfile{\number\wd\marxupbox}\immediate\write\boxesfile{\number\ht\marxupbox}\immediate\write\boxesfile{\number\dp\marxupbox}\box\marxupbox};
\path[-,line width=0.4pt,line cap=butt,line join=miter,dash pattern=](-26.6316pt,21.758pt)--(-20pt,21.758pt)--(-20pt,13.0788pt)--(-26.6316pt,13.0788pt)--cycle;
\path[-,line width=0.4pt,line cap=butt,line join=miter,dash pattern=](-25.8617pt,13.0788pt)--(-22pt,13.0788pt)--(-22pt,6.9212pt)--(-25.8617pt,6.9212pt)--cycle;
\node[anchor=north west,inner sep=0] at (-25.8617pt,13.0788pt){\savebox{\marxupbox}{{\({\scriptstyle j}\)}}\immediate\write\boxesfile{274}\immediate\write\boxesfile{\number\wd\marxupbox}\immediate\write\boxesfile{\number\ht\marxupbox}\immediate\write\boxesfile{\number\dp\marxupbox}\box\marxupbox};
\path[-,line width=0.4pt,line cap=butt,line join=miter,dash pattern=](-27.8617pt,15.0788pt)--(-20pt,15.0788pt)--(-20pt,4.9212pt)--(-27.8617pt,4.9212pt)--cycle;
\path[-,line width=0.4pt,line cap=butt,line join=miter,dash pattern=](-26.3398pt,2.4911pt)--(-22pt,2.4911pt)--(-22pt,-2.4911pt)--(-26.3398pt,-2.4911pt)--cycle;
\node[anchor=north west,inner sep=0] at (-26.3398pt,2.4911pt){\savebox{\marxupbox}{{\({\scriptstyle k}\)}}\immediate\write\boxesfile{275}\immediate\write\boxesfile{\number\wd\marxupbox}\immediate\write\boxesfile{\number\ht\marxupbox}\immediate\write\boxesfile{\number\dp\marxupbox}\box\marxupbox};
\path[-,line width=0.4pt,line cap=butt,line join=miter,dash pattern=](-28.3398pt,4.4911pt)--(-20pt,4.4911pt)--(-20pt,-4.4911pt)--(-28.3398pt,-4.4911pt)--cycle;
\path[-,line width=0.4pt,line cap=butt,line join=miter,dash pattern=](22pt,12.485pt)--(24.5769pt,12.485pt)--(24.5769pt,7.4334pt)--(22pt,7.4334pt)--cycle;
\node[anchor=north west,inner sep=0] at (22pt,12.485pt){\savebox{\marxupbox}{{\({\scriptstyle l}\)}}\immediate\write\boxesfile{276}\immediate\write\boxesfile{\number\wd\marxupbox}\immediate\write\boxesfile{\number\ht\marxupbox}\immediate\write\boxesfile{\number\dp\marxupbox}\box\marxupbox};
\path[-,line width=0.4pt,line cap=butt,line join=miter,dash pattern=](20pt,14.485pt)--(26.5769pt,14.485pt)--(26.5769pt,5.4334pt)--(20pt,5.4334pt)--cycle;
\end{tikzpicture}}\).
In {\sc{}Roger}, it is even possible to factor the antisymmetrisation operator, and obtain the following concise expression:
\(\allowbreak{}\mathnormal{(}\mskip 0.0mu\mathnormal{∂}\mskip 3.0muΓ\mskip 3.0mu\mathnormal{+}\mskip 3.0muΓ\mskip 3.0mu\allowbreak{}\mathnormal{∘}\allowbreak{}\mskip 3.0mu\allowbreak{}\mathnormal{(}\mskip 0.0mu\mathsf{id}\mskip 3.0mu{⊗}\mskip 3.0muΓ\mskip 0.0mu\mathnormal{)}\allowbreak{}\mskip 0.0mu\mathnormal{)}\allowbreak{}\mskip 3.0mu\allowbreak{}\mathnormal{∘}\allowbreak{}\mskip 3.0muα\mskip 3.0mu\allowbreak{}\mathnormal{∘}\allowbreak{}\mskip 3.0mu\allowbreak{}\mathnormal{(}\mskip 0.0mu\allowbreak{}\mathnormal{(}\mskip 0.0mu\mathsf{id}\mskip 3.0mu\mathnormal{-}\mskip 3.0muσ\mskip 0.0mu\mathnormal{)}\allowbreak{}\mskip 3.0mu{⊗}\mskip 3.0mu\mathsf{id}\mskip 0.0mu\mathnormal{)}\allowbreak{}\).
Despite its concision, this form obscures which index plays which role, and thus is rarely found in the literature.
The above definition can be encoded directly in {\sc{}Albert} as follows:
\begin{list}{}{\setlength\leftmargin{1.0em}}\item\relax
\ensuremath{\begin{parray}\column{B}{@{}>{}l<{}@{}}\column[0em]{1}{@{}>{}l<{}@{}}\column[1em]{2}{@{}>{}l<{}@{}}\column[2em]{3}{@{}>{}l<{}@{}}\column[3em]{4}{@{}>{}l<{}@{}}\column{5}{@{}>{}l<{}@{}}\column{6}{@{}>{}l<{}@{}}\column{7}{@{}>{}l<{}@{}}\column{8}{@{}>{}l<{}@{}}\column{9}{@{}>{}l<{}@{}}\column{10}{@{}>{}l<{}@{}}\column{11}{@{}>{}l<{}@{}}\column{12}{@{}>{}l<{}@{}}\column{13}{@{}>{}l<{}@{}}\column{14}{@{}>{}l<{}@{}}\column{E}{@{}>{}l<{}@{}}%
\>[1]{\mathsf{curvature}\mskip 3.0mu\mathnormal{::}\mskip 3.0mu}\>[14]{\mathsf{CoordinateCategory}\mskip 3.0muz\mskip 3.0mu\mathnormal{\Rightarrow }\mskip 3.0mu\mathsf{P}\mskip 3.0muz\mskip 3.0mu\mathsf{r}\mskip 3.0mu\dual{T_{z}}\mskip 3.0mu\mathnormal{⊸}\mskip 3.0mu\mathsf{P}\mskip 3.0muz\mskip 3.0mu\mathsf{r}\mskip 3.0muT_{z}\mskip 3.0mu\mathnormal{⊸}\mskip 3.0mu\mathsf{P}\mskip 3.0muz\mskip 3.0mu\mathsf{r}\mskip 3.0muT_{z}\mskip 3.0mu\mathnormal{⊸}\mskip 3.0mu\mathsf{P}\mskip 3.0muz\mskip 3.0mu\mathsf{r}\mskip 3.0muT_{z}\mskip 3.0mu\mathnormal{⊸}\mskip 3.0mu\mathsf{R}\mskip 3.0muz\mskip 3.0mu\mathsf{r}}\<[E]\\
\>[1]{\mathsf{curvature}\mskip 3.0mu^{\mathsf{l}}\mskip 3.0mu_{\mathsf{k}}\mskip 3.0mu_{\mathsf{i}}\mskip 3.0mu_{\mathsf{j}}}\<[E]\\
\>[2]{\mathnormal{=}\mskip 3.0mu\mathsf{plus}\mskip 3.0mu\allowbreak{}\mathnormal{(}\mskip 0.0muλ\mskip 3.0mu\mathsf{a}\mskip 3.0mu\mathnormal{\rightarrow }\mskip 3.0mu\mathbf{case}\mskip 3.0mu\mathsf{a}\mskip 3.0mu\mathbf{of}}\<[E]\\
\>[3]{\mathsf{False}\mskip 3.0mu}\>[5]{\mathnormal{\rightarrow }\mskip 3.0mu\mathsf{minus}\mskip 3.0mu\allowbreak{}\mathnormal{(}\mskip 0.0muλ\mskip 3.0mu\mathsf{b}\mskip 3.0mu\mathnormal{\rightarrow }\mskip 3.0mu\mathbf{case}\mskip 3.0mu\mathsf{b}\mskip 3.0mu\mathbf{of}}\<[E]\\
\>[4]{\mathsf{False}\mskip 3.0mu}\>[6]{\mathnormal{\rightarrow }\mskip 3.0mu\mathsf{partial}\mskip 3.0mu}\>[7]{_{\mathsf{i}}\mskip 3.0mu}\>[9]{\allowbreak{}\mathnormal{(}\mskip 0.0mu\mathsf{christoffel}\mskip 3.0mu_{\mathsf{j}}\mskip 3.0mu}\>[12]{_{\mathsf{k}}\mskip 3.0mu^{\mathsf{l}}\mskip 0.0mu\mathnormal{)}\allowbreak{}}\<[E]\\
\>[4]{\mathsf{True}\mskip 3.0mu}\>[6]{\mathnormal{\rightarrow }\mskip 3.0mu\mathsf{partial}\mskip 3.0mu}\>[7]{_{\mathsf{j}}\mskip 3.0mu}\>[9]{\allowbreak{}\mathnormal{(}\mskip 0.0mu\mathsf{christoffel}\mskip 3.0mu_{\mathsf{i}}\mskip 3.0mu}\>[12]{_{\mathsf{k}}\mskip 3.0mu^{\mathsf{l}}\mskip 0.0mu\mathnormal{)}\allowbreak{}\mskip 0.0mu\mathnormal{)}\allowbreak{}}\<[E]\\
\>[3]{\mathsf{True}\mskip 3.0mu}\>[5]{\mathnormal{\rightarrow }\mskip 3.0mu\mathsf{contract}\mskip 3.0mu\allowbreak{}\mathnormal{(}\mskip 0.0muλ\mskip 3.0mu^{\mathsf{m}}\mskip 3.0mu_{\mathsf{m}}\mskip 3.0mu\mathnormal{\rightarrow }\mskip 3.0mu\mathsf{minus}\mskip 3.0mu\allowbreak{}\mathnormal{(}\mskip 0.0muλ\mskip 3.0mu\mathsf{b}\mskip 3.0mu\mathnormal{\rightarrow }\mskip 3.0mu\mathbf{case}\mskip 3.0mu\mathsf{b}\mskip 3.0mu\mathbf{of}}\<[E]\\
\>[4]{\mathsf{False}\mskip 3.0mu}\>[6]{\mathnormal{\rightarrow }\mskip 3.0mu\mathsf{christoffel}\mskip 3.0mu_{\mathsf{i}}\mskip 3.0mu}\>[8]{_{\mathsf{m}}\mskip 3.0mu^{\mathsf{l}}\mskip 3.0mu}\>[10]{{\tikzstar{0.11}{0.25}{5}{-18}{fill=black}}\mskip 3.0mu}\>[11]{\mathsf{christoffel}\mskip 3.0mu_{\mathsf{j}}\mskip 3.0mu}\>[13]{_{\mathsf{k}}\mskip 3.0mu^{\mathsf{m}}}\<[E]\\
\>[4]{\mathsf{True}\mskip 3.0mu}\>[6]{\mathnormal{\rightarrow }\mskip 3.0mu\mathsf{christoffel}\mskip 3.0mu_{\mathsf{j}}\mskip 3.0mu}\>[8]{_{\mathsf{m}}\mskip 3.0mu^{\mathsf{l}}\mskip 3.0mu}\>[10]{{\tikzstar{0.11}{0.25}{5}{-18}{fill=black}}\mskip 3.0mu}\>[11]{\mathsf{christoffel}\mskip 3.0mu_{\mathsf{i}}\mskip 3.0mu}\>[13]{_{\mathsf{k}}\mskip 3.0mu^{\mathsf{m}}\mskip 0.0mu\mathnormal{)}\allowbreak{}\mskip 0.0mu\mathnormal{)}\allowbreak{}\mskip 0.0mu\mathnormal{)}\allowbreak{}}\<[E]\end{parray}}\end{list} Unfortunately, the operands of each addition must be written in the branch of a case expression, making the expression verbose as a whole.
Even though it is defined in terms of Christoffel symbols, the Riemann
curvature (as an algebraic object) does not depend on the choice of
coordinates. This is a consequence of \cref{277}.

\begin{theorem}[Ricci identity]For every covector field \(\mathsf{u}\),
\(∇{_i}∇{_j}u{^k} - ∇{_j}∇{_i}u{^k}\) = \(R{^k}{_i}{_j}{_l}u{^l}\) 
\label{277}\end{theorem}\begin{proof}We carry out the proof using the diagram notation. To be sure, we don't claim that the diagrammatic proof is novel, it is a mere illustration.
But we feel it is a good example of using the {\sc{}dsl}s: all the steps are defined, type checked, and diagrams are rendered with our library.
We note first the following lemma: \({
}\) which is an instance of \cref{230}.

We then compute symbolically the left-hand-side in the theorem's statement, starting with the expansion of the outer derivative into partial derivatives and affinity terms (\cref{230}):

\({%
}\) 

The middle term is zero, by symmetry of Γ. Expanding the first term using the lemma and linearity of partial derivatives, we get

\({%
}\) 

Because partial derivatives commute, the first term is zero. We expand the middle term using the composition rule (\cref{201}) for partial derivatives:

\({%
}\) 

then use the product law, \cref{223}, on the middle term, and obtain

\({%
}\) 

The partial derivative of the identity morphism is zero so the whole second term can be simplified away

\({%
}\) 

then we commute swap and antisymmetrisation, so the middle term changes sign:

\({%
}\) 

then we use the lemma again to expand \(∇u\) in the last term

\({%
}\) 

and note that the middle two terms cancel to yield:

\({%
}\) 

To get to the intended result, it remains to use the structural laws of {\sc{}smc}s.\end{proof} 
\subsection{Code for \cref{266}}\label{391} A contraction of the Riemann curvature
occurs twice in \cref{266}, and as such deserves to be extracted as an
intermediate definition. It is called the Ricci tensor.
\begin{list}{}{\setlength\leftmargin{1.0em}}\item\relax
\ensuremath{\begin{parray}\column{B}{@{}>{}l<{}@{}}\column[0em]{1}{@{}>{}l<{}@{}}\column{2}{@{}>{}l<{}@{}}\column{3}{@{}>{}l<{}@{}}\column{E}{@{}>{}l<{}@{}}%
\>[1]{\mathsf{ricci}\mskip 0.0mu\mathnormal{,}\mskip 3.0mu\mathsf{grLhs}\mskip 3.0mu\mathnormal{::}\mskip 3.0mu}\>[2]{\allowbreak{}\mathnormal{(}\mskip 0.0mu\mathsf{Additive}\mskip 3.0muz\mskip 0.0mu}\>[3]{\mathnormal{,}\mskip 3.0mu\mathsf{CoordinateCategory}\mskip 3.0muz\mskip 0.0mu\mathnormal{)}\allowbreak{}\mskip 3.0mu\mathnormal{\Rightarrow }\mskip 3.0mu\mathsf{P}\mskip 3.0muz\mskip 3.0mu\mathsf{r}\mskip 3.0muT_{z}\mskip 3.0mu\mathnormal{⊸}\mskip 3.0mu\mathsf{P}\mskip 3.0muz\mskip 3.0mu\mathsf{r}\mskip 3.0muT_{z}\mskip 3.0mu\mathnormal{⊸}\mskip 3.0mu\mathsf{R}\mskip 3.0muz\mskip 3.0mu\mathsf{r}}\<[E]\\
\>[1]{\mathsf{ricci}\mskip 3.0mu_{\mathsf{j}}\mskip 3.0mu_{\mathsf{k}}\mskip 3.0mu\mathnormal{=}\mskip 3.0mu\mathsf{contract}\mskip 3.0mu\allowbreak{}\mathnormal{(}\mskip 0.0muλ\mskip 3.0mu^{\mathsf{i}}\mskip 3.0mu_{\mathsf{i}}\mskip 3.0mu\mathnormal{\rightarrow }\mskip 3.0mu\mathsf{curvature}\mskip 3.0mu^{\mathsf{i}}\mskip 3.0mu_{\mathsf{j}}\mskip 3.0mu_{\mathsf{i}}\mskip 3.0mu_{\mathsf{k}}\mskip 0.0mu\mathnormal{)}\allowbreak{}}\<[E]\end{parray}}\end{list} This tensor can be contracted one more time to obtain what is called the scalar (or Gaussian) curvature:
\begin{list}{}{\setlength\leftmargin{1.0em}}\item\relax
\ensuremath{\begin{parray}\column{B}{@{}>{}l<{}@{}}\column[0em]{1}{@{}>{}l<{}@{}}\column{2}{@{}>{}l<{}@{}}\column{3}{@{}>{}l<{}@{}}\column{4}{@{}>{}l<{}@{}}\column{E}{@{}>{}l<{}@{}}%
\>[1]{\mathsf{gaussian}\mskip 3.0mu\mathnormal{::}\mskip 3.0mu}\>[3]{\allowbreak{}\mathnormal{(}\mskip 0.0mu\mathsf{Additive}\mskip 3.0muz\mskip 0.0mu}\>[4]{\mathnormal{,}\mskip 3.0mu\mathsf{CoordinateCategory}\mskip 3.0muz\mskip 0.0mu\mathnormal{)}\allowbreak{}\mskip 3.0mu\mathnormal{\Rightarrow }\mskip 3.0mu\mathsf{R}\mskip 3.0muz\mskip 3.0mu\mathsf{r}}\<[E]\\
\>[1]{\mathsf{gaussian}\mskip 3.0mu\mathnormal{=}\mskip 3.0mu\mathsf{constant}\mskip 3.0mu\allowbreak{}\mathnormal{(}\mskip 0.0mu\mathrm{1}\mskip 3.0mu\mathnormal{/}\mskip 3.0mu\mathrm{2}\mskip 0.0mu\mathnormal{)}\allowbreak{}\mskip 3.0mu{\tikzstar{0.11}{0.25}{5}{-18}{fill=black}}\mskip 3.0mu\mathsf{contract}\mskip 3.0mu}\>[2]{\allowbreak{}\mathnormal{(}\mskip 0.0muλ\mskip 3.0mu^{\mathsf{i}}\mskip 3.0mu_{\mathsf{i}}\mskip 3.0mu\mathnormal{\rightarrow }\mskip 3.0mu\mathsf{ricci}\mskip 3.0mu_{\mathsf{i}}\mskip 3.0mu\allowbreak{}\mathnormal{(}\mskip 0.0mu\mathsf{lower}\mskip 3.0mu^{\mathsf{i}}\mskip 0.0mu\mathnormal{)}\allowbreak{}\mskip 0.0mu\mathnormal{)}\allowbreak{}}\<[E]\end{parray}}\end{list} From there, the left-hand-side of \cref{266} is
defined as follows: \begin{list}{}{\setlength\leftmargin{1.0em}}\item\relax
\ensuremath{\begin{parray}\column{B}{@{}>{}l<{}@{}}\column[0em]{1}{@{}>{}l<{}@{}}\column[1em]{2}{@{}>{}l<{}@{}}\column{3}{@{}>{}l<{}@{}}\column{E}{@{}>{}l<{}@{}}%
\>[1]{\mathsf{grLhs}\mskip 3.0mu_{\mathsf{i}}\mskip 3.0mu_{\mathsf{j}}\mskip 3.0mu\mathnormal{=}\mskip 3.0mu\mathsf{plus}\mskip 3.0mu\allowbreak{}\mathnormal{(}\mskip 0.0muλ\mskip 3.0mu\mathsf{c}\mskip 3.0mu\mathnormal{\rightarrow }\mskip 3.0mu\mathbf{case}\mskip 3.0mu\mathsf{c}\mskip 3.0mu\mathbf{of}}\<[E]\\
\>[2]{\mathsf{False}\mskip 3.0mu}\>[3]{\mathnormal{\rightarrow }\mskip 3.0mu\mathsf{ricci}\mskip 3.0mu_{\mathsf{i}}\mskip 3.0mu_{\mathsf{j}}}\<[E]\\
\>[2]{\mathsf{True}\mskip 3.0mu}\>[3]{\mathnormal{\rightarrow }\mskip 3.0mu\mathsf{gaussian}\mskip 3.0mu{\tikzstar{0.11}{0.25}{5}{-18}{fill=black}}\mskip 3.0mu\mathsf{metric}\mskip 3.0mu_{\mathsf{i}}\mskip 3.0mu_{\mathsf{j}}\mskip 0.0mu\mathnormal{)}\allowbreak{}}\<[E]\end{parray}}\end{list} We can then convert it to a
morphism: \begin{list}{}{\setlength\leftmargin{1.0em}}\item\relax
\ensuremath{\begin{parray}\column{B}{@{}>{}l<{}@{}}\column[0em]{1}{@{}>{}l<{}@{}}\column{2}{@{}>{}l<{}@{}}\column{3}{@{}>{}l<{}@{}}\column{E}{@{}>{}l<{}@{}}%
\>[1]{\mathsf{grLhsM}\mskip 3.0mu\mathnormal{::}\mskip 3.0mu}\>[2]{\allowbreak{}\mathnormal{(}\mskip 0.0mu\mathsf{Additive}\mskip 3.0muz\mskip 0.0mu}\>[3]{\mathnormal{,}\mskip 3.0mu\mathsf{CoordinateCategory}\mskip 3.0muz\mskip 0.0mu\mathnormal{)}\allowbreak{}\mskip 3.0mu\mathnormal{\Rightarrow }\mskip 3.0mu\allowbreak{}\mathnormal{(}\mskip 0.0muT_{z}\mskip 3.0mu\mathnormal{⊗}\mskip 3.0muT_{z}\mskip 0.0mu\mathnormal{)}\allowbreak{}\mskip 1.0mu\overset{z}{\leadsto }\mskip 1.0mu\mathbf{1}}\<[E]\\
\>[1]{\mathsf{grLhsM}\mskip 3.0mu\mathnormal{=}\mskip 3.0mu\mathsf{tensorEval}_{1}\mskip 3.0mu\allowbreak{}\mathnormal{(}\mskip 0.0muλ\mskip 3.0mu\mathsf{k}\mskip 3.0mu\mathnormal{\rightarrow }\mskip 3.0mu\mathsf{split}\mskip 3.0mu\mathsf{k}\mskip 3.0mu\mathnormal{\&}\mskip 3.0muλ\mskip 3.0mu\allowbreak{}\mathnormal{(}\mskip 0.0mu_{\mathsf{i}}\mskip 0.0mu\mathnormal{,}\mskip 3.0mu_{\mathsf{j}}\mskip 0.0mu\mathnormal{)}\allowbreak{}\mskip 3.0mu\mathnormal{\rightarrow }\mskip 3.0mu\mathsf{grLhs}\mskip 3.0mu_{\mathsf{i}}\mskip 3.0mu_{\mathsf{j}}\mskip 0.0mu\mathnormal{)}\allowbreak{}}\<[E]\end{parray}}\end{list} In the above, the operator \(\allowbreak{}\mathnormal{(}\mskip 0.0mu\mathnormal{\&}\mskip 0.0mu\mathnormal{)}\allowbreak{}\) is the linear post-fix application:
\begin{list}{}{\setlength\leftmargin{1.0em}}\item\relax
\ensuremath{\begin{parray}\column{B}{@{}>{}l<{}@{}}\column[0em]{1}{@{}>{}l<{}@{}}\column{2}{@{}>{}l<{}@{}}\column{E}{@{}>{}l<{}@{}}%
\>[1]{\allowbreak{}\mathnormal{(}\mskip 0.0mu\mathnormal{\&}\mskip 0.0mu\mathnormal{)}\allowbreak{}\mskip 3.0mu\mathnormal{::}\mskip 3.0mu}\>[2]{\mathsf{a}\mskip 3.0mu\mathnormal{⊸}\mskip 3.0mu\allowbreak{}\mathnormal{(}\mskip 0.0mu\mathsf{a}\mskip 3.0mu\mathnormal{⊸}\mskip 3.0mu\mathsf{b}\mskip 0.0mu\mathnormal{)}\allowbreak{}\mskip 3.0mu\mathnormal{⊸}\mskip 3.0mu\mathsf{b}}\<[E]\\
\>[1]{\mathsf{x}\mskip 3.0mu\mathnormal{\&}\mskip 3.0mu\mathsf{f}\mskip 3.0mu\mathnormal{=}\mskip 3.0mu\mathsf{f}\mskip 3.0mu\mathsf{x}}\<[E]\end{parray}}\end{list} 
\subsection{Point-mass example (the Schwarzschild metric)}\label{392} 
To complete our test case for {\sc{}Albert},
in this section we define the metric which describes the
gravitational effects of a point-sized mass: the Schwarzschild metric.
We then verify that this metric satisfies of \cref{266} by evaluating both sides and checking
that they match.

The first step is to define the coordinate system.
Roughly speaking, Schwarzschild coordinates are spherical coordinates with an extra component for time.
\begin{list}{}{\setlength\leftmargin{1.0em}}\item\relax
\ensuremath{\begin{parray}\column{B}{@{}>{}l<{}@{}}\column[0em]{1}{@{}>{}l<{}@{}}\column{E}{@{}>{}l<{}@{}}%
\>[1]{\mathbf{data}\mskip 3.0mu\mathsf{Spherical}\mskip 3.0mu\mathnormal{=}\mskip 3.0mu\mathsf{Time}\mskip 3.0mu\mathnormal{|}\mskip 3.0mu\mathsf{Rho}\mskip 3.0mu\mathnormal{|}\mskip 3.0mu\mathsf{Theta}\mskip 3.0mu\mathnormal{|}\mskip 3.0mu\mathsf{Phi}}\<[E]\end{parray}}\end{list} The point mass will
be located at the origin (ρ=0) at every point in time. In other words,
the mass is at rest in this coordinate system.
Hence, because we have a point-mass, the \(T{_i}{_j}\) tensor is zero everywhere except at the origin, where it is infinite.

As for all coordinate representations, the compact-closed category
structure for \(\mathsf{Matrix}\mskip 3.0mu\mathsf{Spherical}\) falls out directly from
\cref{110}. Likewise for the partial derivative. The only missing piece of the structure
is the metric.
The Schwarzschild metric is defined in terms of the considered mass
\(M\) or alternatively by the Schwarzschild radius \(r_s\),
the two being connected by the equation \(r_s = κMc^2/4π\).
  We use the parameter \(r_s\) in the rest of the section.
The metric is then given by tabulating the following function as a matrix.
\begin{list}{}{\setlength\leftmargin{1.0em}}\item\relax
\ensuremath{\begin{parray}\column{B}{@{}>{}l<{}@{}}\column[0em]{1}{@{}>{}l<{}@{}}\column{2}{@{}>{}l<{}@{}}\column{3}{@{}>{}l<{}@{}}\column{4}{@{}>{}l<{}@{}}\column{5}{@{}>{}l<{}@{}}\column{6}{@{}>{}l<{}@{}}\column{E}{@{}>{}l<{}@{}}%
\>[1]{\mathsf{schwarzschild}\mskip 3.0mu\mathnormal{::}\mskip 3.0mu\mathsf{Spherical}\mskip 3.0mu\mathnormal{\rightarrow }\mskip 3.0mu\mathsf{Spherical}\mskip 3.0mu\mathnormal{\rightarrow }\mskip 3.0mu\mathsf{S}_{\mathsf{Spherical}}}\<[E]\\
\>[1]{\mathsf{schwarzschild}\mskip 3.0mu}\>[3]{\mathsf{Time}\mskip 3.0mu}\>[4]{\mathsf{Time}\mskip 3.0mu}\>[5]{\mathnormal{=}\mskip 3.0mu\mathnormal{-}\mskip 3.0mu}\>[6]{\allowbreak{}\mathnormal{(}\mskip 0.0mu\mathrm{1}\mskip 3.0mu\mathnormal{-}\mskip 3.0mur_s\mskip 3.0mu\mathnormal{/}\mskip 3.0mu\mathsf{rho}\mskip 0.0mu\mathnormal{)}\allowbreak{}\mskip 3.0mu\mathnormal{*}\mskip 3.0mu\allowbreak{}\mathnormal{(}\mskip 0.0mu\mathsf{c}\mskip 3.0mu\string^\mskip 3.0mu\mathrm{2}\mskip 0.0mu\mathnormal{)}\allowbreak{}}\<[E]\\
\>[1]{\mathsf{schwarzschild}\mskip 3.0mu}\>[3]{\mathsf{Rho}\mskip 3.0mu}\>[4]{\mathsf{Rho}\mskip 3.0mu}\>[5]{\mathnormal{=}\mskip 3.0mu}\>[6]{\allowbreak{}\mathnormal{(}\mskip 0.0mu\mathrm{1}\mskip 3.0mu\mathnormal{-}\mskip 3.0mur_s\mskip 3.0mu\mathnormal{/}\mskip 3.0mu\mathsf{rho}\mskip 0.0mu\mathnormal{)}\allowbreak{}\mskip 3.0mu\string^\mskip 3.0mu\allowbreak{}\mathnormal{(}\mskip 0.0mu\mathnormal{-}\mskip 3.0mu\mathrm{1}\mskip 0.0mu\mathnormal{)}\allowbreak{}}\<[E]\\
\>[1]{\mathsf{schwarzschild}\mskip 3.0mu}\>[3]{\mathsf{Theta}\mskip 3.0mu}\>[4]{\mathsf{Theta}\mskip 3.0mu}\>[5]{\mathnormal{=}\mskip 3.0mu\mathsf{rho}\mskip 3.0mu\string^\mskip 3.0mu\mathrm{2}}\<[E]\\
\>[1]{\mathsf{schwarzschild}\mskip 3.0mu}\>[3]{\mathsf{Phi}\mskip 3.0mu}\>[4]{\mathsf{Phi}\mskip 3.0mu}\>[5]{\mathnormal{=}\mskip 3.0mu\allowbreak{}\mathnormal{(}\mskip 0.0mu\mathsf{rho}\mskip 3.0mu\mathnormal{*}\mskip 3.0mu\mathsf{sin}\mskip 3.0mu\mathsf{theta}\mskip 0.0mu\mathnormal{)}\allowbreak{}\mskip 3.0mu\string^\mskip 3.0mu\mathrm{2}}\<[E]\\
\>[1]{\mathsf{schwarzschild}\mskip 3.0mu}\>[3]{\mathnormal{\_}\mskip 3.0mu}\>[4]{\mathnormal{\_}\mskip 3.0mu}\>[5]{\mathnormal{=}\mskip 3.0mu\mathrm{0}}\<[E]\\
\>[1]{\mathsf{rho}\mskip 3.0mu}\>[2]{\mathnormal{=}\mskip 3.0mu\mathsf{variable}\mskip 3.0mu\mathsf{Rho}}\<[E]\\
\>[1]{\mathsf{theta}\mskip 3.0mu}\>[2]{\mathnormal{=}\mskip 3.0mu\mathsf{variable}\mskip 3.0mu\mathsf{Theta}}\<[E]\end{parray}}\end{list} We refer the reader to a course in general relativity for the physical meaning of it.
We can then define the \(\mathsf{Matrix}\mskip 3.0mu\mathsf{Spherical}\) instance as follows:
\begin{list}{}{\setlength\leftmargin{1.0em}}\item\relax
\ensuremath{\begin{parray}\column{B}{@{}>{}l<{}@{}}\column[0em]{1}{@{}>{}l<{}@{}}\column[1em]{2}{@{}>{}l<{}@{}}\column{E}{@{}>{}l<{}@{}}%
\>[1]{\mathbf{instance}\mskip 3.0mu\mathsf{MetricCategory}\mskip 3.0muM_{\mathsf{Spherical}}\mskip 3.0mu\mathbf{where}}\<[E]\\
\>[2]{\mathbf{type}\mskip 3.0muT_{M_{\mathsf{Spherical}}}\mskip 3.0mu\mathnormal{=}\mskip 3.0mu\mathsf{Atom}\mskip 3.0mu\mathsf{Spherical}}\<[E]\\
\>[2]{g\mskip 3.0mu\mathnormal{=}\mskip 3.0mu\mathsf{Tab}\mskip 3.0mu\allowbreak{}\mathnormal{(}\mskip 0.0muλ\mskip 3.0mu\allowbreak{}\mathnormal{(}\mskip 0.0mu\mathsf{Atom}\mskip 3.0mu\mathsf{i}\mskip 0.0mu\mathnormal{,}\mskip 3.0mu\mathsf{Atom}\mskip 3.0mu\mathsf{j}\mskip 0.0mu\mathnormal{)}\allowbreak{}\mskip 3.0mu\mathnormal{\_}\mskip 3.0mu\mathnormal{\rightarrow }\mskip 3.0mu\mathsf{schwarzschild}\mskip 3.0mu\mathsf{i}\mskip 3.0mu\mathsf{j}\mskip 0.0mu\mathnormal{)}\allowbreak{}}\<[E]\end{parray}}\end{list} 
At this point, we can directly evaluate \(\mathsf{grLhsM}\) with \(z\mskip 3.0mu\mathnormal{=}\mskip 3.0mu\mathsf{Matrix}\mskip 3.0mu\mathsf{Spherical}\), and
obtain a 4 by 4 matrix of symbolic expressions depending on \(\mathsf{Spherical}\) coordinate variables.
We find that it simplifies to zero everywhere it is defined.\footnote{The metric has two singularities, one at the origin and one at \(ρ=r_s\), the event horizon. It satisfies \cref{266} everywhere else.\label{393}} Thus we can verify that the Schwarzschild metric satisfies the general relativity equation.

\section{Implementation of {\sc{}Albert}}\label{394} 
In this section we explain the implementation of the combinators of
{\sc{}Albert} (\cref{263} and \cref{264}).
As outlined in \cref{1} and in first approximation, it
is provided by the library for symmetric
monoidal categories of \citet{bernardy_evaluating_2021}.  The key idea
is that every linear function \(∀\mskip 3.0mu\mathsf{r}\mskip 1.0mu.\mskip 3.0mu\mathsf{P}\mskip 3.0muz\mskip 3.0mu\mathsf{r}\mskip 3.0mu\mathsf{a}\mskip 3.0mu\mathnormal{⊸}\mskip 3.0mu\mathsf{P}\mskip 3.0muz\mskip 3.0mu\mathsf{r}\mskip 3.0mu\mathsf{b}\) can be converted (naturally) to a morphism of type \(\mathsf{a}\mskip 1.0mu\overset{z}{\leadsto }\mskip 1.0mu\mathsf{b}\),
and back:
\begin{list}{}{\setlength\leftmargin{1.0em}}\item\relax
\ensuremath{\begin{parray}\column{B}{@{}>{}l<{}@{}}\column[0em]{1}{@{}>{}l<{}@{}}\column{2}{@{}>{}l<{}@{}}\column{3}{@{}>{}l<{}@{}}\column{4}{@{}>{}l<{}@{}}\column{5}{@{}>{}l<{}@{}}\column{6}{@{}>{}l<{}@{}}\column{7}{@{}>{}l<{}@{}}\column{E}{@{}>{}l<{}@{}}%
\>[1]{\mathsf{encode}\mskip 3.0mu}\>[2]{\mathnormal{::}\mskip 3.0mu}\>[3]{\mathsf{SymmetricMonoidal}\mskip 3.0muz\mskip 3.0mu\mathnormal{\Rightarrow }\mskip 3.0mu}\>[4]{\allowbreak{}\mathnormal{(}\mskip 0.0mu\mathsf{a}\mskip 1.0mu\overset{z}{\leadsto }\mskip 1.0mu\mathsf{b}\mskip 0.0mu\mathnormal{)}\allowbreak{}\mskip 3.0mu\mathnormal{\rightarrow }\mskip 3.0mu\allowbreak{}\mathnormal{(}\mskip 0.0mu∀\mskip 3.0mu\mathsf{r}\mskip 1.0mu.\mskip 3.0mu}\>[6]{\mathsf{P}\mskip 3.0muz\mskip 3.0mu\mathsf{r}\mskip 3.0mu\mathsf{a}\mskip 3.0mu\mathnormal{⊸}\mskip 3.0mu\mathsf{P}\mskip 3.0muz\mskip 3.0mu\mathsf{r}\mskip 3.0mu\mathsf{b}\mskip 0.0mu\mathnormal{)}\allowbreak{}}\<[E]\\
\>[1]{\mathsf{decode}\mskip 3.0mu}\>[2]{\mathnormal{::}\mskip 3.0mu}\>[5]{\mathsf{SymmetricMonoidal}\mskip 3.0muz\mskip 3.0mu\mathnormal{\Rightarrow }\mskip 3.0mu\allowbreak{}\mathnormal{(}\mskip 0.0mu∀\mskip 3.0mu\mathsf{r}\mskip 1.0mu.\mskip 3.0mu}\>[7]{\mathsf{P}\mskip 3.0muz\mskip 3.0mu\mathsf{r}\mskip 3.0mu\mathsf{a}\mskip 3.0mu\mathnormal{⊸}\mskip 3.0mu\mathsf{P}\mskip 3.0muz\mskip 3.0mu\mathsf{r}\mskip 3.0mu\mathsf{b}\mskip 0.0mu\mathnormal{)}\allowbreak{}\mskip 3.0mu\mathnormal{\rightarrow }\mskip 3.0mu\allowbreak{}\mathnormal{(}\mskip 0.0mu\mathsf{a}\mskip 1.0mu\overset{z}{\leadsto }\mskip 1.0mu\mathsf{b}\mskip 0.0mu\mathnormal{)}\allowbreak{}}\<[E]\end{parray}}\end{list} 
In fact, any {\sc{}smc} \(z\) is isomorphic to the category of Haskell
linear functions between corresponding ports (whose hom-set family is
\(\mathsf{Hom}\mskip 3.0mu\allowbreak{}\mathnormal{(}\mskip 0.0mu\mathsf{a}\mskip 0.0mu\mathnormal{,}\mskip 3.0mu\mathsf{b}\mskip 0.0mu\mathnormal{)}\allowbreak{}\mskip 3.0mu\mathnormal{=}\mskip 3.0mu\mathnormal{∀}\mskip 3.0mu\mathsf{r}\mskip 1.0mu.\mskip 3.0mu\mathsf{P}\mskip 3.0muz\mskip 3.0mu\mathsf{r}\mskip 3.0mu\mathsf{a}\mskip 3.0mu\mathnormal{⊸}\mskip 3.0mu\mathsf{P}\mskip 3.0muz\mskip 3.0mu\mathsf{r}\mskip 3.0mu\mathsf{b}\)). This works even if
\(z\) has additional structure, and in particular if it is a
tensor category.

So, if we were to choose \(\mathsf{R}\mskip 3.0muz\mskip 3.0mu\mathsf{r}\) to be \(\mathsf{P}\mskip 3.0muz\mskip 3.0mu\mathsf{r}\mskip 3.0mu\mathbf{1}\), the
\(\mathsf{encode}\) and \(\mathsf{decode}\) functions would take care of most of
the respective tasks of \(\mathsf{tensorEmbed}\) and \(\mathsf{tensorEval}\). It
would remain to dualise the codomain and apply unitors appropriately,
as shown in \cref{91}.

With this simple setup, we can implement the embedding of tensors
(including δ), as well as multiplication. Unfortunately, there are several complications.

\subsection{Complication: supporting derivatives}\label{395} 
The main issue with the above implementation sketch arises when trying
to implement the covariant derivative of a tensor expression \(\mathsf{t}\) as a sum of its partial derivative and affinity terms.  Indeed, the
affinity terms of a tensor expression \(\mathsf{t}\) depends on the number
(and type) of free index variables in it. But this information is not
made available by the library of \citet{bernardy_evaluating_2021} which we base our work upon. Simply put, the type \(\mathsf{P}\mskip 3.0muz\mskip 3.0mu\mathsf{r}\mskip 3.0mu\mathbf{1}\) is opaque. To explain the solution that we employ, we need to first
peek inside the implementation of this library.

Its principle is that a port (in our application, an index) is represented as a
morphism in the (free) Cartesian extension of the underlying category
\(z\): \begin{list}{}{\setlength\leftmargin{1.0em}}\item\relax
\ensuremath{\begin{parray}\column{B}{@{}>{}l<{}@{}}\column[0em]{1}{@{}>{}l<{}@{}}\column{2}{@{}>{}l<{}@{}}\column{3}{@{}>{}l<{}@{}}\column{E}{@{}>{}l<{}@{}}%
\>[1]{\mathbf{data}\mskip 3.0mu\mathsf{P}\mskip 3.0muz\mskip 3.0mu\mathsf{r}\mskip 3.0mu\mathsf{a}\mskip 3.0mu}\>[2]{\mathbf{where}\mskip 3.0mu\mathsf{P}\mskip 3.0mu\mathnormal{::}\mskip 3.0mu\mathsf{CartesianExt}\mskip 3.0muz\mskip 3.0mu}\>[3]{\mathsf{r}\mskip 3.0mu\mathsf{a}\mskip 3.0mu\mathnormal{\rightarrow }\mskip 3.0mu\mathsf{P}\mskip 3.0muz\mskip 3.0mu\mathsf{r}\mskip 3.0mu\mathsf{a}}\<[E]\end{parray}}\end{list} The \(\mathsf{encode}\) function simply embeds the morphism in the Cartesian extension. The
\(\mathsf{decode}\) function takes a linear function \(\mathsf{f}\mskip 3.0mu\mathnormal{:}\mskip 3.0mu\mathsf{P}\mskip 3.0muz\mskip 3.0mu\mathsf{r}\mskip 3.0mu\mathsf{a}\mskip 3.0mu\mathnormal{⊸}\mskip 3.0mu\mathsf{P}\mskip 3.0muz\mskip 3.0mu\mathsf{r}\mskip 3.0mu\mathsf{b}\) and applies it to the identity morphism (\(\mathsf{id}\mskip 3.0mu\mathnormal{:}\mskip 3.0mu\mathsf{P}\mskip 3.0muz\mskip 3.0mu\mathsf{a}\mskip 3.0mu\mathsf{a}\)),
obtaining \(\mathsf{f}\mskip 3.0mu\allowbreak{}\mathnormal{(}\mskip 0.0mu\mathsf{P}\mskip 3.0mu\mathsf{id}\mskip 0.0mu\mathnormal{)}\allowbreak{}\mskip 3.0mu\mathnormal{:}\mskip 3.0mu\mathsf{CartesianExt}\mskip 3.0muz\mskip 3.0mu\mathsf{a}\mskip 3.0mu\mathsf{b}\).  The non-obvious
property proven by \citet{bernardy_evaluating_2021} is that, because
\(\mathsf{f}\) is linear, \(\mathsf{f}\mskip 3.0mu\allowbreak{}\mathnormal{(}\mskip 0.0mu\mathsf{P}\mskip 3.0mu\mathsf{id}\mskip 0.0mu\mathnormal{)}\allowbreak{}\) is always equivalent to a morphism
in the underlying category \(z\); it never needs to refer to projections or duplications.
The function which recovers this morphism has the following signature, but remember that it will crash if the input morphism makes essential use of the Cartesian structure:
\begin{list}{}{\setlength\leftmargin{1.0em}}\item\relax
\ensuremath{\begin{parray}\column{B}{@{}>{}l<{}@{}}\column[0em]{1}{@{}>{}l<{}@{}}\column{2}{@{}>{}l<{}@{}}\column{3}{@{}>{}l<{}@{}}\column{4}{@{}>{}l<{}@{}}\column{E}{@{}>{}l<{}@{}}%
\>[1]{\mathsf{cartesianToMonoidal}\mskip 3.0mu}\>[2]{\mathnormal{::}\mskip 3.0mu}\>[3]{\mathsf{CartesianExt}\mskip 3.0muz\mskip 3.0mu}\>[4]{\mathsf{a}\mskip 3.0mu\mathsf{b}\mskip 3.0mu\mathnormal{\rightarrow }\mskip 3.0mu\mathsf{a}\mskip 1.0mu\overset{z}{\leadsto }\mskip 1.0mu\mathsf{b}}\<[E]\end{parray}}\end{list} To be able to support tensor derivatives, we cannot rely on \(\mathsf{decode}\), and we will have to access the internal function \(\mathsf{cartesianToMonoidal}\) directly.

With this in mind, we can solve our problem, namely finding an implementation for \(\mathsf{R}\mskip 3.0muz\mskip 3.0mu\mathsf{r}\) which lets us track free variables.  We do this by embedding
a pair of morphisms \(\mathsf{t}\mskip 3.0mu\mathnormal{:}\mskip 3.0mu\mathsf{x}\mskip 1.0mu\overset{z}{\leadsto }\mskip 1.0mu\mathbf{1}\) and
\(\mathsf{p}\mskip 3.0mu\mathnormal{:}\mskip 3.0mu\mathsf{CartesianExt}\mskip 3.0muz\mskip 3.0mu\mathsf{r}\mskip 3.0mu\mathsf{x}\) (instead of just \(\mathsf{p}\)).
Note that the type \(\mathsf{x}\) is existentially bound, not a parameter to \(\mathsf{R}\).
\begin{list}{}{\setlength\leftmargin{1.0em}}\item\relax
\ensuremath{\begin{parray}\column{B}{@{}>{}l<{}@{}}\column[0em]{1}{@{}>{}l<{}@{}}\column{2}{@{}>{}l<{}@{}}\column{3}{@{}>{}l<{}@{}}\column{4}{@{}>{}l<{}@{}}\column{E}{@{}>{}l<{}@{}}%
\>[1]{\mathbf{data}\mskip 3.0mu\mathsf{R}\mskip 3.0muz\mskip 3.0mu\mathsf{r}\mskip 3.0mu\mathbf{where}\mskip 3.0mu}\>[2]{\mathsf{Compose}\mskip 3.0mu\mathnormal{::}\mskip 3.0mu}\>[3]{\allowbreak{}\mathnormal{(}\mskip 0.0mu\mathsf{x}\mskip 1.0mu\overset{z}{\leadsto }\mskip 1.0mu\mathbf{1}\mskip 0.0mu\mathnormal{)}\allowbreak{}\mskip 3.0mu\mathnormal{\rightarrow }\mskip 3.0mu\mathsf{CartesianExt}\mskip 3.0muz\mskip 3.0mu}\>[4]{\mathsf{r}\mskip 3.0mu\mathsf{x}\mskip 3.0mu\mathnormal{\rightarrow }\mskip 3.0mu\mathsf{R}\mskip 3.0muz\mskip 3.0mu\mathsf{r}}\<[E]\end{parray}}\end{list} Here, \(\mathsf{x}\) is the context used by the tensor
\(\mathsf{t}\), without any spurious component. The tensor \(\mathsf{t}\) carries the payload of the expression. Because
it inhabits the non-Cartesian category \(z\), we know that it uses the whole context
\(\mathsf{x}\). Even though the type declares that the morphism \(\mathsf{p}\) is built from the Cartesian extension
of \(z\), the implementation will arrange that it will only
use the sub-category of \(z\) whose derivative is zero. This is enforced by
hiding the \(\mathsf{encode}\) function on \(\mathsf{P}\mskip 3.0muz\mskip 3.0mu\mathsf{r}\mskip 3.0mu\mathsf{a}\) from the user: we only
provide the \(\mathsf{raise}\) and \(\mathsf{lower}\) functions to
manipulate this type, and they are safe because \(∇g=0\):
\begin{list}{}{\setlength\leftmargin{1.0em}}\item\relax
\ensuremath{\begin{parray}\column{B}{@{}>{}l<{}@{}}\column[0em]{1}{@{}>{}l<{}@{}}\column{2}{@{}>{}l<{}@{}}\column{E}{@{}>{}l<{}@{}}%
\>[1]{\mathsf{raise}\mskip 3.0mu}\>[2]{\mathnormal{=}\mskip 3.0mu\mathsf{encode}\mskip 3.0mu\mathsf{juggleUp}}\<[E]\\
\>[1]{\mathsf{lower}\mskip 3.0mu}\>[2]{\mathnormal{=}\mskip 3.0mu\mathsf{encode}\mskip 3.0mu\mathsf{juggleDown}}\<[E]\end{parray}}\end{list} 
With this setup the derivative can be computed correctly on a
pair \(\mathsf{Compose}\mskip 3.0mu\mathsf{t}\mskip 3.0mu\mathsf{p}\).
Indeed, the morphism-level derivative
(\ensuremath{\begin{parray}\column{B}{@{}>{}l<{}@{}}\column[0em]{1}{@{}>{}l<{}@{}}\column{2}{@{}>{}l<{}@{}}\column{3}{@{}>{}l<{}@{}}\column{E}{@{}>{}l<{}@{}}%
\>[1]{∇\mskip 3.0mu}\>[2]{\mathnormal{::}\mskip 3.0mu}\>[3]{\allowbreak{}\mathnormal{(}\mskip 0.0mu\mathsf{a}\mskip 1.0mu\overset{z}{\leadsto }\mskip 1.0mu\mathsf{b}\mskip 0.0mu\mathnormal{)}\allowbreak{}\mskip 3.0mu\mathnormal{\rightarrow }\mskip 3.0mu\allowbreak{}\mathnormal{(}\mskip 0.0mu\allowbreak{}\mathnormal{(}\mskip 0.0mu\mathsf{v}\mskip 3.0mu\mathnormal{⊗}\mskip 3.0mu\mathsf{a}\mskip 0.0mu\mathnormal{)}\allowbreak{}\mskip 1.0mu\overset{z}{\leadsto }\mskip 1.0mu\mathsf{b}\mskip 0.0mu\mathnormal{)}\allowbreak{}}\<[E]\end{parray}}) need only
be applied to the \(\mathsf{t}\) component of the pair. As for the \(\mathsf{p}\) component of the pair, the use of the
Cartesian fork operator \(\allowbreak{}\mathnormal{(}\mskip 0.0mu\mathnormal{▵}\mskip 0.0mu\mathnormal{)}\allowbreak{}\) ensures that the new index is
passed to the right component of \(∇\mskip 3.0mu\mathsf{t}\).

\begin{list}{}{\setlength\leftmargin{1.0em}}\item\relax
\ensuremath{\begin{parray}\column{B}{@{}>{}l<{}@{}}\column[0em]{1}{@{}>{}l<{}@{}}\column{2}{@{}>{}l<{}@{}}\column{E}{@{}>{}l<{}@{}}%
\>[1]{\mathsf{deriv}\mskip 3.0mu\mathnormal{::}\mskip 3.0mu}\>[2]{\allowbreak{}\mathnormal{(}\mskip 0.0mu\mathsf{ConnectionCategory}\mskip 3.0muz\mskip 0.0mu\mathnormal{)}\allowbreak{}\mskip 3.0mu\mathnormal{\Rightarrow }\mskip 3.0mu\mathsf{P}\mskip 3.0muz\mskip 3.0mu\mathsf{r}\mskip 3.0mu\mathsf{v}\mskip 3.0mu\mathnormal{⊸}\mskip 3.0mu\mathsf{R}\mskip 3.0muz\mskip 3.0mu\mathsf{r}\mskip 3.0mu\mathnormal{⊸}\mskip 3.0mu\mathsf{R}\mskip 3.0muz\mskip 3.0mu\mathsf{r}}\<[E]\\
\>[1]{\mathsf{deriv}\mskip 3.0mu\allowbreak{}\mathnormal{(}\mskip 0.0mu\mathsf{P}\mskip 3.0mu\mathsf{i}\mskip 0.0mu\mathnormal{)}\allowbreak{}\mskip 3.0mu\allowbreak{}\mathnormal{(}\mskip 0.0mu\mathsf{Compose}\mskip 3.0mu\mathsf{t}\mskip 3.0mu\mathsf{p}\mskip 0.0mu\mathnormal{)}\allowbreak{}\mskip 3.0mu\mathnormal{=}\mskip 3.0mu\mathsf{Compose}\mskip 3.0mu\allowbreak{}\mathnormal{(}\mskip 0.0mu∇\mskip 3.0mu\mathsf{t}\mskip 0.0mu\mathnormal{)}\allowbreak{}\mskip 3.0mu\allowbreak{}\mathnormal{(}\mskip 0.0mu\mathsf{i}\mskip 3.0mu\mathnormal{▵}\mskip 3.0mu\mathsf{p}\mskip 0.0mu\mathnormal{)}\allowbreak{}}\<[E]\end{parray}}\end{list} \begin{list}{}{\setlength\leftmargin{1.0em}}\item\relax
\ensuremath{\begin{parray}\column{B}{@{}>{}l<{}@{}}\column[0em]{1}{@{}>{}l<{}@{}}\column{2}{@{}>{}l<{}@{}}\column{3}{@{}>{}l<{}@{}}\column{4}{@{}>{}l<{}@{}}\column{E}{@{}>{}l<{}@{}}%
\>[1]{\allowbreak{}\mathnormal{(}\mskip 0.0mu\mathnormal{▵}\mskip 0.0mu\mathnormal{)}\allowbreak{}\mskip 3.0mu\mathnormal{::}\mskip 3.0mu}\>[2]{\allowbreak{}\mathnormal{(}\mskip 0.0mu\mathsf{Cartesian}\mskip 3.0muz\mskip 0.0mu}\>[3]{\mathnormal{)}\allowbreak{}\mskip 3.0mu\mathnormal{\Rightarrow }\mskip 3.0mu}\>[4]{\allowbreak{}\mathnormal{(}\mskip 0.0mu\mathsf{a}\mskip 1.0mu\overset{z}{\leadsto }\mskip 1.0mu\mathsf{b}\mskip 0.0mu\mathnormal{)}\allowbreak{}\mskip 3.0mu\mathnormal{\rightarrow }\mskip 3.0mu\allowbreak{}\mathnormal{(}\mskip 0.0mu\mathsf{a}\mskip 1.0mu\overset{z}{\leadsto }\mskip 1.0mu\mathsf{c}\mskip 0.0mu\mathnormal{)}\allowbreak{}\mskip 3.0mu\mathnormal{\rightarrow }\mskip 3.0mu\mathsf{a}\mskip 1.0mu\overset{z}{\leadsto }\mskip 1.0mu\allowbreak{}\mathnormal{(}\mskip 0.0mu\mathsf{b}\mskip 3.0mu\mathnormal{⊗}\mskip 3.0mu\mathsf{c}\mskip 0.0mu\mathnormal{)}\allowbreak{}}\<[E]\end{parray}}\end{list} 
The transcoding functions between {\sc{}Albert} and its categorical
semantics then become: \begin{list}{}{\setlength\leftmargin{1.0em}}\item\relax
\ensuremath{\begin{parray}\column{B}{@{}>{}l<{}@{}}\column[0em]{1}{@{}>{}l<{}@{}}\column{2}{@{}>{}l<{}@{}}\column{3}{@{}>{}l<{}@{}}\column{E}{@{}>{}l<{}@{}}%
\>[1]{\mathsf{embedTensor}_{1}\mskip 3.0mu}\>[2]{\mathnormal{::}\mskip 3.0mu}\>[3]{\allowbreak{}\mathnormal{(}\mskip 0.0mu\mathsf{a}\mskip 1.0mu\overset{z}{\leadsto }\mskip 1.0mu\mathbf{1}\mskip 0.0mu\mathnormal{)}\allowbreak{}\mskip 3.0mu\mathnormal{\rightarrow }\mskip 3.0mu\mathsf{P}\mskip 3.0muz\mskip 3.0mu\mathsf{r}\mskip 3.0mu\mathsf{a}\mskip 3.0mu\mathnormal{⊸}\mskip 3.0mu\mathsf{R}\mskip 3.0muz\mskip 3.0mu\mathsf{r}}\<[E]\\
\>[1]{\mathsf{embedTensor}_{1}\mskip 3.0mu\mathsf{t}\mskip 3.0mu\mathsf{p}\mskip 3.0mu\mathnormal{=}\mskip 3.0mu\mathsf{Compose}\mskip 3.0mu\mathsf{t}\mskip 3.0mu\mathsf{p}}\<[E]\end{parray}}\end{list} \begin{list}{}{\setlength\leftmargin{1.0em}}\item\relax
\ensuremath{\begin{parray}\column{B}{@{}>{}l<{}@{}}\column[0em]{1}{@{}>{}l<{}@{}}\column[1em]{2}{@{}>{}l<{}@{}}\column{3}{@{}>{}l<{}@{}}\column{4}{@{}>{}l<{}@{}}\column{5}{@{}>{}l<{}@{}}\column{6}{@{}>{}l<{}@{}}\column{E}{@{}>{}l<{}@{}}%
\>[1]{\mathsf{evalTensor}_{1}\mskip 3.0mu\mathnormal{::}\mskip 3.0mu\allowbreak{}\mathnormal{(}\mskip 0.0mu}\>[5]{\mathsf{SymmetricMonoidal}\mskip 3.0muz\mskip 0.0mu\mathnormal{)}\allowbreak{}\mskip 3.0mu\mathnormal{\Rightarrow }\mskip 3.0mu\allowbreak{}\mathnormal{(}\mskip 0.0mu∀\mskip 3.0mu\mathsf{r}\mskip 1.0mu.\mskip 3.0mu}\>[6]{\mathsf{P}\mskip 3.0muz\mskip 3.0mu\mathsf{r}\mskip 3.0mu\mathsf{a}\mskip 3.0mu\mathnormal{⊸}\mskip 3.0mu\mathsf{R}\mskip 3.0muz\mskip 3.0mu\mathsf{r}\mskip 0.0mu\mathnormal{)}\allowbreak{}\mskip 3.0mu\mathnormal{\rightarrow }\mskip 3.0mu\mathsf{a}\mskip 1.0mu\overset{z}{\leadsto }\mskip 1.0mu\mathbf{1}}\<[E]\\
\>[1]{\mathsf{evalTensor}_{1}\mskip 3.0mu\mathsf{f}\mskip 3.0mu\mathnormal{=}\mskip 3.0mu\mathsf{decoder}\mskip 3.0mu\allowbreak{}\mathnormal{(}\mskip 0.0mu\mathsf{f}\mskip 3.0mu\allowbreak{}\mathnormal{(}\mskip 0.0mu\mathsf{P}\mskip 3.0mu\mathsf{id}\mskip 0.0mu\mathnormal{)}\allowbreak{}\mskip 0.0mu\mathnormal{)}\allowbreak{}\mskip 3.0mu\mathbf{where}}\<[E]\\
\>[2]{\mathsf{decoder}\mskip 3.0mu\mathnormal{::}\mskip 3.0mu\allowbreak{}\mathnormal{(}\mskip 0.0mu}\>[4]{\mathsf{SymmetricMonoidal}\mskip 3.0muz\mskip 0.0mu\mathnormal{)}\allowbreak{}\mskip 3.0mu\mathnormal{\Rightarrow }\mskip 3.0mu\mathsf{R}\mskip 3.0muz\mskip 3.0mu\mathsf{a}\mskip 3.0mu\mathnormal{\rightarrow }\mskip 3.0mu\mathsf{a}\mskip 1.0mu\overset{z}{\leadsto }\mskip 1.0mu\mathbf{1}}\<[E]\\
\>[2]{\mathsf{decoder}\mskip 3.0mu\allowbreak{}\mathnormal{(}\mskip 0.0mu\mathsf{Compose}\mskip 3.0mu\mathsf{t}\mskip 3.0mu\mathsf{p}\mskip 0.0mu\mathnormal{)}\allowbreak{}\mskip 3.0mu}\>[3]{\mathnormal{=}\mskip 3.0mu\mathsf{t}\mskip 3.0mu\allowbreak{}\mathnormal{∘}\allowbreak{}\mskip 3.0mu\mathsf{cartesianToMonoidal}\mskip 3.0mu\mathsf{p}}\<[E]\end{parray}}\end{list} 
Multiplication remains straightforward to implement:
\begin{list}{}{\setlength\leftmargin{1.0em}}\item\relax
\ensuremath{\begin{parray}\column{B}{@{}>{}l<{}@{}}\column[0em]{1}{@{}>{}l<{}@{}}\column{E}{@{}>{}l<{}@{}}%
\>[1]{\mathsf{Compose}\mskip 3.0mu\mathsf{t}_{1}\mskip 3.0mu\mathsf{p}_{1}\mskip 3.0mu{\tikzstar{0.11}{0.25}{5}{-18}{fill=black}}\mskip 3.0mu\mathsf{Compose}\mskip 3.0mu\mathsf{t}_{2}\mskip 3.0mu\mathsf{p}_{2}\mskip 3.0mu\mathnormal{=}\mskip 3.0mu\mathsf{Compose}\mskip 3.0mu\allowbreak{}\mathnormal{(}\mskip 0.0mu\bar{ρ}\mskip 3.0mu\allowbreak{}\mathnormal{∘}\allowbreak{}\mskip 3.0mu\allowbreak{}\mathnormal{(}\mskip 0.0mu\mathsf{t}_{1}\mskip 3.0mu{⊗}\mskip 3.0mu\mathsf{t}_{2}\mskip 0.0mu\mathnormal{)}\allowbreak{}\mskip 0.0mu\mathnormal{)}\allowbreak{}\mskip 3.0mu\allowbreak{}\mathnormal{(}\mskip 0.0mu\mathsf{p}_{1}\mskip 3.0mu\mathnormal{▵}\mskip 3.0mu\mathsf{p}_{2}\mskip 0.0mu\mathnormal{)}\allowbreak{}}\<[E]\end{parray}}\end{list} 
\subsection{Complication: supporting addition of tensors}\label{396} 
The type of the addition operator (\(\mathsf{plus}\mskip 3.0mu\mathnormal{::}\mskip 3.0mu\allowbreak{}\mathnormal{(}\mskip 0.0mu\mathsf{Bool}\mskip 3.0mu\mathnormal{\rightarrow }\mskip 3.0mu\mathsf{R}\mskip 3.0muz\mskip 3.0mu\mathsf{r}\mskip 0.0mu\mathnormal{)}\allowbreak{}\mskip 3.0mu\mathnormal{⊸}\mskip 3.0mu\mathsf{R}\mskip 3.0muz\mskip 3.0mu\mathsf{r}\)) poses a problem for the above implementation. The issue is
that the semantics of \(\mathsf{plus}\mskip 3.0mu\mathsf{f}\) depends on both \(\mathsf{f}\mskip 3.0mu\mathsf{False}\) and
\(\mathsf{f}\mskip 3.0mu\mathsf{True}\). However, \(\mathsf{plus}\) is linear in \(\mathsf{f}\), and thus
we can call \(\mathsf{f}\) only once in the implementation of
\(\mathsf{plus}\). The way to work around this problem is to embed a
non-linear implementation inside a linear datatype. That is, it suffices
to add a \emph{constructor} \(\mathsf{Plus}\mskip 3.0mu\mathnormal{::}\mskip 3.0mu\allowbreak{}\mathnormal{(}\mskip 0.0mu\mathsf{Bool}\mskip 3.0mu\mathnormal{\rightarrow }\mskip 3.0mu\mathsf{R}\mskip 3.0muz\mskip 3.0mu\mathsf{r}\mskip 0.0mu\mathnormal{)}\allowbreak{}\mskip 3.0mu\mathnormal{⊸}\mskip 3.0mu\mathsf{R}\mskip 3.0muz\mskip 3.0mu\mathsf{r}\) to the \(\mathsf{R}\) datatype.  Operationally, if \(\mathsf{f}\) is the argument
to \(\mathsf{plus}\), it is \emph{stored once} in the \(\mathsf{R}\) data
structure, and it is only at the moment of evaluating \(\mathsf{plus}\mskip 3.0mu\mathsf{f}\) to a
morphism that we invoke \(\mathsf{f}\). But, at the evaluation point, there is
no linearity restriction on the tensor expression of type \(\mathsf{R}\mskip 3.0muz\mskip 3.0mu\mathsf{r}\): it can be
used an arbitrary number of times, and, according to the typing rules
of Linear Haskell, it means that its payload (\(\mathsf{f}\)) has itself no
usage restrictions.  The fact that the evaluation function has no linearity restriction
on its tensor expression argument (\(\mathsf{t}\)) means that \(\mathsf{t}\) cannot have any free index variables in
it.  At first sight, this is a drawback.  However there is a good reason for restricting evaluation to
closed tensor expressions: it ensures that index variables do not escape
the scope of the tensor expression where they belong.

The rest of the implementation of {\sc{}Albert} needs to be modified to
support this new constructor, by defining a case for sums. Fortunately
such a case is never difficult to handle: we simply distribute every
operation over the operands of the sum.

\subsection{Complication: contraction}\label{397} 
Unfortunately, our definition for \(\mathsf{R}\) so far does not in fact
support contraction.  This is because all indices are inputs of the
expression, and to connect inputs together we need to use the \(η\) combinator. To do so, we need an additional \(\mathbf{1}\) input--- but
no such input is explicitly available in the \(\mathsf{R}\) type. The
workaround for the problem is to add such a unit input
explicitly. Hence the final implementation of the \(\mathsf{R}\) type is:
\begin{list}{}{\setlength\leftmargin{1.0em}}\item\relax
\ensuremath{\begin{parray}\column{B}{@{}>{}l<{}@{}}\column[0em]{1}{@{}>{}l<{}@{}}\column[1em]{2}{@{}>{}l<{}@{}}\column{3}{@{}>{}l<{}@{}}\column{4}{@{}>{}l<{}@{}}\column{5}{@{}>{}l<{}@{}}\column{6}{@{}>{}l<{}@{}}\column{E}{@{}>{}l<{}@{}}%
\>[1]{\mathbf{data}\mskip 3.0mu\mathsf{S}\mskip 3.0muz\mskip 3.0mu\mathsf{r}\mskip 3.0mu\mathbf{where}}\<[E]\\
\>[2]{\mathsf{Compose}\mskip 3.0mu}\>[4]{\mathnormal{::}\mskip 3.0mu}\>[5]{\allowbreak{}\mathnormal{(}\mskip 0.0mu\mathsf{x}\mskip 1.0mu\overset{z}{\leadsto }\mskip 1.0mu\mathbf{1}\mskip 0.0mu\mathnormal{)}\allowbreak{}\mskip 3.0mu\mathnormal{\rightarrow }\mskip 3.0mu\mathsf{CartesianExt}\mskip 3.0muz\mskip 3.0mu}\>[6]{\mathsf{r}\mskip 3.0mu\mathsf{x}\mskip 3.0mu\mathnormal{\rightarrow }\mskip 3.0mu\mathsf{S}\mskip 3.0muz\mskip 3.0mu\mathsf{r}}\<[E]\\
\>[2]{\mathsf{Plus}\mskip 3.0mu}\>[3]{\mathnormal{::}\mskip 3.0mu\allowbreak{}\mathnormal{(}\mskip 0.0mu\mathsf{Bool}\mskip 3.0mu\mathnormal{\rightarrow }\mskip 3.0mu\mathsf{S}\mskip 3.0muz\mskip 3.0mu\mathsf{r}\mskip 0.0mu\mathnormal{)}\allowbreak{}\mskip 3.0mu\mathnormal{⊸}\mskip 3.0mu\mathsf{S}\mskip 3.0muz\mskip 3.0mu\mathsf{r}}\<[E]\\
\>[1]{\mathbf{type}\mskip 3.0mu\mathsf{R}\mskip 3.0muz\mskip 3.0mu\mathsf{r}\mskip 3.0mu\mathnormal{=}\mskip 3.0mu\mathsf{P}\mskip 3.0muz\mskip 3.0mu\mathsf{r}\mskip 3.0mu\mathbf{1}\mskip 3.0mu\mathnormal{⊸}\mskip 3.0mu\mathsf{S}\mskip 3.0muz\mskip 3.0mu\mathsf{r}}\<[E]\end{parray}}\end{list} The implementation of \(\mathsf{contract}\) is then: \begin{list}{}{\setlength\leftmargin{1.0em}}\item\relax
\ensuremath{\begin{parray}\column{B}{@{}>{}l<{}@{}}\column[0em]{1}{@{}>{}l<{}@{}}\column{E}{@{}>{}l<{}@{}}%
\>[1]{\mathsf{contract}\mskip 3.0mu\mathsf{f}\mskip 3.0mu\mathsf{u}\mskip 3.0mu\mathnormal{=}\mskip 3.0mu\mathsf{uncurry}\mskip 3.0mu\allowbreak{}\mathnormal{(}\mskip 0.0mu\mathsf{uncurry}\mskip 3.0mu\mathsf{f}\mskip 0.0mu\mathnormal{)}\allowbreak{}\mskip 3.0mu\allowbreak{}\mathnormal{(}\mskip 0.0mu\mathsf{encode}\mskip 3.0mu\allowbreak{}\mathnormal{(}\mskip 0.0mu\allowbreak{}\mathnormal{(}\mskip 0.0muη\mskip 3.0mu{⊗}\mskip 3.0mu\mathsf{id}\mskip 0.0mu\mathnormal{)}\allowbreak{}\mskip 3.0mu\allowbreak{}\mathnormal{∘}\allowbreak{}\mskip 3.0muρ\mskip 0.0mu\mathnormal{)}\allowbreak{}\mskip 3.0mu\mathsf{u}\mskip 0.0mu\mathnormal{)}\allowbreak{}}\<[E]\end{parray}}\end{list} The other combinators can simply ignore or thread this unit input. The core of the final implementation is shown in \cref{398}.

\begin{figure}[] 
\textbf{-- -- Part of the implementation borrowed from \citet{bernardy_evaluating_2021} -- --} 
\begin{list}{}{\setlength\leftmargin{1.0em}}\item\relax
 \ensuremath{\begin{parray}\column{B}{@{}>{}l<{}@{}}\column[0em]{1}{@{}>{}l<{}@{}}\column{2}{@{}>{}l<{}@{}}\column{3}{@{}>{}l<{}@{}}\column{4}{@{}>{}l<{}@{}}\column{E}{@{}>{}l<{}@{}}%
\>[1]{\mathsf{cartesianToMonoidal}\mskip 3.0mu}\>[2]{\mathnormal{::}\mskip 3.0mu}\>[3]{\mathsf{CartesianExt}\mskip 3.0muz\mskip 3.0mu}\>[4]{\mathsf{a}\mskip 3.0mu\mathsf{b}\mskip 3.0mu\mathnormal{\rightarrow }\mskip 3.0mu\mathsf{a}\mskip 1.0mu\overset{z}{\leadsto }\mskip 1.0mu\mathsf{b}}\<[E]\end{parray}} 

\ensuremath{\begin{parray}\column{B}{@{}>{}l<{}@{}}\column[0em]{1}{@{}>{}l<{}@{}}\column{2}{@{}>{}l<{}@{}}\column{3}{@{}>{}l<{}@{}}\column{4}{@{}>{}l<{}@{}}\column{5}{@{}>{}l<{}@{}}\column{6}{@{}>{}l<{}@{}}\column{7}{@{}>{}l<{}@{}}\column{E}{@{}>{}l<{}@{}}%
\>[1]{\mathsf{encode}\mskip 3.0mu}\>[2]{\mathnormal{::}\mskip 3.0mu}\>[3]{\mathsf{SymmetricMonoidal}\mskip 3.0muz\mskip 3.0mu\mathnormal{\Rightarrow }\mskip 3.0mu}\>[4]{\allowbreak{}\mathnormal{(}\mskip 0.0mu\mathsf{a}\mskip 1.0mu\overset{z}{\leadsto }\mskip 1.0mu\mathsf{b}\mskip 0.0mu\mathnormal{)}\allowbreak{}\mskip 3.0mu\mathnormal{\rightarrow }\mskip 3.0mu\allowbreak{}\mathnormal{(}\mskip 0.0mu∀\mskip 3.0mu\mathsf{r}\mskip 1.0mu.\mskip 3.0mu}\>[6]{\mathsf{P}\mskip 3.0muz\mskip 3.0mu\mathsf{r}\mskip 3.0mu\mathsf{a}\mskip 3.0mu\mathnormal{⊸}\mskip 3.0mu\mathsf{P}\mskip 3.0muz\mskip 3.0mu\mathsf{r}\mskip 3.0mu\mathsf{b}\mskip 0.0mu\mathnormal{)}\allowbreak{}}\<[E]\\
\>[1]{\mathsf{decode}\mskip 3.0mu}\>[2]{\mathnormal{::}\mskip 3.0mu}\>[5]{\mathsf{SymmetricMonoidal}\mskip 3.0muz\mskip 3.0mu\mathnormal{\Rightarrow }\mskip 3.0mu\allowbreak{}\mathnormal{(}\mskip 0.0mu∀\mskip 3.0mu\mathsf{r}\mskip 1.0mu.\mskip 3.0mu}\>[7]{\mathsf{P}\mskip 3.0muz\mskip 3.0mu\mathsf{r}\mskip 3.0mu\mathsf{a}\mskip 3.0mu\mathnormal{⊸}\mskip 3.0mu\mathsf{P}\mskip 3.0muz\mskip 3.0mu\mathsf{r}\mskip 3.0mu\mathsf{b}\mskip 0.0mu\mathnormal{)}\allowbreak{}\mskip 3.0mu\mathnormal{\rightarrow }\mskip 3.0mu\allowbreak{}\mathnormal{(}\mskip 0.0mu\mathsf{a}\mskip 1.0mu\overset{z}{\leadsto }\mskip 1.0mu\mathsf{b}\mskip 0.0mu\mathnormal{)}\allowbreak{}}\<[E]\end{parray}} 

\ensuremath{\begin{parray}\column{B}{@{}>{}l<{}@{}}\column[0em]{1}{@{}>{}l<{}@{}}\column{2}{@{}>{}l<{}@{}}\column{3}{@{}>{}l<{}@{}}\column{E}{@{}>{}l<{}@{}}%
\>[1]{\mathbf{data}\mskip 3.0mu\mathsf{P}\mskip 3.0muz\mskip 3.0mu\mathsf{r}\mskip 3.0mu\mathsf{a}\mskip 3.0mu}\>[2]{\mathbf{where}\mskip 3.0mu\mathsf{P}\mskip 3.0mu\mathnormal{::}\mskip 3.0mu\mathsf{CartesianExt}\mskip 3.0muz\mskip 3.0mu}\>[3]{\mathsf{r}\mskip 3.0mu\mathsf{a}\mskip 3.0mu\mathnormal{\rightarrow }\mskip 3.0mu\mathsf{P}\mskip 3.0muz\mskip 3.0mu\mathsf{r}\mskip 3.0mu\mathsf{a}}\<[E]\end{parray}} 

\ensuremath{\begin{parray}\column{B}{@{}>{}l<{}@{}}\column[0em]{1}{@{}>{}l<{}@{}}\column{2}{@{}>{}l<{}@{}}\column{3}{@{}>{}l<{}@{}}\column{4}{@{}>{}l<{}@{}}\column{5}{@{}>{}l<{}@{}}\column{E}{@{}>{}l<{}@{}}%
\>[1]{\mathsf{curry}\mskip 3.0mu}\>[2]{\mathnormal{::}\mskip 3.0mu\allowbreak{}\mathnormal{(}\mskip 0.0mu\mathsf{SymmetricMonoidal}\mskip 3.0muz\mskip 0.0mu}\>[3]{\mathnormal{)}\allowbreak{}\mskip 3.0mu}\>[4]{\mathnormal{\Rightarrow }\mskip 3.0mu}\>[5]{\allowbreak{}\mathnormal{(}\mskip 0.0mu\mathsf{P}\mskip 3.0muz\mskip 3.0mu\mathsf{r}\mskip 3.0mu\allowbreak{}\mathnormal{(}\mskip 0.0mu\mathsf{a}\mskip 3.0mu\mathnormal{⊗}\mskip 3.0mu\mathsf{b}\mskip 0.0mu\mathnormal{)}\allowbreak{}\mskip 3.0mu\mathnormal{⊸}\mskip 3.0mu\mathsf{k}\mskip 0.0mu\mathnormal{)}\allowbreak{}\mskip 3.0mu\mathnormal{⊸}\mskip 3.0mu\allowbreak{}\mathnormal{(}\mskip 0.0mu\mathsf{P}\mskip 3.0muz\mskip 3.0mu\mathsf{r}\mskip 3.0mu\mathsf{a}\mskip 3.0mu\mathnormal{⊸}\mskip 3.0mu\mathsf{P}\mskip 3.0muz\mskip 3.0mu\mathsf{r}\mskip 3.0mu\mathsf{b}\mskip 3.0mu\mathnormal{⊸}\mskip 3.0mu\mathsf{k}\mskip 0.0mu\mathnormal{)}\allowbreak{}}\<[E]\\
\>[1]{\mathsf{curry}\mskip 3.0mu\mathsf{f}\mskip 3.0mu\mathsf{a}\mskip 3.0mu\mathsf{b}\mskip 3.0mu\mathnormal{=}\mskip 3.0mu\mathsf{f}\mskip 3.0mu\allowbreak{}\mathnormal{(}\mskip 0.0mu\mathsf{merge}\mskip 3.0mu\allowbreak{}\mathnormal{(}\mskip 0.0mu\mathsf{a}\mskip 0.0mu\mathnormal{,}\mskip 3.0mu\mathsf{b}\mskip 0.0mu\mathnormal{)}\allowbreak{}\mskip 0.0mu\mathnormal{)}\allowbreak{}}\<[E]\end{parray}} 

\ensuremath{\begin{parray}\column{B}{@{}>{}l<{}@{}}\column[0em]{1}{@{}>{}l<{}@{}}\column{2}{@{}>{}l<{}@{}}\column{3}{@{}>{}l<{}@{}}\column{4}{@{}>{}l<{}@{}}\column{5}{@{}>{}l<{}@{}}\column{E}{@{}>{}l<{}@{}}%
\>[1]{\mathsf{uncurry}\mskip 3.0mu}\>[2]{\mathnormal{::}\mskip 3.0mu\allowbreak{}\mathnormal{(}\mskip 0.0mu\mathsf{SymmetricMonoidal}\mskip 3.0muz\mskip 0.0mu}\>[3]{\mathnormal{)}\allowbreak{}\mskip 3.0mu}\>[4]{\mathnormal{\Rightarrow }\mskip 3.0mu}\>[5]{\allowbreak{}\mathnormal{(}\mskip 0.0mu\mathsf{P}\mskip 3.0muz\mskip 3.0mu\mathsf{r}\mskip 3.0mu\mathsf{a}\mskip 3.0mu\mathnormal{⊸}\mskip 3.0mu\mathsf{P}\mskip 3.0muz\mskip 3.0mu\mathsf{r}\mskip 3.0mu\mathsf{b}\mskip 3.0mu\mathnormal{⊸}\mskip 3.0mu\mathsf{k}\mskip 0.0mu\mathnormal{)}\allowbreak{}\mskip 3.0mu\mathnormal{⊸}\mskip 3.0mu\allowbreak{}\mathnormal{(}\mskip 0.0mu\mathsf{P}\mskip 3.0muz\mskip 3.0mu\mathsf{r}\mskip 3.0mu\allowbreak{}\mathnormal{(}\mskip 0.0mu\mathsf{a}\mskip 3.0mu\mathnormal{⊗}\mskip 3.0mu\mathsf{b}\mskip 0.0mu\mathnormal{)}\allowbreak{}\mskip 3.0mu\mathnormal{⊸}\mskip 3.0mu\mathsf{k}\mskip 0.0mu\mathnormal{)}\allowbreak{}}\<[E]\end{parray}} \\ \ensuremath{\begin{parray}\column{B}{@{}>{}l<{}@{}}\column[0em]{1}{@{}>{}l<{}@{}}\column{E}{@{}>{}l<{}@{}}%
\>[1]{\mathsf{uncurry}\mskip 3.0mu\mathsf{f}\mskip 3.0mu\mathsf{p}\mskip 3.0mu\mathnormal{=}\mskip 3.0mu\mathbf{case}\mskip 3.0mu\mathsf{split}\mskip 3.0mu\mathsf{p}\mskip 3.0mu\mathbf{of}\mskip 3.0mu\allowbreak{}\mathnormal{(}\mskip 0.0mu\mathsf{a}\mskip 0.0mu\mathnormal{,}\mskip 3.0mu\mathsf{b}\mskip 0.0mu\mathnormal{)}\allowbreak{}\mskip 3.0mu\mathnormal{\rightarrow }\mskip 3.0mu\mathsf{f}\mskip 3.0mu\mathsf{a}\mskip 3.0mu\mathsf{b}}\<[E]\end{parray}}  \end{list} 
\textbf{-- -- Part of the implementation specific to the present work -- --} \begin{list}{}{\setlength\leftmargin{1.0em}}\item\relax
\ensuremath{\begin{parray}\column{B}{@{}>{}l<{}@{}}\column[0em]{1}{@{}>{}l<{}@{}}\column[1em]{2}{@{}>{}l<{}@{}}\column[2em]{3}{@{}>{}l<{}@{}}\column{4}{@{}>{}l<{}@{}}\column{5}{@{}>{}l<{}@{}}\column{6}{@{}>{}l<{}@{}}\column{7}{@{}>{}l<{}@{}}\column{8}{@{}>{}l<{}@{}}\column{9}{@{}>{}l<{}@{}}\column{10}{@{}>{}l<{}@{}}\column{11}{@{}>{}l<{}@{}}\column{12}{@{}>{}l<{}@{}}\column{13}{@{}>{}l<{}@{}}\column{14}{@{}>{}l<{}@{}}\column{15}{@{}>{}l<{}@{}}\column{16}{@{}>{}l<{}@{}}\column{17}{@{}>{}l<{}@{}}\column{18}{@{}>{}l<{}@{}}\column{19}{@{}>{}l<{}@{}}\column{E}{@{}>{}l<{}@{}}%
\>[1]{\mathsf{zeroTensor}\mskip 3.0mu\mathnormal{=}\mskip 3.0mu\mathsf{tensorEmbed}_{1}\mskip 3.0mu0}\<[E]\\
\>[1]{\mathsf{constant}\mskip 3.0mu\mathsf{s}\mskip 3.0mu\allowbreak{}\mathnormal{(}\mskip 0.0mu\mathsf{P}\mskip 3.0mu\mathsf{u}\mskip 0.0mu\mathnormal{)}\allowbreak{}\mskip 3.0mu\mathnormal{=}\mskip 3.0mu\mathsf{Compose}\mskip 3.0mu\allowbreak{}\mathnormal{(}\mskip 0.0mu\mathsf{s}\mskip 3.0mu\smalltriangleleft \mskip 3.0mu\mathsf{id}\mskip 0.0mu\mathnormal{)}\allowbreak{}\mskip 3.0mu\mathsf{u}}\<[E]\\
\>[1]{\mathsf{plus}\mskip 3.0mu\mathsf{f}\mskip 3.0mu\mathsf{u}\mskip 3.0mu\mathnormal{=}\mskip 3.0mu\mathsf{Plus}\mskip 3.0mu\allowbreak{}\mathnormal{(}\mskip 0.0muλ\mskip 3.0mu\mathsf{b}\mskip 3.0mu\mathnormal{\rightarrow }\mskip 3.0mu\mathsf{f}\mskip 3.0mu\mathsf{b}\mskip 3.0mu\mathsf{u}\mskip 0.0mu\mathnormal{)}\allowbreak{}}\<[E]\\
\>[1]{\allowbreak{}\mathnormal{(}\mskip 0.0mu\mathsf{f}\mskip 3.0mu{\tikzstar{0.11}{0.25}{5}{-18}{fill=black}}\mskip 3.0mu\mathsf{g}\mskip 0.0mu\mathnormal{)}\allowbreak{}\mskip 3.0mu\mathsf{u}\mskip 3.0mu\mathnormal{=}\mskip 3.0mu\mathsf{dupU}\mskip 3.0mu\mathsf{u}\mskip 3.0mu\mathnormal{\&}\mskip 3.0muλ\mskip 3.0mu\allowbreak{}\mathnormal{(}\mskip 0.0mu\mathsf{u}_{1}\mskip 0.0mu\mathnormal{,}\mskip 3.0mu\mathsf{u}_{2}\mskip 0.0mu\mathnormal{)}\allowbreak{}\mskip 3.0mu\mathnormal{\rightarrow }\mskip 3.0mu\mathsf{f}\mskip 3.0mu\mathsf{u}_{1}\mskip 3.0mu\mathnormal{⋆⋆}\mskip 3.0mu\mathsf{g}\mskip 3.0mu\mathsf{u}_{2}\mskip 3.0mu\mathbf{where}}\<[E]\\
\>[2]{\allowbreak{}\mathnormal{(}\mskip 0.0mu\mathnormal{⋆⋆}\mskip 0.0mu\mathnormal{)}\allowbreak{}\mskip 3.0mu}\>[5]{\mathnormal{::}\mskip 3.0mu\allowbreak{}\mathnormal{(}\mskip 0.0mu\mathsf{SymmetricMonoidal}\mskip 3.0muz\mskip 0.0mu}\>[14]{\mathnormal{)}\allowbreak{}\mskip 3.0mu\mathnormal{\Rightarrow }\mskip 3.0mu\mathsf{S}\mskip 3.0muz\mskip 3.0mu\mathsf{r}\mskip 3.0mu\mathnormal{⊸}\mskip 3.0mu\mathsf{S}\mskip 3.0muz\mskip 3.0mu\mathsf{r}\mskip 3.0mu\mathnormal{⊸}\mskip 3.0mu\mathsf{S}\mskip 3.0muz\mskip 3.0mu\mathsf{r}}\<[E]\\
\>[2]{\mathsf{Compose}\mskip 3.0mu\mathsf{t}_{1}\mskip 3.0mu\mathsf{p}_{1}\mskip 3.0mu}\>[6]{\mathnormal{⋆⋆}\mskip 3.0mu\mathsf{Compose}\mskip 3.0mu\mathsf{t}_{2}\mskip 3.0mu\mathsf{p}_{2}\mskip 3.0mu}\>[12]{\mathnormal{=}\mskip 3.0mu\mathsf{Compose}\mskip 3.0mu\allowbreak{}\mathnormal{(}\mskip 0.0mu\bar{ρ}\mskip 3.0mu\allowbreak{}\mathnormal{∘}\allowbreak{}\mskip 3.0mu\allowbreak{}\mathnormal{(}\mskip 0.0mu\mathsf{t}_{1}\mskip 3.0mu{⊗}\mskip 3.0mu\mathsf{t}_{2}\mskip 0.0mu\mathnormal{)}\allowbreak{}\mskip 0.0mu\mathnormal{)}\allowbreak{}\mskip 3.0mu\allowbreak{}\mathnormal{(}\mskip 0.0mu\mathsf{p}_{1}\mskip 3.0mu\mathnormal{▵}\mskip 3.0mu\mathsf{p}_{2}\mskip 0.0mu\mathnormal{)}\allowbreak{}}\<[E]\\
\>[2]{\mathsf{Plus}\mskip 3.0mu\mathsf{f}\mskip 3.0mu}\>[6]{\mathnormal{⋆⋆}\mskip 3.0mu\mathsf{t}\mskip 3.0mu}\>[12]{\mathnormal{=}\mskip 3.0mu\mathsf{Plus}\mskip 3.0mu\allowbreak{}\mathnormal{(}\mskip 0.0muλ\mskip 3.0mu\mathsf{c}\mskip 3.0mu\mathnormal{\rightarrow }\mskip 3.0mu\mathsf{f}\mskip 3.0mu\mathsf{c}\mskip 3.0mu\mathnormal{⋆⋆}\mskip 3.0mu\mathsf{t}\mskip 0.0mu\mathnormal{)}\allowbreak{}}\<[E]\\
\>[2]{\mathsf{t}\mskip 3.0mu}\>[6]{\mathnormal{⋆⋆}\mskip 3.0mu\mathsf{Plus}\mskip 3.0mu\mathsf{f}\mskip 3.0mu}\>[12]{\mathnormal{=}\mskip 3.0mu\mathsf{Plus}\mskip 3.0mu\allowbreak{}\mathnormal{(}\mskip 0.0muλ\mskip 3.0mu\mathsf{c}\mskip 3.0mu\mathnormal{\rightarrow }\mskip 3.0mu\mathsf{t}\mskip 3.0mu\mathnormal{⋆⋆}\mskip 3.0mu\mathsf{f}\mskip 3.0mu\mathsf{c}\mskip 0.0mu\mathnormal{)}\allowbreak{}}\<[E]\\
\>[1]{\mathsf{deriv}\mskip 3.0mu\mathsf{i}\mskip 3.0mu\mathsf{r}\mskip 3.0mu\mathsf{u}\mskip 3.0mu\mathnormal{=}\mskip 3.0mu\mathsf{deriv'}\mskip 3.0mu\mathsf{i}\mskip 3.0mu\allowbreak{}\mathnormal{(}\mskip 0.0mu\mathsf{r}\mskip 3.0mu\mathsf{u}\mskip 0.0mu\mathnormal{)}\allowbreak{}\mskip 3.0mu\mathbf{where}}\<[E]\\
\>[2]{\mathsf{deriv'}\mskip 3.0mu\mathnormal{::}\mskip 3.0mu}\>[19]{\allowbreak{}\mathnormal{(}\mskip 0.0mu\mathsf{ConnectionCategory}\mskip 3.0muz\mskip 0.0mu\mathnormal{)}\allowbreak{}\mskip 3.0mu\mathnormal{\Rightarrow }\mskip 3.0mu\mathsf{P}\mskip 3.0muz\mskip 3.0mu\mathsf{r}\mskip 3.0muT_{z}\mskip 3.0mu\mathnormal{⊸}\mskip 3.0mu\mathsf{S}\mskip 3.0muz\mskip 3.0mu\mathsf{r}\mskip 3.0mu\mathnormal{⊸}\mskip 3.0mu\mathsf{S}\mskip 3.0muz\mskip 3.0mu\mathsf{r}}\<[E]\\
\>[2]{\mathsf{deriv'}\mskip 3.0mu}\>[4]{\allowbreak{}\mathnormal{(}\mskip 0.0mu\mathsf{P}\mskip 3.0mu\mathsf{i}\mskip 0.0mu\mathnormal{)}\allowbreak{}\mskip 3.0mu}\>[6]{\allowbreak{}\mathnormal{(}\mskip 0.0mu\mathsf{Compose}\mskip 3.0mu\mathsf{t}\mskip 3.0mu\mathsf{p}\mskip 0.0mu\mathnormal{)}\allowbreak{}\mskip 3.0mu}\>[11]{\mathnormal{=}\mskip 3.0mu\mathsf{Compose}\mskip 3.0mu\allowbreak{}\mathnormal{(}\mskip 0.0mu∇\mskip 3.0mu\mathsf{t}\mskip 0.0mu\mathnormal{)}\allowbreak{}\mskip 3.0mu\allowbreak{}\mathnormal{(}\mskip 0.0mu\mathsf{i}\mskip 3.0mu\mathnormal{▵}\mskip 3.0mu\mathsf{p}\mskip 0.0mu\mathnormal{)}\allowbreak{}}\<[E]\\
\>[2]{\mathsf{deriv'}\mskip 3.0mu}\>[4]{\mathsf{p}\mskip 3.0mu}\>[6]{\allowbreak{}\mathnormal{(}\mskip 0.0mu\mathsf{Plus}\mskip 3.0mu\mathsf{f}\mskip 0.0mu\mathnormal{)}\allowbreak{}\mskip 3.0mu}\>[11]{\mathnormal{=}\mskip 3.0mu\mathsf{Plus}\mskip 3.0mu\allowbreak{}\mathnormal{(}\mskip 0.0muλ\mskip 3.0mu\mathsf{c}\mskip 3.0mu\mathnormal{\rightarrow }\mskip 3.0mu\mathsf{deriv'}\mskip 3.0mu\mathsf{p}\mskip 3.0mu\allowbreak{}\mathnormal{(}\mskip 0.0mu\mathsf{f}\mskip 3.0mu\mathsf{c}\mskip 0.0mu\mathnormal{)}\allowbreak{}\mskip 0.0mu\mathnormal{)}\allowbreak{}}\<[E]\\
\>[1]{\mathsf{tensorEval}_{1}\mskip 3.0mu\mathsf{f}\mskip 3.0mu}\>[10]{\mathnormal{=}\mskip 3.0mu\mathsf{tensorEval'}_{0}\mskip 3.0mu\allowbreak{}\mathnormal{(}\mskip 0.0mu\mathsf{uncurry}\mskip 3.0mu\mathsf{f}\mskip 3.0mu\allowbreak{}\mathnormal{(}\mskip 0.0mu\mathsf{P}\mskip 3.0mu\allowbreak{}\mathnormal{(}\mskip 0.0mu\mathsf{FreeCart.Embed}\mskip 3.0muρ\mskip 0.0mu\mathnormal{)}\allowbreak{}\mskip 0.0mu\mathnormal{)}\allowbreak{}\mskip 0.0mu\mathnormal{)}\allowbreak{}\mskip 3.0mu\mathbf{where}}\<[E]\\
\>[2]{\mathsf{tensorEval'}_{0}\mskip 3.0mu\mathnormal{::}\mskip 3.0mu\allowbreak{}\mathnormal{(}\mskip 0.0mu}\>[15]{\mathsf{SymmetricMonoidal}\mskip 3.0muz\mskip 0.0mu\mathnormal{,}\mskip 3.0mu\mathsf{Additive}\mskip 3.0mu}\>[18]{z\mskip 0.0mu\mathnormal{)}\allowbreak{}\mskip 3.0mu\mathnormal{\Rightarrow }\mskip 3.0mu\mathsf{S}\mskip 3.0muz\mskip 3.0mu\mathsf{a}\mskip 3.0mu\mathnormal{\rightarrow }\mskip 3.0mu\mathsf{a}\mskip 1.0mu\overset{z}{\leadsto }\mskip 1.0mu\mathbf{1}}\<[E]\\
\>[2]{\mathsf{tensorEval'}_{0}\mskip 3.0mu\mathsf{u}\mskip 3.0mu\mathnormal{=}\mskip 3.0mu\mathbf{case}\mskip 3.0mu\mathsf{u}\mskip 3.0mu\mathbf{of}}\<[E]\\
\>[3]{\allowbreak{}\mathnormal{(}\mskip 0.0mu\mathsf{Compose}\mskip 3.0mu\mathsf{t}\mskip 3.0mu\mathsf{p}\mskip 0.0mu\mathnormal{)}\allowbreak{}\mskip 3.0mu}\>[7]{\mathnormal{\rightarrow }\mskip 3.0mu\mathsf{t}\mskip 3.0mu\allowbreak{}\mathnormal{∘}\allowbreak{}\mskip 3.0mu\mathsf{cartesianToMonoidal}\mskip 3.0mu\mathsf{p}}\<[E]\\
\>[3]{\mathsf{Plus}\mskip 3.0mu\mathsf{f}\mskip 3.0mu}\>[7]{\mathnormal{\rightarrow }\mskip 3.0mu\mathsf{tensorEval'}_{0}\mskip 3.0mu\allowbreak{}\mathnormal{(}\mskip 0.0mu\mathsf{f}\mskip 3.0mu\mathsf{False}\mskip 0.0mu\mathnormal{)}\allowbreak{}\mskip 3.0mu\mathnormal{+}\mskip 3.0mu\mathsf{tensorEval'}_{0}\mskip 3.0mu\allowbreak{}\mathnormal{(}\mskip 0.0mu\mathsf{f}\mskip 3.0mu\mathsf{True}\mskip 0.0mu\mathnormal{)}\allowbreak{}}\<[E]\\
\>[1]{\mathsf{tensorEval}\mskip 3.0mu\mathsf{f}\mskip 3.0mu}\>[8]{\mathnormal{=}\mskip 3.0mu\bar{ρ}\mskip 3.0mu\allowbreak{}\mathnormal{∘}\allowbreak{}\mskip 3.0muσ\mskip 3.0mu\allowbreak{}\mathnormal{∘}\allowbreak{}\mskip 3.0mu\allowbreak{}\mathnormal{(}\mskip 0.0mu\mathsf{tensorEval}_{1}\mskip 3.0mu\allowbreak{}\mathnormal{(}\mskip 0.0mu\mathsf{Interface.uncurry}\mskip 3.0mu\mathsf{f}\mskip 0.0mu\mathnormal{)}\allowbreak{}\mskip 3.0mu{⊗}\mskip 3.0mu\mathsf{id}\mskip 0.0mu\mathnormal{)}\allowbreak{}\mskip 3.0mu\allowbreak{}\mathnormal{∘}\allowbreak{}\mskip 3.0mu\bar{α}\mskip 3.0mu\allowbreak{}\mathnormal{∘}\allowbreak{}\mskip 3.0mu\allowbreak{}\mathnormal{(}\mskip 0.0mu\mathsf{id}\mskip 3.0mu{⊗}\mskip 3.0muη\mskip 0.0mu\mathnormal{)}\allowbreak{}\mskip 3.0mu\allowbreak{}\mathnormal{∘}\allowbreak{}\mskip 3.0muρ}\<[E]\\
\>[1]{\mathsf{tensorEmbed}_{1}\mskip 3.0mu\mathsf{f}\mskip 3.0mu\allowbreak{}\mathnormal{(}\mskip 0.0mu\mathsf{P}\mskip 3.0mu\mathsf{q}\mskip 0.0mu\mathnormal{)}\allowbreak{}\mskip 3.0mu\allowbreak{}\mathnormal{(}\mskip 0.0mu\mathsf{P}\mskip 3.0mu\mathsf{u}\mskip 0.0mu\mathnormal{)}\allowbreak{}\mskip 3.0mu\mathnormal{=}\mskip 3.0mu\mathsf{Compose}\mskip 3.0mu\allowbreak{}\mathnormal{(}\mskip 0.0mu\mathsf{f}\mskip 3.0mu\allowbreak{}\mathnormal{∘}\allowbreak{}\mskip 3.0mu\bar{ρ}\mskip 0.0mu\mathnormal{)}\allowbreak{}\mskip 3.0mu\allowbreak{}\mathnormal{(}\mskip 0.0mu\mathsf{q}\mskip 3.0mu\mathnormal{▵}\mskip 3.0mu\mathsf{u}\mskip 0.0mu\mathnormal{)}\allowbreak{}}\<[E]\\
\>[1]{\mathsf{tensorEmbed}\mskip 3.0mu\mathsf{t}\mskip 3.0mu\mathsf{i}\mskip 3.0mu\mathsf{j}\mskip 3.0mu}\>[9]{\mathnormal{=}\mskip 3.0mu\mathsf{tensorEmbed}_{1}\mskip 3.0mu\allowbreak{}\mathnormal{(}\mskip 0.0muϵ\mskip 3.0mu\allowbreak{}\mathnormal{∘}\allowbreak{}\mskip 3.0mu\allowbreak{}\mathnormal{(}\mskip 0.0mu\mathsf{t}\mskip 3.0mu{⊗}\mskip 3.0mu\mathsf{id}\mskip 0.0mu\mathnormal{)}\allowbreak{}\mskip 0.0mu\mathnormal{)}\allowbreak{}\mskip 3.0mu\allowbreak{}\mathnormal{(}\mskip 0.0mu\mathsf{merge}\mskip 3.0mu\allowbreak{}\mathnormal{(}\mskip 0.0mu\mathsf{i}\mskip 0.0mu\mathnormal{,}\mskip 3.0mu\mathsf{j}\mskip 0.0mu\mathnormal{)}\allowbreak{}\mskip 0.0mu\mathnormal{)}\allowbreak{}}\<[E]\\
\>[1]{\mathsf{delta}\mskip 3.0mu\mathsf{i}\mskip 3.0mu\mathsf{j}\mskip 3.0mu\mathsf{u}\mskip 3.0mu\mathnormal{=}\mskip 3.0mu\mathsf{eatU}\mskip 3.0mu\mathsf{u}\mskip 3.0mu\allowbreak{}\mathnormal{(}\mskip 0.0mu\mathsf{delta'}\mskip 3.0mu\mathsf{i}\mskip 3.0mu\mathsf{j}\mskip 0.0mu\mathnormal{)}\allowbreak{}\mskip 3.0mu\mathbf{where}}\<[E]\\
\>[2]{\mathsf{delta'}\mskip 3.0mu\mathnormal{::}\mskip 3.0mu\mathsf{CompactClosed}\mskip 3.0muz\mskip 3.0mu\mathnormal{\Rightarrow }\mskip 3.0mu}\>[17]{\mathsf{P}\mskip 3.0muz\mskip 3.0mu\mathsf{r}\mskip 3.0mu\mathsf{a}\mskip 3.0mu\mathnormal{⊸}\mskip 3.0mu\mathsf{P}\mskip 3.0muz\mskip 3.0mu\mathsf{r}\mskip 3.0mu\dual{\mathsf{a}}\mskip 3.0mu\mathnormal{⊸}\mskip 3.0mu\mathsf{S}\mskip 3.0muz\mskip 3.0mu\mathsf{r}}\<[E]\\
\>[2]{\mathsf{delta'}\mskip 3.0mu\allowbreak{}\mathnormal{(}\mskip 0.0mu\mathsf{P}\mskip 3.0mu\mathsf{i}\mskip 0.0mu\mathnormal{)}\allowbreak{}\mskip 3.0mu\allowbreak{}\mathnormal{(}\mskip 0.0mu\mathsf{P}\mskip 3.0mu\mathsf{j}\mskip 0.0mu\mathnormal{)}\allowbreak{}\mskip 3.0mu\mathnormal{=}\mskip 3.0mu\mathsf{Compose}\mskip 3.0muϵ\mskip 3.0mu\allowbreak{}\mathnormal{(}\mskip 0.0mu\mathsf{i}\mskip 3.0mu\mathnormal{▵}\mskip 3.0mu\mathsf{j}\mskip 0.0mu\mathnormal{)}\allowbreak{}}\<[E]\\
\>[1]{\mathsf{eatU}\mskip 3.0mu\mathnormal{::}\mskip 3.0mu\allowbreak{}\mathnormal{(}\mskip 0.0mu}\>[13]{\mathsf{SymmetricMonoidal}\mskip 3.0muz\mskip 0.0mu\mathnormal{)}\allowbreak{}\mskip 3.0mu\mathnormal{\Rightarrow }\mskip 3.0mu\mathsf{P}\mskip 3.0muz\mskip 3.0mu\mathsf{r}\mskip 3.0mu\mathbf{1}\mskip 3.0mu\mathnormal{⊸}\mskip 3.0mu\mathsf{S}\mskip 3.0muz\mskip 3.0mu\mathsf{r}\mskip 3.0mu\mathnormal{⊸}\mskip 3.0mu\mathsf{S}\mskip 3.0muz\mskip 3.0mu\mathsf{r}}\<[E]\\
\>[1]{\mathsf{eatU}\mskip 3.0mu\allowbreak{}\mathnormal{(}\mskip 0.0mu\mathsf{P}\mskip 3.0mu\mathsf{f}\mskip 0.0mu\mathnormal{)}\allowbreak{}\mskip 3.0mu\allowbreak{}\mathnormal{(}\mskip 0.0mu\mathsf{Compose}\mskip 3.0mu\mathsf{t}\mskip 3.0mu\mathsf{p}\mskip 0.0mu\mathnormal{)}\allowbreak{}\mskip 3.0mu\mathnormal{=}\mskip 3.0mu\mathsf{Compose}\mskip 3.0mu\allowbreak{}\mathnormal{(}\mskip 0.0mu\mathsf{t}\mskip 3.0mu\allowbreak{}\mathnormal{∘}\allowbreak{}\mskip 3.0mu\bar{ρ}\mskip 0.0mu\mathnormal{)}\allowbreak{}\mskip 3.0mu\allowbreak{}\mathnormal{(}\mskip 0.0mu\mathsf{p}\mskip 3.0mu\mathnormal{▵}\mskip 3.0mu\mathsf{f}\mskip 0.0mu\mathnormal{)}\allowbreak{}}\<[E]\\
\>[1]{\mathsf{eatU}\mskip 3.0mu\mathsf{p}\mskip 3.0mu\allowbreak{}\mathnormal{(}\mskip 0.0mu\mathsf{Plus}\mskip 3.0mu\mathsf{f}\mskip 0.0mu\mathnormal{)}\allowbreak{}\mskip 3.0mu\mathnormal{=}\mskip 3.0mu\mathsf{Plus}\mskip 3.0mu\allowbreak{}\mathnormal{(}\mskip 0.0muλ\mskip 3.0mu\mathsf{b}\mskip 3.0mu\mathnormal{\rightarrow }\mskip 3.0mu\mathsf{eatU}\mskip 3.0mu\mathsf{p}\mskip 3.0mu\allowbreak{}\mathnormal{(}\mskip 0.0mu\mathsf{f}\mskip 3.0mu\mathsf{b}\mskip 0.0mu\mathnormal{)}\allowbreak{}\mskip 0.0mu\mathnormal{)}\allowbreak{}}\<[E]\\
\>[1]{\mathsf{dupU}\mskip 3.0mu\mathnormal{::}\mskip 3.0mu\mathsf{SymmetricMonoidal}\mskip 3.0muz\mskip 3.0mu\mathnormal{\Rightarrow }\mskip 3.0mu}\>[16]{\mathsf{P}\mskip 3.0muz\mskip 3.0mu\mathsf{r}\mskip 3.0mu\mathbf{1}\mskip 3.0mu\mathnormal{⊸}\mskip 3.0mu\allowbreak{}\mathnormal{(}\mskip 0.0mu\mathsf{P}\mskip 3.0muz\mskip 3.0mu\mathsf{r}\mskip 3.0mu\mathbf{1}\mskip 0.0mu\mathnormal{,}\mskip 3.0mu\mathsf{P}\mskip 3.0muz\mskip 3.0mu\mathsf{r}\mskip 3.0mu\mathbf{1}\mskip 0.0mu\mathnormal{)}\allowbreak{}}\<[E]\\
\>[1]{\mathsf{dupU}\mskip 3.0mu\mathnormal{=}\mskip 3.0mu\mathsf{split}\mskip 3.0mu\allowbreak{}\mathnormal{∘}\allowbreak{}\mskip 3.0mu\allowbreak{}\mathnormal{(}\mskip 0.0mu\mathsf{encode}\mskip 3.0muρ\mskip 0.0mu\mathnormal{)}\allowbreak{}}\<[E]\end{parray}}\end{list} \begin{list}{}{\setlength\leftmargin{1.0em}}\item\relax
\ensuremath{\begin{parray}\column{B}{@{}>{}l<{}@{}}\column[0em]{1}{@{}>{}l<{}@{}}\column{2}{@{}>{}l<{}@{}}\column{E}{@{}>{}l<{}@{}}%
\>[1]{\mathsf{raise}\mskip 3.0mu}\>[2]{\mathnormal{=}\mskip 3.0mu\mathsf{encode}\mskip 3.0mu\mathsf{juggleUp}}\<[E]\\
\>[1]{\mathsf{lower}\mskip 3.0mu}\>[2]{\mathnormal{=}\mskip 3.0mu\mathsf{encode}\mskip 3.0mu\mathsf{juggleDown}}\<[E]\end{parray}}\end{list} \caption{Final implementation}\label{398}\end{figure} 
\newpage{} \section{Related work}\label{399} \subsection{Tensor presentations in introductory texts}\label{400} 
We expect many of our readers not to be already familiar with tensors,
and therefore they will need to read pedagogical introductions to the topic,
as the authors did when preparing this paper.
We believe that a warning is in order, because,
typically, introductory texts lean heavily
on the representations of tensors as (generalised) arrays of
coefficients. Consequently, undue importance is enthused to what
happens under change of coordinates in the manifold--- even though
from an algebraic perspective, coordinate systems do not even enter
the picture. All the introductory texts that we could find take this
approach \citep{grinfeld_introduction_2013,rowland_tensor_2023,fleisch_student_2012,porat_gentle_2014,dullemond_introduction_2010}.
The quotes below are by
no means intended as singling out particular authors:  this approach is pervasive.
These are the kinds of definition that we find:

\begin{quote}[Tensors] are geometrical objects over vector
spaces, whose coordinates obey certain laws of transformation under
change of basis.{\unskip\nobreak\hfil\penalty50 \hskip2em\hbox{}\nobreak\hfil\citep{porat_gentle_2014}\parfillskip=0pt \finalhyphendemerits=0 \par}\end{quote} 
\begin{quote}A tensor of rank \(n\) is an array of \(3^n\) values (in 3-D space) called “tensor
components” that combine with multiple directional indicators (basis vectors)
to form a quantity that does not vary as the coordinate system is changed.{\unskip\nobreak\hfil\penalty50 \hskip2em\hbox{}\nobreak\hfil\citep{fleisch_student_2012}\parfillskip=0pt \finalhyphendemerits=0 \par}\end{quote} 
\begin{quote}An nth-rank tensor in m-dimensional space is a mathematical
object that has n indices and \(m^n\) components and obeys certain
transformation rules.{\unskip\nobreak\hfil\penalty50 \hskip2em\hbox{}\nobreak\hfil\citep{rowland_tensor_2023}\parfillskip=0pt \finalhyphendemerits=0 \par}\end{quote} \noindent{} At the end of the document one finds out that the transformation rules
are the multiplication by Jacobians.
Other sources take a more pedagogical approach and start with vectors and covectors:
\begin{quote}\(y^α\) : contravariant vector{\unskip\nobreak\hfil\penalty50 \hskip2em\hbox{}\nobreak\hfil\citep[][p. 13]{dullemond_introduction_2010}\parfillskip=0pt \finalhyphendemerits=0 \par}\end{quote} \noindent{} Here \(y\) is considered to be some array of numbers, enthused with
a property (covariance), which refers to the fact that the inverse of
the Jacobian should multiply the coefficients when changing coordinate
systems. (Furthermore \(α\) is implicitly lambda bound here--- a
kind of liberty that most textbooks take in all areas of mathematics.)

For the student already familiar with linear algebra, this can be
particularly confusing, because \emph{every} vector or matrix
transforms this way. So why bother highlighting this property? This
presentation continues with

\begin{quote}The object \(g_{μν}\) is a kind of tensor that is neither a
matrix nor a vector or covector.{\unskip\nobreak\hfil\penalty50 \hskip2em\hbox{}\nobreak\hfil\citep[][p. 17]{dullemond_introduction_2010}\parfillskip=0pt \finalhyphendemerits=0 \par}\end{quote} 
In algebraic terms, it is not hard to state that the metric is a
linear function of two arguments, but the representational approaches
incites the authors to beat around the bush.  Besides, the
representational approaches defines the metric as the Gram matrix of
the base vectors, and it being a tensor requires some proof.

In the algebraic definition this kind of pitfall is avoided.  Because
of the awkward definition of tensors in the representational view,
there is much discussion about what is and what isn't a tensor. One
can find statements like the following:

\begin{quote}\(v^μ + w_μ\) is not a tensor.{\unskip\nobreak\hfil\penalty50 \hskip2em\hbox{}\nobreak\hfil\citep[][p. 30]{dullemond_introduction_2010}\parfillskip=0pt \finalhyphendemerits=0 \par}\end{quote} Algebraically, this
is attempting to add objects of different type together. In {\sc{}Albert}, the above expression is not well-typed
either. (And yet, to confuse matters even more, this expression \emph{can} be made sense of with the pervasive index juggling conventions.)

But there is a more subtle way in which something might be ``not a tensor'':

\begin{quote}The Christoffel symbol is not a tensor because it
contains all the information about the curvature of the coordinate
system and can therefore be transformed entirely to zero if the
coordinates are straightened. Nevertheless we treat it as any ordinary
tensor in terms of the index notation.{\unskip\nobreak\hfil\penalty50 \hskip2em\hbox{}\nobreak\hfil\citep[][p. 36]{dullemond_introduction_2010}\parfillskip=0pt \finalhyphendemerits=0 \par}\end{quote} 
We believe that this kind of statement is puzzling to the
novice. Indeed, the first sentence states that because \(Γ\) (the Christoffel symbol) is zero
in some bases, then it cannot be a tensor (implicit to the above is
that no Jacobian can transform a zero tensor to a nonzero tensor). The
student might still wonder if this statement applies for a manifold
that cannot be straightened (i.e. one that is not flat), and why it can
be handled like a tensor anyway.

In the algebraic view that we employ, \(Γ\) is a morphism like any
other. So is it a tensor after all?  In fact, the value of \(Γ\) for any given basis is a tensor. But \(Γ\) is defined as an
expression which explicitly references the basis, and changes with it.
The confusing aspect is that the coefficient representation of
\emph{any} tensor changes when the basis changes. But \(Γ\) changes
\emph{as a geometric object} as the coordinate system changes, which means that
it cannot be converted to another basis by the usual jacobian-based transformation
while retaining the properties of the Christoffel symbol.

What we find puzzling is that the choice of the representational approach
appears to be a conscious one:

\begin{quote}[We] have used the coordinate approach to tensors, as opposed to the formal
geometrical approach. Although this approach is a bit old fashioned, I still
find it the easier to comprehend on first learning, especially if the learner is
not a student of mathematics or physics.{\unskip\nobreak\hfil\penalty50 \hskip2em\hbox{}\nobreak\hfil\citep{porat_gentle_2014}\parfillskip=0pt \finalhyphendemerits=0 \par}\end{quote} 
It is not to say that the algebraic approach is absent from the
literature. While it appears to be chiefly geared towards specialists,
it also can be found in textbooks, but we could not find one that
explicitly makes the link between all notations.  For instance, \citet{bowen_introduction_1976} take
an algebraic approach, and as such only manipulate the category-oriented (point-free) notations.
Using only this notation is sufficient for them because
they do not manipulate any complicated tensor expressions, but it also
means that readers will have a hard time connecting other notations to the point-free language.
\citet{jeevanjee_introduction_2011} does better by describing the tensor algebras, and discussing various representations.
However the equivalence between the various languages is not mentioned.

As mentioned earlier \citet{thorne15:ModernClassicalPhysics} do
mention this equivalence informally. We refer the more advanced reader to \citet{bleecker2005gauge} for
a full development in the language of linear algebra and differential forms, without any connection
to the underlying categorical structures.

 \subsection{Einstein notation}\label{401} 
The Einstein notation appears to arise as a generalisation of the
notation to denote elements of matrices. This indexing notation is so
well established that we could not trace where it originates.

However, not every expression which involves accessing indices can be
mapped to a tensor. In a nutshell, indices must be treated abstractly
and used linearly. Thus, under these conditions, which are exactly
those of {\sc{}Albert}, there is an equivalence
between Einstein notation and the algebraic specification of tensors as
morphisms.

Despite our best efforts, some distance remains between
{\sc{}Albert} and standard Einstein notation. First the Einstein notation leaves
the binders of indices implicit, but because {\sc{}Albert} is
embedded in a functional language, every index variable must be lambda
bound. In particular, contraction is written as \(\mathsf{contract}\mskip 3.0mu\allowbreak{}\mathnormal{(}\mskip 0.0muλ\mskip 3.0mu\mathsf{ᵏ}\mskip 3.0mu\mathsf{ₖ}\mskip 3.0mu\mathnormal{\rightarrow }\mskip 3.0mu\mathnormal{…}\mskip 0.0mu\mathnormal{)}\allowbreak{}\).  One could imagine a pre-processing step to shorten the
notation and more closely approach the Einstein notation, but we find that sticking to
the lambda calculus convention avoids a source of confusion.

Second,
it should be noted that the use of standalone subscript and
superscript variables is a small liberty that we took in
typesetting the paper: while the Haskell compiler accepts super and subscript characters in
variable names, it disallows them as the first character.  Third, the
operands of an addition are \(\mathbf{case}\) branches. We believe that an
alternative syntax should be provided by {\sc{}ghc}, because it
is generally useful in presence of linear types.  Fourth, a pattern
syntax for \(\mathsf{merge}\) and \(\mathsf{split}\) would shorten index
manipulation. Pattern synonyms are already available in
{\sc{}ghc}, but are not currently compatible with linear
types. This is a technical shortcoming that could be addressed in Haskell implementations.

While this paper covers all the principles of the tensor notations,
we have not aimed for an exhaustive coverage. In particular, we did not discuss the
Levi-Civita \emph{tensor}, which has many uses and is related to antisymmetrisation.

\subsection{{\sc{}dsl}s for array programming and scientific programming}\label{402} 
There is overlap between the present work and languages oriented to
scientific programming: they can both be used to describe
array-oriented computations.  Array-oriented programming languages
have a long history, perhaps starting with
{\sc{}apl} \citep{iverson_programming_1962}. Notable standalone
languages include Single-Assignment C \citep{scholz_single_1994} or
{\sc{}sisal} \citep{feo_report_1990} and even Matlab \citep{gilat_matlab_2004}. When it comes to
array {\sc{}edsl}s, there is an abundance of libraries available.
Limiting ourselves to the Haskell ecosystem, notable examples include
 RePa \citep{lippmeier_regular_2010} and accelerate
\citep{chakravarty_accelerating_2011}.

An important difference between this work and most array-oriented
languages is that they focus on representations first, and semantics come
only as a means to support optimisations.  In contrast, our approach
puts categorical semantics at the core.  This means that
constructing a dataflow representation (typically as a free {\sc{}smc}) is
natural. This representation can then be optionally optimised and
interpreted as operations on matrices and arrays, perhaps relying
on the aforementioned libraries or a dedicated
compiler \citep{kjolstad_tensor_2017} as backend.

The implementation model that we suggest is to construct a dataflow
graph (typically as a free categorical representation), optimise it,
and then interpret it as operations on matrices. This kind of model
was already at the heart of {\sc{}sisal}, but has been
popularised recently in machine-learning applications by the
TensorFlow library \citep{abadi_tensorflow_2016}. We note, however, that
TensorFlow (and similar machine learning packages) do not offer an
index-based notation.\footnote{There is a function called
\(\text{einsum}\) in both PyTorch and Tensorflow but it is not typed, and
not a general syntax for the Einstein notation. It only applies to one
term using contraction or shape transformation.\label{403}} Instead, the
programmer must keep track of dimensions using indices. The situation
is similar to de Bruijn index representations for lambda terms.
Our work, on the other hand, rejects non-tensor (non-linear) primitives,
and as such would preclude many useful array operations. These would need
to be added as an additional layer.

Another salient feature of the present work is its specific ability to
support calculus. Most scientific programming languages provide
arrays, but let the user figure out how to model tensor calculus
operations. However, there are exceptions. \citet{sussman2013functional} develop
a lisp-based {\sc{}dsl} for differential geometry. It is a point-free language close in spirit to {\sc{}Roger}, even though
the categorical structures remain implicit. Diderot \citep{chiw_diderot_2012} is specifically oriented towards tensor calculus, with explicit support for indices, as in Einstein notation.
Cadabra \citep{peeters2006field} is another computer-algebra system supporting tensor calculus expressions.

\subsection{Categorical Semantics}\label{404} 
It is generally more convenient to provide instances of categorical
structures rather than handling lambda terms directly. In this paper we have used matrix
and diagram instances, but many other applications exist
\citep{elliott_compiling_2017}.  A major selling point of categorical
semantics is that they avoid the need of
manipulating variables explicitly (perhaps as de Bruijn indices).  This
advantage is already identifiable in
the work of \citet{cousineau_categorical_1985}, but
\citet{elliott_compiling_2017} has shown how to leverage them in
{\sc{}edsl}s.  He does so by providing a compiler plugin that translates
lambda terms of the {\sc{}edsl} to morphisms in a closed Cartesian
category. This last characteristic means that no support for linear
types or specific support of {\sc{}smc}s is available.

These shortcomings have been addressed by
\citet{bernardy_evaluating_2021}, who show how to evaluate linear
functions to morphisms in an {\sc{}smc}, with no compiler modification. While
they lay down the foundation for the present work, their library is
unfortunately not sufficient for our purposes.  The technical
additions provided by this paper include the ability to represent tensor
derivatives (\cref{395}), tensor addition (\cref{396}), and contraction (\cref{397}).

Categorical semantics are particularly well suited to perform automatic
differentiation (AD), as \citet{elliott_simple_2018} has shown. We
have shown that derivatives can be represented, and we expect their symbolic
computation can be done with standard techniques. We have not shown
how AD can be performed, but because we use categorical semantics, Elliott-style AD
is the natural approach to implement our interface for derivatives. In fact, it is the method
that we have used to implement the Schwarzschild metric example (\cref{392}).

\subsection{Penrose diagram notation}\label{405} 
The diagram notation that we have used is a (graphical) {\sc{}dsl} in its own right. It
is particularly suited for morphisms in {\sc{}smc}s (and their extensions;
such as compact closed categories). \citet{selinger2011survey} provides a survey for various kinds of categories.
The correspondence between definitional
and topological equivalence is their main advantage: it means that
topological intuitions can be leveraged for formal proofs
\citep{hinze_kan_2012,blinn_using_2002}. \citet{kissinger_pictures_2012} even built a tool on this premise.

To our knowledge, the diagram notation was in fact first developed for
tensor calculus by \citet{penrose_applications_1971}. Later it was
generalised to represent morphisms in monoidal categories
\citep{joyal_geometry_1991}.  For tensors, there is a considerable
amount of variation between the diagrammatic notation used by various
authors.  We have adhered to a standardised subset of this
notation. Furthermore, we have ensured that each diagram is built
strictly using well-defined building blocks, combined strictly using
sequential \(\allowbreak{}\mathnormal{(}\mskip 0.0mu\allowbreak{}\mathnormal{∘}\allowbreak{}\mskip 0.0mu\mathnormal{)}\allowbreak{}\) and parallel \(\allowbreak{}\mathnormal{(}\mskip 0.0mu{⊗}\mskip 0.0mu\mathnormal{)}\allowbreak{}\) composition.

In contrast, the literature on tensor applications
does not prescribe a
direction for reading diagrams (whereas ours are always read
left to right). As long as the various input and outputs of atomic
morphisms are clearly identified, there is no problem in not
specifying a direction for diagrams of compact closed categories, because all
directions are equivalent thanks to the snake laws.

In his seminal work, Penrose additionally does not make a graphical
difference between the dual \(\dual{\mathsf{a}}\) and \(\mathsf{a}\). This means, for instance, that metrics
are (graphically) indistinguishable from \(η\) and
\(ϵ\). Most of the time the difference is inconsequential,
however it becomes important when expressing covariant derivatives in
terms of partial derivatives and affinities. Indeed, the affinities
for \(\mathsf{a}\) and \(\dual{\mathsf{a}}\) are not the same.

\section{Conclusion}\label{406} 
We have studied three equally expressive notations for tensor algebra
and tensor calculus, using the {\sc{}dsl} methodology: a point-free morphism
language ({\sc{}Roger}) based directly on the categorical semantics; an
index-based language ({\sc{}Albert}) based on the Einstein notation in wide
use in the literature; and a streamlined version of the diagram
notation by Penrose.

{\sc{}Roger} is ideal to define semantics (as instances of the
categorical structures). Its main drawback is that its syntax make it
difficult to track connections between components. In particular it is
difficult to recognise if two expressions are equivalent.

The diagrammatic notation is good at representing connections between
building blocks of complex expressions, and makes it easy to check for
equivalence.  Therefore it is ideal for presenting morphisms and
proofs of equivalence. Its main drawback is that it lacks an
established way to deal with polymorphism.

The Einstein notation is a natural extension of matrix element
notation, and its programming language counterpart is a natural
extension of array indexing.  Indeed its textual, compact
nature makes it easy to input into computers and easy to typeset in books.
The connection between
the building blocks of a complex Einstein notation expression is done by inspecting
indices and checking their repeated occurrences. It is not as obvious  as in
diagram notation, but much less arduous than in point-free notation.
This notation
is chosen in most of the literature on tensors, even though we find
that diagram notation is often pedagogically superior.

We formalised it as an {\sc{}edsl}, {\sc{}Albert},
where every index is represented by a (linear) lambda bound variable.
Expressions in {\sc{}Albert} are very close to standard Einstein notation, and they evaluate to representations in any of the
instances of the abstract categorical structures.
Following this workflow, one can build a program which can
simultaneously be used to manipulate tensors as matrices, and produce
diagrams in Penrose notation for the same computations. This means
that {\sc{}Albert} users enjoy most of the benefits of all representations.

The connection between the categorical structures, Einstein notation and diagrams is not a new one.
However, despite our best
efforts, we have not seen it precisely documented anywhere before
this work.  In one direction, it is not difficult to see that Einstein notation is an instance of the abstract structure.
However, the other
direction (from Einstein notation to categories) is not so obvious.
To the best of our understanding, the connection
relies essentially on the isomorphism between linear functions
(between indices) and morphisms in an {\sc{}smc}. This isomorphism is known
in the functional programming community \citep{benton_lnl_1995} but we
could not find any presentation of tensors which points out this fact.
Furthermore, on its own, this isomorphism is not sufficient to account
for all aspects of tensor calculus: the semantics of sums and derivatives
need careful treatment (\cref{394}).

Another difficulty that we faced when studying tensor calculus using
mathematics textbooks is that they mix concepts and notations from the
abstract algebraic level with those at representational level, without
warning. This kind of freedom can be disturbing for someone used
to the rigid conventions of lambda calculi and programming languages
in general. We hope that this paper helps the growing crowd of (functional)
programmers to approach tensor notation and its applications.

\paragraph*{Conflicts of Interest}\hspace{1.0ex}\label{407} None

\bibliography{PaperTools/bibtex/jp.bib,local.bib}
\bibliographystyle{jfplike} 
\label{lastpage01} 
\end{document}